\documentclass[a4paper]{article}
\usepackage{graphicx}%
\usepackage{multirow}%
\usepackage{amsmath,amssymb,amsfonts}%
\usepackage{amsthm}%
\usepackage{mathrsfs}%
\usepackage[title]{appendix}% 
\usepackage{xcolor}%
\usepackage{textcomp}%
\usepackage{manyfoot}%
\usepackage{booktabs}%
\usepackage{algorithm}%
\usepackage{algorithmicx}% 
\usepackage{algpseudocode}%
\usepackage{listings}%
\usepackage{hyperref}
\usepackage{pdflscape}
\usepackage{adjustbox}
\usepackage[authoryear]{natbib}
\newtheorem{theorem}{Theorem}%  meant for continuous numbers
\newtheorem{proposition}{Proposition}% to get separate numbers for theorem and proposition etc.

\newtheorem{definition}{Definition}%
\newtheorem{example}{Example}%

\newcommand{\ignore}[1]{} % 
\usepackage{mdframed} 

\newcommand{\mybf}[1]{#1}
\newcommand{\myem}[1]{\emph{#1}}
\newcommand{\bigO}{\mathcal{O}}
\newcommand{\red}[1]{\color{red}#1}
\newcommand{\R}{\mathbb{R}}

\newcommand{\lap}{{\rm Lap}}
\newcommand{\norm}[1]{\lVert#1\rVert}
\newcommand{\dpns}{{DP}} % DP without space after, Use when small case first letter would be needed
\newcommand{\dps}{{DP }} % DP with space after, Use when small case first letter would be needed "differential privacy"
\newcommand{\DP}{{DP }} % DP with space after, Use when capitalized first letter would be needed: "Differential Privacy"
\newcommand{\ml}{{ML }} % DP with space after, Use when small case first letter would be needed "differential privacy"
\DeclareMathAlphabet\mathbfcal{OMS}{cmsy}{b}{n}
\mdfsetup{
    linewidth=1.10pt
}

\begin{document}
\begin{sloppy}

\title{A Comprehensive Guide to Differential Privacy:  From Theory to User Expectations}
\author{
  Napsu Karmitsa\textsuperscript{1*},\;
  Antti Airola\textsuperscript{1},\;
  Tapio Pahikkala\textsuperscript{1},\;
  Tinja Pitk\"am\"aki\textsuperscript{1}\\[1.5ex]
  \textsuperscript{1}\small Department of Computing, University of Turku, FI-20014 Turku, Finland
}

\date{}

\maketitle

\vspace{-2em}

\begin{center}
{$^*$Corresponding author. E-mail:} \href{mailto:napsu@karmitsa.fi}{napsu@karmitsa.fi};

{Contributing authors:}
\href{mailto:ajairo@utu.fi}{ajairo@utu.fi}; \href{mailto:aatapa@utu.fi}{aatapa@utu.fi}; \href{mailto:tievpi@utu.fi}{tievpi@utu.fi};
\end{center}

\vspace{1.5em}

\begin{abstract}
The increasing availability of personal data has enabled significant advances in fields such as machine learning, healthcare, and cybersecurity. However, this data abundance also raises serious privacy concerns, especially in light of powerful re-identification attacks and growing legal and ethical demands for responsible data use. Differential privacy (DP) has emerged as a principled, mathematically grounded framework for mitigating these risks. This review provides a comprehensive survey of DP, covering its theoretical foundations, practical mechanisms, and real-world applications. It explores key algorithmic tools and domain-specific challenges --- particularly in privacy-preserving machine learning and synthetic data generation. The report also highlights usability issues and the need for improved communication and transparency in DP systems. Overall, the goal is to support informed adoption of DP by researchers and practitioners navigating the evolving landscape of data privacy.\bigskip

\noindent \textbf{Keywords:} Differential Privacy, Machine Learning, Synthetic Data, Federated Learning, Cybersecurity, User Expectations

\end{abstract}

\section{Introduction} \label{cha_intro}
{Differential Privacy} (DP), introduced by %Dwork et al.\ 
\cite{Dwork2006,DworkRoth2014}, is a rigorous mathematical framework for protecting individual privacy in data analysis, including machine learning. At its core, DP ensures that the result of an analysis --- or the output of an algorithm --- does not significantly change depending on whether any one individual’s data is included or not. This means that someone seeing the result \textbf{cannot confidently tell if a particular individual’s data was used} in the computation. 
This property makes DP particularly effective against re-identification attacks, even when an adversary has some prior knowledge about the individuals in the dataset. Such background knowledge, known as \emph{auxiliary information}, may include partial information --- such as age, ZIP code, or health condition --- or access 
to other related datasets that, when combined with non-private outputs, could otherwise enable re-identification. As a result, DP is now widely regarded as the "gold standard" for privacy protection, offering strong, quantifiable, and mathematically proven guarantees --- rather than relying on ad-hoc anonymization techniques like removing names or masking values (\cite{Ponomareva2023,wagh2021}). 
 
In what follows, we present an in-depth survey of DP, covering its foundational principles, evolving methodologies, and diverse practical applications. Particular attention is given to challenges in usability and communication, as well as the need for interdisciplinary collaboration to ensure that DP systems are not only secure, but also understandable and adoptable by researchers, developers, and the general public. The document is structured to introduce foundational concepts and progressively explore more advanced techniques and applications, culminating in a discussion of user expectations and future directions in the field.

\medskip 

\noindent
\textbf{Section \ref{moti}: Motivation} 
gives the motivation for DP by highlighting real-world privacy risks and failures of traditional anonymization methods. It presents examples such as linkage and membership inference attacks on public datasets, and illustrates why stronger guarantees are essential for protecting individual data.
\medskip 

\noindent
\textbf{Section \ref{cha2}: Definitions and Properties} % ok
  lays the theoretical foundation for the report, covering formal definitions of DP and core properties such as composition and post-processing. It also introduces different trust models --- central, local, and distributed --- which reflects different assumptions about
trust and control over raw data, and noise mechanisms like the Laplace and Gaussian mechanisms, which form the basis for private data analysis.
\medskip 

\noindent
\textbf{Section \ref{chaML}: Differential Privacy in Machine Learning} 
shifts focus to machine learning, explaining how DP can be applied during model training. The chapter discusses the widely used DP-SGD algorithm and explores challenges such as balancing utility and privacy, hyperparameter tuning under privacy constraints, and the privacy accounting methods used to track cumulative privacy loss during training. As machine learning systems increasingly rely on sensitive data, integrating DP into training workflows is a critical step toward building trustworthy and ethically responsible AI models.
\medskip 

\noindent
\textbf{Section \ref{section_privacy_budget}: Understanding and Defining the Privacy Budget} 
explores how to select and manage the privacy budget. It presents strategies for budgeting in practice, emphasizing the trade-offs between privacy and utility.
In addition the chapter presents examples of privacy budgets
used in real-world applications. A clear understanding of the privacy budget is essential for designing effective DP systems, as it directly influences both the level of protection offered and the practical usability of the resulting data or models.
\medskip 

\noindent
\textbf{Section \ref{sec_synthetic_data}: Privacy-Preserving Synthetic Data} % OK
addresses synthetic data generation using DP. It introduces a taxonomy of DP-based methods, including histogram and marginal-based techniques, probabilistic graphical models, and deep generative models. The chapter also examines how to evaluate the utility and privacy of synthetic datasets and highlights key challenges such as representation bias and fairness concerns that may arise when deploying DP-generated synthetic data. Despite these challenges, DP-based synthetic data holds strong potential for enabling safe data sharing and analysis, particularly in domains where direct access to real data is restricted.
\medskip

\noindent
\textbf{Section \ref{cha_enhancing}: Enhancing Differential Privacy}
examines the combination of DP with complementary technologies to meet both privacy and utility requirements. We focus on three key directions: the integration of {DP with cryptographic techniques}, the application of {DP in federated learning (FL)}, and hybrid approaches that jointly leverage {FL, cryptographic techniques, and DP}. These combinations are especially promising, as they open new pathways for building robust, privacy-preserving systems that remain practical and effective in real-world, distributed environments.
\medskip 

\noindent
\textbf{Section \ref{chaUseCases}: Use Cases of Differential Privacy}
 explores a range of scenarios in which DP plays a vital role in enabling privacy-preserving  data sharing and computation. The chapter begins with applications in cybersecurity, with a focus on cyber-physical systems (CPS), anomaly detection, and face recognition, where DP reduces risks in sensitive and high-stakes environments. It then turns to the healthcare and finance sectors --- both highly data-intensive and privacy-sensitive --- and examines how DP supports secure analytics while meeting regulatory and ethical requirements. Collectively, these case studies demonstrate the versatility and growing relevance of DP in real-world systems.
\vspace{-0.4mm}
 
\noindent
\textbf{Section \ref{chaUserExpectations}: Practicalities and User Expectations}
examines how users and practitioners perceive and interact with DP in practical contexts.  It covers insights from empirical studies on communication strategies, implementation challenges, the risks of creating a false sense of privacy ("privacy theater"), and uncertainties around regulation. The chapter emphasizes the importance of bridging the gap between formal guarantees and real-world deployment to build trust, enhance usability, and support effective, privacy-conscious decision-making. To promote responsible communication in practice, we conclude the section by introducing a practical DP workflow and  a concise DP disclosure checklist that operationalizes these considerations.

 \bigskip 

\noindent
\textbf{Section \ref{chaFuture}: Future Trends and Research Directions}
 outlines key directions for future research in DP, emphasizing the need for scalable and efficient DP-training methods, especially for complex machine learning models. It highlights emerging areas such as the integration of DP with federated learning and cryptographic techniques, the development of advanced privacy mechanisms and accounting methods, and personalized privacy frameworks. The chapter also identifies challenges in applying DP to modern architectures like transformers and large language models, and stresses the importance of robust privacy auditing, explainability, and fairness-aware design. Together, these themes point toward a more adaptable, transparent, and practically deployable future for DP.\medskip

The papers reviewed in this study were primarily collected from \emph{Scopus}, \emph{IEEE Explorer}, and \emph{Google Scholar} using search strings such as 'differential privacy \& machine learning', 'differential privacy \& syntetic data', 'differential privacy \& cybersecurity', and other related keywords. These search terms were chosen to ensure comprehensive coverage of the relevant literature in the field.

%In summary, this paper synthesizes current research trends, identifies critical challenges, and outlines promising future directions for DP. It emphasizes DP's capacity to provide strong privacy protections, encouraging its broader adoption by clearly communicating both theoretical insights and practical guidelines for effective implementation. %The paper aims to make the concept of DP more accessible and to support its effective deployment in data-driven domains, especially in machine learning. 
\newpage

\section{Motivation}\label{moti}
In an era defined by data, organizations across all sectors --- government agencies, healthcare providers, research institutions, and technology companies --- routinely collect and analyze vast amounts of personal information. This data, which includes census and health records, behavioral logs, social media activity, sensor data from connected devices, and cybersecurity telemetry, enables insights that drive innovation, improve services, combat fraud, and inform policy decisions.%\medskip

However, the abundance of such data brings growing concerns over individual privacy. High-profile data breaches, misuse of personal information, and increasing public awareness have raised questions about how to responsibly extract value from data without compromising the privacy of individuals it represents. At the same time, regulations such as the General Data Protection Regulation (GDPR, \cite{gdpr2016}) and other data protection laws worldwide have introduced stricter requirements for privacy, transparency, and accountability.%\medskip 

Alongside legal pressures and growing public concern, there is a parallel demand for reproducible science, transparent decision-making, and --- in some cases --- open or controlled data access, especially in data-driven domains such as healthcare, finance, and machine learning. These goals often conflict with the need to protect privacy, creating a pressing need for techniques that can reconcile utility with confidentiality. As a result, privacy-preserving data analysis is no longer a theoretical luxury --- it is a practical necessity. Striking a balance between data utility and individual privacy has become a central challenge for organizations tasked with releasing or analyzing sensitive data.

\subsection{Privacy Risks and Attacks}
Traditional approaches to protecting privacy, such as \emph{de-identification} and \emph{access control}, have proven inadequate to face modern adversaries equipped with powerful computational tools and vast auxiliary datasets. De-identification typically refers to the removal or masking of direct identifiers (such as names or social security numbers) from datasets in an effort to anonymize them. However, this process often retains quasi-identifiers --- like ZIP code, birth date, and gender --- that can still be used for re-identification when combined with external data. Access control, by contrast, limits who can view or query a dataset, often through authentication, authorization, or secure data enclaves. While effective in some contexts, access control does not fundamentally mitigate privacy risks once data is released or aggregated statistics are shared. Multiple  high-profile re-identification attacks have demonstrated that seemingly anonymous data can be traced back to individuals when combined with publicly available information. These attacks --- each illustrated with representative examples --- include the following:

\begin{itemize}
    \item  \textbf{Linkage Attacks} combine anonymized data with external auxiliary information to re-identify individuals. \cite{sweeney1997weaving} famously re-identified medical records using voter registration data, demonstrating how quasi-identifiers like ZIP code, birth date, and gender can be used to re-link de-identified datasets to individuals. Similarly, the Netflix Prize dataset was shown to be vulnerable to linkage attacks using auxiliary information from IMDb~(\cite{narayanan2008robust}).

\item \textbf{Reconstruction Attacks} aim to infer the original sensitive dataset by exploiting the outputs of a data release or model, often by querying it many times. \cite{10.1145/773153.773173} demonstrated that releasing too many highly accurate statistics allows an adversary to reconstruct a large portion of the original dataset. This foundational result has been further strengthened and generalized in subsequent work, including~\cite{10.1145/1250790.1250804}, \cite{10.1007/978-3-540-85174-5-26}, and \cite{10.1145/2213556.2213570}. A striking real-world example occurred with the 2010 U.S. Census, where researchers at the Census Bureau were able to reconstruct individual-level records from published statistical summaries, prompting the agency to adopt DP for future data releases (\cite{10.1145/3287287}). These examples illustrate what has come to be known informally as the \emph{Fundamental Law of Information Recovery} (\cite{DworkRoth2014}) --- the principle that overly accurate answers to too many queries can destroy privacy.

    \item \textbf{Differencing Attacks} infer sensitive information by comparing the outputs of aggregate queries that differ by only a single individual. A classic method is the “tracker attack” described by \cite{denning1979tracker}, where carefully constructed queries bypassed suppression rules to isolate individual values.
    A modern example was demonstrated on Facebook’s advertising platform, where an attacker could infer sensitive traits --- such as sexual orientation or medical conditions --- by crafting highly specific target audiences and comparing differences in ad reach statistics~(\cite{korolova2010privacy,venkatadri2018privacy}).

    \item \textbf{Membership Inference Attacks} aim to determine whether a specific individual’s data is included in a dataset or has been used to train a machine learning model. An early example by \cite{homer2008resolving} demonstrated that it is possible to identify whether an individual contributed DNA to a mixture used in genetic studies, using statistical distance measures to assess their presence. Since then, membership inference attacks have been observed in a variety of contexts, including genomic research (\cite{GWAS2}), aggregate location data (\cite{zang2011anonymization}), and deep learning models (\cite{shokri2017membership}).
\end{itemize}
These examples reveal that privacy breaches are not merely theoretical possibilities --- they pose concrete threats in real-world systems. Moreover, traditional anonymization techniques and ad hoc privacy safeguards often lack rigorous methods for quantifying cumulative privacy risk across multiple analyses or data releases. They often fail to account for an adversary’s background knowledge and offer limited protection against adaptive or future attacks.

\subsection{Differential Privacy as a Response}
In response to these risks, DP has emerged as a rigorous, future-proof framework for quantifying and controlling privacy loss (\cite{Dwork2006,DworkRoth2014}).  DP formalizes the idea that the outcome of an analysis should be (nearly) equally likely whether or not any single individual's data is included. This ensures that no individual can be harmed --- or benefit --- significantly from participating in a dataset, offering a strong form of privacy regardless of auxiliary information available to an adversary\footnote{Note that harm may still result from the outcome of the analysis (e.g., increased insurance rates for smokers), but such harm does not depend on whether any particular individual participated in the dataset --- a classic example being the "smoking man" scenario (\cite{DworkRoth2014}).}. %\medskip

DP is particularly well-suited to the challenges of modern data release and machine learning:
\begin{itemize}
    \item It provides \textbf{strong theoretical privacy guarantees}, quantifying worst-case disclosure risk in a mathematically rigorous way.
    \item It provides \textbf{composability}, allowing the cumulative privacy risk across multiple queries to be formally bounded.

    \item It is \textbf{resilient to post-processing}, meaning that further manipulation of the results does not compromise privacy guarantees.

    \item It is \textbf{future-proof}, offering protection even as external datasets grow and attack methods evolve.
\end{itemize} 

Due to these properties, DP has been adopted by major organizations such as Google (\cite{google2023distributed}), Apple (\cite{apple2017,apple}), Microsoft  (\cite{ding2017telemetry}), and the U.S. Census Bureau  (\cite{UScensus4}). These real-world deployments underscore the importance of DP as a practical tool for enabling data analytics and machine learning while safeguarding personal information. 
However, the path to practical deployment remains challenging, with issues including the privacy-utility trade-off, parameter setting, user understanding, and communication of guarantees. %\enlargethispage{\baselineskip}

\section{Definitions and Properties}\label{cha2}
In this section, we present the formal foundations of differential privacy (\dpns), along with its most important theoretical properties. We begin by introducing the standard definition of \dps and then examine several  well-established relaxations that extend its applicability in practical settings. The section also introduces the concept of the {\em privacy budget}, which quantifies the cumulative privacy loss  (and will be further analysed in Section \ref{section_privacy_budget}), and examines various \emph{trust models} that influence where and how \dps mechanisms are deployed within a system. Finally, we discuss commonly used randomization mechanisms that satisfy \dpns, highlighting their mathematical guarantees.

\subsection{Differential Privacy Definitions}
\label{sec_DPdefinition}
%At its core, differential privacy ensures that the probability of obtaining any particular output does not change significantly when any single individual’s data is added or removed from the dataset. This guarantee is usually parameterized by a privacy loss parameter, commonly denoted as 
%$\varepsilon$, where smaller values of  
%$\varepsilon$ indicate stronger privacy guarantees.
A differentially private algorithm provides a formal guarantee that limits how much its behavior can change when a single individual’s data is added to or removed from the dataset. This guarantee holds universally for every individual and every dataset. In practice, this means that, under reasonably strong privacy limits, even if a malicious adversary gains access to the output of an algorithm, they cannot determine with high confidence whether a particular individual’s data was included in the dataset. In other words, the presence or absence of any single record has  only a bounded impact on the overall output, which provides strong plausible deniability for individuals. 
%A differentially private algorithm guarantees that its behavior hardly changes when a single individual’s data is added to or removed from the dataset. This guarantee holds universally for every individual and every dataset. In practice, this means that even if a malicious adversary gains access to the output of an algorithm, they cannot determine with high confidence whether a particular individual’s data was included in the dataset. In other words, the presence or absence of any single record has a minimal impact on the overall output, which provides strong plausible deniability for individuals. 
An attacker --- even with substantial auxiliary information --- will find it difficult to infer sensitive details about any one person, because the algorithm’s output remains nearly the same regardless of that person’s data. %\medskip

At its core, DP provides \emph{privacy by process}, achieved by introducing randomness into computations. %(\cite{DworkRoth2014}). 
In what follows, we present the relevant notation and formal definitions of DP.
%DP provides \emph{privacy by process}: in particular, it introduces randomness to computations \cite{DworkRoth2014}.
%In what follows, we introduce the relevant notation and provide the formal definitions for \dpns.
\bigskip 

\noindent
\textbf{Mathematical Notations and Definitions:}
Let $\mathcal{D}$ be a potentially infinite universe of datasets and let $D \subset \mathcal{D}$ be a dataset consisting of $n$ records. Let $f$ be a (query) function that takes a dataset as input and outputs a quantity of interest. For example, $f$ may compute the mean of a specific feature or, more complexly, train a deep neural network and return the network's weights.    We next define the mechanism, neighboring datasets, and the notion of $\varepsilon$-\dpns. In addition, the notations used in this paper are collected in Table \ref{math_notations}.%\medskip

\begin{definition}[\textbf{Mechanism} $\mathbfcal{A}$] 
%\begin{definition}[Mechanism $\mathcal{A}$] 
Mechanism $\mathcal{A}:\mathcal{D}\rightarrow \mathcal{R}$ is a randomized algorithm that approximates the result of the function $f$. It operates
on a dataset $D \subset \mathcal{D}$ and returns a value in the range $R$. %$\range(\mathcal{A})$.
\end{definition}%\vspace{-0.3cm}

\begin{mdframed}%\vspace{-0.5cm}
\begin{example}
\textbf{Randomized Algorithm} (\cite{Ponomareva2023}) \medskip

\noindent
The mechanism $\mathcal{A}$ may, for example, be a noisy version of the query function $f$, defined as:
$$
\mathcal{A}(D) = f(D) + \text{noise},
$$
where the noise is sampled from a specific probability distribution (e.g., Laplace or Gaussian). %\smallskip

Note, however, that 
the mechanism needs not to be additive as in this example; some mechanisms achieve privacy through randomized response, subsampling, or more complex transformations that are not expressible as simply adding noise to a function's output.
\end{example}\smallskip
\end{mdframed}

\begin{table}[ht]
\caption{Mathematical Notations}\smallskip
\label{math_notations}
\resizebox{\textwidth}{!}{\begin{tabular}{@{}lll@{}}
\toprule
\textbf{Symbol} && \textbf{Description} \\
\midrule
       	$\varepsilon,\, \delta, \, \alpha, \, \rho, \, \mu$ && Privacy parameters (budget) \\ 
       	$\mathcal{D}$&& Universe of all datasets \\ 
        $D,D'$ && Datasets  \\ 
     	$D \sim D'$ && Neighboring datasets (i.e., datasets differing by one record)  \\ 
  $N$   	&& Set of all neighboring datasets $N = \{(D,D') \mid D \text{ and }D' \text{ are neighbors}\}$ \\ 
     	$n$ && Number of records in $D$ \\ 
     	$f$ && The (non-noisy) query function \\ 
     	     $\Delta_f^p$ 	&& Sensitivity of function $f$ measured using $\ell_p$-norm\\ 
$\mathcal{A},\mathcal{M}$ && Mechanisms or randomized algorithms \\ 
     	$R$&& Set of all possible outputs of $\mathcal{A}$  \\ 
%     	$\range(\mathcal{A})$& Set of all possible outputs of $\mathcal{A}$  \\ \hline
     	$S$ && Set of outputs of $\mathcal{A}$, $S \subseteq R$\\ 
%    $S$ & outputs of $\mathcal{A}$, $S \subseteq \range(\mathcal{A})$\\ \hline$\varepsilon, \delta$& Privacy parameters (privacy budgets)  \\ \hline
     	$P[\cdot]$&& Probability  \\ 
     	
${\lap}(b)$ && Laplace distribution with parameter $b$\\ 
 $\mathcal{N}(\mu,\sigma^2)$    	&& Gaussian (normal) distribution with mean $\mu$ and standard deviation $\sigma$\\ 
  $u$    	&& Utility function, $u:\mathcal{D} \times R \rightarrow \R$ \\ 
$A \propto B$ && $A$ is proportional to $B$ \\ 
$R_{ERM}(\theta)$     	&& Excess empirical risk with parameters $\theta$\\ 
$\mathcal{L}(\theta;D)$     	&& Loss function with parameters (weights) $\theta$ on dataset $D$ \\ 
$\mathcal{R}(\theta)$ && Regularization term  \\ 
$\eta$ && Learning rate \\ 
$\sigma$     	&& Noise level \\ 
$C$   	&& Clipping norm \\ 
 $g$, $\tilde{g}$   	&& Gradient, noisy gradient \\ 
$B$     	&& Batch size \\ 
$D_\alpha(P||Q)$ && R\'enyi Divergence for probability distributions $P$ and $Q$ \\ 
\bottomrule
\end{tabular}} %\vspace{-7mm}
\end{table} 

\begin{definition}[\textbf{Neighboring Datasets}] Two datasets  $D,D' \subset \mathcal{D}$ are said to be neighbors, denoted $D \sim D'$, if $D\subset D'$ and $|D'| = |D|+1$ (or vice versa).
\label{neighboring_datasets}
\end{definition}%\vspace{-2mm}
\noindent
This means that $D$ and $D'$ are neighbors, if $D'$ is a copy of $D$ with one record added or removed (the \emph{add-or-remove} criterion). Note that this definition changes the size of the dataset, which may cause some complications when applying it (\cite{Ponomareva2023}). Alternatively, some definitions adopt the \emph{replace-one} criterion (\cite{Vadhan2017}) --- where one record is replaced with a different record --- or the \emph{zero-out} criterion (\cite{erlingsson2020encodeshuffleanalyzeprivacy,Kairouz2021}) --- where one record is set to a default "{\em zero}" value --- both of which keep the dataset size unchanged but result in slightly different privacy guarantees. The formal definition of DP works with any of these neighboring relations; here, we adopt the \emph{add-or-remove} notion for consistency with the original formulation by \cite{Dwork2006}, and we will return to the implications of the alternative definitions later. %\smallskip

\begin{definition}[\textbf{Pure $\boldsymbol{\varepsilon}$-Differential Privacy}, \cite{Dwork2006}]
For $\varepsilon > 0$, a mechanism $\mathcal{A}:\mathcal{D}\rightarrow R$ satisfies $\varepsilon$-\dps if, for
any two neighboring datasets $D \sim D' \subset \mathcal{D}$ and any set
$S \subseteq R$ of possible outputs, the following holds:
$$
    P[\mathcal{A}(D) \in S] \leq \exp(\varepsilon) \cdot P[\mathcal{A}(D') \in S].
$$
Here, $P[\cdot]$ denotes the probability measure over the randomness inherent in the mechanism $\mathcal{A}$.

The quantity 
$$
    \ln \frac{P[\mathcal{A}(D) \in S]}{ P[\mathcal{A}(D') \in S]}
$$
is called the {\em privacy loss}.
%; that is, the probabilities are computed with respect to the random choices made by $\mathcal{A}$ during its execution.
\label{pureDP}
\end{definition}\vspace{-2mm}
\noindent
In other words, $\mathcal{A}$  is $\varepsilon$-differentially private if the inclusion or exclusion of any single individual's data in the dataset changes the likelihood of any particular output by no more than a multiplicative factor of 
 $\exp(\varepsilon)$. A smaller $\varepsilon$ 
 indicates {\em stronger privacy} --- meaning that the output of $\mathcal{A}$ is less sensitive to the data of any individual --- while a larger $\varepsilon$ 
 implies {\em weaker privacy} guarantees 
  (\cite{DworkRoth2014}).
In particular, when $\varepsilon=0$, the output becomes completely independent of the input, rendering the analysis meaningless. The optimal choice of $\varepsilon$ depends on the application: for common statistical queries (e.g., computing a mean), 
$\varepsilon$
 is typically chosen to be less than 1, whereas in deep learning scenarios, this choice is often relaxed to $\varepsilon=10$ (\cite{Ponomareva2023}). We discuss the selection of the privacy budget in Section~\ref{section_privacy_budget}.  \vspace{-1mm}
 %\medskip
 
%\medskip
\begin{mdframed}%\vspace{-5.5mm}
 \begin{example}\label{ex_RR}
\textbf{Randomized Response for a Yes-or-No Question}\medskip \enlargethispage{\baselineskip}

\noindent
Suppose a survey asks a sensitive yes-or-no question, such as:
\begin{quote}
\emph{"Have you ever cheated on an exam?"}
\end{quote}
To protect individual privacy, instead of answering directly, each participant uses the following \emph{randomized response} procedure (\cite{Warner1965}):\vspace{-1mm}

\begin{itemize}
    \item {\small Flip a fair coin.}\vspace{-1.5mm}
    \item {\small If the coin lands \emph{heads}, answer truthfully.}\vspace{-1.5mm}
    \item {\small If the coin lands \emph{tails}, flip the coin again:}\vspace{-1.5mm}
    \begin{itemize}
        \item {\small If the second flip is heads, answer "yes."}\vspace{-1.5mm}
        \item {\small If the second flip is tails, answer "no."}
    \end{itemize}
\end{itemize}\vspace{-2mm}
This setup ensures that each participant's answer is randomized and private: the surveyor cannot tell with certainty whether the answer reflects the true response or was generated randomly.\medskip

\noindent
\textbf{Privacy Analysis:}
The privacy loss for a given output measures how much more likely it is to observe that output depending on the true underlying response.
Specifically:
\begin{itemize}
    \item {\small If the true answer is "yes", the propability of reporting "yes" is 
    
    $\qquad\qquad  0.5$ (truthful heads)$+ 0.25$ (random yes after tails) $= 0.75$.}
    \item {\small If the true answer is "no",  the probability of reporting "yes" is  
    
    $\qquad\qquad  0.25$ (random yes after tails).}
\end{itemize}

%\begin{itemize}
%    \item If the true answer is ``yes,'' the probability of reporting ``yes'' is 
%    $0.75$.
%    \item If the true answer is ``no,'' the probability of reporting ``yes'' is $0.25$.
%\end{itemize}

\noindent
Thus, the privacy loss for observing "yes" is:
\[
\text{Privacy loss} = \ln\left(\frac{0.75}{0.25}\right) = \ln(3).
\]
Similarly, for observing "no," the privacy loss is:
\[
\text{Privacy loss} = \ln\left(\frac{0.25}{0.75}\right) = \ln\left(\frac{1}{3}\right) = -\ln(3).
\]

\noindent
The maximum privacy loss across all outputs defines the DP parameter $\varepsilon$. Here:
\[
\varepsilon = \log(3) \approx 1.1.
\]

%The mechanism guarantees $\varepsilon$-DP for a specific $\varepsilon$.

%The probability of answering ``yes'' given the true answer:
%\begin{itemize}
 %   \item If the true answer is ``yes'': $0.5$ (truthful heads) $+ 0.25$ (random yes after tails) $= 0.75$.
 %   \item If the true answer is ``no'': $0.25$ (random yes after tails).
%\end{itemize}

%Thus, the ratio of probabilities is:
%\[
%\frac{0.75}{0.25} = 3,
%\]
%and therefore,
%\[
%\varepsilon = \ln(3) \approx 1.1.
%\]
\noindent
This means the randomized response mechanism satisfies $(\varepsilon = \ln(3))$-DP.\medskip
 
\noindent
\textbf{Intuitive Explanation:} 
In this example, a participant's answer is up to three times more likely to be "yes" if they actually cheated than if they did not. This looseness is quantified by $\varepsilon$, which in this case is fixed at $\ln(3)$.  While smaller values of 
$\varepsilon$ generally correspond to stronger privacy (but noisier data), the classic randomized response mechanism provides a moderate privacy level by design, offering meaningful protection while still enabling useful aggregate insights.

%\bigskip
%\noindent
%\textbf{Summary Intuition:}
%
%Randomized response hides an individual's true answer behind randomness.  
%The smaller the value of $\varepsilon$, the more private the mechanism (but potentially noisier the aggregate results).

 \end{example}\smallskip
\end{mdframed}%\smallskip
%\smallskip
%\bigskip

\noindent
\textbf{Unit of Privacy:}
The choice of what constitutes a "\emph{record}" in Definition \ref{neighboring_datasets} is central to interpreting Definition \ref{pureDP}. The level at which the privacy guarantee is applied is referred to as the \emph{unit of privacy}. In \emph{example-level} (\emph{sample-level} or \emph{instance-level}) \emph{\dpns}, each individual data entry (e.g., a row in the dataset) is treated as sensitive, ensuring that the addition or removal of a single entry does not significantly affect the output of the algorithm.%\medskip

In contrast, {\em user-level \dps} (often called \emph{client-level} privacy in federated learning) protects all data contributed by a single user --- even if it consists of multiple entries --- by ensuring that the collective impact of that user’s data is limited. 
This distinction is crucial in scenarios where a user contributes multiple data points, as user-level privacy offers a stronger and more comprehensive protection compared to example-level guarantees.
%\medskip

Additional units of privacy appear in more complex settings. For instance, \emph{user-day-level} privacy applies when data is collected repeatedly over time, treating each user's data for a single day as the sensitive unit. \emph{Attribute-level} privacy, on the other hand, protects specific features or columns in the dataset rather than entire records. The chosen unit of privacy significantly affects both the strength of the guarantee and the amount of noise needed to satisfy it. %\medskip

\bigskip

\noindent
\textbf{Alternative Neighboring Criteria:}
As mentioned before, the add-or-remove approach given in Definition \ref{neighboring_datasets} can be replaced with alternative neighboring criteria  that preserve the dataset's size.
The zero-out criterion involves zeroing out a record by replacing all its entries with a designated "{\em zero}" record --- often exactly zero for numeric data --- rather than removing it entirely  (\cite{erlingsson2020encodeshuffleanalyzeprivacy,Kairouz2021}). In this case, the privacy parameter $\varepsilon$ in Definition \ref{pureDP} remains semantically equivalent to the add-or-remove approach %used in Definition \ref{neighboring_datasets} 
(\cite{Ponomareva2023}).%\medskip

Alternatively, the replace-one criteria that replaces a record with an arbitrary different record (\cite{Vadhan2017}) is effectively equivalent to both adding and removing a record simultaneously. As a result, this definition can be considered to provide privacy guarantees that are roughly twice as strong as those of the previous two approaches. %\newpage
 
In addition, the definition of neighboring datasets may vary depending on the specific privacy notion adopted. For example, in {\em label \dpns} (\cite{Chaudhuri2011LabelDP}), only the labels (i.e., the target outputs in supervised learning, such as class labels)  are considered sensitive, so neighboring datasets differ in the label of a single instance while keeping features fixed. In contrast, {\em group privacy} (\cite{DworkRoth2014,Vadhan2017}) extends the standard definition by allowing neighboring datasets to differ in up to $k$ records, providing protection for groups rather than just individuals. It is important to note that group privacy and user‑level DP discussed above are distinct concepts: group privacy may use various units of privacy, including user‑level DP.

\bigskip

\noindent
\textbf{Approximate DP:} 
In many practical settings, the strict guarantee provided by pure differential privacy is relaxed by allowing a small probability $\delta$ of failure in the privacy protection. This relaxation leads to the notion of $(\varepsilon, \delta)$-\dpns, commonly referred to as {\em approximate \dpns}. It is particularly relevant in machine learning applications, where enforcing exact privacy guarantees can be infeasible due to the complexity of the models and training procedures. 

\begin{definition}[\textbf{($\boldsymbol{\varepsilon}$, $\boldsymbol{\delta}$)-Differential Privacy}, \cite{Dwork2006}]
\label{approximateDP}
For $\varepsilon > 0$ and $\delta \in [0,1]$, a mechanism $\mathcal{A}:\mathcal{D}\rightarrow R$ satisfies ($\varepsilon,\delta$)-\dps if, for
any two neighboring datasets $D \sim D' \subset \mathcal{D}$ and any set
$S \subseteq R$ of possible outputs, the following holds: \vspace{-1mm}
$$
    P[\mathcal{A}(D) \in S] \leq \exp(\varepsilon) \cdot P[\mathcal{A}(D') \in S] + \delta.
$$
\end{definition}
\noindent
If $\delta = 0$, the mechanism satisfies pure $\varepsilon$-\dpns, providing a strong guarantee for every possible output. When $\delta > 0$, we have approximate \dpns, which is slightly weaker but often necessary for mechanisms such as those adding Gaussian noise. 
 Intuitively, $\delta$ represents the probability of a \emph{"bad event"} in which the privacy guarantee may fail to hold. For example, a mechanism that selects a single individual from a dataset of $n$ people and releases their data without any noise can still satisfy ($\varepsilon,\delta$)-DP, as long as $\delta > \frac{1}{n}$. To guard against such \emph{catastrophic failures}, the value of $\delta$ is typically set extremely small in practice (\cite{DworkRoth2014}). A common guideline is to choose $\delta \ll \frac{1}{n}$ --- the value supported by the worst case analysis.
 \medskip
 
\begin{mdframed}%\vspace{-5mm}
\begin{example}
\label{ex_worse_case}
\textbf{Worst-Case Analysis}~(\cite{Ponomareva2023}) \smallskip

\noindent
Consider the following {\em worst-case assumption} for each individual record: 

If a record $r$ is included in the dataset, a mechanism satisfying $(\varepsilon, \delta)$-\dps may produce a specific output $S_r$ with probability $\delta$, and crucially, $S_r$ cannot occur if $r$ is not present. 
If an attacker observes $S_r$, they can conclusively infer that $r$ is in the dataset.

Under this assumption, each record carries a probability $\delta$ of being successfully identified.
For a dataset of size $n$, the expected number of successful attacks is $\delta n$. To keep this expected leakage negligible, one must choose $\delta \ll 1/n$, ensuring that the expected number of successful identifications is significantly less than 1.
\end{example}\smallskip
\end{mdframed}\medskip

\subsection{Properties of Differential Privacy}
\label{sec_DPproperties}
The parameters $\varepsilon$ and $\delta$ are called the {\em privacy budget} --- they quantify how much privacy loss is allowed. Each query or computation, whether a statistical query or a machine learning algorithm, spends some privacy budget; the more queries you perform --- or the higher the accuracy or utility you demand --- the more privacy budget is consumed. This section discusses key properties of \dpns, which are essential for understanding and applying \dps in practical settings. First, we introduce the concept of {\em composition}, which explains how the privacy loss accumulates when multiple differentially private mechanisms are applied either sequentially or in parallel. 
Then, we examine the {\em post-processing immunity property}, which ensures that any additional computations applied to the output of a differentially private mechanism do not weaken its privacy guarantees.
Finally, we present the notion of {\em group privacy}, which extends \dps to protect collections of records (or groups) by bounding the cumulative effect of changes in multiple records. These properties together provide a robust framework for analyzing the overall privacy risk when designing systems that employ \dpns.

\bigskip

\noindent
\textbf{Sequential Composition:}
%Applying multiple \dps mechanisms to the same dataset remains differentially private but with some degradation in the privacy parameters. 
In many practical scenarios, multiple \dps mechanisms are applied sequentially to the same dataset, for example, in iterative algorithms in machine learning. While the result of such composition remains differentially private, the overall privacy guarantee degrades. The following proposition quantifies the cumulative privacy loss incurred by sequential composition.

\begin{proposition}[\textbf{Sequential Composition}, \cite{Dwork2006b, Dwork2009}]\label{Prop_SequentialComposition}
Let 
 $\mathcal{A}_1(D),\ldots,\mathcal{A}_k(D)$ be a collection of $k$ mechanisms, where each mechanism $\mathcal{A}_i$ is $(\varepsilon_i,\delta_i)$-differentially private for $i=1,\ldots,k$. Then the combined mechanism
$$\mathcal{M}(D)=(\mathcal{A}_1(D),\ldots,\mathcal{A}_k(D))$$
satisfies $(\varepsilon_\mathcal{M},\delta_\mathcal{M})$-\dps with $$\varepsilon_\mathcal{M} = \sum_{i=1}^k\varepsilon_i \qquad \text{and} \qquad \delta_\mathcal{M} = \sum_{i=1}^k\delta_i.$$
\end{proposition}
\noindent
This result shows that the privacy losses of individual mechanisms add up when they are applied sequentially, meaning that the overall privacy budget must be managed carefully to ensure that the cumulative $\varepsilon$ and $\delta$ remain within acceptable bounds. \cite{DworkRoth2014} showed that this simple linear composition bound on $\varepsilon$ can be improved by allowing a slight increase in $\delta$. Specifically, they consider the linear composition of the {\em expected} privacy loss across mechanisms, which can be converted into a cumulative privacy budget $\varepsilon$ with high probability. This result is formalized in the \emph{advanced composition theorem}, which is proved to apply to any differentially private mechanism.\medskip 
\begin{theorem}[\textbf{Advanced Composition}, {\cite{DworkRoth2014}}]
\label{thm:advanced_composition}
Let $\mathcal{A}_1, \mathcal{A}_2, \ldots, \mathcal{A}_k$ be a collection of $k$ mechanisms, where each $\mathcal{A}_i$ satisfies $(\varepsilon, \delta)$-\dpns. Then the combined mechanism $\mathcal{M}$
that applies all $k$ mechanisms sequentially satisfies $(\varepsilon', k\delta + \delta')$-DP for any $\delta' > 0$, where
\[
\varepsilon' = \sqrt{2k \ln(1/\delta')} \, \varepsilon + k\varepsilon (e^\varepsilon - 1).
\]
In particular, when $\varepsilon$ is small (e.g., $\varepsilon \leq 1$), the bound simplifies approximately to
\[
\varepsilon' \approx \sqrt{2k \ln(1/\delta')} \, \varepsilon + k\varepsilon^2.
\]
\end{theorem}

\noindent
The advanced composition theorem applies to \emph{$k$-fold adaptive composition}, where each mechanism in the sequence may depend on the outputs of the previous ones. This captures real-world scenarios such as iterative model training or sequential data analyses. The theorem ensures that even in this adaptive setting, the cumulative privacy loss can be tightly bounded.
%\medskip

While the advanced composition theorem provides significantly improved bounds over naive linear composition, it can still be overly conservative in practice --- especially in settings involving many training iterations, such as deep learning. To obtain much tighter and more practical privacy estimates, \cite{abadi2016deep} introduced the \emph{Moments Accountant} method, which tracks higher moments of the privacy loss random variable, rather than just its expectation. This approach anticipates the later formalization of \emph{Rényi differential privacy} (Rényi \dpns) by %by Mironov 
\cite{RDP2017}, which offers a more refined and composable framework for privacy accounting. We return to these sharper cumulative privacy guarantees in the context of DP in machine learning in Section~\ref{sec_Accounting}. 
%\medskip

A key consequence of the composition property is that the total desired $\varepsilon$ can be treated as a "privacy budget". This budget is analogous to a financial budget --- it can be split into portions and spent gradually. 
It is worth noting that exceeding the allocated budget, for instance,  by releasing additional statistics, does not immediately expose the dataset to attack, but it does increase the overall $\varepsilon$, thereby weakening the privacy guarantee in a measurable way.\bigskip

\begin{mdframed} %\vspace{-5mm}
\begin{example}
\textbf{Privacy Budget Allocation}\medskip

\noindent
Suppose an analyst wants to publish a set of statistics about a sensitive dataset while maintaining an overall privacy loss of $\varepsilon=0.5$. They could choose to perform five analyses, each with $\varepsilon=0.1$, such that the cumulative privacy loss remains exactly at $0.5$. %\medskip

Alternatively, if some analyses require more precise results, the analyst could allocate the budget unevenly --- for instance, three analyses at $\varepsilon=0.15$ and two analyses at $\varepsilon=0.05$.  %\medskip

In both cases, the total 
$\varepsilon$ is the sum of the individual privacy losses, ensuring that the overall risk is bounded.
However, in the latter case, the total privacy loss would be $3 \times 0.15 + 2 \times 0.05 = 0.55$, which exceeds the original budget of $0.5$
 and results in a slightly weakened privacy guarantee. 

\end{example}\smallskip
\end{mdframed}%\medskip
%\medskip
\bigskip
\bigskip

\noindent
\textbf{Parallel Composition:} In sequential composition all mechanisms are applied to same data set $D$.  
 In contrast, parallel composition assumes that the dataset can be partitioned into disjoint subsets, with each mechanism applied to one unique subset. In this case, the overall privacy loss is determined solely by the worst-case loss among the individual mechanisms.
 The following proposition formalizes this parallel composition property.

\begin{proposition}[\textbf{Parallel Composition}, \cite{Ponomareva2023}]
Let 
 $\mathcal{A}_1,\ldots,\mathcal{A}_k$ be a collection of $\,k$ mechanisms, where each mechanism $\mathcal{A}_i$ is $(\varepsilon_i,\delta_i)$-differentially private for $i=1,\ldots,k$. 
  Suppose the dataset $D$ is partitioned into $k$ mutually disjoint subsets  $D_1,\ldots,D_k$, and each mechanism $\mathcal{A}_i$ is applied exclusively to its corresponding subset $D_i$
 Then the combined mechanism
$$\mathcal{M}(D)=(\mathcal{A}_1(D_1),\ldots,\mathcal{A}_k(D_k))$$
satisfies $(\max_{i=1,\ldots,k} \varepsilon_i, \max_{i=1,\ldots,k} \delta_i)$-\dpns.
\end{proposition}

\ignore{
\begin{proposition}[\textbf{Parallel Composition} \cite{Ponomareva2023}]
Let 
 $\mathcal{A}_1,\ldots,\mathcal{A}_t$ be a collection of $\,t$ mechanisms, where the $i$-th mechanism satisfies $(\varepsilon_i,\delta_i)$-\dps for $i=1,\ldots,t$. 
  Suppose the dataset $D$ is partitioned into $t$ mutually disjoint subsets  $D_1,\ldots,D_t$, and each mechanism $\mathcal{A}_i$ is applied exclusively to its corresponding subset $D_i$
 Then the combined mechanism
$$\mathcal{M}(D)=(\mathcal{A}_1(D_1),\ldots,\mathcal{A}_t(D_t))$$
satisfies $(\max_{i=1,\ldots,t} \varepsilon_i, \max_{i=1,\ldots,t} \delta_i)$-\dpns.
\end{proposition}
} % tässä oli t kun aiemmin k. En nyt yhtään keksi oliko tälle joku järkevä syy.

\noindent
The parallel composition guarantee is stronger than the sequential composition guarantee because the overall privacy loss does not compound when mechanisms operate on non-overlapping subsets of the dataset. \bigskip

\begin{mdframed} %\vspace{-5mm}
\begin{example}
\textbf{Parallel Composition}\medskip

\noindent
Consider a hospital database partitioned into $k$ disjoint subsets by department (e.g., cardiology, oncology, pediatrics). Suppose each subset $D_i$
is analyzed with a mechanism $\mathcal{A}_i$
that is $(\varepsilon_i,\delta_i)$-differentially private. %\medskip

The combined release 
$(\mathcal{A}_1(D_1),\ldots,\mathcal{A}_k(D_k))$
satisfy $(\max_{i=1,\ldots,k} \varepsilon_i, \max_{i=1,\ldots,k} \delta_i)$-DP rather than just $(\sum_{i=1}^k \varepsilon_i, \sum_{i=1}^k \delta_i)$-DP, since no single individual’s data appears in more than one subset $D_i$. %\medskip

In contrast, if the same individual’s data appeared in multiple subsets, the privacy losses would compound as in sequential composition.
This illustrates why parallel composition offers a stronger guarantee than sequential composition.\smallskip
\end{example}
\end{mdframed}%\medskip

\bigskip

\noindent
\textbf{Invariance to Post Processing:}
Differentially private mechanisms are immune to post-processing. That is, any function applied to the output of a differentially private mechanism --- whether by a data analyst or an adversary --- cannot degrade its privacy guarantee, provided the original dataset is not accessed. This property ensures that no additional sensitive information can be extracted beyond what is already protected by the mechanism itself. The following proposition formalizes this principle.

\begin{proposition}[\textbf{Post Processing}, \cite{DworkRoth2014}]
\label{postProcessing}
Let $\mathcal{A}:\mathcal{D} \rightarrow R$ be a randomized mechanism that satisfies $(\varepsilon,\delta)$-\dpns.  Then, for any (possibly randomized) function $f:R \rightarrow R'$, the composed mechanism $f \circ \mathcal{A}$, defined by  
$(f \circ \mathcal{A})(D) = f(\mathcal{A}(D))$, also satisfies $(\varepsilon,\delta)$-\dpns.
\end{proposition}
% Sekä Dwork+Roth että Vadhan sanoo "randomized" ei pelkästään "possible". Eli onko tuo randomized vaatimus?
% En äkikseltään keksi miksi sen pitäisi olla.

\noindent
This result guarantees that further computations on the output of a differentially private mechanism do not incur additional privacy loss. It enables flexible downstream analysis while preserving the privacy of individuals in the dataset.\medskip

\begin{mdframed}%\vspace{-5mm}
\begin{example}
\textbf{Post Processing}\medskip

\noindent
Suppose $\mathcal{A}$ is a mechanism that returns a noisy count of the number of users in a dataset 
$D$ who have a certain property. For instance, 
$\mathcal{A}(D)$ might output a value $x$ that is the true count plus some added noise --- making it 
$\varepsilon$-\dpns.%\medskip

Now, define a function 
$f$ that doubles its input, i.e., $f(y)=2y$. Even if an analyst applies $f$ to the output of $\mathcal{A}$, $f(\mathcal{A}(D))=2\times\mathcal{A}(D)=2x$, the overall mechanism $f \circ \mathcal{A}$ remains $\varepsilon$-\dpns.%\medskip

This demonstrates post-processing immunity: the additional computation (multiplying by 2) does not weaken the \dps guarantee provided by 
$\mathcal{A}$, so no extra sensitive information is revealed by the post-processing step.
\end{example}\smallskip
\end{mdframed}%\medskip

\bigskip

\noindent
\textbf{Group Privacy:}
Group privacy allows us to extend the privacy guarantee from a smaller unit of privacy (i.e., each individual record or example) to a larger group. Specifically, \cite{DworkRoth2014} have shown that if an algorithm is $\varepsilon$-\dps at the example-level, then for any two datasets that differ in at most $k$
 records, the overall guarantee scales to $k\varepsilon$. This demonstrates that \dps is primarily designed to protect each individual record --- the fundamental unit of privacy --- rather than concealing the effect of large, coordinated changes. Nevertheless, this group privacy bound is useful, as a very small per-example $\varepsilon$ effectively controls the influence of small groups of records.\medskip

\begin{proposition}[\textbf{Group Privacy under $\boldsymbol{\varepsilon}$-\DP}, \cite{DworkRoth2014}] \label{groupDP1}
Suppose a mechanism 
$\mathcal{A}:\mathcal{D} \rightarrow R$ satisfies $\varepsilon$-\dpns. Then for any two datasets $D$ and $D'$ that differ in at most $k$ records, $\mathcal{A}$
 satisfies 
$k\varepsilon$-\dpns. That is, for every measurable set $S \subseteq R$,
$$
    P[\mathcal{A}(D) \in S] \leq \exp(k\varepsilon) \cdot P[\mathcal{A}(D') \in S].
$$
\end{proposition}

\noindent
This guarantee means that if multiple records are changed in the dataset, the privacy loss under $\varepsilon$-\dps scales linearly with the number of records affected, ensuring that the impact of any group of individuals is appropriately bounded. In contrast, when working with $(\varepsilon,\delta)$-\dpns, the parameter $\delta$ does not scale linearly with the group size; its accumulation is more complex and requires a distinct analysis. 
\cite{Vadhan2017} extended Proposition \ref{groupDP1} to  $(\varepsilon,\delta)$-\dpns. %Vadhan and Lindell.

\begin{proposition}[\textbf{Group Privacy under $\boldsymbol{(\varepsilon,\delta)}$-\DP}, \cite{Vadhan2017}] Suppose a mechanism 
$\mathcal{A}:\mathcal{D} \rightarrow R$ satisfies $(\varepsilon,\delta)$-\dpns. Then for any two datasets $D$ and $D'$ that differ in at most $k$ records, $\mathcal{A}$
 satisfies 
$(k\varepsilon,k\exp(k\varepsilon)\,\delta)$-\dpns.
\label{groupDP2}
\end{proposition}

\subsection{Variants of Differential Privacy}
\label{sec_DPvariants}
In addition to $(\varepsilon, \delta)$-\dps (Definition  \ref{approximateDP}), several alternative formulations have been introduced to modify the strictness of the original, pure \dps definition (Definition \ref{pureDP}). 
These variants are designed to better accommodate practical scenarios and to offer more flexible trade-offs between privacy and utility. They play a crucial role in facilitating the effective deployment of \dps in real-world applications. Below, we summarize the most prominent ones. The overview of these \dps variants is given in Table \ref{table_DPvariants}.\medskip

\noindent
\textbf {R\'enyi Differential Privacy}, introduced by %(R\'enyi \dpns) %, {\red introduced by Mironov (2017)} 
\cite{RDP2017}, defines privacy in terms of the Rényi divergence between output distributions. R\'enyi \dps offers tighter bounds than $(\varepsilon,\delta)$-\dps on the cumulative privacy loss under composition, making it well-suited for iterative algorithms like {\em Differentially Private Stochastic Gradient Descent} (DP-SGD, see Section \ref{sec_DP-SGD}) and deep learning scenarios (\cite{abadi2016deep}).
R\'enyi \dps parameters can be converted to $(\varepsilon,\delta)$-\dps bounds, making it more intuitive and interoperable with the standard framework. We will formally define R\'enyi \dps in Section \ref{sec_DP-SGD}. \medskip

\noindent \enlargethispage{\baselineskip}
\textbf {Zero-Concentrated Differential Privacy} (zCDP) by \cite{BunMark2016CDPS} is another relaxation that quantifies privacy loss using Rényi divergence. It provides a more convenient framework for analyzing composition and subsampling, and often yields better utility-privacy trade-offs in theoretical guarantees.
zCDP has been effectively applied, for instance, in federated learning scenarios (\cite{KairouzPeter2021TDDG,Gboard2023}, see also Section \ref{sec_FL}), 
\ignore{
Concentrated Differentially Private Federated Learning with Performance Analysis
Hu, R.,Guo, Y.,Gong, Y.
IEEE Open Journal of the Computer Society
 2021}
where it helps balance the trade-off between privacy and utility without relying on a fully trusted server. %This is achieved through local gradient perturbation and secure aggregation.
In addition, zCDP supports conversion into $(\varepsilon,\delta)$-\dps bounds for compatibility. 
\ignore{
In some implementations, such as per-record zero concentrated differential privacy (PzCDP), the privacy loss is a function of each record's value, making it particularly useful for datasets with skewed or heavy-tailed distributions.
}
\medskip

\begin{landscape}
\begin{table}[ph!]
\caption{Overview  of Differential Privacy Variants}
\label{table_DPvariants}
\begin{tabular}{@{}llllll@{}}
\toprule
\textbf{Variant} & \textbf{Core Idea} & \textbf{Metric} & \textbf{Advantage} & \textbf{Limitation} & \textbf{Use Case} \\
\midrule
\textbf{Pure DP} & Strict $\varepsilon$-bounded privacy & $\varepsilon$  & Strong, clear guarantee &Low utility in practice &Theoretical studies,\\
&&& && Basic mechanisms \\ \midrule
\textbf{Approx. DP} & Allows small probability $\delta$ & $(\varepsilon, \delta)$ & More flexible, & Weaker than pure DP& ML-training, DP-SGD\\
&of privacy leakage && Allows Gaussian noise &Risk of catastrophic failure &\\ \midrule
\textbf{R\'enyi DP} & Tracks loss via Rényi & $(\alpha, \varepsilon)$ & Improved composition, & Needs $(\varepsilon,\delta)$ conversion &   ML training, DP-SGD \\
&divergence &&  No catastrophic failure & \quad for interpretability \\ \midrule
\textbf{zCDP} &Rényi divergence and & $\rho$ & Improved composition, &  Needs $(\varepsilon,\delta)$ conversion &  Subsampling,  \\
& concentrated privacy via $\rho$&&  No catastrophic failure & \quad for interpretability & Theoretical analysis \\ \midrule
\textbf{GDP} & Privacy as Gaussian & $\mu$ & Improved composition, & Needs $(\varepsilon,\delta)$ conversion &  ML training, DP-SGD \\
& hypothesis testing && Intuitive,  & \quad for interpretability \\
&&& No catastrophic failure
&& \\ \midrule
\textbf{Local DP} &  Privacy enforced locally & Local $\varepsilon$  & No need for trusted  &  High noise, low utility & Telemetry, surveys \\
& before data leaves device && \quad aggregator
&&\\ \midrule
\textbf{PDP} & Per-user privacy settings & $\varepsilon_i$ & Fine-tuned privacy &  Complex to manage &  User-controlled sharing \\ \midrule
\textbf{Heterog.\ DP}      	& Varying $\varepsilon$ by context &  Context- 
&  Flexible trade-offs &  Risk of unfairness &  Federated learning \\ 
 & & \quad specific $\varepsilon$ \\
\midrule
\textbf{SP}     	& Varying $\varepsilon$ by outlierness & Data- & Utility for rare events &  Weaker protection for &  Anomaly detection \\ 
&&\quad aware $\varepsilon$ && \quad outliers \\
\bottomrule
\end{tabular}
\end{table}
\end{landscape}

\noindent
\textbf{Gaussian Differential Privacy} (GDP), proposed by \cite{DongJinshuo2022Gdp}, is a refined variant of DP that redefines privacy guarantees using a hypothesis testing interpretation. A mechanism satisfies $\mu$-GDP if an adversary cannot distinguish between two neighboring datasets better than distinguishing between two normal distributions $\mathcal{N}(0,1)$ and $\mathcal{N}(\mu,1)$. This formulation offers a more interpretable privacy guarantee than classical $(\varepsilon, \delta)$-DP by linking the privacy loss to statistical testing power. GDP also satisfies strong composition properties and admits a central limit theorem for cumulative privacy loss, making it conceptually appealing for repeated analyses.  Additionally, GDP guarantees can be converted into corresponding $(\varepsilon, \delta)$-DP guarantees, allowing compatibility with existing privacy accounting frameworks and regulatory standards.
\medskip

\noindent
In \textbf{Local Differential Privacy} %(Local \dpns) 
formalized in \cite{Kasi2011}, the privacy guarantee is enforced before data leaves the client device. Each individual perturbs their data locally before aggregation, removing the need for a trusted server. Local \dps is used in applications like telemetry collection (e.g., Google’s RAPPOR (\cite{rappor}) and Apple’s \dps deployment (\cite{apple2017})). While it offers strong individual privacy, it typically results in lower utility due to high noise levels. We will consider local \dps more in Section \ref{sec_PrivacySettings}.\medskip

\noindent
\textbf{Personalized Differential Privacy} (PDP, %(Personalized \dpns)  
\cite{boenisch2024wayindividualizedprivacyassignment,Edabi2015,Jorgensen2015}) 
 is an advanced privacy protection mechanism that tailors privacy guarantees to individual users' preferences and data sensitivity. This approach is particularly useful in scenarios where users have varying privacy requirements, allowing for more flexible and accurate data privacy
 management. 
 %\cite{Jorgensen2015} löytyy myös määritelmä, jos sellaisia tänne halutaan
 %Tästä olisi jotain kehittyneempiä mekanismejä olemassa  
 % Jos löytyy use caseista tms jotain, niin tähän olisi mukava saada viite johonkin lukkun ja kongreettinen esimerkki missä käytetään
 % Ehkä täältä voi viitata user expectations lukuun???
 \ignore{
Utility-aware Personalized Exponential Mechanism (UPEM): This mechanism distinguishes between different possible results with the same personalized score, enhancing the utility of the data while maintaining privacy.

Utility-aware Exponential Mechanism for Personalized Differential Privacy,
Niu B.,Chen Y.,Wang B., (...),Li F.
IEEE Wireless Communications and Networking Conference, WCNC 2020

Fine-Grained Personalized Differential Privacy (FG-PDP): This mechanism reduces sampling errors and boosts data utility by performing approximating and filtering processes on the unsampled dataset before noise processing.

Enhancing Data Utility in Personalized Differential Privacy: A Fine-Grained Processing Approach,
Liu Z.,Wang W.,Liang H.,Yuan Y.
Lecture Notes in Computer Science (including subseries Lecture Notes in Artificial Intelligence and Lecture Notes in Bioinformatics) 2025

} 
 \medskip 

\noindent
\textbf{Heterogeneous Differential Privacy}, %(Heterogeneous \dpns) 
 first introduced by \cite{AlagganMohammad2017HDP},
is closely related to personalized \dpns, and in many contexts, they refer to the same general idea --- assigning different privacy levels to different individuals or data points. However, heterogeneous \dps is typically interpreted more broadly to encompass system-level implementations, where privacy budgets are determined by factors such as data sensitivity, model roles, or organizational policies --- rather than being explicitly specified by users. This broader notion of heterogeneous \dps has been applied in practical settings such as federated learning~(\cite{10032626,LingJie2024Eflp}, see also Section \ref{sec_FL}).
\medskip

\noindent
\textbf{Sensitive Privacy} (SP) presented in the work of  \cite{sp2019,9623509} is a recent extension proposed to address challenges in outlier detection.
It provides \dps levels based on each record’s "outlierness". That is, each record 
adapts the privacy level based on the degree of deviation from the norm. 
SP is used in tasks like outlier identification and anomaly detection (see Section \ref{sec_anomaly}), where achieving both meaningful privacy and useful accuracy under traditional \dps is essentially impossible without relaxing assumptions. SP improves utility in data-dependent tasks while preserving strong protections for the majority of the dataset.\medskip

Choosing the appropriate \dps variant is often dictated by the task, data distribution, and system constraints, especially in sensitive domains like anomaly detection, where rare events must be accurately identified without compromising individual privacy. %\newpage

\subsection{Privacy Settings: Central, Local, and Distributed}
\label{sec_PrivacySettings}
% Trust models???
\DP can be implemented under different trust models 
depending on where the noise is added and who handles the raw data. \DP comes in three main models: the standard model --- often called {\em central DP} --- assumes a trusted data curator holds the raw data and applies the \dps algorithm before releasing results. There are also {\em local DP}, where each individual randomizes their own data before sending it to an aggregator, ensuring privacy even if the server is untrusted, and {\em distributed DP}, which combines local noise addition and secure aggregation or anonymizing shuffle to improve utility while reducing central trust requirements (see Figure \ref{fig:DPsettings}). %\medskip
\begin{figure}[htb]
    \centering 
    \includegraphics[width=0.95\linewidth]{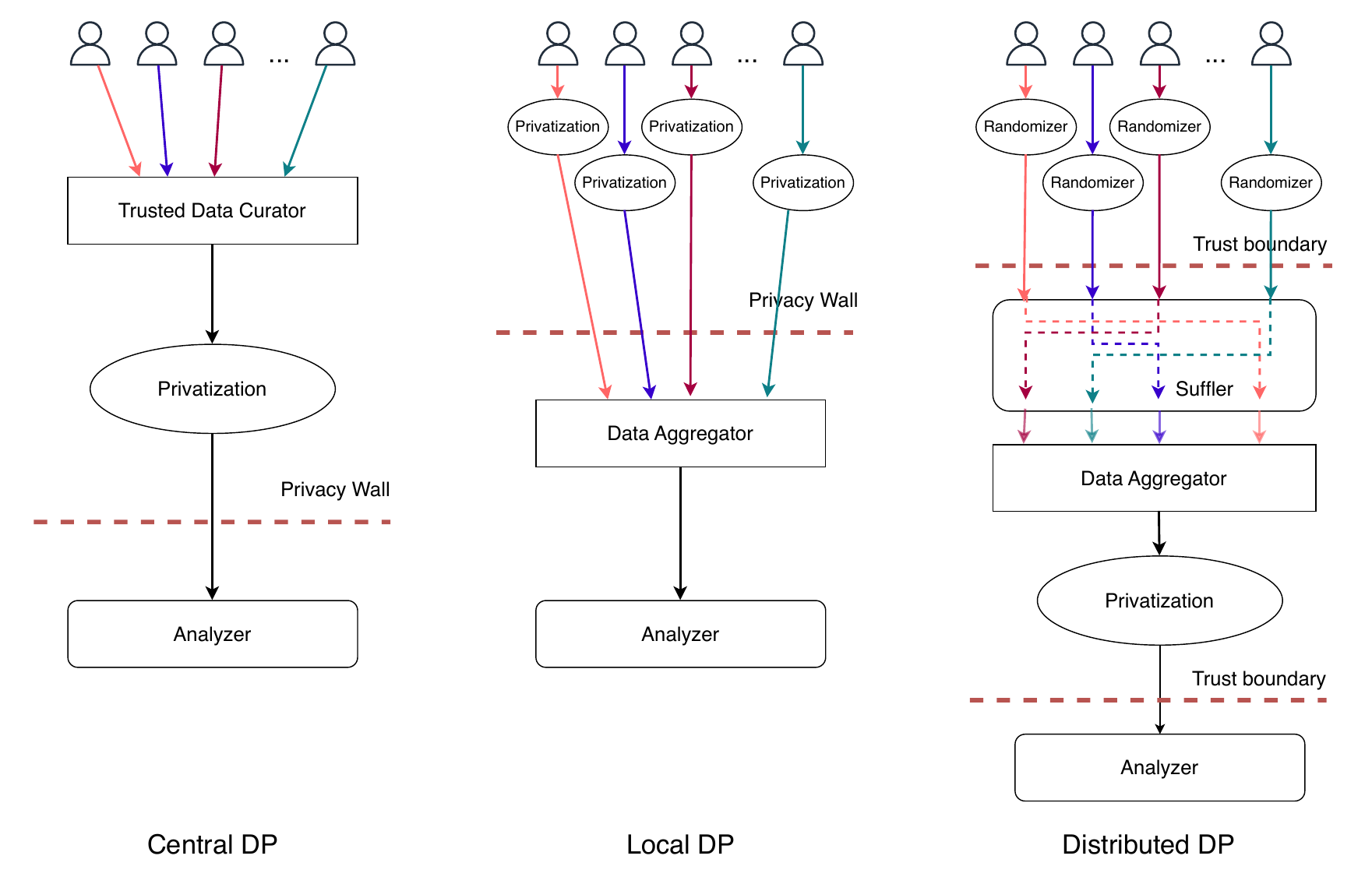}
    \caption{Different DP Settings: In distributed \dpns, a trusted suffler is used before aggregation (see, e.g., \cite{CheuAlbert2019DDPv,WeiYu2024DDPv}). } % Local DP
    \label{fig:DPsettings}
\end{figure}

This section reviews each of these \dps settings, highlights their key characteristics, and outlines their respective strengths, limitations, and typical use cases. To illustrate the differences, we use a simple real-world example: a survey asking individuals a sensitive yes/no question.  In addition, a summary comparison is provided in Table~\ref{table:DPsettings}.%\bigskip

%While this report mainly focuses on the central model, which underpins most \dps mechanisms like Laplace or Gaussian noise addition, we also briefly address scenarios where local and distributed \dps techniques (e.g., randomized response and differentially private federated learning) are particularly relevant, such as certain cybersecurity applications (see Section~\ref{sec_cybersecurity}).
%\medskip 

\bigskip

\noindent
\textbf{Central Differential Privacy:} 
{\em Central \dps} (often referred to as {\em global \dpns}) represents the classical \dps setting introduced in the foundational work by \cite{Dwork2006}. In this setting, a {\em trusted service provider} --- commonly called {\em trusted curator} or {\em aggregator} --- collects raw, unperturbed data from users.
The \dps mechanism (e.g., Laplace or Gaussian noise addition, or a privacy-aware machine learning algorithm) is then applied {\em centrally}, ensuring that the released output satisfies formal \dps guarantees. 
%\medskip

In the central \dps model, users must trust the aggregator to both {\em safeguard the raw data} and {\em correctly implement the privacy mechanism}, ensuring that the released output does not reveal sensitive information about any individual. The adversary is assumed to observe only the final output produced by the trusted aggregator.
%\medskip
%
In the context of machine learning, this setting corresponds to a scenario where multiple users contribute their raw data to a centralized dataset, curated by the aggregator, and this data is then used to train and release a machine learning model.

 Central \dps typically offers {\em high utility}, due to the availability of unperturbed data and the ability to add relatively low amounts of noise while preserving privacy.

\ignore{
Key Features:

Data collection: Raw data is sent to a trusted server.

Noise addition: Performed by the server, after query computation.

Utility: Typically high, due to the global view and low noise requirements.

Trust model: Requires users to trust the data curator.}

Typical use cases of central \dps include:
\begin{itemize} 
    \item \textbf{Official statistics} --- for example, U.S. census data (\cite{UScensus4,UScensus2}).
    \item \textbf{Centralized machine learning and analytics} --- for example, cardiovascular risk prediction (\cite{orabe2025}).
    \item \textbf{Research data analysis under strong governance} --- for example, protecting patient information in healthcare (\cite{Dankar2013}).
\end{itemize}

\begin{mdframed} %\vspace{-5mm}
\begin{example}
{\textbf{Health Survey with Central \dps}}\medskip

\noindent
Imagine a health survey asking 1,000 people if they have a certain condition. 
Under central DP, everyone {\em confidentially reports the truthful answer} (Yes/No) to the central health agency. The agency’s server now has the exact count of "Yes" answers (say 600 out of 1,000). 
To preserve privacy, the agency {\em adds a small random perturbation} to this count before publishing it. They might report "about 598" or "603" instead of exactly 600, by adding a little noise. This noise is carefully calibrated using a privacy budget $\varepsilon$ so that any single person’s presence or absence {\em does not noticeably change the overall result}. In essence, an individual’s data has only a tiny influence on the noisy output providing plausible deniability.\smallskip
\end{example}
\end{mdframed}  
\medskip

In the central \dps model, the service provider is assumed to be trusted. However, in practice, this trust assumption can be difficult to uphold — even large, reputable companies have failed in protecting users’ privacy as noted in Example \ref{ex7}. These incidents highlight the risks of centralizing raw user data and suggest that the assumption of a fully trusted third party is often unrealistic.\medskip

\begin{mdframed} %  \vspace{-5mm}
\begin{example} \textbf{Real-World Failures of the Trusted Curator  Assumption}\label{ex7} \medskip

\noindent
In 2018, hundreds of thousands of Google+ users had their private data leaked {(\cite{macmillan2018google})}, followed later that year by a bug in the Google+ API that exposed the data of 52.5 million users (\cite{newman2018googleplus}). In 2019, another breach exposed IDs, phone numbers, and names of hundreds of millions of Facebook users (\cite{fisher2019facebook}). These cases illustrate the potential dangers of storing unprotected user data in centralized systems.
\end{example}\smallskip
\end{mdframed}  
%https://www.sciencedirect.com/science/article/pii/S0920548923001083?casa_token=VwdXvSYDzjMAAAAA:okG3Rioh4IYeMfjmLxAhEupuwCNJLsSzNu28qv6r6AmUgxbd8X6gW3cMd8XkB-ItCgTZJTPaGcQ#b7
\medskip 

\bigskip

\noindent
\textbf{Local  Differential Privacy:}
Local DP originally formulated by \cite{Kasi2011} offers a stronger privacy model that eliminates the need for a trusted curator. In this setting, each user applies a predefined randomized mechanism --- such as in the classic randomized response technique --- to perturb their data {\em locally} before sending it to the server. As a result, the server never sees the original data, and privacy is guaranteed regardless of the aggregator’s behavior. 
Formally, local \dps is defined as follows:

\begin{definition}[\textbf{Local Differential Privacy}, \cite{Kasi2011}]
For $\varepsilon > 0$, a mechanism $\mathcal{A}:\mathcal{D}\rightarrow R$ satisfies $\varepsilon$-local \dps if, for
any two values $x$ and $x'$, and for any set
$S \subseteq R$ of possible outputs, the following holds:
$$
    P[\mathcal{A}(x) \in S] \leq \exp(\varepsilon) \cdot P[\mathcal{A}(x') \in S].
$$
\label{localDP}
\end{definition}\vspace{-5mm}

\noindent
Compared to the pure $\varepsilon$-DP definition (see Definition~\ref{pureDP}), which considers neighboring datasets differing by a single record, {\em local DP applies the privacy guarantee at the level of individual data points}. It requires the mechanism to produce statistically indistinguishable outputs for any pair of possible inputs, even if they are arbitrarily different.
%\medskip

In the local model, {\em all noise is added on the user’s device}, and only the privatized data is shared. This eliminates the need for trust in the server, making local DP particularly attractive in sensitive or large-scale environments. However, since each user acts independently, local DP mechanisms typically require {\em stronger noise}, which can lead to {\em lower utility} compared to central DP, especially for small or moderate-sized datasets. %\medskip
  
To address this trade-off in practical applications, \cite{Schein2018} have proposed {\em limited-precision local privacy} (LPLP) which weakens the strict indistinguishability requirements of local DP by only requiring privacy guarantees for inputs within a specified distance threshold. This relaxation improves utility in tasks like count data analysis, where exact local DP would be prohibitively noisy. 
%\newpage

\ignore{
Key Features:

Data collection: Users only send privatized (noisy) data.

Noise addition: Performed locally on the user device.

Utility: Generally lower, due to stronger noise and lack of global structure.

Trust model: No trust in the server is required.
}

Typical use cases of local \dps and LPLP include:
\begin{itemize}
  \item \textbf{Privacy-preserving surveys and polling} --- for example, collecting browser configuration data in Google Chrome using RAPPOR (\cite{rappor}).
  \item \textbf{User behavior telemetry} --- for example, Apple's collection of emoji usage and typing behavior across iOS devices (\cite{apple,apple2017}).
  \item \textbf{Simple analytics with very large user bases} --- for example, Apple's local DP system used for Safari domain popularity measurements (\cite{apple,apple2017}).
\end{itemize}\bigskip

\begin{mdframed} %\vspace{-5mm}
\begin{example}\textbf{Health Survey with Local \dps}
\medskip

\noindent
Imagine again a health survey asking 1,000 people whether they have a certain condition. %\medskip
With local DP, each person will {\em randomize their answer} themselves using a simple procedure --- for example, the classic randomized response method (\cite{Warner1965}, see also Example \ref{ex_RR}):
\begin{itemize}
    \item {\small Flip a coin. If it comes up heads, {\em answer truthfully} ("Yes" or "No").}
\item {\small If the first coin comes up tails, {\em flip a second coin}. Answer "Yes" if the second coin is heads, and "No" if the second coin is tails (in this case, the answer is purely random).}
\end{itemize}
Each participant follows the above steps privately. This means there is a 50\% chance they report their true answer, and a 50\% chance their answer is independent of the truth (random). {\em From the coordinator’s perspective}, any individual “Yes” could either be truthful or just due to the coin flips. Thus, the individual’s true response remains hidden. %\medskip

However, because the randomization process is known, the survey analyst can still {\em estimate} the overall "Yes" rate. For instance, if 550 out of 1,000 responses came back "Yes", the analyst knows that approximately 250 of those are likely due to the random process. Therefore, the estimated number of truthful "Yes" answers is 300 out of the 500 non-random responses. Based on this, the analyst can estimate that about 60\% of respondents --- roughly 600 out of 1,000 --- genuinely have the condition. %\medskip

In general, with a sufficiently large number of participants, the random noise averages out, allowing a reasonably accurate estimate of the population statistics --- while preserving the privacy of each individual.\smallskip

\end{example}
\end{mdframed}  %\bigskip

\bigskip

\noindent
\textbf{Distributed Differential Privacy:} {\em Distributed \dps} (see, e.g., \cite{CheuAlbert2019DDPv,Dwork2006b,KairouzPeter2021TDDG,WeiYu2024DDPv}) offers a middle ground between central and local \dpns.  The goal is to achieve accuracy closer to the central model without relying on a single trusted curator. This is typically accomplished through the use of {\em multiple semi-trusted parties} or {\em cryptographic techniques}. Users participate in a {\em collaborative protocol} that guarantees \dps while ensuring that no single party ever sees all the raw data. In effect, {\em trust is distributed}: as long as not all parties collude, individual data remains private, and the final output is equivalent to what a trusted curator would produce --- noise included.%\medskip  

A key advantage of distributed DP is its approach to noise addition. Instead of each user adding large, independent noise (as in local \dpns), each contributes a small amount of noise, and the protocol ensures that the final aggregate satisfies \dpns. This shared responsibility enables significantly {\em higher utility than local \dpns}, often approaching the accuracy of central \dpns. \bigskip

\begin{mdframed} %\vspace{-5mm}
\begin{example}\textbf{Health Survey with Distributed \DP} 
\medskip

\noindent
Once again imagine a health survey asking 1,000 people if they have a certain condition. This time, instead of one survey coordinator, there are {\em two independent coordinators} --- say, Alice and Bob --- working together. 
In addition, each participant is advised to divide their answer into two "shares," such that neither share reveals any information on its own.
%
% Tässä on virhe XOR ei voi käyttää summaan
%For a simple yes/no question, this can be done by selecting a random bit for Alice and a second bit --- computed as the {\em XOR of the random bit and the actual answer} -- for Bob. Effectively, the two bits sum to participant's actual answer.
%For example, if the true answer is "Yes" (1), a participant might send Alice a 0 and Bob a 1 (since 0 $\oplus$ 1 = 1), or alternatively, send Alice a 1 and Bob a 0. For a "No" (0) answer, both bits are 0.\smallskip
%
For example, for this simple yes/no question, each participant can use {\em additive secret sharing} over a finite field to split their binary response in a privacy-preserving way. Specifically, a random integer $r \in \{0,\ldots,q-1\}$ is generated and sent to Alice, while the value $(x-r) \mod q$ is sent to Bob, where $x\in \{0,1\}$ is the participant's actual answer and $q \geq n = 1000$. 
%\medskip

Neither Alice nor Bob can infer the participant’s answer individually, as both shares appear uniformly random. However, by summing their shares and combining the results modulo $q$, they can accurately compute the total number of "Yes" responses --- without ever seeing any individual's true input. Before releasing the result, a small amount of random noise is jointly added to the total to ensure \dpns.%\medskip 

In this {\em two-party distributed \dps scheme}, no single entity has access to any participant’s full answer. Alice holds one "half" of the data and Bob the other. As long as Alice and Bob do not collude and share their half-datasets, an individual’s true response remains private. Yet, together they are still able to compute an accurate aggregate count (plus noise). This is analogous to secure multi-party computation ensuring privacy, combined with a \dps noise addition. 
 The final output satisfies \dpns, but with distributed trust: privacy remains intact as long as at least one of the two parties behaves honestly.
\end{example}\smallskip
\end{mdframed}  \bigskip

Another approach to distributed \dps is using an {\em anonymizing shuffle} by \cite{CheuAlbert2019DDPv} (see Figure \ref{fig:DPsettings}). Each user sends a randomized report through an semi-trusted shuffler that strips any user identifiers and mixes all the reports before they reach the aggregator. The aggregator then adds noise to the aggregated, shuffled data. Since the data is anonymized by the shuffle, the aggregator/analyst can not link records to individuals, and only a small amount of noise is needed for privacy. 
 \cite{CheuAlbert2019DDPv}  show that this "shuffled model" can achieve accuracy close to the central model while removing the need to trust a single server.%\medskip

%The {\em privacy guarantee} in distributed \dps typically relies on an "honest-but-curious" assumption: for example, a shuffler correctly anonymizes messages, or in a multi-party setup, at least one server does not collude. Under such assumptions, individuals receive privacy guarantees similar to those in the central model: any single user's data has only a minimal impact on the final result (characterized by the privacy parameter $\varepsilon$), and no adversary can confidently isolate an individual’s input.

In summary, distributed \dps provides strong privacy guarantees with minimal central trust, by distributing both data and computation. It can achieve utility close to the central model while reducing reliance on a trusted curator. However, this comes at the cost of increased system complexity and the need for partial trust assumptions, such as ensuring that not all involved parties collude or the shuffler correctly anonymizes messages.
%\medskip
%\newpage
%The distributed model spreads data and computation across multiple parties or devices. It offers privacy with minimal central trust – approximating the accuracy of central \dps while approaching the trustlessness of local DP. The trade-off is increased system complexity and the need for partial trust assumptions --- for instance, "not everyone colludes".

\ignore{
Key Features:
\begin{itemize}
    \item Data collection: Users may send encrypted or masked data.
    \item Noise addition: Shared among participants; less per-user noise.
    \item Utility: Better than local DP, close to central DP in some cases. 

    \item Trust model: Requires partial trust or secure coordination.
\end{itemize}}

Typical use cases of distributed \dps include:
\begin{itemize}
 % Tämä on ok
    \item \textbf{Federated learning with privacy guarantees} --- for example, Google’s deployment of distributed DP with secure aggregation in Smart Text Selection to reduce data memorization (\cite{google2023distributed}), and privacy-preserving model training for disease prediction (\cite{Khanna2022}).
    
    \item \textbf{Privacy-preserving data sharing without centralization} --- for example, distributed census or survey data collection where individual data contributors apply noise locally and aggregate results through secure protocols (\cite{CheuAlbert2019DDPv}), and sparse location heatmap generation using distributed DP and secure aggregation (\cite{Bagdasaryan2022SparseFA}). 
\end{itemize}

\ignore{
 Data is held by multiple parties.  
 
 Noise is generated collaboratively (e.g., via secure aggregation or distributed noise generation) across the parties.  
 
 Offers a balance between privacy and utility without requiring a single trusted curator, yet it is more complex to implement. \medskip}
\medskip

%%%%%%%%%%%%%%%

Each of these \dps settings serves a different set of privacy and deployment needs. Choosing the appropriate setting depends on the specific context, including data sensitivity, infrastructure, trust assumptions, and desired accuracy. A comparison of the \dps settings is given in Table~\ref{table:DPsettings}.\medskip

\begin{table}[ht]
\caption{Comparison of Differential Privacy Settings}
\label{table:DPsettings}
\resizebox{\textwidth}{!}{\begin{tabular}{@{}llll@{}}
\toprule
\textbf{Feature} & \textbf{Central DP}  & \textbf{Local DP}  & \textbf{Distributed DP}\\
\midrule
\textbf{Who sees raw data?}      & Central server     & No one              & No one \\ 
\textbf{Where is noise added?}   & At the server      & On the user device  & Users + secure protocol \\ 
\textbf{Utility}                 & High               & Low                 & Medium to High \\ 
\textbf{Trust requirement}       & Trusted curator    & None                & Partial trust or cryptography \\ 
\textbf{Typical applications}    & Statistics, ML     & Telemetry, polling  & Federated learning,  \\ 
 &    &   & Multi-party analytics \\ 
\bottomrule
\end{tabular}}
\end{table}

\subsection{Mechanisms for Differential Privacy}
Achieving DP in practice requires introducing randomness into computations. In complex systems, \dps is implemented in a modular fashion using building-block mechanisms. The key challenge is to add just enough noise to protect sensitive information while still preserving the usefulness (utility) of the results.%\medskip

Various standard mechanisms have been developed, each suited to different types of queries or data. 
Designing a \dps mechanism requires 
careful consideration of the query’s sensitivity and a thoughtful allocation of the privacy budget  $\varepsilon$ across queries. 
 In this section, we first review the fundamental mechanisms --- Laplace, Gaussian, and Exponential --- along with an explanation of sensitivity. Then, we briefly recall some other widely-used \dps mechanisms and techniques in the literature.

%randomized responce?

\subsubsection{Sensitivity of Function} 
Before introducing the noise mechanisms, it is important to define sensitivity. The global sensitivity of a function measures how much the function’s output can change when a single individual’s data is modified.
\begin{definition}[\textbf{Global $\boldsymbol{\ell_p}$-sensitivity}, \cite{Ponomareva2023}]
Given a query function $f:\mathcal{D} \rightarrow \R^d$ and a set of all possible pairs of neighboring datasets 
$N=\{(D,D') \mid D \text{ and }D' \text{ are neighbors}\}$, we define the global $\ell_p$-sensitivity of $f$ as 
$$
\Delta_f^p =\max_{D,D'\in N} \norm{f(D)-f(D')}_p,
$$
where $\norm{\cdot}_p$ is the $\ell_p$-norm.
\end{definition}\enlargethispage{\baselineskip}
\noindent
 Intuitively, $\Delta_f^p$ quantifies the worst-case contribution any single individual can have on the function’s value. %\bigskip
It plays a crucial role in differential privacy, as the amount of random noise that must be added is directly proportional to $\Delta_f^p$. Queries with higher sensitivity require more noise to sufficiently obscure the effect of any single individual.
\smallskip

 \begin{mdframed}\vspace{-1mm}
 \begin{example}
 \textbf{ Sensitivity of a function} \smallskip

\noindent
If $f$  counts the number of records satisfying a certain property, then adding or removing a single individual can change the count by at most 1. Thus, the sensitivity of such a counting query is $\Delta_f^1=1$.
%\medskip

If $f$ returns the sum of ages of all individuals in the dataset, then the sensitivity $\Delta_f^1$ equals the maximum possible age, as a single individual could contribute that amount to the total.%\medskip
\end{example}\smallskip
 \end{mdframed}
\smallskip

In practice, sensitivity may be unbounded or difficult to estimate. A common solution is to apply {\em clipping}, either to individual data entries or to the output of the query, ensuring that values lie within a bounded range (or have bounded norm). The choice of this range is a critical design parameter, as it introduces a bias–variance trade-off (\cite{Amin19}).
For standard queries --- such as computing a sum, mean, or gradient --- clipping ensures a bounded sensitivity that is both interpretable and easy to compute.\smallskip

\begin{mdframed}%\vspace{-5mm}
 \begin{example}
 \textbf{ Unbounded Sensitivity of a function} \medskip

\noindent
If $f$ returns the average income of individuals in a dataset with no upper bound on individual income, then adding or removing a single person with an extremely high income can significantly change the average. In this case, the sensitivity $\Delta_f^p$ is unbounded (potentially infinite), making it unsuitable for direct use in \dps without additional constraints such as clipping or bounding the income values.
%\smallskip
%This highlights the importance of ensuring that the function’s output is appropriately bounded when designing privacy-preserving algorithms.
\end{example}\smallskip
 \end{mdframed}\medskip

\subsubsection{Basic Mechanisms}
We are now ready to review the fundamental mechanisms — Laplace, Gaussian, and Exponential. We note, however, that while these mechanisms form the theoretical foundation of DP, they are no longer sufficient on their own for many modern applications. Contemporary machine learning tasks often require more specialized mechanisms that account for iterative training, complex model structures, or distributed data settings (these will be considered in Sections \ref{sec_otherDPmechanisms} and \ref{chaML}).

\bigskip

\noindent
\textbf{Laplace Mechanism:} 
One of the basic tools in \dps is the Laplace mechanism, used for numeric results. The idea is to add noise drawn from a Laplace distribution (see, Figure \ref{fig:laplace}) to the true answer of a query.
\begin{definition}[\textbf{Laplace Distribution}]
The standard Laplace distribution, centered at zero, is a continuous probability distribution defined on 
 $\R$ 
 with the probability density function
$$
g(u) = \frac{1}{2b}\exp{(-\frac{|u|}{b})}, 
$$
where $u \in \R$, $b>0$ is the {\em scale parameter}, and the variance is given by $\sigma^2=2b^2$. We denote by $\lap(b)$ the Laplace distribution with parameter $b$.\medskip

\end{definition}
\begin{figure}[ht]
    \centering
    \includegraphics[width=0.7\linewidth]{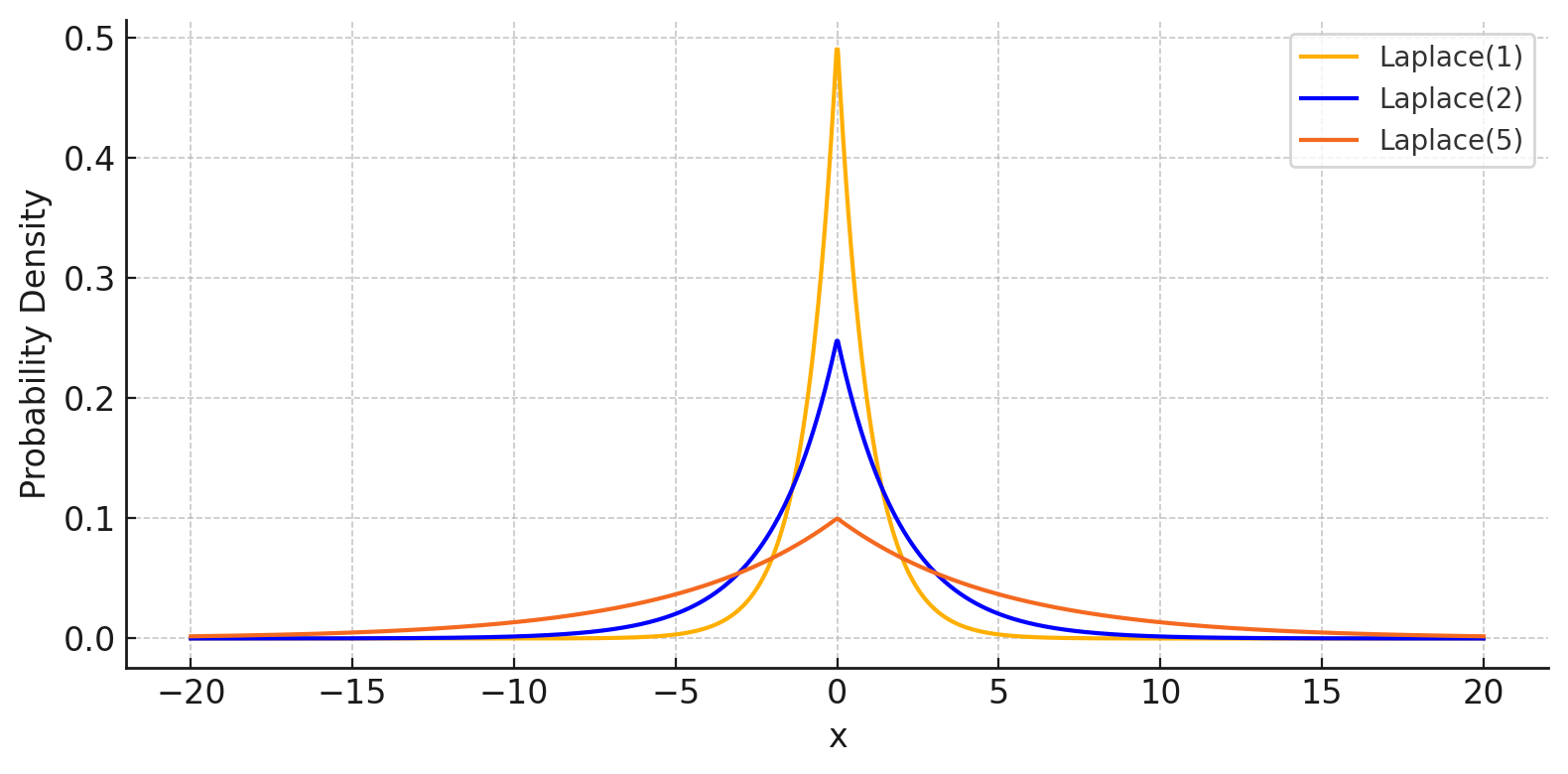}
    \caption{Laplace Distribution with Different Scale Parameters.} % better figure is needed
    \label{fig:laplace}
\end{figure} %\medskip

\noindent
For a query with output $f(D) \in \R^d$, the Laplace mechanism adds noise sampled from the Laplace distribution to each of the $d$ dimensions in the output, with the variance of the noise calibrated to the $\ell_1$-sensitivity of the query, ensuring that the effect of any single record is adequately masked.  
% Näistä voisi tehdä myös lauseita, joissa yhdistetään tuo että se takaa DP:n
%\begin{definition}[Laplace Mechanism %\cite{Dwork2006,DworkRoth2014}]
%Given a query function $f:\mathcal{D} \rightarrow \R^d$, we define the Laplace mechanism as
%$$
%\mathcal{A}_L(D;f,\varepsilon) = f(D) + %(Y_1,Y_2,\ldots,Y_d), 
%$$
%where $Y_i$ is independent and identically distributed %(i.i.d.) random variables drawn from $\lap(\Delta_f^1 / %\varepsilon)$.
%\end{definition}
\begin{theorem}[\textbf{Laplace Mechanism}, \cite{Dwork2006,DworkRoth2014}]
Given a query function $f:\mathcal{D} \rightarrow \R^d$,  the Laplace mechanism
$$
\mathcal{A}_L(D;f,\varepsilon) = f(D) + (Y_1,Y_2,\ldots,Y_d), 
$$
where each $Y_i$ is independent and identically distributed (i.i.d.) random variable drawn from $\lap(\Delta_f^1 / \varepsilon)$, is guaranteed to satisfy $\varepsilon$-\dpns.
\end{theorem}
%The Laplace mechanism $\mathcal{A}_L(D;f,\varepsilon)$ is guaranteed to satisfy $\varepsilon$-\dps \cite{DworkRoth2014}.
The scale parameter (and hence the variance) of the Laplace noise increases with the query's sensitivity $\Delta_f^1$. 
 %This is intuitive: queries with higher sensitivity can change more drastically with the addition or removal of a single record, and therefore require more noise to mask those changes. 
 Conversely, the scale parameter decreases as $\varepsilon$ increases, meaning that stronger (tighter) \dps guarantees (i.e., smaller 
$\varepsilon$) require the addition of more noise.
 In practice, the Laplace mechanism is straightforward and widely used for releasing aggregated statistics like sums, counts, and averages with \dps guarantees.\medskip
 
 \begin{mdframed}%\vspace{-5.5mm}
 \begin{example}
 
\textbf{Laplace Mechanism}\medskip

\noindent
Consider a query $f$ counting how many incidents in a log involve a certain malware --- adding or removing one log entry changes the count by at most $1$, so the sensitivity is $\Delta_f^1=1$. The Laplace mechanism would add noise proportional to $1/\varepsilon$ to obscure that $\pm 1$ difference: 
$$
\mathcal{A}_L(D;f,\varepsilon)=f(D)+ Y \qquad \text{ with } Y \sim \lap \left(\frac{1}{\varepsilon}\right).
$$        
\noindent
With appropriate choice of $\varepsilon$, the query result is noisy but statistically close to the real count, and an adversary cannot tell if any single log entry was included or not.
\end{example}
\end{mdframed}%\vspace{-6mm}

\begin{mdframed}%\vspace{-5.5mm}
\begin{example}\label{example_PatientPrivacy} 
\textbf{Protecting Patient Privacy} \medskip

\noindent
\textbf{Scenario: }  
A hospital wishes to publish the number of patients diagnosed with a particular disease without compromising individual privacy.  Let $D$ represent a dataset where entries indicate whether a patient has the disease (1) or not (0). For example, $D = [1, 0, 1, \mathbf{1}, 0]$ corresponds to 3 patients with the disease. A neighboring dataset $D'$ differs by one individual: $D' = [1, 0, 1, \mathbf{0}, 0]$, where only 2 patients have the disease. \medskip

\noindent
\textbf{Intuition: }  
By adding noise to the output, the hospital reduces the influence of any single individual's data, making it difficult for an adversary to determine whether a specific patient is included. \medskip

\noindent
\textbf{Procedure: }  
To preserve privacy, the hospital applies the Laplace mechanism:
$$
    \mathcal{A}_L(D; f, \varepsilon) = f(D) + \text{Noise}, 
$$
where $\text{Noise} \sim \lap(1/\varepsilon)$ and $f(D)$ is the count of patients with the disease. This mechanism,  with $\varepsilon = 1$, is illustrated in Figure \ref{fig:exampleLap}. %\medskip

The true count is $3$ for $D$ and $2$ for $D'$. Suppose the mechanism outputs $S=4$. Then, for $\varepsilon = 1$, the probabilities of this output are:
\begin{align*}
	            &P[\mathcal{A}_L(D;f,1)=4] = \tfrac{1}{2} \exp(-|4-3|)=\tfrac{1}{2}e^{-1}      \quad \text{and}      \\
	            &P[\mathcal{A}_L(D';f,1)=4] = \tfrac{1}{2} \exp(-|4-2|)=\tfrac{1}{2}e^{-2}.
\end{align*}
The ratio of these probabilities is: 
$$
\frac{P[\mathcal{A}(D)=4]}{ P[\mathcal{A}(D')=4]} = \frac{e^{-1} }{e^{-2} } = e^1, 
$$
which satisfies the 1-\dps condition. \medskip

\noindent
\textbf{Conclusion: }  
Even if an attacker observes the noisy output $S = 4$, they cannot confidently determine whether a specific patient's data was included. The privacy guarantee ensures that the presence or absence of any one individual has only a limited effect on the output distribution, thereby preserving patient confidentiality.
\end{example}\smallskip
\end{mdframed}%\medskip

\begin{figure}[hbt!]
    \centering
    \includegraphics[width=0.7\linewidth]{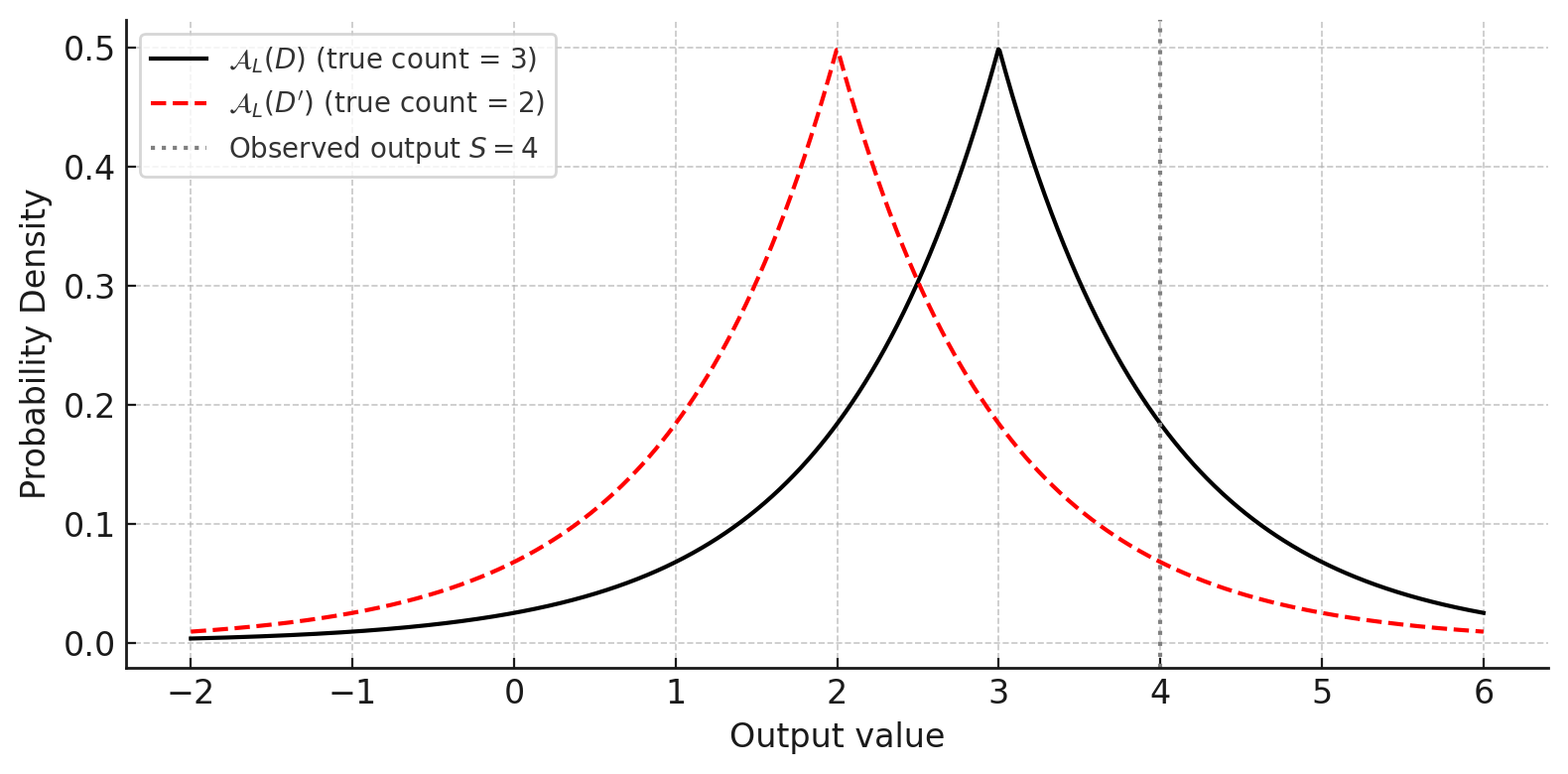}
    \caption{Laplace Mechanism Output for Neighboring Datasets in Example \ref{example_PatientPrivacy}.} % better figure is needed
    \label{fig:exampleLap}
\end{figure}

\medskip

\noindent \enlargethispage{\baselineskip}
\textbf{Gaussian Mechanism:} The Gaussian mechanism is similar in spirit to the Laplace mechanism but uses noise from the Gaussian (normal) distribution (see Figure \ref{fig:gaussian}). Gaussian noise is often used when we allow a small $\delta$ probability of failure, that is, for $(\varepsilon,\delta)$-\dps (Definition \ref{approximateDP}). Unlike the Laplace mechanism --- which relies on $\ell_1$-sensitivity --- the Gaussian mechanism calibrates the noise based on the $\ell_2$-sensitivity of the query and the desired privacy parameters $\varepsilon$ and $\delta$.
 
 \begin{definition}[\textbf{Gaussian Distribution}] The Gaussian distribution with mean $0$ and standard deviation $\sigma$ is defined on $\R$ by the probability density function
 $$ 
 g(u)=\frac{1}{\sqrt{1 \pi \sigma^2}} \exp \left(-\frac{u^2}{2 \sigma^2} \right)
 $$
for $u \in \R$. We denote this distribution by $\mathcal{N}(0,\sigma^2)$. \end{definition}
 \begin{figure}[ht!]
    \centering
    \includegraphics[width=0.7\linewidth]{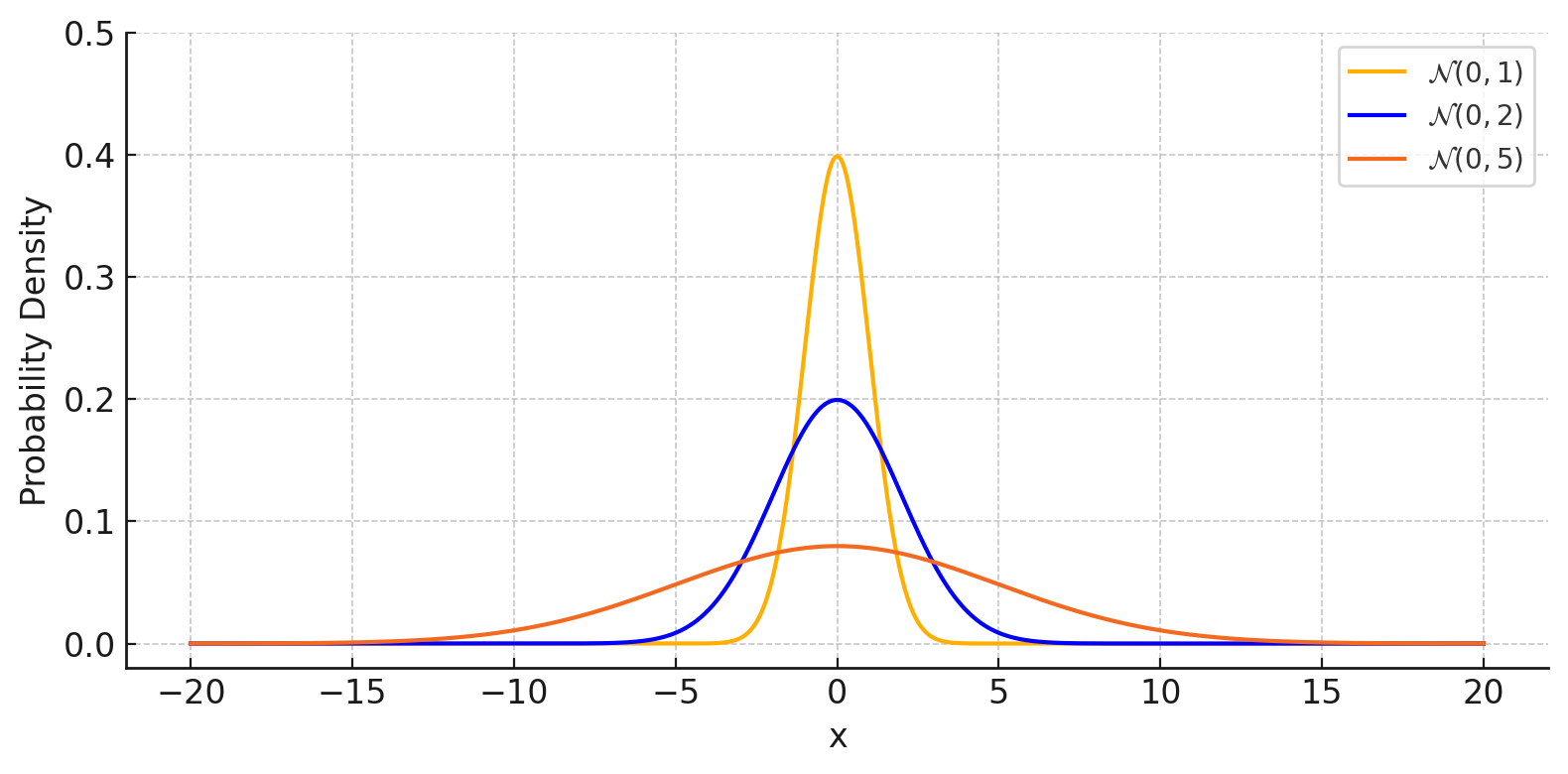}
    \caption{Gaussian Distribution with Different Deviations.} % better figure is needed
    \label{fig:gaussian}
\end{figure}

\noindent 
For a query with output $f(D) \in \R^d$, the Gaussian mechanism adds independent noise to each of the $d$ dimensions. A common choice is to set
\begin{align}
\label{sigma_in_gaussian}
\sigma = \frac{\Delta_f^2 \sqrt{2 \ln (1.25/\delta)}}{\varepsilon}.
\end{align}
This choice ensures that the mechanism satisfies $(\varepsilon,\delta)$-\dpns. 

\begin{theorem}[Gaussian Mechanism, \cite{DworkRoth2014}] Given a query function $f:\mathcal{D} \rightarrow \R^d$, the Gaussian mechanism defined as
$$
\mathcal{A}_G(D;f,\varepsilon,\delta) = f(D) + (Y_1,Y_2,\ldots,Y_d), 
$$
where $Y_i \sim \mathcal{N}(0,\sigma^2)$ ($i=1,\ldots,d$)
with 
$\sigma$ chosen as in \eqref{sigma_in_gaussian}, satisfies $(\varepsilon,\delta)$-\dpns. \end{theorem}

%\begin{definition}[Gaussian Mechanism %\cite{DworkRoth2014}] Given a query function %$f:\mathcal{D} \rightarrow \R^d$, the Gaussian mechanism %is defined as
%$$
%\mathcal{A}_G(D;f,\varepsilon,\delta) = f(D) + %(Y_1,Y_2,\ldots,Y_d), 
%$$
%where $Y_i \sim \mathcal{N}(0,\sigma^2)$ %($i=1,\ldots,d$)
%with 
%$\sigma$ chosen as in \eqref{sigma_in_gaussian}. %\end{definition}
%\vspace{-6mm}

 \begin{mdframed}%\vspace{-5.5mm}
 \begin{example}
 \textbf{Noisy Average BMI (Gaussian Mechanism)}\medskip

\noindent 
Suppose a hospital wants to publish the average body mass index (BMI) of 100 individuals in a dataset. Let \( f(D) \) denote the average BMI. To compute the sensitivity of the query, assume individual BMI values are bounded between 10 and 60, covering the realistic physiological range. The maximum change in the average due to the addition, removal, or replacement of a single individual is:
\[
\Delta_f = \frac{60 - 10}{100} = 0.5,
\]
so the \(\ell_2\)-sensitivity of the function is \( \Delta^2_f = 0.5 \). %\medskip

To satisfy \((\varepsilon, \delta)\)-\dps with parameters \( \varepsilon = 1 \) and \( \delta = 10^{-5} \), the noise scale is set as:
\[
\sigma = \frac{\Delta^2_f \sqrt{2 \ln (1.25/\delta)}}{\varepsilon} = \frac{0.5 \cdot \sqrt{2 \ln (1.25 / 10^{-5})}}{1}.
\]%\smallskip
The mechanism then returns the perturbed output:
\[
\mathcal{A}_G(D;f,\varepsilon,\delta) = f(D) + Y,
\] where $Y \sim \mathcal{N}(0, \sigma^2)$,
ensuring that the contribution of any one individual's BMI remains hidden in the released average.%\bigskip

As a rule of thumb, about 99.7\% of Gaussian noise lies within three standard deviations of the mean. Thus, outputs typically fall in the interval \( f(D) \pm 3\sigma \), giving a practical sense of the expected perturbation. For example, if the true average BMI is \( f(D) = 24.8 \) and \( \sigma \approx 2.5 \), most outputs would lie between approximately 17.3 and 32.3.
\end{example}\smallskip
  \end{mdframed}\smallskip\medskip

In summary, while the Laplace mechanism is effective for pure $\varepsilon$-\dpns, the Gaussian mechanism --- providing 
 $(\varepsilon,\delta)$-\dps --- is often favored in high-dimensional and complex settings. It benefits from lower noise, since $\ell_2$-sensitivity is generally smaller than $\ell_1$-sensitivity, and the faster decay of its tails, which reduces the likelihood of extreme noise values. Although it allows a small $\delta$ probability of privacy loss, its flexibility and compatibility with advanced composition theorems (see Theorem \ref{thm:advanced_composition} and Section \ref{sec_Accounting}) make it well-suited for tasks like privacy-preserving machine learning, where achieving a balance between accuracy and privacy is essential.
\bigskip

\noindent
\textbf{Exponential Mechanism:} 
In contrast to the Laplace and Gaussian mechanisms, which can only handle numerical query results, the exponential mechanism is designed for situations where the output is non-numeric or when a quality score can be assigned to potential outputs. 
Rather than perturbing the output, the exponential mechanism selects an output from a discrete set by weighing each candidate according to a utility function $u:D \times R \rightarrow \R$ that measures how desirable the output 
$r \in R$ is given the dataset 
$D$. Note that the set $R$ and the utility function $u$ are assumed to be public, while the dataset $D$ is private. The goal is to make sure that releasing some $r \in R$ does not reveal sensitive information about the records in $D$.
\vspace{-1mm}

\begin{definition}[\textbf{Exponential Mechanism}, \cite{McSherry2007}]
\label{exp_mechanism} For the utility function $u:\mathcal{D} \times R \rightarrow \R$,
the exponential mechanism $\mathcal{A}_E(D,u,R)$ selects and  outputs an element
$r \in R$ with probability proportional to
$$
\exp\left( \frac{\varepsilon u(D,r)}{2 \Delta_u}\right),
$$
where $\Delta_u$
 is the sensitivity of the utility function $u$, reflecting the maximum change in 
$u(D,r)$ when a single record in 
$D$ is modified. 
\end{definition} %\vspace{-5mm}

\begin{theorem}[\textbf{Exponential Mechanism: Privacy Guarantee}, \cite{McSherry2007}]
The exponential mechanism $\mathcal{A}_E(D,u,R)$ given in Definition \ref{exp_mechanism} satisfies $\varepsilon$-\dpns.
\end{theorem}%\vspace{-3mm}

The exponential mechanism is particularly useful for queries where directly adding noise to the result is not feasible, such as selecting the best candidate from a set of options. While the Laplace and Gaussian mechanisms are ideal for numeric outputs, the exponential mechanism provides a flexible and effective way to achieve DP for a wide range of outputs by favoring those with higher utility while still protecting individual privacy.
\medskip

\begin{mdframed}%\vspace{-5mm} 
\begin{example}\textbf{{Exponental Mechanism}}\medskip

\noindent
Suppose a platform wants to recommend one of three movies --- A, B, or C --- based on user ratings collected from a large number of individuals. Each movie receives an aggregated score (e.g., average rating):
\begin{itemize}
    \item {\small Movie A: 6;}
    \item {\small Movie B: 9;}
    \item {\small Movie C: 4.}
\end{itemize}

To protect the privacy of individual users while favoring higher-rated options, the exponential mechanism $\mathcal{A}_E$ is used. The utility function $u(m)$ is defined as the aggregated score of movie $m$, and the mechanism selects a movie with probability proportional to
\[
P[\mathcal{A}_E(u) = m] \propto \exp\left(\frac{\varepsilon u(m)}{2 \Delta_u}\right),
\]
where $\Delta_u$ is the sensitivity --- that is, the maximum change in utility when a single user’s ratings are added or removed. Assuming $\varepsilon = 1$ and $\Delta u = 1$, we compute:
\[
P[\mathcal{A}_E(u) =\text{A}] \propto e^3,\quad P[\mathcal{A}_E(u) =\text{B}] \propto e^{4.5},\quad P[\mathcal{A}_E(u) =\text{C}] \propto e^2.
\]
After normalization, this yields approximate selection probabilities:
\[
P[\mathcal{A}_E(u) =\text{A}] \approx 17\%,\quad P[\mathcal{A}_E(u) =\text{B}] \approx 77\%,\quad P[\mathcal{A}_E(u) =\text{C}] \approx 6\%.
\]

\medskip

\noindent
\textbf{Interpretation.}
The exponential mechanism selects a high-utility output while ensuring \dps with respect to individual contributions. Even if one user’s ratings are added or removed, the probability of selecting any specific movie changes only slightly (bounded by $\varepsilon$). This guarantees that no output reveals too much about any single user’s preferences.
\end{example}\smallskip
\end{mdframed}

%\medskip

% Tästä voisi tehdä summary laatikon
%\begin{itemize}
%\item Laplace Mechanism: Adds Laplace noise to outputs for pure $\varepsilon$-\dpns.
%\item Gaussian Mechanism: Adds Gaussian noise for $(\varepsilon,\delta)$-\dpns, commonly used in ML due to lower noise requirements in high-dimensional settings.
%\item Exponential Mechanism: Supports non-numeric outputs by assigning selection probabilities based on a scoring function.
%\end{itemize}
\bigskip

\subsubsection{Other Mechanisms and Techniques} 
\label{sec_otherDPmechanisms}
Next we list some other popular mechanisms and techniques for achieving \dpns. 
Each of these mechanisms is designed for different types of queries and applications, and the choice of the mechanism depends on the specific privacy and utility requirements of the task at hand. Table \ref{table_DPmechanisms} summarizes the mechanisms considered here.
For more discussion and surveys of additional techniques we refer to \cite{DworkRoth2014},
\cite{Jayaraman2019}, and
\cite{Ponomareva2023}.

\begin{table}[htp]
\caption{Overview  of Differential Privacy Mechanisms}
\label{table_DPmechanisms}
\resizebox{\textwidth}{!}{\begin{tabular}{@{}lllll@{}}
\toprule
\textbf{Mechanism} & \textbf{Core Idea} & \textbf{Metric} & \textbf{Key Feature} & \textbf{Use Case} \\
\midrule
\textbf{Laplace} & Adds Laplace & $\varepsilon$-DP & Scales noise to &  Count queries \\
 & noise to numeric &  & $\ell_1$-sensitivity &  Simple statistics,  \\
 & query outputs &  &  & \\ \midrule

\textbf{Gaussian} & Adds Gaussian &  ($\varepsilon,\delta$)-DP & Scales noise to & ML training, \\ 
 & noise to numeric &  & $\ell_2$-sensitivity & DP-SGD  \\ 
 & outputs &  &  & \\ \midrule

\textbf{Exponential} & Selects among & $\varepsilon$-DP or & Probability & Private selection, \\ 
 & discrete (non-numeric) & ($\varepsilon,\delta$)-DP  & weighted by & Maximum likelihood  \\ 
 & outputs using a utility &  & utility score & estimation \\ 
& function &  &  &  \\ \midrule

\textbf{RR} & Provides local \dps & $\varepsilon$-local DP & Responds truthfully &  Surveys,\\ 
 & for categorical or &  & or randomly with&  Telemetry \\ 
 & binary attributes &  &  probability & \\ \midrule

\textbf{RNM} & Choose the best & ($\varepsilon,\delta$)-DP & Adds noise to each & Classification, \\ 
 & option under DP & & score, returns max & Ranking \\  \midrule

\textbf{SVT} & Privately answer & ($\varepsilon,\delta$)-DP & Adds noise only & Interactive query  \\ 
 & a sequence of &  &  to comparisons & systems \,\, \\ 
 & threshold queries &  &  & \\ \midrule

\textbf{S\&A} & Aggregates private & ($\varepsilon,\delta$)-DP & Divide-aggregate &  Model training,\\
& outputs from & & design for complex &  Histograms\\
 & data partitions &  &   tasks &  \\ \midrule

\textbf{PATE} & Aggregates votes & ($\varepsilon,\delta$)-DP & Ensures privacy via & Deep learning, \\
& from teacher &  & noisy majority & Semi-supervised \\
 & models with noise &  &  vote &  learning \\ \midrule

\textbf{OutP} & Adds noise to model & ($\varepsilon,\delta$)-DP & Post-optimization  &  Convex ERM \\ 
 &  output after training & &  noise &  problems \\  \midrule

\textbf{ObjP} & Adds noise inside  & ($\varepsilon,\delta$)-DP & Optimizes a  & Smooth, strongly\\ 
 &  the loss function & &  perturbed  & convex ERM \\ 
 & &  & \quad objective &   \\ \midrule

\textbf{Gradient} & Private training & ($\varepsilon,\delta$)-DP & Clips gradients and & Deep learning,\\ 
\textbf{Perturbation} & via noisy gradient & (tracked via & adds noise per step & Nonconvex models, \\ 
\textbf{(DP-SGD)} & updates & R\'enyi \dpns) &  & Large-scale ML \\ 
\midrule

\textbf{Stateful DP} & Maintains and  & ($\varepsilon,\delta$)-\dps & Reduces total noise & Federated learning, \\ 
\textbf{(DP-FTRL)}& perturbs a  & ($\varepsilon,\delta$)-\dps & via correlated & Small or full-batch \\ 
& cumulative sum of & 
&  updates, &  settings \\
& gradients  &  &  Avoids subsampling &  \\ \midrule

\textbf{DP via } & Privacy test filters  &$(\varepsilon,\delta)$-DP &Separates privacy & Synthetic data \\ 
\textbf{Plausible} & synthetic data from  && enforcement from   & release  \\ 
\textbf{Deniability} & a utility-focused & & data generation &  \\ 
& generative model & & &  \\ 
\bottomrule
\end{tabular}}
\end{table}

\medskip
\noindent
\textbf{Randomized Response} (RR) by \cite{Warner1965} is one of the earliest techniques, where respondents randomly choose to answer truthfully or provide an alternative answer, thus protecting individual responses. It provides Local \dps guarantee for categorical or binary attributes. \medskip
    %S. L. Warner. Randomized response: A survey technique for eliminating evasive answer bias. Journal of the American Statistical Association, 60(309):63–69, 1965.

\noindent
 \textbf{Report-Noisy-Max} (RNM), described in \cite{DworkRoth2014}, adds noise (e.g., from Laplace distribution) to the object scores and then 
    selects the item with the highest (or lowest) score, ensuring that the selection is differentially private.\medskip

\noindent    
\textbf{Sparse Vector Technique} (SVT), also descriped in \cite{DworkRoth2014},
 is designed to efficiently answer a sequence of threshold queries. Instead of adding noise to every query, SVT adds noise to a private threshold and --- up to a fixed number of positive responses --- only responds to queries that exceed this threshold. This selective reporting allows for improved utility in interactive or adaptive query settings, where only a few significant results are expected. SVT plays a key role in enabling differentially private monitoring, statistical testing, and iterative algorithms. \medskip
\ignore{
The core idea of the Sparse Vector Technique (SVT) is to privately answer a sequence of threshold queries by:

Allowing "yes" answers only when a query result exceeds a noisy threshold, and

Limiting the number of such "yes" answers to control privacy loss.

It adds noise to the threshold once and to individual queries as needed, which makes it more efficient than adding noise to every answer. It’s especially useful for interactive settings where only a few results are expected to cross the threshold.
}
   
\noindent
\textbf{Sample and Aggregate} (S\&A, see, e.g., \cite{Bassily2018,Dwork2006,Nissim2007}) % 
is a general framework for achieving DP in complex or non-smooth computations. It works by dividing the dataset into multiple disjoint subsets (sampling), applying a non-private function to each subset independently, and then aggregating the results using a differentially private method. This approach transforms a potentially sensitive computation into one that can be privatized via stable aggregation. S\&A is particularly useful for extending \dps to black-box functions or models that are not easily analyzed for sensitivity.
%Bassily laajentanut niin, että käytetään saman tyyppistä julkista opetusdataa kuin PATEssa
\medskip

% \textbf{Private Multiplicative Weights}: A more advanced mechanism that iteratively updates an estimate of the data distribution while ensuring \dpns, often used when a large number of queries must be answered accurately.

 \noindent
\textbf{Private Aggregation of Teacher Ensembles} (PATE),
 developed by \cite{papernot2017semi,papernot2018scalable},  
  provides a privacy-preserving mechanism designed for machine learning. It trains multiple \emph{teacher} models on disjoint subsets of private data, and a \emph{student} model learns from the aggregated predictions of the teachers, with noise added to the aggregation to ensure \dpns. PATE is particularly well-suited for classification tasks and has been shown to offer strong empirical privacy guarantees in deep learning applications.\medskip

% ok
\noindent
\textbf{Objective and Output Perturbation} are two closely related mechanisms designed to enforce \dps in optimization-based machine learning models, particularly for empirical risk minimization (ERM).
\begin{itemize}
\item {\em Output Perturbation} (OutP), introduced by \cite{chaudhuri2011differentially} and later extended by \cite{Wu2016}, adds noise directly to the final model parameters after training on the unmodified objective function. It is simple to implement but typically requires strong convexity assumptions, convergence to global optimum, and may result in utility loss due to post-hoc noise. % Wu et.al. bold-on approach
%A notable work by Wu et al. (2016) proposed a “bolt-on” differential privacy approach for generic models: they run standard SGD to train a model and then add noise to the final weights, claiming \dps under certain assumptions about convexity and smoothness. This method is easy to implement but, in practice, tends to yield either high $\varepsilon$ (if noise is small) or poor accuracy (if noise is large) for complex models.

\item {\em Objective Perturbation} (ObjP), proposed by \cite{chaudhuri2011differentially} and further developed by \cite{pmlr-v23-kifer12} and \cite{ZhangJun2012FmRa}, in contrast, injects noise into the loss function itself by adding a random linear term before optimization.  This approach can offer better utility guarantees under certain smoothness and convexity conditions, as the noise is accounted for during the optimization process. However, also objective pertrubation mechanism require convergence to a global optimum.
%ZhangJun2012FmRa Functional mechanism
%Chaudhuri et al. (2011) introduced this method for logistic regression, showing that it can often add less noise than output perturbation because it’s baked into the optimization. The idea is related to regularization: the random term can be seen as a form of random regularization that does not overly hurt utility if done carefully. Kifer et al. (2012) and subsequent works extended this to other settings. An interesting instance is the Functional Mechanism by Zhang et al. (2012), which adds noise to the objective’s polynomial representation.

\end{itemize}
Both objective and output perturbation mechanisms provide \dps guarantees, but they differ in where the noise is introduced --- after versus within the optimization process.
\medskip

%\textbf{\dpns-Stochastic Gradient Descent} (DP-SGD)
% is a differentially private mechanism, specifically tailored for training models via stochastic gradient descent. It’s widely used in practice (e.g., in TensorFlow Privacy, Opacus {\red \cite{}}).

\noindent  
\textbf{Gradient Perturbation} methods enforce \dps by modifying the training process --- typically some form of stochastic gradient descent --- such that each update step limits the influence of individual data points. This is done by {\em clipping per-example gradients} to a fixed norm and {\em adding random noise} to the aggregated gradients before updating model parameters.

The strength of this approach lies in its generality --- it works with any model trained via gradient-based methods, including non-convex models like deep neural networks --- making it the most widely adopted technique for \dps training today. A key advantage is that it does not require convergence to a global optimum, unlike output or objective perturbation methods.

The most widely used gradient perturbation method is the \emph{Differentially Private Stochastic Gradient Descent} (DP-SGD) introduced by \cite{abadi2016deep} (see also
\cite{BassilyRaef2014PERM,ShuangSong2013Sgdw}). 
It modifies the standard SGD algorithm by clipping individual gradients to bound sensitivity and adding calibrated Gaussian noise to the aggregated gradients at each training step. The privacy loss is tracked over time using advanced accounting techniques such R\'enyi \dpns. 
   Recent enhancements to DP-SGD include {\em adaptive clipping}  (\cite{andrew2021adaptive, pichapati2019adaptive}), where the clipping norm is dynamically adjusted during training based on gradient statistics, improving utility without compromising privacy guarantees.  
   In addition, other iterative training algorithms such as Adam and Adagrad can be adapted for \dps by ensuring per-example gradients are appropriately clipped before noise is added (\cite{McMahan2018}).
   %For instance, McMahan and Andrew (2018) describe how to extend gradient perturbation methods to work with these optimizers, though care must be taken to maintain correct accounting when using accumulated gradient statistics.
   %
   DP-SGD and its variants are particularly well-suited for large-scale models, including deep neural networks, and it is a core component of many practical \dps frameworks in modern machine learning frameworks. Therefore we will consider it in more details in Section \ref{sec_DP-SGD}.\medskip    

\noindent  
\textbf{Differentially Private  Follow-The-Regularized-Leader} (DP-FTRL) by \cite{Kairouz2021} is a training algorithm using a \textbf{stateful \dps mechanism}. Instead of adding independent noise to each gradient (as in DP-SGD and related gradient perturbation methods), it maintains a running sum (or state) of gradients and adds noise only to this cumulative state. The noise is introduced in a correlated manner across training steps, carefully structured to preserve \dps while substantially reducing the total noise accumulated in the final model.%\medskip
DP-FTRL avoids the need for random subsampling (used for privacy amplification in DP-SGD) and tends to preserve more utility, especially in federated or small-batch learning scenarios where sampling may be impractical or undesirable.
\medskip

\noindent
Finally, \textbf{Differential Privacy via Plausible Deniability}, introduced by \cite{bindschaedler2017plausibledeniabilityprivacypreservingdata}, is a framework %introduced by Bindschaedler et al.\ \cite{bindschaedler2017plausibledeniabilityprivacypreservingdata} 
for privacy-preserving \emph{synthetic data generation}. Instead of directly adding noise to the model or its outputs, this approach applies a post-generation privacy test to candidate synthetic records to determine whether they can be safely released. By randomizing this test, the mechanism achieves formal $(\varepsilon,\delta)$-differential privacy guarantees. This decouples the data generation process from privacy enforcement, allowing the use of high-utility generative models while still ensuring strong privacy protection for the released synthetic data.

\ignore{
DP-SGD is a mechanized approach to training machine learning models (especially neural networks) by:

Clipping per-example gradients (to bound sensitivity),

Adding Gaussian noise to the aggregated gradients,

And tracking privacy loss using accounting methods like RDP or Moments Accountant.}
\medskip

In the next section, we turn our attention to DP in machine learning, where these mechanisms are adapted and extended to protect individual data during model training while preserving predictive performance. 

\section{Differential Privacy in Machine Learning}\label{chaML}
Modern machine learning (ML) models often handle sensitive data and can inadvertently memorize or reveal information about their training examples. This has raised concerns as attackers can exploit trained models via membership inference or extraction attacks to determine if a particular record --- for example, a patient in a medical dataset --- was part of the training data. Differential privacy (DP) enables a principled way to quantify and limit privacy leakage in ML. 
This section provides an overview of how and where DP can be integrated into the ML pipeline, with a particular focus on training-time methods. It introduces the motivation for applying DP to ML models, outlines integration points across the pipeline, and presents a detailed treatment of the widely used \emph{Differentially Private Stochastic Gradient Descent} (DP-SGD) algorithm --- its foundations, privacy guarantees, utility trade-offs, as well as practical implementation aspects. Other training-time approaches are also briefly discussed. For a more comprehensive and practice-oriented guide of how to DP-fly ML, we refer to the excellent survey by \cite{Ponomareva2023}.

\subsection{Applying DP to ML Models: Introduction}
DP is increasingly recognized as a key framework for formalizing and enforcing privacy guarantees in data analysis and ML.
%DP has emerged as the {\em de facto} standard for protecting individual privacy in data analysis and ML. 
At a high level, DP ensures that the outcome of a computation --- the trained model in our case --- is {\em insensitive to any single training example’s presence or absence}, thereby limiting what an adversary can learn about that example from the output.  However, applying DP to ML (especially to complex deep learning models) is non-trivial. Naively training a model with DP often entails a trade-off: as we strengthen privacy (smaller $\varepsilon$), model accuracy can degrade due to the noise added or other modifications necessary to enforce privacy. 
Key challenges include choosing appropriate privacy parameters, modifying training algorithms to enforce DP, tuning hyperparameters under the added noise, and sometimes even adjusting model architectures --- all while ensuring acceptable model performance.
%\medskip

Despite these challenges, the motivation to integrate DP into ML is strong. In many domains (healthcare, finance, cybersecurity, user behavior modeling, etc.), protecting training data privacy is not only ethically important but often legally required. Fortunately, recent research demonstrates that with sufficiently large training data, sufficient computation, and careful tuning, it is often possible to get models that are nearly as accurate as non-private models while providing meaningful privacy guarantees. Further, even in settings with limited data or computation, adding even a relatively weak DP guarantee can significantly mitigate worst-case privacy risks and provide value (\cite{Ponomareva2023}).

\subsection{Stages of Differential Privacy Integration}
Due to the post-processing property of DP (see Proposition \ref{postProcessing}), DP can be introduced at various stages of the ML workflows. Broadly, there are three main points of intervention: input/data level, during the training process, and output/prediction level.    Table \ref{table_where_to_apply} summarizes these stages. \bigskip  %\vspace{-2mm}%\medskip

\begin{table}[ht]
\caption{Where to Apply Differential Privacy}
\label{table_where_to_apply} 
\resizebox{\textwidth}{!}{\begin{tabular}{@{}llll@{}}
\toprule
\textbf{Stage} & \textbf{What It Protects} & \textbf{How It Is Applied} & \textbf{Use Cases} \\
\midrule
\textbf{Input / } & Individual data records & Noise is added before & Data sharing (e.g., \\
\textbf{Data Level} &  & training, e.g., via  & medical or demographic \\
 &  & synthetic data generation  & data), \\
 &  & or local DP mechanisms & Public data releases \\
\midrule
\textbf{Training Level} & Learned model  & Noise is injected during & Federated learning, \\
 & parameters & training, e.g., through & Private deep learning,\\
 & & gradient perturbation & Private deep learning,\\
&  & (DP-SGD) & Secure ML deployment \\
\midrule
\textbf{Prediction / } & Individual predictions & Noise is added to model & Private APIs,\\
\textbf{Output Level} & & outputs or query  & Interactive analytics,\\
 & & responses & Recommendation  \\
 & &  &  systems \\
\bottomrule
\end{tabular}}
\bigskip
\end{table}%\vspace{-5.5mm}

\noindent
\textbf{Input/Data Level:} If the input data is made differentially private --- for example, by applying local DP techniques such as LPLP by {\cite{Schein2018}}, or by generating differentially private synthetic data (see Section \ref{sec_synthetic_data}) --- then
any model trained on that data will also be differentially private, as will be all outputs produced by the model. Thus, input-level integration of DP ensures privacy guarantees for shared datasets and downstream uses.%\medskip
Moreover, once the privacy budget has been spent to produce the differentially private data, it does not continue to accumulate in subsequent use of that data, such as training multiple models, comparing them, or tuning hyperparameters.

This means that 
introducing DP at the input/data level offers the strongest and most comprehensive privacy guarantees.
However, input/data level is often the most challenging place to introduce DP, and training a model with local DP techniques or on truly differentially private synthetic data may severely limit utility (\cite{Perez2024,Ponomareva2023}).
%\begin{itemize}
%    \item Guarantees privacy for shared datasets
%    \item Applies DP to dataset sharing.
%    \item Challenges: High noise requirement.
%    \item Methods: Synthetic data generation (see Section \ref{sec_synthetic_data}) and local DP techniques like the {\em limited-precision local privacy} (LPLP) \cite{Shein2019}.
%\end{itemize}
%

\medskip\vspace{1mm}

\noindent
\textbf{Training Level:}
Applying DP during model training is the most common and effective approach for achieving DP in ML. If the training algorithm is differentially private, so will be the resulting model and its outputs. %\medskip

Depending on which parts of the data are considered sensitive, modifications to the training process can support different types of protection. In {\em label-only protection}~(see, e.g., \cite{Chaudhuri2011LabelDP,pmlr-v151-esfandiari22a,esmaeili2021antipodes,Ghazi2022LabelDP}), only the labels are treated as private, while features are assumed to be public. This setting arises, for instance, in scenarios such as online advertising, recommendation systems, and user surveys and analytics.
%\medskip

By contrast, {\em full training model protection} --- often referred to as {\em DP-training} --- treats  both features and labels as sensitive. The goal in DP-training is to modify the training algorithm such that the final model is differentially private with respect to its entire training set. 
\ignore{
\begin{itemize}
    \item {\em Label-only protection} \cite{Chaudhuri2011LabelDP,pmlr-v151-esfandiari22a,esmaeili2021antipodes,Ghazi2022LabelDP}, considers only the labels of the data to be sensitive. This is the case, for instance, in {\red nice example}.
\item Full training model protection often termed {\em DP-training}.
\end{itemize}}
This ensures that the learned parameters do not reveal information about any individual data point. DP-training is typically implemented by injecting noise into the optimization process, for instance, into the gradients. The most prominent method used in DP-training is DP-SGD, which we discuss in detail in Section~\ref{sec_DP-SGD}.

\bigskip

\noindent
\textbf{Output/Prediction Level:}  
Applying DP at the model's output or prediction may suffice when the model itself does not need to be released (\cite{DworkFeldman2018}). One example is a privacy-preserving prediction API: given a trained model, each time it is queried with input data, the prediction is perturbed with noise to satisfy DP, ensuring that the output does not reveal too much about any individual training example. Another example is the \emph{sample and aggregate} framework (\cite{Bassily2018,Dwork2006,Nissim2007}),
where multiple models (or model instances trained on different data partitions) are aggregated with added noise to provide DP-guaranteed responses to queries.%\medskip

Privacy preserving prediction methods are particularly useful when only inference-time privacy is required. However,  \cite{vandermaaten2020tradeoffsprivateprediction} argue that DP-training typically offers a better privacy-accuracy trade-off than output-level DP, especially when a large inference budget is needed or when the training dataset is large. Moreover, DP-training simultaneously provides stronger access guarantees by protecting also the model parameters.
%\medskip

In summary, integrating DP in ML can happen at multiple levels, but DP-training is by far the most prominent approach. Therefore, the rest of this section focuses on methods for incorporating DP into the training process, with particular emphasis on the standard DP-SGD algorithm. %\medskip
 
\subsection{Differentially Private Stochastic Gradient Descent}
\label{sec_DP-SGD} 
One of the most influential developments in private ML is the adaptation of stochastic gradient descent to satisfy DP. DP-SGD refers to a family of training algorithms based on SGD that incorporate noise to provide DP guarantees. 
The approach was first outlined for simple models by 
\cite{ShuangSong2013Sgdw} and refined in a seminal work by  
\cite{abadi2016deep}, who applied it to deep neural networks together with a technique called the \emph{moments accountant} for tight privacy loss tracking. 
%\smallskip

DP-SGD (and its variants) has since become the workhorse of private deep learning, being implemented in many DP libraries --- for instance, in Tensor Flow Privacy\footnote{\url{https://www.tensorflow.org/responsible_ai/privacy/ }}, Diffprivlib\footnote{\url{https://github.com/IBM/differential-privacy-library }}, and PyTorch Opacus\footnote{\url{https://opacus.ai/ }} --- and used in most state-of-the-art results on DP model training {(see, e.g., \cite{abadi2016deep,andrew2021adaptive,pichapati2019adaptive})}. 
Next, we recall the basic idea and properties of DP-SGD. %\medskip

\subsubsection{Core Idea} 
A standard (non-private) SGD update computes the gradient of the loss on a batch of training examples, and takes a step in that direction. DP-SGD introduces two key modifications to make this process differentially private:

\begin{enumerate}
    \item \textbf{Gradient Clipping:} For each individual training example in a batch, the gradient vector is clipped to a maximum length $C$. That is, if $g_i$ is the gradient for a record $i$, we set
    $$
    g_i \leftarrow \frac{g_i}{\max(1,\norm{g_i}_2/ C)}.
    $$
This ensures that no single example's gradient has an outsized influence on the update. In other words, it limits the sensitivity of the gradient computation.
 
 \item \textbf{Noise Addition:} After computing and clipping per-example gradients, we average them to get the batch gradient as usual, then add random noise to this average. Typically one uses Gaussian noise:
 $$
 \tilde{g}  \leftarrow  \frac{1}{B}\sum_{i =1}^B g_i + \mathcal{N}(0,\sigma^2 C^2 I), 
 $$ 
 where $B$ is the batch size, $\sigma$ is the noise level (see, eq. 
 \eqref{sigma_in_gaussian}), and $I$ is the identity matrix.
 This noisy gradient $\tilde{g}$  is then used to update the model parameters. The noise proportional to the clipping norm $C$ masks the contribution of any single example to the batch gradient.
\end{enumerate}
By repeating these steps for each iteration of SGD, every update satisfies DP for the batch, and with careful accounting (to be discussed next), the whole training process can be shown to satisfy approximate DP with parameters $\varepsilon$ and $\delta$ that depend on the noise level, batch size, and number of iterations. 
Figure~\ref{fig:DP-SGD} illustrates the main idea behind the DP‑SGD algorithm, including gradient clipping, noise addition, and basic privacy accounting. In the figure, the training data is first divided into mini‑batches, typically sampled using Poisson subsampling (i.e., each record is included independently with a fixed probability per iteration, \cite{abadi2016deep}). For each of the $T$ training iterations, individual gradients are computed for the current batch, clipped to a fixed norm, and aggregated. Gaussian noise is then added to the sum before updating model parameters. A privacy accountant tracks the cumulative privacy loss over the course of training. In practice, other sampling strategies --- such as uniform sampling without replacement --- are also used, with the privacy accounting adapted accordingly (see, e.g., TensorFlow Privacy and Opacus).
\bigskip

\begin{figure}[hbt!] 
    \centering 
    \includegraphics[width=0.99\linewidth]{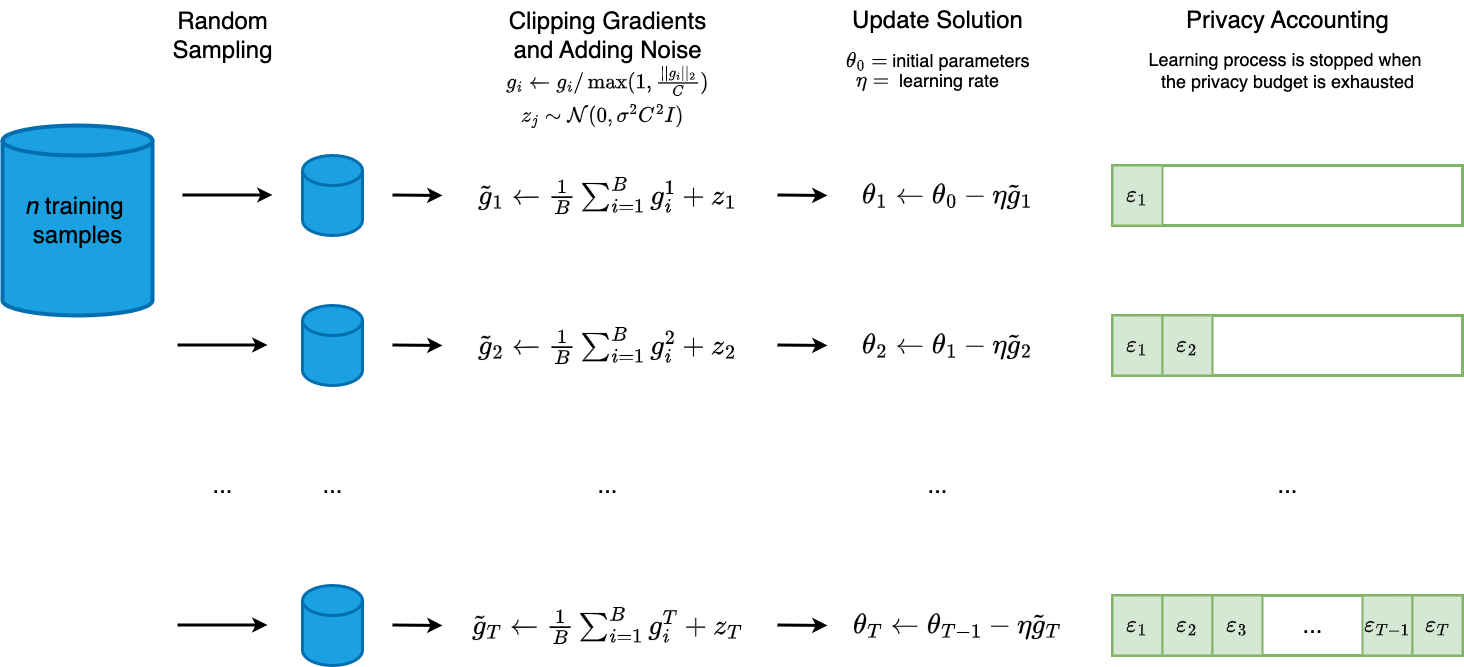}
    \caption{
    {Main Idea of the DP‑SGD Algorithm.} 
    }
    \label{fig:DP-SGD}
\end{figure}
 
\subsubsection{Privacy Accounting}\label{sec_Accounting}
 When training models with DP-SGD, we need to track the total privacy loss across many iterations. The basic rules for combining privacy guarantees --- given in Proposition~\ref{Prop_SequentialComposition} and illustrated in Figure~\ref{fig:DP-SGD} --- and even the stronger composition theorems (Theorem \ref{thm:advanced_composition}) introduced by \cite{DworkRoth2014}, and later refined by  \cite{KairouzPeter2015TCTf}, tend to be loose when applied to hundreds or thousands of training iterations. To obtain much tighter --- and more practical --- privacy guarantees, \cite{abadi2016deep} introduced the \emph{moments accountant} method. Rather than bounding the worst-case privacy loss, this approach tracks how the privacy loss accumulates statistically over the course of training.%\medskip 
\vspace{3mm}

\noindent
\textbf{R\'enyi Differential Privacy:} 
At the heart of the moments accountant is R\'enyi DP introduced by \cite{RDP2017}. R\'enyi DP measures how different the outputs of a mechanism are on neighboring datasets, using a smoother and more flexible tool than traditional DP: the {\em R\'enyi divergence}.\vspace{-2.5mm}

\begin{definition}[\textbf{R\'enyi Divergence}, \cite{RDP2017}]
For two probability distributions $P$ and $Q$, the R\'enyi  divergence of order $\alpha > 1$ is defined as:
$$
{D_\alpha(P||Q)}=\frac{1}{1-\alpha}\log \mathbb{E}_{x\sim Q}\left[\left(\frac{P(x)}{Q(x)}\right)^\alpha\right].
$$% lisää taulukkoon
\end{definition}\vspace{-2mm}

\noindent
This quantity tells us how distinguishable $P$ and $Q$ are from one another, with higher values meaning they are easier to tell apart. \vspace{-2.5mm} %\smallskip
 
\begin{definition}[\textbf{($\boldsymbol{\alpha}$,$\boldsymbol{\varepsilon}$)-R\'enyi Differential Privacy}, \cite{RDP2017}]
 A randomized mechanism $\mathcal{A}:\mathcal{D}\rightarrow R$ satisfies ($\alpha$,$\varepsilon$)-R\'enyi DP if, for
any two neighboring datasets $D \sim D' \subset \mathcal{D}$, the following holds:
$$
D_\alpha(\mathcal{A}(D)||\mathcal{A}(D')) \leq \varepsilon.
$$
\end{definition}\vspace{-2mm}

\noindent
This definition gives us a flexible privacy guarantee that composes additively over multiple steps making it much easier to keep track of cumulative loss.\vspace{-1mm}%\medskip

\bigskip

\noindent
\textbf{Moments Accountant Method:} 
The {moments accountant} uses R\'enyi DP to track how privacy loss builds up over time. It focuses on the log moments of the privacy loss, which
%compose linearly \cite{Adabi2016}.
grow roughly linearly with the number of training iterations (assuming the mechanism is stable). 
This allows us to {\em sum up the moments}, and then convert them into a final ($\varepsilon$,$\delta$)-DP guarantee at the end of training.
%\medskip

The conversion is based on a {\em tail bound}:
\begin{align}
    \label{tailBound}
\delta = \min_{\lambda} \exp(\alpha_{\mathcal{M}}(\lambda) - \lambda \varepsilon),
\end{align}
where 
%$\alpha_{\mathcal{M}}(\lambda)$ is the log of the expected value of the exponential of the privacy loss at moment $\lambda$. Intuitively, this can be interpreted as asking: “What’s the probability that the cumulative privacy loss exceeds a certain threshold?”
\(\alpha_{\mathcal{M}}(\lambda)\) represents the log moment of the privacy loss at order \(\lambda\); that is,
\[
\alpha_{\mathcal{M}}(\lambda) = \log \mathbb{E}_{o \sim \mathcal{M}(D)} \left[ \exp\left( \lambda \cdot c(o; \mathcal{M}, D, D') \right) \right],
\]
where \(c(o; \mathcal{M}, D, D')\) is the privacy loss random variable for output \(o\) under datasets \(D\) and \(D'\).
Intuitively, the tail bound captures the probability that the cumulative privacy loss exceeds a specified threshold.%\medskip

In DP-SGD, each gradient is clipped to a fixed norm \( C \), then noise is added from a Gaussian distribution with scale \( \sigma \). The noise ensures privacy, while the clipping ensures bounded sensitivity.
\newpage

To analyze the total privacy loss, we follow a three-step process:\enlargethispage{\baselineskip}
\begin{enumerate}
  \item \textbf{Compute the moment of the privacy loss} for each training iteration using known formulas (e.g., for subsampled Gaussian mechanisms).%\vspace{-2mm}
  \item \textbf{Sum the moments} across all training iterations.%\vspace{-2mm}
  \item \textbf{Convert the total moment} into an \((\varepsilon, \delta)\)-DP guarantee using the tail bound~\eqref{tailBound}.
\end{enumerate}

\cite{abadi2016deep} gives a commonly used approximation for the moment. That is,
%One commonly used approximation for the moment is \cite{abadi2016deep}:
\[
\alpha(\lambda) \leq \frac{q^2 \lambda (\lambda + 1)}{(1 - q)\sigma^2} + \mathcal{O}\left( \frac{q^3}{\sigma^3} \right),
\]
where \( q = B/n \) is the sampling rate (i.e., batch size divided by dataset size), and \( \lambda \) is the moment order.
This approach allows us to tune the noise level \( \sigma \) and the batch size $B$ to stay within a desired privacy budget --- while still training a useful model.\medskip
 
\begin{mdframed} %\vspace{-5mm}
\begin{example}\textbf{Comparing Privacy Loss Under Different Accounting Methods} (\cite{abadi2016deep})\medskip
 \label{example_moments}

\noindent
Consider a DP-SGD training setup with the following parameters:
\begin{itemize}
    \item {\small Sampling ratio $q=0.01$ (i.e., each batch is 1\% of the dataset),}
    \item {\small Noise level $\sigma = 4$,}
    \item {\small Privacy parameter $\delta = 10^{-5}$,}
    \item {\small Varying number of training epochs $E$,}
    \item {\small For each epoch, the number of training steps is $T=E/q=100E$.}
\end{itemize}
We compare two methods for estimating total privacy loss $\varepsilon$ as a function of the number of training epochs~$E$:
\begin{itemize}
    \item {\small
The \textbf{strong composition theorem} (Theorem \ref{thm:advanced_composition}), which adds privacy loss more conservatively},

\item {\small The \textbf{moments accountant}, which provides tighter bounds through careful tracking of the privacy loss distribution.}
\end{itemize} %\medskip

\noindent
Figure \ref{fig:exampleMom} shows that the moments accountant yields significantly tighter estimates. For example:

\begin{itemize}
    \item {\small 
At $E=100$, strong composition gives $\varepsilon=9.34$, while moments accountant gives only $\varepsilon = 1.26$.}

\item {\small At $E=400$, strong composition gives $\varepsilon=24.22$, compared to just $\varepsilon = 2.55$ with the moments accountant.}
\end{itemize}

\noindent
This highlights the practical importance of advanced accounting techniques for deploying DP in deep learning. 
\end{example}
\end{mdframed}\bigskip

\begin{figure}[hbt!]
    \centering
%\begin{tcolorbox}
    \includegraphics[width=0.7\linewidth]{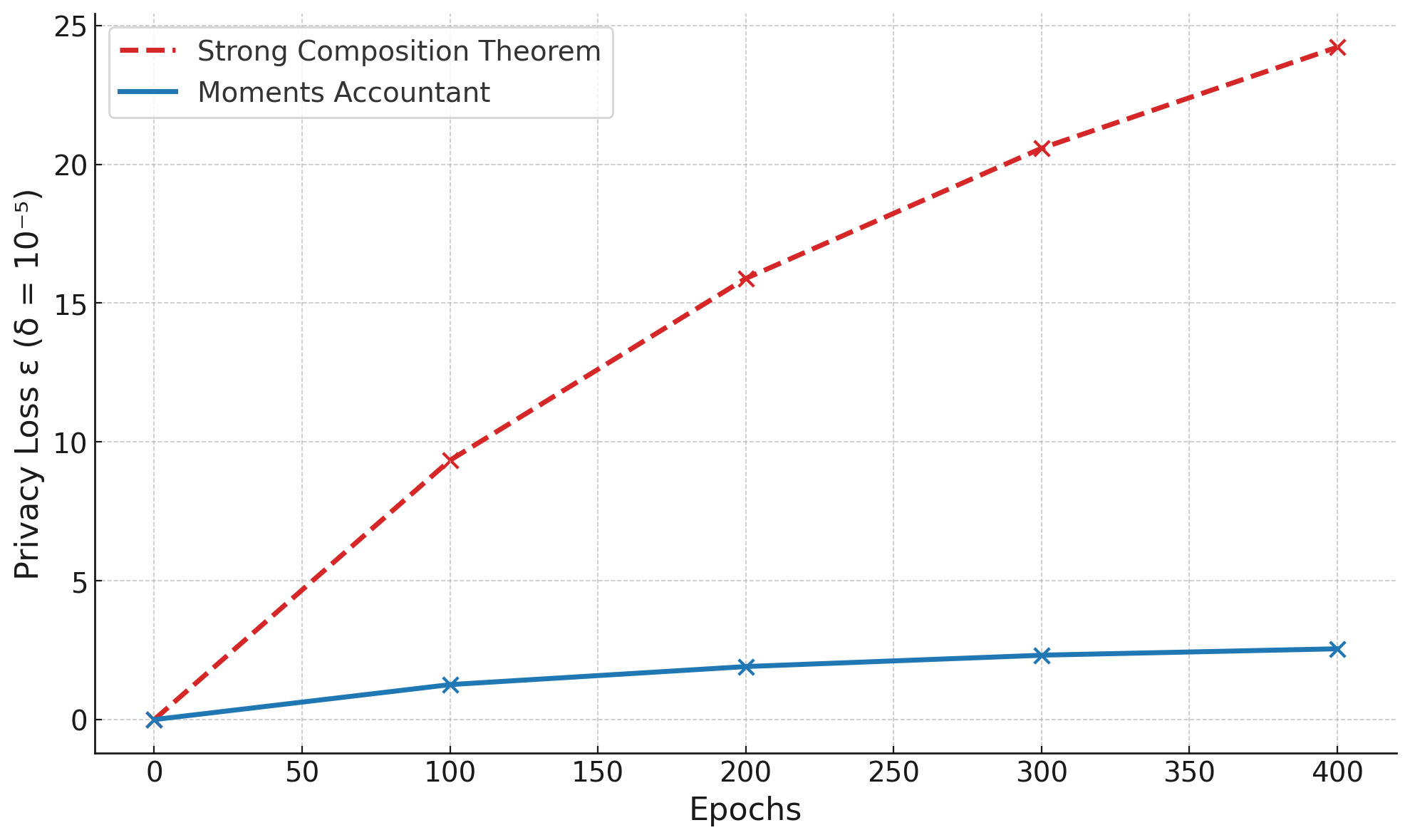}
%    \includegraphics[angle=270,width=1\linewidth]{DP_T1.png}
%\end{tcolorbox}
    \caption{Comparing Privacy Loss under Different Accounting Methods in Example \ref{example_moments}.} % better figure is needed
    \label{fig:exampleMom}
\end{figure} 

\bigskip

\newpage
\noindent
\textbf{Recent Advances in Privacy Accounting:} A number of new techniques have improved the accuracy of privacy accounting beyond the original moments accountant and standard Rényi DP-to-($\varepsilon$,$\delta$)-DP conversions. Notably, Asoodeh et al.\ \cite{AsoodehShahab2020ABBG} proposed a tightest known conversion from Rényi DP to 
 ($\varepsilon$,$\delta$)-DP, which achieves provably optimal bounds for a given R\'enyi DP guarantee. This conversion refines the privacy guarantees obtained from DP-SGD and helps avoid overestimating the cumulative privacy loss. %\medskip

Table~\ref{tab:dp-training-bounds}, adapted from \cite{Ponomareva2023}, summarizes the evolution of privacy bounds under various moments accounting schemes. It assumes that each of $T$ iterations of DP-SGD achieves the same $(\mathcal{O}(q(e^{\varepsilon_s} - 1)), \mathcal{O}(q\delta))$ guarantee, where $q=B/n$ is a sampling ratio and $\varepsilon_s = \sqrt{2 \log 1.25} / \sigma$. %\vspace{-1mm}

\begin{table}[ht]
\caption{Evolution of DP-Training Bounds at a Glance (\cite{Ponomareva2023})}
\label{tab:dp-training-bounds}
\resizebox{\textwidth}{!}{\begin{tabular}{@{}lll@{}}
\toprule
\textbf{DP-Type} & \textbf{DP-Training Bound} & \textbf{Comments} \\
\midrule
$(\varepsilon, \delta)$ & $\left(\mathcal{O}(q(e^{\varepsilon_s} - 1)T), \mathcal{O}(q\delta T)\right)$ & Straightforward application of composition, \\
 &  &  Very loose bounds \\
\midrule
$(\varepsilon, \delta)$ & $\left(\mathcal{O}(q(e^{\varepsilon_s} - 1)\sqrt{T \log(1/\delta)}), \mathcal{O}(q\delta T)\right)$ & Application of strong composition by \\
 & & \cite{DworkRoth2014} \\
\midrule
Rényi & $\left(\mathcal{O}(q(e^{\varepsilon_s} - 1)\sqrt{T}), \delta\right)$ & Conversion from R\'enyi DP to $(\varepsilon, \delta)$-DP by \\
 &  & by \cite{abadi2016deep} (Theorem 2) \\
\midrule
Rényi & $\left(\mathcal{O}(q(e^{\varepsilon_s} - 1)T^k), \delta\right)$, & Conversion from R\'enyi DP to $(\varepsilon, \delta)$-DP by\\
 &$k$ slightly less than $1/2$ &  \cite{AsoodehShahab2020ABBG} (Lemma 1)\\
\bottomrule
\end{tabular}}
\end{table}\medskip

\ignore{
\begin{table}[ht]
\caption{Evolution of DP-Training Bounds at a Glance (\cite{Ponomareva2023})}
\label{tab:dp-training-bounds}
\begin{tabular}{@{}lll@{}}
\toprule
\textbf{DP-Type} & \textbf{DP-Training Bound} & \textbf{Comments} \\
\midrule
$(\varepsilon, \delta)$ & $\left(\mathcal{O}(q(e^{\varepsilon_s} - 1)T), \mathcal{O}(q\delta T)\right)$ & Straightforward application of composition, \\
 &  &  Very loose bounds \\
\\
$(\varepsilon, \delta)$ & $\left(\mathcal{O}(q(e^{\varepsilon_s} - 1)\sqrt{T \log(1/\delta)}), \mathcal{O}(q\delta T)\right)$ & Application of strong composition by \\
 & & \cite{DworkRoth2014} \\
\\
Rényi & $\left(\mathcal{O}(q(e^{\varepsilon_s} - 1)\sqrt{T}), \delta\right)$ & Conversion from R\'enyi DP to $(\varepsilon, \delta)$-DP by \\
 &  & by \cite{abadi2016deep} (Theorem 2) \\
\\
Rényi & $\left(\mathcal{O}(q(e^{\varepsilon_s} - 1)T^k), \delta\right)$, & Conversion from R\'enyi DP to $(\varepsilon, \delta)$-DP by\\
 &$k$ slightly less than $1/2$ &  \cite{AsoodehShahab2020ABBG} (Lemma 1)\\
\bottomrule
\end{tabular}\vspace{-5mm}
\end{table}}

As an alternative, \cite{DongJinshuo2022Gdp} proposed \emph{Gaussian DP} (GDP) as a more natural and analytically tight framework for tracking cumulative privacy loss, especially in iterative mechanisms like DP-SGD. Instead of bounding the privacy loss via $(\varepsilon,\delta)$-DP directly, GDP expresses it through a hypothesis testing interpretation using the $\mu$ parameter. This framework admits a central limit theorem and composes privacy guarantees more tightly than traditional methods. In practice, GDP-based accounting yields more accurate estimates of cumulative privacy loss and can be translated into $(\varepsilon,\delta)$ bounds for reporting, offering a practical and theoretically grounded alternative to moments-based accounting.%\medskip

In a parallel line of work, \cite{KoskelaAntti2019CTDP} introduced privacy accounting based on the full distribution of the privacy loss, rather than bounding its moments. This approach --- known as \emph{privacy loss distribution (PLD) accounting} --- can yield significantly tighter privacy bounds than R\'enyi DP-based methods, especially when combined with subsampling and the Gaussian mechanism. 
PLD accounting tracks the exact behavior of the privacy loss over many iterations, enabling more accurate privacy guarantees without sacrificing utility. 
Further, \cite{KoskelaAntti2022IPAw} extended the approach to support \emph{individual privacy accounting}, which enables bounding DP loss individually for each participant involved in the analysis. 

\subsubsection{Convergence and Utility}
A natural question is how DP-SGD affects model convergence and performance. From a theoretical perspective, the addition of noise and the clipping of gradients introduce bias and variance into the optimization process. Next we study the convergence and utility aspects of DP-SGD.
\bigskip

\noindent
\textbf{Convergence:}
For convex optimization problems, {\cite{BassilyRaef2020SoSG,BassilyRaef2019PSCO,BassilyRaef2014PERM}}  
have shown that DP-SGD can achieve {\em optimal} or {\em near-optimal convergence rates}, up to multiplicative factors that depend on the privacy parameters. In particular, for convex and Lipschitz continuous loss functions, the {\em excess empirical risk} $$R_{ERM}(\theta)= \mathcal{L}(\theta;D)- \min_\theta \mathcal{L}(\theta;D) $$ 
can be bounded by $\bigO(\frac{\sqrt{p}}{\varepsilon n})$ (ignoring logarithmic factors), where $p$ is the number of parameters, $\theta \in \R^p$ denotes the model weights, $\mathcal{L}(\theta; D)$ is the empirical loss on dataset $D$, and $n$ is the number of training examples.%\medskip

This result indicates that the error grows with the model’s dimensionality $p$ and scales inversely with the privacy parameter $\varepsilon$, quantifying the cost of privacy in terms of sample complexity. For strongly convex loss functions, even faster rates of $\bigO(\frac{p}{\varepsilon^2 n^2})$ are achievable. These bounds are essentially tight and show that, as the dataset size $n$ increases (for fixed $p$ and $\varepsilon$), the impact of privacy on performance can become negligible.%\medskip

In contrast, theoretical guarantees are more limited in the nonconvex setting --- common in deep learning --- where DP-SGD is not guaranteed to converge to a global or even local minimum. However, this limitation is not unique to DP-SGD; it applies to non-private SGD as well. In both cases, convergence guarantees typically focus on reaching a stationary point --- a point where the gradient vanishes --- under standard smoothness assumptions. For DP-SGD, achieving such convergence may require additional assumptions, such as the gradient distributions being symmetric (\cite{10.5555/3495724.3496879}) or having bounded heavy tails (\cite{Das2023}). A recent study by \cite{BassilyRaef2021DPSO} establish new lower bounds for differentially private nonconvex stochastic optimization and show that standard algorithms such as DP-SGD with Gaussian noise match these bounds up to polylogarithmic factors. This is the first result to demonstrate that DP-SGD is (nearly) optimal for finding stationary points in nonconvex settings under DP.%\medskip

%Empirically, DP-SGD has been shown to successfully train large models to competitive accuracy, albeit often with some degradation in final performance compared to non-private training {\red \cite{good citation}}. {\red For example, models may require more epochs to reach comparable accuracy, and the added noise can lead to convergence to slightly suboptimal solutions.}\medskip
% tähän voisi laittaa konkreettisia esimerkkejä 
\medskip

\noindent
\textbf{Utility:} Two key factors that influence the utility of DP-SGD in practice are the noise scale $\sigma$ and the clipping norm $C$. The noise scale $\sigma$ affects directly to the trade-off between privacy and utility. Larger values of $\sigma$ offer stronger privacy by obscuring individual gradient contributions but introduce more variance, which can slow convergence and degrade performance. Conversely, smaller values reduce noise and improve utility but weaken privacy. The choice of $\sigma$ is typically guided by the desired privacy budget $(\varepsilon, \delta)$, the number of training iterations, and the batch size, as determinated by a privacy accountant (\cite{Ponomareva2023}). In practice, $\sigma$ is often tuned alongside the clipping norm $C$ to achieve a good balance between privacy guarantees and model accuracy.%\medskip{}
% Ponomareva sanoo, että Batch size ja iteraatiot paikalleen ja sitten määritellään sigma budjetin avulla (eli ei tuota C:tä). Voisi tarkastaa nuo muut, jotka käsitteli hyperparametrien tuunausta.

The clipping norm $C$ must be carefully tuned to strike a balance between privacy and learning dynamics. If $C$ is set too low, many gradients will be excessively clipped, introducing bias and degrading model accuracy. Conversely, if $C$ is too high, individual examples can exert a stronger influence on the update, requiring larger noise to maintain the same privacy guarantee --- thereby increasing variance and also reducing accuracy. \cite{Ponomareva2023} recommend running a {sweep over different values of the clipping norm $C$} and selecting the smallest value for which model utility degrades only slightly compared to the utility of a non-private model. 
Alternatively, rather than fixing the clipping norm a priori, one can use {\em adaptive clipping}, as proposed by \cite{andrew2021adaptive}, which adjusts the clipping threshold dynamically during training based on the observed gradient norms. Another approach is {\em per-layer clipping}, introduced by \cite{McMahan2018b,McMahan2018}, where separate clipping norms are applied to each layer to account for differences in gradient scale across the network. These techniques aim to improve utility by tailoring the clipping strategy to the training dynamics.
%\medskip

\ignore{
The noise scale $\sigma$ is usually chosen based on a target privacy budget $\varepsilon$: a higher $\sigma$ (more noise) means stronger privacy (smaller $\varepsilon$) at the cost of accuracy. {\red
There is a point of diminishing returns --- beyond a certain noise level, the model won’t train effectively. Research has shown that for a fixed dataset size, increasing batch size can be a more "noise-efficient" way to improve privacy-utility trade-off than simply increasing $\sigma$.}} 

In summary, DP-SGD can converge to good solutions, but careful hyperparameter tuning is critical. In addition to the noise scale and clipping norm, other hyperparameters also directly affect the privacy and utility of DP-SGD. Tuning of these parameters will be discussed in the next section.

\subsubsection{Hyperparameter Tuning under DP}\label{sec_hyperparameters}
Hyperparameter tuning plays a crucial role in obtaining good performance for DP models. However, when tuning is performed using sensitive data, it contributes to the overall privacy loss and must be carefully managed and accounted for. This section summarizes strategies and best practices for effective hyperparameter tuning in the DP setting, for more details see, for example, \cite{Ponomareva2023}, \cite{kurakin2022trainingimagenetscaledifferential}, \cite{li2022largelanguagemodelsstrong}, and \cite{papernot2022hyperparametertuningrenyidifferential}.

\bigskip

\noindent
\textbf{Important Hyperparameters and Tuning Strategy:}
The main DP-specific hyperparameters in DP-SGD are the gradient clipping norm $C$ and the noise multiplier $\sigma$, but others --- such as learning rate, batch size, and number of training epochs --- also strongly influence utility. %\cite{kurakin2022trainingimagenetscaledifferential,li2022largelanguagemodelsstrong,Ponomareva2023}. \medskip

\cite{Ponomareva2023} recommend the following practical tuning procedure:
\begin{enumerate}
\item \textbf{Use non-private training to initialize.} Start by selecting hyperparameters such as learning rate, batch size and number of epochs that perform well in a non-private setting.
\item \textbf{Search the clipping norm $\boldsymbol{C}$.} With the other parameters fixed, tune the clipping norm $C$ with $\sigma=0$ (i.e., without noise), identifying the smallest $C$ that only slightly degrades utility.
\item \textbf{Determine $\boldsymbol{\sigma}$.} Once $C$ is chosen, compute the noise multiplier $\sigma$ that satisfies the target privacy budget $(\varepsilon, \delta)$ using a privacy accountant.
    \item \textbf{Tune learning rate.} With $C$ and $\sigma$ fixed, perform a sweep over learning rates to maximize utility. Optionally, refine $C$ further if performance suggests a better trade-off.
\end{enumerate}
This approach helps isolate the contribution of each parameter to utility and keeps tuning efficient.

\bigskip

\noindent
\textbf{Reducing Privacy Cost and Privacy Accounting for Tuning:}
Each tuning run on private data consumes part of the privacy budget. The ideal approach to reduce tuning cost is to use a \emph{separate hold-out dataset} or \emph{public data} for tuning, but that is not always feasible. Therefore, several accounting strategies and hyperparameter selection mechanisms have been proposed to mitigate the tuning cost:

\begin{itemize}
\item \textbf{Sequential Composition:} The simplest approach  (see Proposition \ref{Prop_SequentialComposition}) --- add up the privacy cost of each tuning run. This method is straightforward but can be overly pessimistic, especially when many trials are conducted.

\item \textbf{Advanced Composition, R\'enyi DP, and PLD Accounting:} These methods provide tighter bounds by considering the statistical nature of privacy loss accumulation. In particular, PLD accounting, introduced by \cite{KoskelaAntti2019CTDP}, tracks the full privacy loss distribution and often yields better bounds than R\'enyi DP, especially under subsampling and Gaussian noise.

\item \textbf{Hyperparameter Selection via the Exponential Mechanism:} In this selection method $M$ models are trained using different hyperparameter configurations sampled from a pool of $K$ options, and the best-performing one is chosen using the exponential mechanism. This reduces privacy cost by focusing the budget on the final selection rather than all trials, but may return a model that is not strictly optimal (\cite{Ponomareva2023}).

\item \textbf{Hyperparameter Selection via Randomized Number of Trials:} The number of tuning trials $M$ is itself a random variable drawn from a distribution like Poisson or truncated negative binomial. This clever trick allows for privacy costs that grow only logarithmically with the number of tuning runs — significantly improving the privacy-utility trade-off in practice  (\cite{papernot2022hyperparametertuningrenyidifferential}).
\end{itemize}

\noindent
The last two techniques are especially useful when only the best model is released, as they allow for more exploration of the hyperparameter space without linearly increasing the privacy cost.

Regardless of the tuning strategy or accounting method used, it is essential to clearly document which parts of the pipeline accessed private data and how privacy was accounted for. This includes specifying whether tuning was included in the privacy budget, which mechanisms were applied, and any assumptions made --- such as subsampling or public data usage --- that affect the interpretation of the privacy guarantees.

\subsection{Other Approaches}
In addition to DP-SGD, a range of alternative methods have been proposed for training  ML models with DP. These approaches differ primarily in where the noise is introduced during the training process (see Example \ref{ex_perturpation_methods}). Broadly, they fall into the following categories:

\begin{enumerate}
    \item \textbf{Output perturbation}, which adds noise to the final trained model parameters (\cite{chaudhuri2011differentially,Wu2016});
    \item \textbf{Objective perturbation}, which injects noise into the optimization objective itself (\cite{chaudhuri2011differentially,pmlr-v23-kifer12,ZhangJun2012FmRa});
    \item \textbf{Gradient perturbation}, such as modifications of  DP-SGD and other gradient based optimization methods, where noise is added to clipped gradients at each training step
(\cite{andrew2021adaptive,BassilyRaef2021DPSO,BassilyRaef2021NDPS,boenisch2024wayindividualizedprivacyassignment,McMahan2018,pichapati2019adaptive}); and
    \item \textbf{Hybrid or advanced techniques}, such as teacher-student frameworks (e.g., PATE, \cite{papernot2017semi,papernot2018scalable}) and stateful methods like DP-FTRL (\cite{Kairouz2021}).
\end{enumerate} \vspace{-0.5mm}

%In addition, Bassily and Raef \cite{BassilyRaef2021DPSO} study differentially private stochastic optimization in convex and
%non-convex, smooth and nonsmooth settings. 
\noindent
While these techniques have already been introduced in Section~\ref{sec_otherDPmechanisms}, we briefly revisit them here to highlight their complementary strengths. 

DP-SGD remains the most widely used method for training deep models, thanks to its flexibility and applicability to non-convex settings. In contrast, output and objective perturbation methods are particularly effective in convex settings with well-characterized optimization behavior. Teacher-student methods, like PATE developed by \cite{papernot2017semi,papernot2018scalable}, are attractive when public data and disjoint training subsets are available. More recent developments, including stateful methods like DP-FTRL by \cite{Kairouz2021}, provide promising alternatives that may offer improved performance in specific regimes, such as federated or small-batch learning. Additionally, advanced gradient perturbation techniques --- such as those introduced by \cite{BassilyRaef2021DPSO,BassilyRaef2021NDPS} --- extend DP to structured nonsmooth and non-convex optimization  problems. These methods leverage smoothed or variance-reduced updates to achieve tight utility bounds and efficient convergence under both $\ell_1$ and $\ell_2$ geometries\footnote{ 
$\ell_1$ and $\ell_2$ geometries refer to optimization problems constrained or regularized using $\ell_1$ or $\ell_2$ norms, respectively. $\ell_1$ geometry is common in sparse models (e.g., Lasso), while $\ell_2$ corresponds to Euclidean constraints or regularization, as in ridge regression or many deep learning applications.
}. Further, IDP-SGD, proposed by \cite{boenisch2024wayindividualizedprivacyassignment}, introduces individualized privacy budgets for each data point and adapts the sampling and noise mechanisms in DP-SGD accordingly. This approach improves utility while respecting diverse user privacy preferences.

\ignore{
When training a ML model with DP, one sets a privacy
budget. This uniform budget represents an overall maximal privacy violation that
any user is willing to face by contributing their data to the training set. We argue
that this approach is limited because different users may have different privacy
expectations. Thus, setting a uniform privacy budget across all points may be
overly conservative for some users or, conversely, not sufficiently protective for
others. In this paper, we capture these preferences through individualized privacy
budgets. To demonstrate their practicality, we introduce a variant of Differentially
Private Stochastic Gradient Descent (DP-SGD) which supports such individualized
budgets. DP-SGD is the canonical approach to training models with differential
privacy. We modify its data sampling and gradient noising mechanisms to arrive at
our approach, which we call Individualized DP-SGD (IDP-SGD). Because IDPSGD
provides privacy guarantees tailored to the preferences of individual users
and their data points, we empirically find it to improve privacy-utility trade-offs.}

%\medskip

A summary of these approaches --- including their core ideas, privacy guarantees, and typical use cases --- can be found in Table~\ref{table_DPmechanisms} in Section~\ref{sec_otherDPmechanisms}.
In addition, Example \ref{ex_perturpation_methods}, adapted from  \cite{Jayaraman2019}, highlights the three main points at which noise can be injected in different perturbation methods.\medskip

\begin{mdframed}%\vspace{-5mm}
 \begin{example}\label{ex_perturpation_methods}
 \textbf{Privacy Mechanisms in Perturbation Methods} \medskip

\noindent 
The following pseudocode outlines a generic training procedure for a ML model and highlights the three main points at which noise can be injected to achieve DP. The training process is defined in terms of a loss function $\mathcal{L}(\theta,X_i, y_i)$, a regularization term $\lambda \mathcal{R}(\theta)$, model parameters $\theta$, and learning rate $\eta$. Gaussian noise terms {$\mathcal{N}_1$, $\mathcal{N}_2$, and $\mathcal{N}_3$} 
are used to represent the added privacy noise at each respective stage. 
\begin{description}
\item[\textbf{Machine Learning Model}] {\phantom{m} }

\begin{description}
\item[\textbf{Data:}] Training dataset \((X, y)\)
\item[\textbf{Result:}] Model parameters \(\theta\)
\item[\(\theta_0 \leftarrow {\rm Init(0)}\)]\vspace{1mm}

\item[{\color{red} \normalfont \#1. {\em Objective perturbation:} Add noise here}]\vspace{1mm}

\item[\(J(\theta) = \frac{1}{n} \sum_{i=1}^{n} \mathcal{L}(\theta, X_i, y_i) + \lambda \mathcal{R}(\theta) {\color{red} + \mathcal{N}_1}\)]
\item[]\vspace{-1.5mm}
\item[\textbf{for}] {\em epoch} \textbf{in} {\em epochs} \textbf{do} \smallskip \newline %\\
     {\phantom{mm} }{\color{blue} \normalfont \#2. {\em Gradient perturbation:} Add noise here} \vspace{0mm} \newline 
     {\phantom{mm} }\(\theta \leftarrow \theta - \eta (\nabla J(\theta) {\color{blue} + \mathcal{N}_2})\)
\item[\textbf{end}]
\item[]\vspace{-1.5mm}
 \item[{\color{teal} \normalfont \#3. {\em Output perturbation:} Add noise here}]
 \item[\textbf{return}] \(\theta {\color{teal} +\mathcal{N}_3}\) \medskip%\vspace{-0.3cm}
\end{description} 
\end{description} 
 
\end{example}
\end{mdframed}\medskip\medskip
 
For completeness, we note that in scenarios where model inputs are public and only labels are considered sensitive, {\em label privacy} schemes have been proposed (seee, e.g., \cite{Chaudhuri2011LabelDP,pmlr-v151-esfandiari22a,esmaeili2021antipodes,Ghazi2022LabelDP}).  
% Nämä kaikki on DPfly paperista
While promising, such methods are outside the scope of this review, which focuses on protecting the privacy of the training data more broadly. 

\subsection{Challenges with DP-Training} 
Training ML models under DP introduces unique challenges that go beyond standard model development. While DP provides strong theoretical guarantees, achieving practical utility without compromising privacy remains difficult. The key challenges --- as pointed out in
 \cite{Cummings2024Advancing} and \cite{Ponomareva2023} --- include:
\begin{itemize}
    \item \textbf{Privacy–Utility Trade-off:} Achieving strong privacy guarantees often leads to a significant reduction in model accuracy. This trade-off is particularly pronounced in complex models or tasks requiring high precision.

    \item \textbf{Computational Overhead:} DP training, especially with DP-SGD, introduces significant computational and memory overhead. Per-example gradient computation, gradient clipping, and noise addition limit hardware acceleration and increase memory consumption, often requiring smaller batch sizes or specialized implementations.

\item \textbf{Hyperparameter Sensitivity and Training Stability:} 
DP algorithms are highly sensitive to hyperparameter settings, such as learning rates, clipping norms, and noise multipliers. 
Selecting suitable values is challenging because traditional tuning procedures can consume part of the privacy budget (see Section~\ref{sec_hyperparameters}). 
{Moreover, gradient clipping introduces bias in gradient estimates, and together with noise injection can destabilize optimization, particularly in early training phases, often requiring smaller learning rates, warm-up schedules, or gradient-norm monitoring to avoid divergence and maintain stable training.}

    \item \textbf{Data Requirements:} DP methods typically require large datasets to maintain utility. In scenarios with limited data, the added noise for privacy can overwhelm the signal, degrading performance.

    \item \textbf{Model Architecture Constraints:} Certain model architectures are more amenable to DP. For instance, models with bounded activations or smaller capacities tend to perform better under DP constraints, limiting the choice of architectures (see Section \ref{sec_archit} for more details). 

    \item \textbf{Implementation Challenges:} Integrating DP into existing ML pipelines is non-trivial. It requires careful accounting of privacy loss and often necessitates custom implementations or adaptations of standard algorithms.
In particular, mismatches between theoretical privacy accounting assumptions (e.g., Poisson subsampling) and the actual training implementation may invalidate reported privacy guarantees if not carefully verified.

    \item \textbf{Evaluation Difficulties:} Assessing the effectiveness of DP mechanisms is complex. Standard evaluation metrics may not capture the nuances introduced by privacy constraints, making it hard to benchmark and compare methods.
\end{itemize}

\noindent
The challenges outlined above underscore the complexity of applying DP in real-world ML systems. Yet despite these hurdles, a growing body of research and practice has demonstrated that DP-training can be both feasible and effective, especially when approached with careful engineering and domain-specific adaptations. In particular,  \cite{Papernot2021Making} suggest selecting model architectures, initializations, and hyperparameters explicitly designed from the start for effective privacy-preserving training.%\medskip
In the following section, we turn from theoretical and systemic challenges to the practical aspects of DP-training, highlighting key design choices and implementation strategies. %These include guidance on hyperparameter tuning, model selection, and evaluation procedures.
 
\subsection{Practicalities of DP-Training} 
\label{sec_archit}

Integrating DP into model training requires more than just choosing a DP algorithm and the privacy budget $\varepsilon$ and $\delta$ --- it also demands attention to how the training process interacts with privacy guarantees. We now highlight some key challenges and design decisions that can influence both the effectiveness and correctness of DP training.
\bigskip

\noindent
\textbf{Avoiding DP Violations from Model Components and Training Strategies: }
Some standard components in modern ML pipelines are incompatible with DP mechanisms like DP-SGD. For example, certain loss functions --- such as \emph{pairwise} or \emph{ranking losses} --- cannot be decomposed into per-example terms, making them incompatible with per-example gradient computation and clipping required by DP-SGD. Similarly, \emph{Batch Normalization} by \cite{Ioffe2015} computes statistics across a mini-batch and can leak information between examples. Because DP-SGD assumes per-example independence, such aggregations violate the required assumptions unless explicitly privatized (see, e.g., \cite{davody2021effectnormalizationlayersdifferentially,10.5555/3600270.3600698}). This issue can also be avoided by using alternative normalization layers such as \emph{Group Normalization} by \cite{10.1007/978-3-030-01261-8_1}, \emph{Layer Normalization} by \cite{ba2016layernormalization}, or \emph{Weight Normalization} by \cite{10.5555/3157096.3157197}, which operate independently across examples and are therefore more compatible with DP training.%\medskip 

Another common pitfall arises from auxiliary procedures like \emph{early stopping}, which --- if based on a non-private validation set --- can leak information unless properly privatized or accounted for. Similarly, caching precomputed statistics or features across the entire training set can violate privacy guarantees if the cached values reflect sensitive data

\bigskip

\noindent
\textbf{Model and Training Design for Utility under DP: } 
Beyond correctness, model architecture and training design must often be modified to preserve utility under privacy constraints. 
For example, \emph{activation functions} can influence the sensitivity of gradients to input perturbations, which directly affects how much noise must be added during DP training. \cite{papernot2020temperedsigmoidactivationsdeep} observed that unbounded functions like \emph{ReLU} can lead to gradient explosions during training, causing more aggressive gradient clipping and, consequently, higher information loss. Bounded and smoother activations --- such as \emph{tempered sigmoid} --- can better control gradient norms, reducing the impact of clipping and noise injection. However, these functions may introduce additional hyperparameters that require tuning. The effectiveness of activation functions in DP training is therefore both function-specific and architecture-dependent
%\medskip

%In addition, large embedding tables --- especially in NLP models --- can yield gradients with high variance, particularly for rare tokens. These outliers are heavily clipped, resulting in disproportionate noise and reduced accuracy. Strategies such as limiting vocabulary size, reducing embedding dimensions, or applying embedding-specific regularization may help mitigate these effects \cite{Denisov2022}.\medskip

The role of regularization in DP training is even more nuanced. 
%Nevertheless, there is no consensus on whether traditional regularization techniques improve utility in DP training.  
Some studies (e.g., \cite{davody2021effectnormalizationlayersdifferentially}) argue that regularization techniques such as \emph{dropout} or \emph{weight decay} can improve model robustness to noise. Others (e.g., \cite{anil-etal-2022-large} and \cite{de2022unlockinghighaccuracydifferentiallyprivate}) contend that DP already serves as a form of implicit regularization --- reducing overfitting --- and that additional regularization may be redundant or even counterproductive. As a result, the effectiveness of regularization techniques under DP appears context-dependent and should be evaluated empirically.

As a summary, \cite{Ponomareva2023} note that complex architectures with many parameters or deep layers may suffer more from noise. Simpler and shallower models are often more robust under DP training. While the role of regularization is debated, techniques like weight decay may help by implicitly controlling gradient magnitudes and clipping behavior in some settings. Additionally, reducing the number of output classes, using smoother loss functions, or limiting the sensitivity of evaluation metrics can improve utility under tight privacy budgets. Finally, to reduce the risk of subtle implementation errors, Ponomareva et al.\ strongly recommend relying on well-audited libraries such as \emph{Opacus}\footnote{\url{https://github.com/pytorch/opacus}} or \emph{TensorFlow Privacy}\footnote{\url{https://github.com/tensorflow/privacy}
}, which offer robust privacy accounting tools and verified implementations of DP-SGD and related mechanisms.

\section{Defining the Privacy Budget}
\label{section_privacy_budget}
In differential privacy (DP), the privacy budget 
$\varepsilon$ quantifies  the maximum privacy loss tolerated by a mechanism.
 In this section, we examine the role of 
$\varepsilon$, explore the trade-off between privacy and utility, provide practical guidelines for selecting 
$\varepsilon$, and present examples of privacy budgets used in real-world applications.
 
\subsection{The Role and Interpretation of \texorpdfstring{\(\boldsymbol{\varepsilon}\)}{e}}
%\subsection{\red The Role of \(\boldsymbol{\varepsilon}\) in Differential Privacy}
Formally, the privacy parameter $\varepsilon$ bounds how much the probability of any output can change due to the presence or absence of a single individual in the dataset (see Definition \ref{pureDP}). A smaller $\varepsilon$ --- that is, more noise added --- means stronger privacy but less accurate output; a larger $\varepsilon$ allows better utility but weaker privacy.
By definition, an $\varepsilon$ of zero would mean perfect privacy but essentially useless data, while very large $\varepsilon$ would mean almost no privacy protection. In practice, $\varepsilon$ is rarely zero or infinite; it is chosen to strike a balance between privacy and accuracy requirements. 
%\medskip

%Importantly, $\varepsilon$ provides a \emph{worst-case guarantee}: it bounds the maximum allowable privacy loss. %One way to interpret $\varepsilon$ is via the bound it places on the likelihood ratio of outputs: 
%An $\varepsilon$-DP mechanism ensures that for \emph{any two datasets differing in one individual}, and for \emph{any outcome}, the probability of that outcome can differ by at most a factor of $e^\varepsilon$. 
%This means an adversary’s confidence in whether a particular person’s data was included can increase by at most $e^\varepsilon$ under the worst circumstances. 
%For example, $\varepsilon = 1$ limits this factor to $e^1 \approx 2.718$, and $\varepsilon = 0.1$ limits it to $e^{0.1} \approx 1.105$ (very little change), whereas $\varepsilon = 10$ allows factor over $22000$, a vastly larger possible change in odds.  \medskip

It is worth noting that $\varepsilon$ provides a \emph{worst-case guarantee}: it bounds the maximum possible change in output probabilities due to the inclusion or removal of a single individual. Specifically,
an $\varepsilon$-DP mechanism ensures that for \emph{any two datasets differing in one individual}, and for \emph{any possible outcome}, the probability of that outcome can differ by at most a factor of $e^\varepsilon$. 
For example, $\varepsilon = 0.1$ limits this factor to $e^{0.1} \approx 1.1$ (a very little change), and $\varepsilon = 1$ limits it to $e^{1} \approx 2.7$ (quite a small change), whereas $\varepsilon = 10$ allows factor exceeding $22,000$, a vastly larger shift in odds, reflecting much weaker privacy.  \smallskip

\newpage
\enlargethispage{1.9\baselineskip}

\begin{mdframed}\vspace{-1mm}
\begin{example}\label{exAlice}
 \textbf{First Name Histogram with Differential Privacy} \smallskip

\noindent
Consider a statistical agency that wishes to publish the number of individuals with each first name in a confidential census dataset. Suppose the dataset contains 5,000 individuals, and the universe of possible first names consists of 10,000 entries (e.g., from a national name registry). Each individual contributes to exactly one entry in the histogram corresponding to their first name.\smallskip 

Assume, for instance, that:\vspace{-0.2cm}
\begin{itemize}
    \item {\small 400 individuals are named Alice,}\vspace{-0.2cm}
    \item {\small 300 individuals are named Bob,}\vspace{-0.2cm}
    \item {\small 1 individual is named Xanthe (a rare name),}\vspace{-0.2cm}
    \item {\small and so on.}\vspace{-0.2cm}
\end{itemize}

\noindent 
The agency aims to release a \emph{differentially private histogram} by applying the \emph{Laplace mechanism} to each count. The histogram query returns a 10,000-dimensional vector, where each component corresponds to the count of one possible first name. Since each person contributes to at most one name, the sensitivity of the query is 1. The Laplace mechanism adds independent noise to each count, drawn from the Laplace distribution with scale $1/\varepsilon$. \vspace{-0.2cm}
\bigskip

\noindent
\textbf{Interpretation of the Privacy Parameter $\boldsymbol{\varepsilon}$:} The parameter $\varepsilon$ controls the privacy-accuracy trade-off. Smaller values of $\varepsilon$ provide stronger privacy guarantees but result in larger noise, whereas larger values yield more accurate results at the cost of weaker privacy. Table \ref{tab:example_eps} shows the effect of different $\varepsilon$ values on the count for "Alice" (with true count = 400) and Xanthe (with true count = 1). These values indicate the typical magnitude of the noise added to each count.
%\footnote{Note: the "95\% Noise Range" in Table \ref{tab:example_eps}  refers to the interval in which the Laplace noise is expected to fall with 95\% probability, computed as approximately $\pm 3$ times the noise scale. In addition, negative noisy counts (e.g., the $–15$ for "Xanthe") can occur with Laplace noise. In practice, differentially private histograms typically clip negative counts to zero or apply post‑processing to ensure meaningful outputs.} \medskip

With $\varepsilon=0.1$, the noise is large enough that even if someone knows that Alice might be in the dataset, they cannot confidently tell if she is — her presence makes almost no noticeable difference. However, with $\varepsilon=10$, the count is almost exact — which is great for accuracy, but reveals much more about individuals in the dataset (poor privacy).\vspace{-0.2cm}

\bigskip

\noindent
\textbf{Common vs. Rare Names:} The same amount of noise affects common and rare names differently: For a common name like Alice (400 occurrences), Laplace noise with scale $1/\varepsilon$ only slightly distorts the count, while for a rare name like Xanthe (only 1 occurrence), the same noise constitutes a much larger relative change. This illustrates why differential privacy calibrates noise to the \emph{worst-case} sensitivity: even rare individuals receive the same level of formal privacy protection as those with common names.\vspace{-0.2cm}

\bigskip

\noindent
\textbf{Understanding $\boldsymbol{\varepsilon}$ through Output Probabilities:}
Differential privacy further guarantees that for any two datasets differing in a single individual, the probability of any given output can change by at most a factor of $e^\varepsilon$ (see Figure \ref{fig_epsilon_effect}). For example, with $\varepsilon = 1$, the maximum multiplicative change in output probabilities is about $e^1 \approx 2.7$, meaning that the presence or absence of any one individual can at most multiply the probability of any given output by 2.7 — a limited but potentially noticeable influence. With $\varepsilon = 0.1$, the bound is $e^{0.1} \approx 1.1$, implying a much smaller influence on the output. \vspace{-0.2cm}

\bigskip

\noindent
\textbf{Impact of Dataset Size on Noise and Accuracy.}
The magnitude of the noise added by the Laplace mechanism depends only on the query's global sensitivity and the privacy parameter~$\varepsilon$; it is independent of the dataset size. However, dataset size affects the relative impact of the noise: in larger datasets, the noise constitutes a smaller proportion of the true counts, resulting in more accurate estimates. In smaller datasets, the same noise can cause greater relative distortion. Thus, while privacy guarantees are independent of dataset size, utility typically improves as the dataset grows.
\end{example}
\end{mdframed}%\vspace{-5mm}

\begin{table}[ht]
\caption{Effect of Different $\varepsilon$ Values on the Differentially Private Counts for "Alice" (true count = 400) and "Xanthe" (true count = 1) in Example~\ref{exAlice}.}
\label{tab:example_eps}
\resizebox{\textwidth}{!}{ \begin{tabular}{@{}llllll@{}}
\toprule
\textbf{$\boldsymbol{\varepsilon}$} & 
\textbf{Laplace} &  
\textbf{95\% Noise} & 
\textbf{Example Noisy} & 
\textbf{Example Noisy} & 
\textbf{Privacy} \\
 & 
\textbf{Scale ($1/\boldsymbol{\varepsilon}$)} &  
\textbf{Range}\footnotemark[1] & 
\textbf{Count for Alice} & 
\textbf{Count for Xanthe} & 
\textbf{Strength} \\
& 
&  
& 
\textbf{(400)} & 
\textbf{(1)} & 
\\
\midrule
$0.1$ & $10$   & $\pm30$   & $428$ or $374$     & $21$ or $-15$\footnotemark[2]    & Very strong \\ 
$1$   & $1$    & $\pm3$    & $402$ or $397$     & $3$ or $0$        & Strong       \\ 
$2$   & $0.5$  & $\pm1.5$  & $400.9$ or $398.7$ & $2.1$ or $0.2$    & Moderate     \\ 
$10$  & $0.1$  & $\pm0.3$  & $400.1$ or $399.8$ & $1.2$ or $0.8$    & Weak         \\
\bottomrule
\multicolumn{6}{l}{$^1$ \footnotesize{The "95\% Noise Range" refers to the interval in which the Laplace noise is expected to fall with 95\% }}\\
\multicolumn{6}{l}{\footnotesize{ probability, computed as approximately $\pm 3$ times the noise scale.}}\\
\multicolumn{6}{l}{$^2$ \footnotesize{Negative noisy counts can occur with Laplace noise. In practice, differentially private histograms}}\\
\multicolumn{6}{l}{\footnotesize{  typically clip negative counts to zero or apply post‑processing to ensure meaningful outputs.}}
\end{tabular}}
\end{table}

%\footnotesize{Note: the "95\% Noise Range" in Table \ref{tab:example_eps}  refers to the interval in which the Laplace noise is expected to fall with 95\% probability, computed as approximately $\pm 3$ times the noise scale. In addition, negative noisy counts (e.g., the $–15$ for "Xanthe") can occur with Laplace noise. In practice, differentially private histograms typically clip negative counts to zero or apply post‑processing to ensure meaningful outputs.} 
%\footnote{Negative noisy counts (e.g., the –15 for "Xanthe") can occur with Laplace noise. In practice, differentially private histograms typically clip negative counts to zero or apply post‑processing to ensure meaningful outputs.}

Figure \ref{fig_epsilon_effect}
illustrates how the Laplace mechanism enforces DP. It compares the output distributions of a query with and without an individual (Alice) in the dataset. For any fixed output, the ratio of these probabilities is bounded by 
$e^\varepsilon$, demonstrating that the presence or absence of a single individual has only a limited influence on the result.\bigskip

\begin{figure}[hbt!]
    \centering 
%\begin{tcolorbox}
    \includegraphics[width=0.99\linewidth]{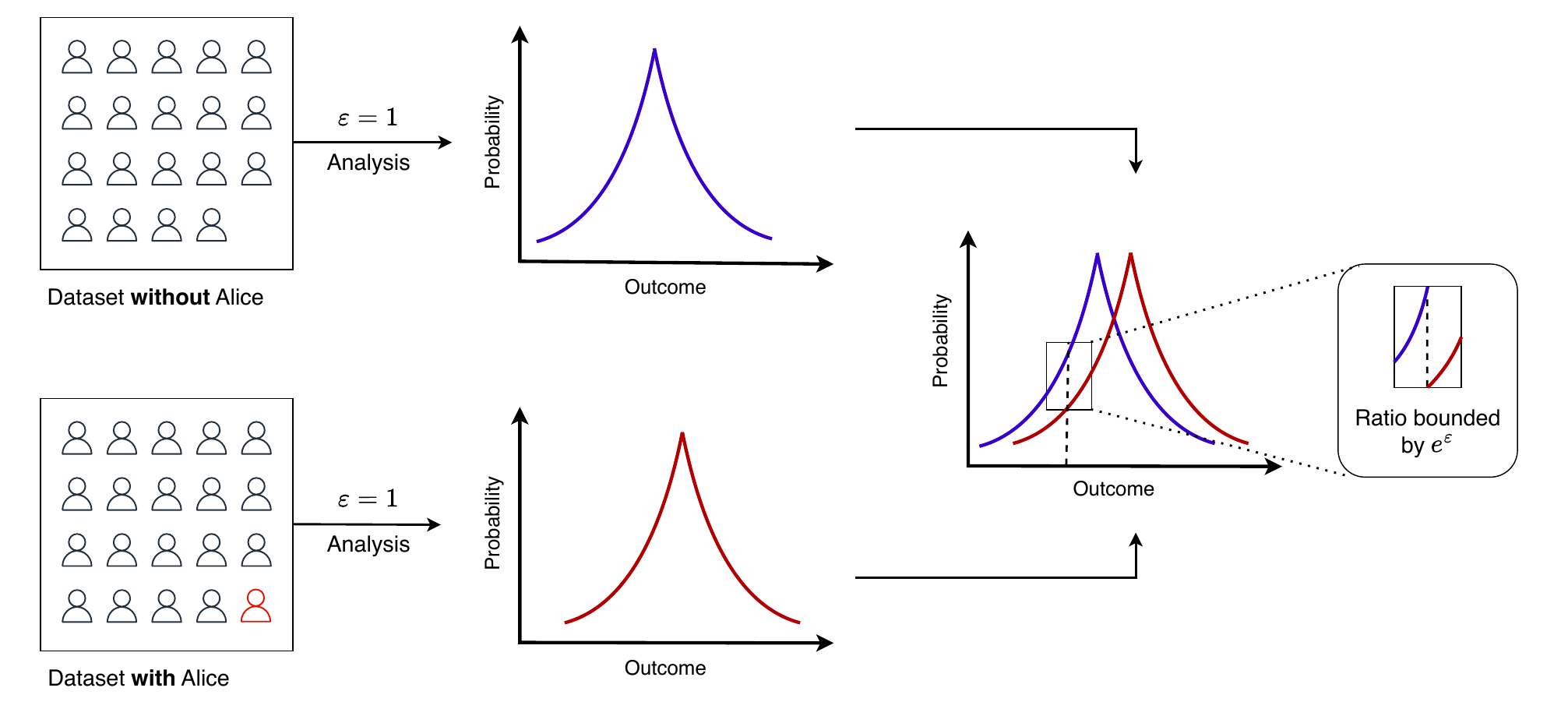}
%    \includegraphics[angle=270,width=1\linewidth]{DP_T1.png}
%\end{tcolorbox}
    \caption{Illustration of the DP Guarantee under the Laplace Mechanism (inspired by~\cite{Franzen2}).}
    %Illustration of the DP Guarantee under the Laplace Mechanism: The figure compares output distributions of a query with and without an individual (Alice) in the dataset. The ratio between the probabilities of any fixed output is bounded by $e^{\varepsilon}$, demonstrating the limited influence of any single individual's data. Figure inspired by~\cite{Franzen2}.} %
    \label{fig_epsilon_effect}
\end{figure}\enlargethispage{\baselineskip}

While $\varepsilon$ is the most visible parameter in \dpns, an equally important --- though often less explicitly discussed --- aspect is the \emph{unit of privacy} (see Section~\ref{sec_DPdefinition}). \DP definitions \ref{pureDP} and \ref{approximateDP} are based on the notion of neighboring datasets, typically assumed to differ by the data of exactly one individual. However, this intuitive assumption does not always align with real-world settings. For example, in mobility data, where each individual may contribute multiple location reports, it is unclear whether neighboring datasets should differ by a single report or by the entire set of reports from one person.
%\medskip

The choice of the unit --- whether per-record, per-user, or another level of granularity --- significantly affects both the interpretation and strength of the resulting privacy guarantees  (\cite{NIST2025}). Crucially, the same \(\varepsilon\) value can correspond to vastly different privacy levels depending on the unit of privacy used as illustrated in Table~\ref{tab:differet_unit_of_privacy}. %\medskip

\begin{table}[ht]
\caption{Two DP Guarantees with the Same $\varepsilon$ and $\delta$ Values but Different Units of Privacy}
\label{tab:differet_unit_of_privacy}
\resizebox{\textwidth}{!}{ \begin{tabular}{@{}lllll@{}}
\toprule
\textbf{Parameter} && \textbf{User‑level DP} && \textbf{Example‑level DP}  \\
\midrule
$\boldsymbol{\varepsilon}$ && $2.5$ && $2.5$  \\
\midrule
$\boldsymbol{\delta}$ && $10^{-25}$ && $10^{-25}$ \\
\midrule
\textbf{Privacy unit} && User level && Example level \\
 &&  (all records per user) &&  (single record) \\
\midrule
\textbf{Concrete example} && Patient  && Single medical visit record \\
 && (all medical visits of one patient) &&  \\
\midrule
\textbf{Practical effect} && Stronger guarantee: && Weaker guarantee: \\
 && Adversary cannot tell whether && Adversary cannot tell whether \\
 && any patient (with all their visits) && a single visit is in the dataset \\
 && is in the dataset &&  \\
\bottomrule
\end{tabular}}
\end{table}

\ignore{
\begin{table}[ht]
\caption{Two DP Guarantees with the Same $\varepsilon$ and $\delta$ Values but Different Units of Privacy}
\label{tab:differet_unit_of_privacy}
\begin{tabular}{@{}lllll@{}}
\toprule
\textbf{Parameter} && \textbf{User‑level DP} && \textbf{Example‑level DP}  \\ \midrule
$\boldsymbol{\varepsilon}$ && $2.5$ && $2.5$  \\
\\
$\boldsymbol{\delta}$ && $10^{-25}$ && $10^{-25}$ \\
\\
\textbf{Privacy unit} && User level && Example level \\
 &&  (all records per user) &&  (single record) \\
\\
\textbf{Concrete example} && Patient  && Single medical visit record \\
 && (all medical visits of one patient) &&  \\
\\
\textbf{Practical effect} && Stronger guarantee: && Weaker guarantee: \\
 && Adversary cannot tell whether && Adversary cannot tell whether \\
 && any patient (with all their visits) && a single visit is in the dataset \\
 && is in the dataset &&  \\
\bottomrule
\end{tabular}
\end{table}}

Finally, $\varepsilon$ is often paired with $\delta$, representing the probability of a rare but unbounded privacy failure. When $\varepsilon$ is small failing to meet $(\varepsilon,0)$-DP may not be catastrophic --- the mechanism may still satisfy, for example, $(2\varepsilon,0)$-DP. 
Moreover, failure to meet $(\varepsilon,0)$-DP simply means that there exists some pair of neighboring datasets $D$ and $D'$ and an output $s\in R$ for which the ratio of probabilities exceeds $e^\varepsilon$. In practice, such worst-case scenarios may involve highly contrived datasets or very unlikely outputs  — a risk that is mitigated by allowing a small $\delta$ in ($\varepsilon,\delta$)-DP. Much like in cryptography — where a weak system might leak anything from a single bit to an entire key --- a deviation from
$(\varepsilon,0)$-DP can range from negligible to severe, depending on the context. A large $\varepsilon$, while not inherently catastrophic, must be interpreted with care.

Worst-case analysis (see Example~\ref{ex_worse_case}) suggests choosing \(\delta \ll 1/n\), where \(n\) is the number of records (\cite{DworkRoth2014}).
Because of the direct relationship between 
$\varepsilon$ and $\delta$, it is generally not meaningful to directly compare two ($\varepsilon,\delta$)-DP guarantees when their $\delta$ values differ (\cite{NIST2025}). This effect is illustrated in Table~\ref{tab:differet_delta}, where the first, smaller
value of $\delta$ allows smaller chance of privacy failure.

\begin{table}[ht]
\caption{Two $(\varepsilon,\delta)$-DP guarantees with the same $\varepsilon$ and unit of privacy but different $\delta$ values. A smaller $\delta$ means a lower chance of a privacy failure.}
\label{tab:differet_delta}
\resizebox{\textwidth}{!}{ \begin{tabular}{@{}lll@{}}
\toprule
\textbf{Parameter} & \textbf{Guarantee A} & \textbf{Guarantee B}  \\
\midrule
$\boldsymbol{\varepsilon}$ & $2.5$ & $2.5$  \\
\midrule
$\boldsymbol{\delta}$ & $10^{-25}$ & $10^{-5}$ \\
\midrule
\textbf{Privacy unit} & User level & User level \\
\midrule
\textbf{Interpretation } & Catastrophic privacy failure allowed & Catastrophic privacy failure allowed\\
\textbf{of $\boldsymbol{\delta}$} & with probability at most $10^{-25}$ & with probability at most $10^{-5}$ \\
 & (extremely unlikely) & (much higher) \\
\bottomrule
\end{tabular}}
\end{table}

This difficulty of comparing $(\varepsilon, \delta)$-DP guarantees across different $\delta$ values has motivated the search for more interpretable reporting standards. 
Recent work by \cite{gomez2025varepsilondeltaconsideredharmful} argues that reporting DP guarantees as a single $(\varepsilon, \delta)$ pair is often incomplete and misleading for practitioners. They recommend Gaussian DP (GDP, \cite{DongJinshuo2022Gdp}) as a better standard for communicating DP guarantees.  GDP uses a single parameter $\mu$, which is directly comparable across algorithms and settings, composes additively like $\varepsilon$, and closely approximates the full privacy profiles of many widely used mechanisms such as DP-SGD. 
Importantly, it also removes the need for choosing an arbitrary $\delta$, simplifying reporting and interpretation.

Table \ref{tab:gdp_conversion}, adapted from \cite{gomez2025varepsilondeltaconsideredharmful},
provides a conversion table between $\mu$-GDP and $(\varepsilon, \delta)$-DP with suggested
replacements for commonly used values. 
As noted in 
\cite{gomez2025varepsilondeltaconsideredharmful}, 
the table also gives a practical illustration of the difficulty
of interpreting $(\varepsilon, \delta)$, as Gaussian mechanism with
$(\varepsilon=8, \delta=10^{-9})$ (blue text) is more private than $(\varepsilon=6, \delta=10^{-5})$ (red text).
 
\begin{table}[ht]
\caption{Values of \(\mu\) Corresponding to Common Values of \((\varepsilon, \delta)\) (\cite{gomez2025varepsilondeltaconsideredharmful})}
\label{tab:gdp_conversion}
\begin{center}
\begin{tabular}{@{}lllllll@{}}
\toprule
\(\boldsymbol{\varepsilon \downarrow}\) \textbackslash{} \(\boldsymbol{\delta \rightarrow}\) &\phantom{mm}& \(\bf{10^{-5}}\) &\phantom{mmm}& \( \bf{10^{-6}}\) &\phantom{mmm}& \(\bf{10^{-9}}\)  \\
\midrule
\textbf{0.1} &&0.03 &&0.03 &&0.02\\ 
\textbf{0.5} &&0.14 &&0.12 &&0.09\\ 
\textbf{1.0} &&0.27 &&0.24 &&0.18\\ 
\textbf{2.0} &&0.50 &&0.45 &&0.35\\ 
\textbf{4.0} &&0.92 &&0.84 &&0.67\\ 
\textbf{6.0} &&\textbf{\color{red} 1.31} &&1.20 &&0.97\\ 
\textbf{8.0} &&1.67 &&1.53 &&\textbf{\color{blue} 1.26}\\ 
\textbf{10.0} &&2.00 &&1.85 &&1.54\\
\bottomrule
\end{tabular} 
\end{center}
\end{table}

%\medskip
%\medskip

\subsection{Choosing \texorpdfstring{$\boldsymbol{\varepsilon}$}{e} in Practice} 
Selecting an appropriate value for 
$\varepsilon$  is one of the most challenging aspects of applying DP in practice. There is \emph{no universally correct value}: the appropriate choice depends on the specific context, including the sensitivity of the data, the type of query or algorithm, and the level of acceptable risk and required utility.
%%but it depends on several factors:
%%\begin{itemize}
%%\item The sensitivity of the data.
%%\item The type of query or algorithm.
%%\item The unit of privacy (e.g., per-user, per-record).
%%\item The expected threat model.
%%\item The acceptable risk and required utility.
%%\end{itemize}
Even within similar domains, $\varepsilon$ values can vary widely due to differing policies, technical constraints, and attitudes toward risk.%\medskip

Therefore, a central responsibility of the data curator (or aggregator) is to determine how much privacy loss is acceptable for a given application. This decision involves balancing the competing interests of data subjects --- who prefer smaller values of 
$\varepsilon$ for stronger privacy --- against those of data analysts, who typically favor larger values to preserve utility by reducing noise.
This balancing act is often more a matter of policy and judgment than exact science. At present, there is no broadly accepted standard or formula for selecting $\varepsilon$. Particularly, 
\cite{NIST2025} noted that setting $\varepsilon$ is still an active area of research and ---
based on interviews with \dps practitioners --- \cite{Dwork_epsilon2019} concluded: \smallskip

\begin{quote}
 \emph{We found no clear consensus on how to choose $\varepsilon$, nor agreement on how to approach this and other key implementation decisions.}
\end{quote}
Furthermore, \cite{Cummings2024Advancing} remarked: 
\begin{quote} 
\emph{While DP has been used in several large-scale deployments, many of these deployments have ended up with privacy loss parameters that provide little if any
meaningful privacy, and the challenge that comes with maintaining a high level of utility while preserving DP
has limited its wider adoption}, 
\end{quote}
thereby acknowledging the persistent difficulty of selecting an appropriate privacy budget and maintaining the utility.
\medskip

%In practice, organizations must weigh privacy risks against utility needs, often through an iterative process of evaluation and refinement.
\subsubsection{Recommended Ranges and Tiered Guidelines for \texorpdfstring{\(\boldsymbol{\varepsilon}\)}{e}}
While the appropriate value of $\varepsilon$ depends heavily on context, use case, and acceptable risk, the DP community has begun to converge on rough practical guidelines. A common framing is to treat $\varepsilon$ values as falling into three broad categories given in Table~\ref{tab:ranges_for_epsilon}.

\begin{table}[ht]
\caption{Recommended Ranges for $\varepsilon$}
\label{tab:ranges_for_epsilon}
\begin{center}
\begin{tabular}{@{}lll@{}}
\toprule
\textbf{Privacy Level} &\phantom{mm}& \textbf{Range of $\boldsymbol{\varepsilon}$} \\
\midrule
Strong privacy && $\varepsilon \leq 1$ \\
Moderate (practical) privacy && $1 < \varepsilon \leq 10$ \\ 
Weak privacy (use with caution) && $\varepsilon > 10$ \\ 
\bottomrule
\end{tabular} 
\end{center}
\end{table}

%\begin{itemize}
%\item 
%\item \textbf{Moderate (practical) privacy:} $1 < \varepsilon \leq 10$ 
%\item \textbf{Weak privacy:} $\varepsilon > 10$ (to be used with caution)
%\end{itemize}

\noindent
This classification is supported by both community consensus and practical experience, although the limit $10$ between moderate and weak privacy is still not entirely settled --- some works use $20$ instead. For instance, \cite{NIST2022} report that $\varepsilon$ values in the low single digits ($0< \varepsilon \leq 5$) are viewed as offering strong privacy protection in many domains, and choosing $\varepsilon \leq 1$ is often viewed as a gold-standard for strict privacy.% \medskip

On the other hand, \cite{NIST2022} observe that deployed systems have used moderately larger values ($5 < \varepsilon < 20$) and still provide robust privacy protection in a variety of settings. In other words, $\varepsilon$ in the high single-digits or low double-digits might still be justified, especially if the data or use case can tolerate a bit more risk or if utility demands are high. %\medskip

If $\varepsilon$ becomes very large (e.g., $> 20$), the privacy guarantee becomes quite loose and we are entering a regime of weak protection. Nonetheless, \cite{NIST2022} note that in some contexts, even $\varepsilon > 20$ may still provide meaningful privacy guarantees when combined with appropriate safeguards, and \cite{Ponomareva2023} remark that any finite $\varepsilon$ is still an improvement over a system with no privacy protections at all.% \medskip

%%%%%%%%%%%%%%%%%%%%%%%
\bigskip

\noindent
\textbf{Ponomareva et al.'s Procedure for Selecting  \texorpdfstring{\(\boldsymbol{\varepsilon}\)}{e}:} A recent comprehensive guide by \cite{Ponomareva2023} provides one of the clearest recommendations for choosing $\varepsilon$ in the context of ML. Their procedure is framed as a set of tiered privacy targets, along with advice to \emph{aim for the strictest tier that still permits a usable model}. In particular, they define three broad tiers for $\varepsilon$ when doing user-level DP (or example-level, where a single user contributes at most one record) under the standard "add-or-remove" adjacency definition (Definition \ref{neighboring_datasets}):
%A more detailed, ML-specific version of these guidelines is proposed by Ponomareva et al.\ \cite{Ponomareva2023}, who introduce a three-tier framework for choosing $\varepsilon$ when doing user-level DP (or example-level, where a single user contributes at most one record) under the standard add-or-remove adjacency definition (Definition \ref{neighboring_datasets}):

\begin{itemize}
\item \textbf{Tier 1 --- Strong privacy guarantee:} $\varepsilon \leq 1$ offers robust protection, directly aligning with the theoretical DP definition \ref{pureDP}. However, this level is often unattainable in modern ML due to substantial utility loss --- it may be practical only for small models or large datasets.

\item \textbf{Tier 2 --- Reasonable privacy–utility trade-off:} $1 < \varepsilon \leq 10$ is the most common setting in practice. This range allows many ML tasks to proceed with acceptable utility while maintaining meaningful privacy protection. Improved accounting techniques for iterative training algorithms may further tighten guarantees within this tier.

\item \textbf{Tier 3 --- Weak privacy:} $\varepsilon > 10$ can still yield empirical privacy benefits --- for instance, by acting as a regularizer and limiting memorization --- but should not be relied upon as a stand-alone privacy solution. When operating in this tier, additional protections such as \emph{privacy auditing} or \emph{attacks} to empirically test what might leak, \emph{data preprocessing} to remove especially sensitive information, \emph{strong regularization} to prevent memorizing specifics, or hybrid approaches are essential.

\end{itemize}
%Ponomareva et al.\  \cite{Ponomareva2023} recommend adopting the strictest feasible tier, relaxing $\varepsilon$ only when necessary to meet utility goals. This tiered strategy complements general guidelines by offering a decision-making framework grounded in both theory and observed practice.

\noindent
Organizations are encouraged to begin by selecting the smallest 
$\varepsilon$  that still allows the data analysis or model to be useful for its intended purpose. This typically involves empirically testing different 
$\varepsilon$  values to observe how model accuracy or statistical error changes as privacy guarantees become stricter. For instance, if a model trained with 
$\varepsilon = 2$ yields acceptable utility, there is no need to increase $\varepsilon$; but if utility degrades too much, one might try $\varepsilon = 4$ or $5$, and so on, until a satisfactory balance is found. This iterative process reflects the fundamental privacy–utility trade-off: either fix a privacy target and accept the resulting accuracy, or fix a minimum utility requirement and find the smallest $\varepsilon$  that meets it. In machine learning (ML), this often means training models with increasing noise (i.e., decreasing $\varepsilon$) until performance drops below an acceptable threshold.

 Finally, it is worth noting that in DP training, other hyperparameters --- such as the clipping strategy, batch size, number of training epochs, and learning rate --- can significantly impact both privacy and utility. The influence of these parameters in the context of DP training is examined, for example, in
 \cite{kurakin2022trainingimagenetscaledifferential},
\cite{li2022largelanguagemodelsstrong}, 
\cite{papernot2022hyperparametertuningrenyidifferential}, and
\cite{Ponomareva2023}
 (see also Section \ref{sec_hyperparameters}).

%\subsubsection{Privacy--Utility Trade-offs}
\subsubsection{Composition and Budget Accounting}
In real-world deployments of DP, a key challenge is managing how the privacy budget is consumed over time. This is governed by the principle of \emph{composition}: when multiple queries or analyses are performed on the same dataset, each consumes a portion of the total available privacy budget. In DP training, for instance, every iteration of the learning algorithm contributes to cumulative privacy loss. Under basic composition rules (see Section~\ref{sec_DPproperties}), the total privacy cost grows approximately as the sum of the individual \(\varepsilon\) values used in each step. As a result, answering too many queries or training for too many iterations can significantly degrade the overall privacy guarantee --- or even eliminate it altogether.

In practice, organizations typically set an overall privacy budget --- e.g., $\varepsilon_{total} = 4$ for the entire project --- and ensure that the sum of the $\varepsilon$ values of all released outputs does not exceed this limit. This necessitates careful planning and may require halting further queries once the budget is exhausted to avoid overexposure. In complex settings, such as training deep neural networks with thousands of iterations, privacy loss accounting tools are used to track the cumulative $(\varepsilon,\delta)$ (see Section \ref{sec_Accounting}). For example, the Moments Accountant technique (\cite{abadi2016deep}) is employed in DP-SGD to compute tight upper bounds on the total privacy loss, often yielding significantly tighter accounting  than naive summation. %\medskip

These challenges become especially pronounced in interactive settings, such as when a dataset is exposed via an API that allows users to make arbitrary queries or retrain models repeatedly. Without strict limits or per-user quotas, the cumulative privacy loss can quickly exceed acceptable bounds. Likewise, when multiple analysts or teams access the same data, their outputs could be combined — intentionally or unintentionally — revealing more than what each individual query would allow. To prevent this, the privacy budget must be considered across all queries and all users collectively, and access mechanisms should enforce shared budget limits.
%\medskip

The privacy budget thus reflects a fundamental trade-off: a smaller total $\varepsilon$ (i.e., stronger privacy) requires either answering fewer queries or tolerating more noise per query, both of which reduce utility. Conversely, a larger $\varepsilon$ enables more accurate outputs or more training iterations, but at the cost of weaker privacy guarantees. 
%\medskip

%Determining how to allocate the budget is ultimately a policy decision, shaped not only by technical considerations but also by legal, ethical, and societal factors, as well as user expectations. In practice, this often involves careful planning, iterative tuning, and the use of privacy accounting tools to ensure that cumulative loss remains within acceptable bounds throughout the data lifecycle.

%
%In many applications, $\varepsilon$ is typically kept in the range $0-10$, often below 5, while  
%$\delta$ is set extremely small (e.g., $10^{-5}$ or lower).
%Intuitively, $\delta$  represents the probability of a "bad event" where the privacy guarantee might be violated and the worst-case analysis (see, Example \ref{ex_worse_case}) suggest to use value $\delta \ll 1/n$, where $n$ is the number of records. \cite{DworkRoth2014}.\medskip
 % DP-ML artikkelissa tästä on esimerkiksi kelpaavaa tekstiä. Halutaanko?

\subsubsection{Alternative Interpretations and Extensions of \texorpdfstring{\(\boldsymbol{\varepsilon}\)}{e}}
While \(\varepsilon\) and \(\delta\) are the most commonly used parameters in DP, they are not the only ones. Researchers have proposed several alternative formulations (see Section~\ref{sec_DPvariants}) that reinterpret or reallocate the privacy budget to address different use cases or improve flexibility. These variants often rely on different mathematical definitions and introduce new parameters, making direct comparisons with standard \((\varepsilon, \delta)\)-DP nontrivial. Even when parameter names overlap, their meanings may differ significantly. For example, the \(\varepsilon\) in R\'enyi  DP is comparable to that of pure DP only when the divergence order \(\alpha\) is very large (\cite{NIST2025}).
%\medskip

Personalized DP is one such variant. It allows each individual data contributor to be protected with their own personal $\varepsilon$ value (\cite{boenisch2024wayindividualizedprivacyassignment,Edabi2015,Jorgensen2015}). 
In such systems, privacy loss is tracked on a per-user basis --- maintaining a separate budget for each individual --- rather than enforcing a single global $\varepsilon$. This enables accommodation of heterogeneous privacy preferences and risk profiles across a population. Such personalized privacy budgets are particularly useful in medical contexts, where patients with rare or highly sensitive conditions may require stricter privacy guarantees than others contributing to the same dataset (\cite{boenisch2024wayindividualizedprivacyassignment}). Similarly, in interactive database systems, personalized budgets allow queries to be answered while ensuring that each user’s individual privacy budget is respected, as demonstrated in \cite{Edabi2015}. %\medskip
 
Another approach, sensitive privacy (SP), introduces a context-aware use of the privacy budget (\cite{sp2019,9623509}).
It is used especially in anomaly detection (see Section \ref{sec_anomaly}).
In SP, the $\varepsilon$  still quantifies the privacy loss, but it is applied selectively based on the sensitivity of each record. Specifically, the \dps guarantee (bounded by $\varepsilon$) is enforced for changes involving $k$-sensitive records ---  
those considered typical or privacy-sensitive --- while relaxed guarantees may be applied to anomalous or less-sensitive records.
This allows strong privacy protection for most individuals while enabling greater utility for tasks such as anomaly detection.
%\medskip

Both personalized DP and SP preserve the core guarantees of DP but redefine how 
$\varepsilon$  is interpreted and controlled --- enabling per-user or context-dependent privacy budgets rather than relying on a one-size-fits-all approach.

\subsubsection{Other Guidelines and Heuristics}

%\begin{itemize}
%    \item \textbf{Risk-based analysis}: Choose \(\varepsilon\) to bound specific identification or inference risks.
%    \item \textbf{Economic models}: Derive \(\varepsilon\) by balancing privacy loss cost and utility gain.
%    \item \textbf{Empirical testing}: Simulate privacy attacks to evaluate effective leakage at a given \(\varepsilon\).
%\end{itemize}

%Beyond Ponomareva et al.’s ML-focused framework \cite{Ponomareva2023}, other researchers and organizations have offered guidance for setting $\varepsilon$, often aligned with specific domains or goals. Here we summarize a few notable guidelines and heuristics from the literature:

While DP provides formal, worst-case privacy guarantees, real-world deployments often involve additional considerations and complementary techniques. In practice, achieving rigorous privacy protection may benefit from a broader perspective that includes empirical privacy auditing, heuristic methods, and context-specific threat modeling. 

\bigskip

\noindent
\textbf{Privacy Auditing and Empirical Evaluation:}
An increasingly studied alternative to purely theoretical guarantees is \emph{privacy auditing}, which uses adversarial techniques to empirically estimate the privacy leakage of ML models. A foundational method is the \emph{membership inference attack}, where an adversary attempts to determine whether a specific data point was part of the training set. Though useful, basic membership inference may underestimate actual privacy risk. More advanced auditing techniques craft worst-case poisoning examples that enhance an adversary’s ability to distinguish neighboring datasets, thereby producing tighter empirical lower bounds on 
$\varepsilon$ than traditional attacks (see, e.g., \cite{JagielskiMatthew2020ADPM,NasrMilad2023TAoD,PillutlaKrishna2023UtPo}); see also recent work on gray-box auditing of DP implementations by \cite{cebere2026privacytheorybugspractice}. We refer to \cite{Cummings2024Advancing} (Chapter 4) and \cite{annamalai2025hitchhikersguideefficientendtoend} for extensive overview of current privacy auditing approaches. 
\bigskip

\noindent
\textbf{Resilience to Attacks Beyond Formal Guarantees:}
Evidence from reconstruction and memorization attacks suggests that even large $\varepsilon$ values --- well beyond what the DP literature traditionally considers acceptable --- can offer non-trivial protection. For example,  \cite{CarliniNicholas2018TSSE} showed that applying DP with  
$\varepsilon=10^9$ still significantly reduced memorization in language models. Similarly,  \cite{BalleBorja2022RTDw} demonstrated that example-level $\varepsilon$ values in the range of 100–10,000 could mitigate training data extraction without harming test performance. These findings indicate that DP mechanisms may provide practical benefits even in high-$\varepsilon$ regimes, though they fall short of the ideal worst-case guarantees. 

On the other hand, \cite{Jayaraman2019} show that different DP accounting methods (e.g., advanced composition, zCDP, Rényi DP) exhibit varying degrees of resistance to membership and attribute inference. Notably, models trained under relaxed DP definitions may achieve theoretical guarantees but still leak significant information under attack --- particularly when $\varepsilon$ is large.

\bigskip

\noindent
\textbf{Heuristic Safeguards and Sensitivity Controls:}
Interestingly, certain non-private practices inspired by DP can also help mitigate privacy risks. For instance, clipping gradient norms, a standard step in DP training (e.g., in DP-SGD), has been shown to defend against various membership inference attacks even without added noise (\cite{Cummings2024Advancing,Jayaraman2019}). This observation suggests that bounding sensitivity --- whether formally or heuristically --- can be an effective privacy-preserving tool on its own.%\medskip
 
% Vaihtoehto
In summary, aside from formal tiered recommendations given by \cite{Ponomareva2023}, the literature suggests using a combination of theory and empirics: start with known "safe" ranges (e.g. $\varepsilon$ in [1, 5] for strong privacy, or up to 10--20 for reasonable privacy, \cite{NIST2022}), consider the specific threats and context of your application, and adjust $\varepsilon$ accordingly, possibly validating the choice via simulations or attacks (\cite{Cummings2024Advancing,Jayaraman2019}). 
It is also advised to be transparent about chosen $\varepsilon$ and the rationale. \cite{Dwork_epsilon2019} have called for an \emph{$\varepsilon$ registry} to compile case studies of $\varepsilon$ choices in different applications. One such registrery is given at  \url{https://registry.oblivious.com/}.

In addition, \cite{NIST2025} encourage vigorous work to inform how to select privacy parameters.
The rationale is that by observing what $\varepsilon$ values have been used (and deemed acceptable) in real-world deployments, future practitioners can make more informed decisions. This leads naturally into examining what values have indeed been used in practice.

\subsection{Examples of Privacy Budgets in Practice}
Since $\varepsilon$ is context-dependent, it is useful to look at concrete examples from both industry deployments and experimental studies. 
%(see also \url{https://registry.oblivious.com/}). 
Below are a few examples illustrating how $\varepsilon$ (and other privacy parameters like $\delta$ and $\rho$)\footnote{Recall that in zCDP, 
$\rho$ serves as a bound on the variance of the privacy loss random variable.} has been set in different scenarios:
%ok
\bigskip

\noindent
\textbf{Statistical Data Release (U.S. Census Bureau):} The 2020 U.S.\ Census is a landmark deployment of DP for national statistics. During the development of the 2020 U.S. Census Disclosure Avoidance System (DAS), the Census Bureau released multiple demonstration data products using a range of privacy budgets to evaluate the impact of \dps on data quality. Among these, a commonly referenced configuration applied full privacy-loss accounting based on zCDP, with a total privacy-loss budget of $\varepsilon = 12.2$. This budget composed of $\varepsilon = 10.3$ (with $\delta = 10^{-10}$ and $\rho \approx 1.05$) for person-level data and $\varepsilon = 1.9$ (with $\delta = 10^{-10}$ and $\rho \approx 0.045$) for housing-unit-level data. These values were used in the May 2020 demonstration data release to solicit public and expert feedback on the trade-off between privacy and accuracy (\cite{UScensus3,UScensus1}).

 However, this was not the final configuration. Following further testing, public consultation, and refinement of the noise injection mechanisms, the Census Bureau ultimately adopted a total privacy-loss budget of $\varepsilon =
19.61$ for the official redistricting data (PL 94-171) release  (\cite{UScensus4,UScensus2}). This final budget was split into $\varepsilon =
17.14$ ($\delta = 10^{-10}$, $\rho=2.56$) for person-level data and $\varepsilon =
2.47$ ($\delta = 10^{-10}$, $\rho=0.07$) for housing-unit-level data.  The increase in the privacy-loss budget reflects the Bureau’s response to concerns about data utility and the legal requirements of redistricting, while still maintaining a formal, finite privacy guarantee under \dpns. The above procedure is a basic example of policy and regularory driven selection of the privacy budget.
 
Nevertheless, the 2020 U.S.\ Census has faced growing criticism from demographers and other social scientists. Scholars have found that while the Census-released test data produce accurate estimates for aggregate total population counts at larger geographic scales (e.g., counties), they introduce substantial discrepancies in estimates for specific subpopulations, such as rural residents and non-white communities  (\cite{UScensus5}).

%ok
\bigskip

\noindent
\textbf{Mobility Data (Facebook COVID-19 Mobility Maps):} During the COVID-19 pandemic, Facebook released aggregated mobility data with \dps to aid researchers and public health officials. In this application, they used $\varepsilon = 2$ with a per-day, per-user privacy accounting to protect location data of users (\cite{facebook}). Here, each user’s day of mobility contributes to aggregates with local DP (user-level privacy per day).

A privacy budget $\varepsilon=2$ per day is relatively strict. The Facebook team likely judged that the public could be uncomfortable with any higher privacy loss on sensitive location traces, and that the aggregates could still be useful with that level of noise. Indeed, $\varepsilon = 2$ means each user’s influence on the mobility metrics is heavily obfuscated, limiting the chance that someone could re-identify a particular user’s movement. In this case high sensitivity data (locations) drove a low $\varepsilon$ choice, accepting some added noise in the published mobility statistics. The utility trade-off was manageable because the data was aggregated at a coarse level (city/county trends), which can tolerate noise.

\bigskip

\noindent
\textbf{Local Telemetry Collection (Apple):}
%ok
Apple has deployed local DP in macOS and iOS to collect usage statistics --- such as emoji usage frequencies and Safari autocomplete domains. According to Apple’s documentation (\cite{apple}), a daily privacy budget per user is enforces in the range $\varepsilon = 2$ to $\varepsilon = 16$. 
Different telemetry items have different $\varepsilon$ allocations within this range, depending on how sensitive the data is considered. In addition, Apple limits the number of times a user's data can be sent per day, thereby bounding the total contribution from each user.

For instance, a highly sensitive data point (such as a health-related signal) is allocated 
$\varepsilon=2$  and limited to one contribution per day, while a less sensitive item (e.g., Safari autoplay intent detection) uses  $\varepsilon=8$ and allows two contributions, resulting in a total budget of $\varepsilon=16$ per user.

Apple’s approach reflects a \emph{privacy-by-design} principle: they fix an upper bound on each user’s daily privacy loss. The specific budget allocated to each query depends on the sensitivity of the data being collected, with more sensitive information receiving tighter privacy constraints. Even at the upper end ($\varepsilon = 16$), the budget remains finite. This ensures that, although noisy reports are collected continuously, \dps guarantees limit the risk of reconstructing individual users' data. This example illustrates how $\varepsilon$  can be managed in ongoing data collection systems --- by distributing a moderate privacy budget across small, bounded contributions, Apple can derive useful aggregate statistics while keeping individual privacy loss controlled on a daily basis.
   
\bigskip

\noindent
\textbf{Differentially Private Machine Learning (Google's Gboard):} 
A notable real-world deployment of differential privacy in ML is Google's \emph{Gboard}, the Android keyboard app (\cite{Gboard2023}). Gboard uses a federated learning framework to train next-word prediction models with device-level DP (for user’s with a single device, device-level DP corresponds directly to user-level DP). In production, Google applies the DP-FTRL algorithm (introduced by \cite{Kairouz2021}) to train language models under zCDP guarantees. The deployed models achieve 
$\rho$-zCDP values ranging from $0.2$ to $2.0$, which correspond to approximate 
$\varepsilon$-DP values in the range 
$\varepsilon \approx 4.5$ to 
$13.7$ when $\delta = 10^{-10}$
  (\cite{Gboard2023}). Models are trained with a report goal of 6500 clients per round and a noise multiplier of 7, and client participation is further controlled through enforced waiting periods between rounds. All next-word prediction and candidate re-ranking models in Gboard now include formal DP guarantees, and future Gboard language models are required to satisfy similar privacy protections. 
  
  This example illustrates how DP is used in a large-scale, user-facing ML system by integrating formal privacy guarantees into the training pipeline while preserving high model utility. In addition, we acknowledge \cite{Gboard2023} and Google for providing a particularly clear and precise presentation of the privacy parameters and accounting methods used in their deployment, which serves as a valuable example of transparency in real-world DP applications.

\bigskip

\noindent
\textbf{Academic ML experiments:} In research settings, $\varepsilon$ varies widely but provides insight into what levels yield acceptable performance.
%{\red Unit of privacy = example-level, neighboring criteria = add or remove.}
%
For instance, the original DP-SGD paper by \cite{abadi2016deep} showed that on simple tasks like MNIST digit classification, one could achieve good accuracy with $\varepsilon$ between $2$ (accuracy $\approx 95\%$) and $8$ (accuracy $\approx 97\%$) and $\delta=10^{-5}$.  These results were made possible by introducing the moments accountant, a technique that provides tighter privacy loss accounting over multiple training steps compared to basic composition, thereby enabling better privacy–utility trade-offs.
Follow-up work by \cite{papernot2017semi}, using the PATE framework, improved the accuracy of a private MNIST model from 97\% to 98\% while simultaneously tightening the privacy bound from $\varepsilon=8$ to $\varepsilon =1.9$, by leveraging public data and an ensemble of teacher models. In both of these works example-level unit of privacy with add or remove neighboring criteria were used.

On the other hand, recent attempts to train large language models from scratch with DP have found that even low-digit $\varepsilon$ %(e.g., $\varepsilon \approx 6$) 
can incur very large accuracy costs. \cite{Ponomareva2022} reported a 34\% relative performance drop  during pretraining of a T5 model at $\varepsilon = 6.06$, though the final task performance after fine-tuning was unaffected. Similarly, 
  \cite{anil-etal-2022-large} observed 14\% performance drop when pre-training Bert with $\varepsilon = 5.36$ using meta-batches. 
These results highlight that the impact of $\varepsilon$ is depends strongly on the model size, data regime and training stage. In contrast, transfer learning --- where a model is pretrained non-privately and then fine-tuned with DP --- can yield much better trade-offs: for example, \cite{yu2024differentially} achieved 
 $\varepsilon = 6.7$ with only a $2.4\%$ relative drop when fine-tuning a RoBERTa-large language model. 

Thus, in academic literature we see that for simpler or well-regularized models, very small $\varepsilon$ can work, whereas for complex models, one might have to settle for higher $\varepsilon$ to get acceptable utility. These results have informed guidelines like the one given in  \cite{Ponomareva2023}: if even state-of-the-art research struggles below $\varepsilon \sim 2$ for large models, expecting a production system to use $\varepsilon = 0.5$ may be unrealistic unless the model or data is fundamentally simpler.\medskip

%\subsection{Conclusion} % Tästä ehkä vois yrittää päästää eroon jos springerin lehti on tavoitteena
Selecting the privacy budget \(\varepsilon\) is a central decision in any DP deployment. 
% tämä uusi conclusionin tilalla tässä
 Ultimately, its allocation is a policy choice --- shaped not only by technical factors, but also by legal, ethical, and societal considerations, as well as user expectations (see Section \ref{chaUserExpectations}). In any case, the chosen value and any assumptions underlying it (such as the unit of privacy, the definition of neighboring dataset, and whether the guarantee covers just the final model or also includes hyperparameter tuning
 together) should be clearly documented to ensure both reproducibility and accountability.

\section{Privacy-Preserving Synthetic Data}
\label{sec_synthetic_data}

Privacy-preserving synthetic data refers to artificially generated data that statistically replicates real-world datasets while providing strong privacy guarantees. The core idea is to learn a model from an original sensitive dataset and then use that model to create new, \emph{synthetic records} that capture statistical patterns of the original data without directly revealing any individual’s information. In the context of differential privacy (DP), the model training or data generation process is designed to satisfy DP, meaning that the presence or absence of any single individual in the training data has a rigorously bounded effect on the generated outputs.
This section explores the generation, usage, and limitations of differentially private synthetic data. 

% HUOM Näitä ei muutettu vielä!!!
\subsection{Introduction and Motivation} 
The motivation for generating privacy-preserving synthetic data arises from the need to balance data utility with confidentiality. Privacy regulations and ethical concerns may restrict organizations from sharing or openly using real datasets containing personal or sensitive information (e.g., medical records, financial transactions).
Synthetic data offer a potential solution: since the released data points are artificially generated rather than exact copies of real individuals’ data, the hope is that sensitive information is not directly exposed. However, using synthetic data is not a privacy panacea --- simply replacing real data with synthetic data does not automatically guarantee privacy or regulatory compliance (\cite{Pitkamaki2024}). In fact, it has been demonstrated that naive synthetic data generation can still leak information: a generative model might even reproduce near-identical copies of original records, or otherwise encode subtle traces of the training data (\cite{jordon2022syntheticdatawhat,Perez2024}). Therefore, formal privacy frameworks, most notably DP, are employed to provide provable privacy guarantees for synthetic data releases.
DP ensures that any single individual's data has only a negligible influence on the synthetic output, thereby limiting disclosure risk. 

The typical use cases of DP synthetic data include: 
\begin{itemize}
\item \textbf{Healthcare}: DP synthetic health data enables privacy-compliant research and testing without access to sensitive patient records (more details are given in Section  \ref{sec_health2}).

\item \textbf{Financial Services:} Synthetic transaction data supports fraud detection algorithm development without exposing client data (more details are given in Section \ref{sec_finance1}).

\item \textbf{Machine Learning Development:} Synthetic datasets facilitate model prototyping and training, especially when real-world data is unavailable or restricted.
\end{itemize}

\subsection{Synthetic Data: Definitions and Distinctions} 
The term synthetic data is sometimes used interchangeably with simulated data, but there are important distinctions in their typical meaning and generation process.  \cite{jordon2022syntheticdatawhat} provide a useful definition: 
\begin{quote}
 \emph{Synthetic data is data that has been generated using a purpose-built mathematical model or algorithm, with the aim of solving a (set of) data science task(s)}.
\end{quote}
This definition is intentionally broad: it does not require the generating model to be derived from real data --- it could, for example, also include an expert-designed simulator. In this work, we adopt a narrower interpretation: we use the term synthetic data to refer specifically to data generated by models that are informed by, or trained on, real-world datasets. With this definition, synthetic data aim to reproduce key statistical properties of the original data while serving a particular purpose (e.g., enabling analysis or training machine learning models). The models used for this purpose can range from simple statistical approaches to complex machine learning algorithms such as generative adversarial networks (GANs) and variational autoencoders (VAEs).

In contrast to synthetic data, \emph{simulated data} typically refers to data generated from a theoretical or mechanistic model of a phenomenon, which may or may not be directly based on real-world data. Such datasets are often produced using prior knowledge or assumptions about the underlying data-generating process --- for example, by sampling from a known probability distribution or running a physics-based simulation --- rather than by learning from a specific collected dataset. While simulations can incorporate parameters estimated from real data, they are not intended to replicate the unique characteristics or irregularities of any particular dataset. %\medskip

Another term, \emph{fabricated data} (sometimes called \emph{dummy data} or \emph{mock data}), refers to completely made-up data with no grounding in real observations. Fabricated data are often used for software testing or examples (e.g., sample names and addresses), and they do not aim to preserve any statistical relationship to real-world data. %\medskip

In summary, data generation approaches can produce different data types, ranging from fabricated data (created without any real-world input), to simulated data (generated from theoretical models or global parameters), to synthetic data (produced by modeling a specific real dataset). In practice, the boundaries between these categories may blur, and the terminology is sometimes applied inconsistently. However, the key distinction is that synthetic data, as discussed here, aims to reproduce the statistical properties of real datasets, while simulated data focuses on modeling underlying processes or systems (e.g., patient flows in hospitals).%\medskip

Figure~\ref{fig:syntheticData} adapted from \cite{Pitkamaki2024} illustrates these distinctions and the relationships between different data types.%\medskip
\begin{figure}[hbt!]
    \centering   
%\begin{tcolorbox}
    \includegraphics[width=0.5500\linewidth]{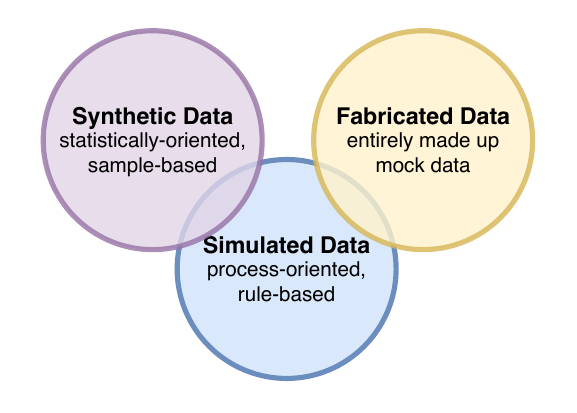}
%    \includegraphics[angle=270,width=1\linewidth]{DP_T1.png}
%\end{tcolorbox}
    \caption{Common Properties of Synthetic, Simulated, and Fabricated Datasets: The
differences are not clear-cut and the terms may be used interchangeably (\cite{Pitkamaki2024}). } % Local DP
    \label{fig:syntheticData}
\end{figure}

It is also useful to clarify subcategories of synthetic data. Depending on how much real data is used in the final release, literature distinguishes fully synthetic data versus partially synthetic data (\cite{Pitkamaki2024}). In \emph{fully synthetic data}, all released records are generated by a model --- none are actual original records. \emph{Partially synthetic data}, on the other hand, might retain some real data attributes or records (for example, releasing a dataset where only certain sensitive columns have been replaced with imputed values, or releasing a mixture of real and fake records).
%\medskip

There is also the notion of \emph{hybrid synthetic data}, in which real and synthetic records are combined in the released dataset. A common application of such data is \emph{data augmentation}, where additional samples—particularly for rare classes or edge cases—are generated to enrich the training set and improve model robustness and generalization (\cite{ALQUDAH2023103809,KIM2023104771}). From a privacy standpoint, however, fully synthetic data remains the main focus of privacy research, as any inclusion of actual data (as in partial or hybrid approaches) means that some real individuals’ information is directly present and thus carries an inherent disclosure risk.
Nevertheless, as already noted, fully synthetic data should not be assumed automatically safe either --- as \cite{ElEmamKhaled2020EIDR} argue, even a dataset that is entirely model-generated can still carry residual information about the original individuals if the model overfits or if certain unique patterns are reproduced. 
This is why \dps needs to be introduced: to formally limit the risk that synthetic data generation inadvertently reproduces or discloses information about any individual.

\subsection{Taxonomy of DP Synthetic Data Generation Methods} 
 A wide range of methods have been proposed for generating synthetic data under \dpns, reflecting different strategies to balance statistical fidelity and privacy (\cite{raisa2023}). These methods can be classified into several broad categories, including approaches based on statistical summaries (e.g. histograms or marginals), probabilistic graphical models, and deep generative models, among others.
 We provide a concise taxonomy below and summarize it in Table~\ref{tab:taxonomy}, highlighting representative techniques in each category. At the end of this section, we include a simplified illustrative example (Example \ref{exampleTaxonomy}) to show how these different approaches could be applied in practice.%\medskip

\begin{table}[h!t]
\caption{Taxonomy of Differentially Private Synthetic Data Generation Methods}
\label{tab:taxonomy}
\resizebox{\textwidth}{!}{\begin{tabular}{@{}lll@{}}
\toprule
\textbf{Category} & \begin{tabular}[c]{@{}l@{}}\textbf{Representative}\\ \textbf{Methods}\end{tabular} 
 & \textbf{Core Characteristics}\\
\midrule
\begin{tabular}[c]{@{}l@{}}\textbf{Histogram /}\\
\textbf{Marginals} \end{tabular} & 
\begin{tabular}[c]{@{}l@{}}MWEM \\ %\cite{HardtMoritz2010Asap} \\ 
NAPSU-MQ \\ %\cite{raisa2023} \\ 
RAP \\ %\cite{AydoreSergul2021DPQR} \\ 
PEP \\ %\cite{LiuTerrance2021IMfP} \\ 
Private-PGM \\ %\cite{McKennaRyan2021WtNC} \\ 
AIM \\ %\cite{McKennaRyan2022Aaaa}
\end{tabular} & 
\begin{tabular}[c]{@{}l@{}}+ High accuracy for selected marginals, \\ 
+ Strong theoretical guarantees for selected \\ \quad queries, \\
+ Well-suited for tabular data (e.g., census, \\ \quad surveys), \\ 
-- Scalability issues for high-dimensional data
\end{tabular} 
\\
\midrule
\begin{tabular}[c]{@{}l@{}}\textbf{Graphical}\\
\textbf{Models}\end{tabular}  & 
\begin{tabular}[c]{@{}l@{}}PrivBayes \\ %\cite{privBayes2014,privBayes2017} \\ 
PrivMRF \\ %\cite{Cai2021} 
\end{tabular} & 
\begin{tabular}[c]{@{}l@{}}+ Captures dependencies between attributes, \\ 
+ Produces interpretable, structured models, \\ 
+ Effective for mixed-type tabular data \\ \quad with relational structure, \\
-- Noise increases with model complexity, \\
-- Risk of oversimplification or overfitting
\end{tabular} \\
\midrule
\begin{tabular}[c]{@{}l@{}}\textbf{Deep}\\
\textbf{Generative}\\
\textbf{Models}\end{tabular}  & 
\begin{tabular}[c]{@{}l@{}}DP-WGAN \\ %\cite{uclanesl_dp_wgan}, \\ 
GS-WGAN \\ %\cite{chen2021gswgangradientsanitizedapproachlearning}, \\ 
PATE-GAN \\ %\cite{yoon2018pategan}, \\ 
G-PATE \\ %\cite{long2021gpatescalabledifferentiallyprivate}, \\ 
DPGM \\ %\cite{Acs2019}, \\ GEM \cite{LiuTerrance2021IMfP}, \\
DPDM \\ %\cite{dockhorn2023dpdm}, \\
DP Diffusion (Pretrained \\ \quad Fine-Tuned)  \\ %\cite{ghalebikesabi2023dpdm}, \\
dp-promise \\ %\cite{wang2023dppromise} 
\end{tabular} & 
\begin{tabular}[c]{@{}l@{}}+ Captures complex, high-dimensional \\ \quad patterns, \\ 
+ Generates realistic, diverse samples, \\ 
+ Suited for unstructured data (e.g., images, \\ \quad text) and high-dimensional tabular data, \\
+ Diffusion models offer stable training,\\ \quad  leverage pre-training, and improved \\ \quad privacy -- utility trade-offs, \\ 
-- Sensitive to noise (GANs) or \\ \quad high computational cost (diffusion), \\ 
-- Requires larger $\varepsilon$ or datasets
\end{tabular} \\
\midrule
%\textbf{Deep Generative Models} & 
%\begin{tabular}[c]{@{}l@{}}DP-WGAN \cite{uclanesl_dp_wgan} \\ 
%GS-WGAN \cite{chen2021gswgangradientsanitizedapproachlearning} \\ 
%PATE-GAN \cite{yoon2018pategan} \\ 
%G-PATE \cite{long2021gpatescalabledifferentiallyprivate} \\ 
%DPGM \cite{Acs2019} \\
%GEM \cite{LiuTerrance2021IMfP} \end{tabular} & 
%\begin{tabular}[c]{@{}l@{}}+ Captures complex, high-dimensional \\ \quad patterns, \\ 
%+ Generates realistic samples, \\ 
%+ Suited for unstructured data (e.g., images, \\ \quad text) and high-dimensional tabular data, \\
%-- Sensitive to noise (mode collapse, \\ \quad degraded quality), \\ 
%-- Requires larger $\varepsilon$ or datasets
%\end{tabular} \\
%\hline
\begin{tabular}[c]{@{}l@{}}\textbf{Other }\\
\textbf{Approaches} \end{tabular} & 
\begin{tabular}[c]{@{}l@{}}Metric Privacy \\ %\cite{BoedihardjoMarch2024Pmrw} \\ 
Plausible Deniability \\ %\cite{bindschaedler2017plausibledeniabilityprivacypreservingdata} 
\end{tabular} & 
\begin{tabular}[c]{@{}l@{}}+ General-purpose utility for diverse tasks, \\ 
%+ Useful in scenarios beyond strict DP \\ \quad (e.g., exploratory analysis, simulations), \\
-- Fewer mature implementations, \\ 
-- Less empirically validated in practice
\end{tabular} \\
\bottomrule
\end{tabular}}
\end{table}

\subsubsection{Histogram and Marginal-Based Methods} 
A simple approach to DP synthetic data is to release noisy statistics of the original data, such as attribute counts or low-dimensional distributions, and then generate synthetic samples consistent with those statistics. For example, a differentially private histogram (see, e.g., \cite{WassermanLarry2010ASFf})  can be constructed for each attribute or attribute combination, and synthetic records are sampled according to the perturbed frequencies. If histograms are built for individual attributes only, this preserves marginal distributions but may break correlations between variables.%\medskip

A more advanced approach ensures that selected joint distributions (marginals) are preserved. The data curator chooses important low-dimensional marginals (e.g., distributions of one or two attributes) and adds DP noise to their values. Synthetic data is then generated to satisfy these noisy marginals as closely as possible.
This can be formulated as a constraint satisfaction or probabilistic inference problem.
An early influential method in this category is the  \emph{Multiplicative Weights Exponential Mechanism} (MWEM) algorithm introduced by \cite{HardtMoritz2010Asap}. MWEM iteratively selects queries (marginals) to privatize, and updates a synthetic data distribution to approximate them, focusing on achieving accuracy for a chosen query workload rather than reconstructing the full data distribution. MWEM was a breakthrough in demonstrating that one can release a synthetic database that is accurate for a large set of statistics under DP, albeit with significant computation. %\medskip

A more recent example is  \emph{Noise-Aware Private Synthetic data Using
Marginal Queries} (NAPSU-MQ), proposed by \cite{raisa2023}. NAPSU-MQ uses noisy marginal queries and the principle of maximum entropy to construct a synthetic data distribution that agrees with them. Crucially, NAPSU-MQ is \emph{noise-aware}: it explicitly models the uncertainty introduced by DP noise when generating data, rather than treating noisy marginals as exact values. This results in synthetic datasets that not only reflect the expected marginal statistics but also properly account for the variability introduced by privacy mechanisms, which is essential for valid statistical inference. 
Compared to earlier methods like MWEM, NAPSU-MQ provides a more principled treatment of uncertainty, making it particularly suitable when downstream analyses require reliable confidence intervals or hypothesis testing. Further, it draws on concepts from multiple imputation to account for uncertainty due to both DP noise and the synthetic data generation process, further improving the validity of statistical inference.
%\medskip 
 
The other recent developments in marginal-based methods have introduced additional refinements:
\emph{Relaxed Adaptive Projection} (RAP) by  \cite{AydoreSergul2021DPQR} applies adaptive projection to produce synthetic data that closely matches noisy answers to selected queries. \cite{LiuTerrance2021IMfP} unified several marginal-based  methods under a common framework, providing a clearer theoretical perspective. In addition, they introduced \emph{Private Entropy
Projection} (PEP) method --- an advanced variant of MWEM that adaptively
reuses past query measurements to boost accuracy. 
 \cite{McKennaRyan2021WtNC} applied a \emph{select-measure-generate} strategy for generating synthetic datasets intended for direct release, rather than using synthetic data solely as an intermediate step for answering queries. Their method, called \emph{Private-PGM}, reconstructs high-dimensional data distributions from noisy marginal queries using 
Probabilistic Graphical Model (PGM)\footnote{Although \emph{Private-PGM} uses PGMs for distribution reconstruction, it is classified here as a marginal-based method because its primary mechanism involves selecting and answering a set of noisy marginals. The PGM serves as a modeling tool rather than defining the privacy mechanism itself.}.
Further,  \cite{McKennaRyan2022Aaaa} proposed the \emph{Adaptive and Iterative Mechanism} (AIM) algorithm, which improves the query selection process in the select-measure-generate strategy by iteratively and greedily selecting queries that most effectively improve the approximation of the original data. %\medskip

Marginal-based methods allow the data curator to prioritize which aspects of the data to preserve and offer theoretical guarantees for those aspects. However, they struggle with high-dimensional data, as the number of possible marginals grows rapidly. Careful selection of key marginals is essential to stay within practical privacy and computational limits.

\subsubsection{Probabilistic Graphical Models} \enlargethispage{\baselineskip}
Another approach to DP synthetic data generation uses probabilistic graphical models, such as \emph{Bayesian networks} or \emph{Markov Random Fields} (MRF), to approximate the joint distribution of the data. These models capture dependencies between attributes while enforcing \dpns, and synthetic data is generated by sampling from the learned distribution.
 %\medskip
  
A prominent example is \emph{PrivBayes}, introduced by \cite{privBayes2014,privBayes2017}. PrivBayes learns a Bayesian network --- a directed acyclic graph where nodes represent attributes and edges represent dependencies --- by incrementally building the structure. At each step, it selects the next attribute and its parents based on criteria like mutual information and adds DP noise to the required frequency counts.
%%\medskip  
%
The result is a noisy but probabilistically structured representation of the data. Once the network structure and noisy conditional probabilities are learned, synthetic data is generated by sampling attributes sequentially according to the network. This process ensures privacy while preserving key dependencies. PrivBayes showed that reasonably accurate synthetic data could be produced for high-dimensional categorical datasets under moderate privacy budgets, outperforming simpler independent attribute models.
%\medskip
 
A related approach is \emph{PrivMRF}, proposed by \cite{Cai2021}, which utilizes a MRF to model dependencies between attributes. Unlike PrivBayes, which imposes constraints on the set of marginals used, PrivMRF allows greater flexibility in selecting marginals to better capture important correlations among attributes. The method first selects a set of low-dimensional marginals and then constructs an MRF that models the dependencies implied by these marginals. Finally, synthetic data is generated by sampling from the MRF model. This approach enables more accurate modeling of attribute correlations while maintaining \dpns. Experiments on benchmark datasets demonstrated that PrivMRF outperformed earlier methods in terms of the accuracy of counting queries and classification tasks performed on the generated synthetic data (\cite{Cai2021}).
%\medskip

Generally, graphical models offer the advantage of explicitly modeling dependencies and providing interpretable structures. However, learning accurate models under DP constraints is challenging, as modeling more dependencies increases noise, while oversimplifying risks losing important correlations. These methods work best when the data has a natural hierarchical or dependency structure; otherwise, the models may become either too complex (exhausting the privacy budget) or too simplistic.

%
%In addition, Diffusion models represent a recent and powerful family of generative models that iteratively transform noise into realistic data samples by learning to reverse a progressive noising process \cite{ho2020denoising,nichol2021improved}. 
%

\subsubsection{Deep Generative Models} %\enlargethispage{\baselineskip}
In recent years, researchers have increasingly applied deep learning to synthetic data generation, using models such as GANs and VAEs under \dpns. 
A GAN consists of two neural networks --- a \emph{generator} and a \emph{discriminator} --- trained together in an adversarial framework (\cite{GANs}). The generator learns to produce realistic synthetic data by using feedback from the discriminator, which is trained to distinguish real from synthetic samples based on features learned from the real data. In contrast, VAEs learn a compressed representation --- often called a \emph{latent space} --- of the data and generate synthetic samples by decoding new points sampled from this learned representation (\cite{kingma2022autoencodingvariationalbayes}).%\medskip

The appeal of these models is their capacity to capture complex, high-dimensional distributions and generate very realistic data samples (as has been shown in non-private settings for images, text, etc.). The challenge is that deep generative models usually require a large number of training iterations and model parameters, which raises the risk of privacy leakage through memorization. To mitigate this, two main strategies have emerged: 
\begin{enumerate}
    \item train the model with a DP training algorithm (e.g., DP-SGD, see Section \ref{sec_DP-SGD}), or
    \item use a student-teacher framework like PATE (\cite{papernot2017semi,papernot2018scalable}) to train models with privacy.
\end{enumerate}
 
Several notable methods illustrate these approaches. DP-WGAN by \cite{uclanesl_dp_wgan} uses a Wasserstein GAN (WGAN) --- a GAN variant known for its stable training --- and applies \dps by adding noise to the stochastic gradient descent updates of both the generator and discriminator networks. By carefully tuning the noise and applying gradient clipping (as per DP-SGD techniques), the model can approximate the real data distribution while limiting the information leakage from each training step.
%\medskip

In addition, \cite{chen2021gswgangradientsanitizedapproachlearning} proposed \emph{Gradient-Sanitized Wasserstein GAN} (GS-WGAN), which enables the release of sanitized synthetic data with formal privacy guarantees. Unlike earlier approaches, GS-WGAN introduces a more precise method for distorting gradient information, allowing for the training of deeper models that can generate higher-quality, more informative synthetic samples. Furthermore, the method is designed to support both centralized and federated (i.e., decentralized) training settings, making it adaptable to diverse data-sharing environments.
%\medskip

Another approach is PATE-GAN, introduced by \cite{yoon2018pategan}, which adapts the PATE framework (\cite{papernot2017semi,papernot2018scalable}). In PATE-GAN, multiple teacher discriminators are trained on disjoint subsets of the real data without requiring privacy during individual teacher training. When training the generator, these teacher discriminators vote on whether generated samples are realistic, and their aggregated (noisy) vote provides a privacy-preserving feedback signal to either a student discriminator or directly to the generator. This ensemble voting mechanism ensures that no single teacher trained on a specific subset of data can disproportionately influence the feedback, as \dps is guaranteed through the added noise.%\medskip

 \cite{long2021gpatescalabledifferentiallyprivate} extended PATE-GAN idea with G-PATE, observing that only the information provided to the generator needs to be privatized, since the generator is the only component released after training. This insight allowed them to train a larger ensemble of discriminators on disjoint data subsets, improving utility while maintaining privacy guarantees. %\medskip

However,  \cite{ganev2025} analyzed and benchmarked six open-source PATE-GAN implementations and found that \emph{all of them leaked more privacy than intended}. This highlights the critical importance of implementation quality and careful privacy accounting: even theoretically sound mechanisms can fall short in practice if not rigorously executed.
%\medskip

While these GAN-based methods have demonstrated the potential to generate high-quality synthetic data in domains such as electronic health records and image datasets, the findings of Ganev et al.\ underscore that realizing this potential in practice requires robust implementation and rigorous auditing to ensure actual privacy guarantees align with theoretical claims.
%\medskip

Beyond GANs, researchers have also explored DP versions of VAEs and recurrent neural networks for sequential data. For example,  \cite{Acs2019} proposed the \emph{Differentially Private Generative Model} (DPGM) for synthetic data generation. This approach first clusters the original dataset into $k$ groups using differentially private kernel $k$-means, after which separate generative models --- such as Restricted Boltzmann Machines or VAEs --- are trained independently on each cluster using differentially private gradient descent.%\medskip

In addition, \cite{LiuTerrance2021IMfP} introduced \emph{Generative Networks with the Exponential Mechanism} (GEM), a deep generative model that directly parameterizes the data distribution using neural networks. While inspired by marginal-based methods like MWEM and PEP, which operate over query answers, GEM circumvents their computational bottlenecks by shifting the focus to optimizing the parameters of a generative model. This allows GEM to capture a richer class of distributions and leverage efficient gradient-based optimization techniques under DP constraints.%\medskip 

Recently, \emph{diffusion models} have emerged as a powerful alternative to GANs and VAEs for differentially private synthetic data generation.\footnote{While diffusion models differ significantly from GANs and VAEs in their training paradigm, they are grouped here under deep generative models as they also rely on deep neural networks to learn complex data distributions.} These models generate data by gradually transforming noise into realistic samples through a learned denoising process (\cite{ho2020denoising}).  \cite{dockhorn2023dpdm} introduced \emph{Differentially Private Diffusion Models} (DPDM), incorporating a tailored training strategy called \emph{noise multiplicity} to reduce gradient variance without additional privacy cost. \cite{ghalebikesabi2023dpdm} demonstrated that large pre-trained diffusion models (e.g., ImageNet-trained) can be fine-tuned under DP to produce high-quality synthetic data, with strong downstream utility on challenging datasets like CIFAR-10 and Camelyon17. \cite{wang2023dppromise} further improved this direction with \emph{dp-promise} (\emph{Differentially Private Diffusion Probabilistic Models for Image Synthesis}), a two-phase framework that leverages the inherent noise in the diffusion forward process as part of the privacy mechanism, reducing the need for additional noise injection and enhancing the privacy -- utility trade-off.%\medskip

Finally, a fundamentally different approach to synthetic data generation with generative models is the plausible deniability framework by \cite{bindschaedler2017plausibledeniabilityprivacypreservingdata}, discussed in Section~\ref{sec:other_approaches}. %\medskip

Overall, deep generative models can capture richer structure than simpler models, but their sensitivity to added noise remains a key limitation: GANs often suffer from mode collapse and reduced sample diversity, VAEs may produce degraded latent representations, and diffusion models, though more stable, can be computationally expensive. As a result, these methods often require larger datasets or higher privacy budgets to reach acceptable utility. Nevertheless, they represent a very active area of development because of their potential payoff: a well-trained DP generative model could enable near real-data fidelity for complex tasks like image recognition or clinical outcome prediction, while offering rigorous privacy guarantees.

\subsubsection{Other Approaches} \label{sec:other_approaches}
Apart from the main categories above, there are other strategies and theoretical frameworks for DP synthetic data. Notably,  \cite{BoedihardjoMarch2024Pmrw} introduce a method based on \emph{metric privacy}, a generalization of DP that provides stronger utility guarantees for a broader class of analyses. They emphasize that most DP mechanisms guarantee utility only for a fixed set of pre-specified queries and provide no assurances for common tasks like clustering or classification.%\medskip

To address this, \cite{BoedihardjoMarch2024Pmrw}  develop a polynomial-time algorithm that constructs a private measure --- a noisy but mathematically faithful approximation of the data distribution in terms of the Wasserstein distance. Synthetic data can then be sampled from this private measure, ensuring accuracy for a wide range of statistical analyses, particularly those based on Lipschitz-continuous functions, rather than just a few predetermined queries.%\medskip

A key technical contribution is a novel \emph{superregular random walk} used to sample from the private measure, ensuring that the synthetic data collectively approximate the true distribution at multiple scales. This work opens a path toward more general-purpose privacy-preserving data generation, with utility preserved across diverse statistical tasks.%\medskip

In addition, as mentioned above, a fundamentally different approach to achieving DP guarantees with generative models is proposed by \cite{bindschaedler2017plausibledeniabilityprivacypreservingdata}, who introduce a generative framework that enforces privacy through \emph{plausible deniability} rather than by directly injecting noise into the model or outputs. Synthetic records are first generated using a standard (non-private) generative model, and then a privacy test is applied to decide whether a candidate record can be safely released. By randomizing this test, the mechanism satisfies formal $(\varepsilon,\delta)$-DP guarantees. This explicit separation of the generative modeling and privacy enforcement stages enables the use of high-utility models while still providing strong privacy protection for the released synthetic data. %\medskip

To illustrate the practical implications of this taxonomy, we present below a simplified scenario demonstrating how each class of methods could be applied to the same dataset.\medskip
\newpage

\begin{mdframed}
 \begin{example}\label{exampleTaxonomy}
 \textbf{Applying DP Synthetic Data Methods to a Hospital Dataset} %\medskip

A hospital wants to release a differentially private synthetic dataset containing patient demographics (age, gender), diagnoses, and treatments.  %\medskip

\textbf{How would different approaches work?}  
\begin{itemize}
    \item \textbf{Histogram/Marginal-based:} Select key low-dimensional marginals (e.g., \{age, gender\}, \{gender, diagnosis\}), add DP noise to these counts, and generate synthetic patients that match the noisy distributions.
    \item \textbf{Graphical Models:} Learn a probabilistic network (e.g., age $\rightarrow$ diagnosis $\rightarrow$ treatment), privatize its conditional probabilities, and sample new patient records preserving these dependencies.
    \item \textbf{Deep Generative Models:} Train a neural network to produce realistic patient records while applying DP techniques (e.g., DP-SGD or teacher-student aggregation) to enforce privacy.
    \item \textbf{Other Approaches:} Use frameworks such as metric privacy, which builds a private approximation of the full data distribution (in terms of Wasserstein distance) to ensure accuracy for a wide range of analyses, or plausible deniability, which generates candidate synthetic records non-privately and applies a randomized privacy test to decide which records can be safely released.
\end{itemize}

%\textbf{Trade-offs:}  
%Marginal-based methods provide accurate low-dimensional statistics but may miss complex dependencies. Graphical models capture interpretable relationships but consume privacy budget during structure learning. Deep generative models can produce highly realistic data but require larger datasets or higher $\varepsilon$. Metric privacy offers broad, task-agnostic utility guarantees, while plausible deniability decouples modeling from privacy enforcement but may require careful tuning of the privacy test.
\end{example}
\end{mdframed}\bigskip

\subsection{Utility and Privacy Evaluation of Synthetic Data} 
Evaluating the quality and utility of privacy-preserving synthetic data is a challenging task. Unlike traditional model evaluation where one can directly compare outputs to ground truth labels, synthetic data quality is a more abstract concept — it involves assessing \emph{how well the synthetic dataset reproduces important properties of the real data} and \emph{how useful and accurate it is for intended downstream uses}. This latter aspect is often referred to as \emph{downstream evaluation}, where the synthetic data is used for training or validating models on tasks of interest to measure its practical utility. Common evaluation criteria include:
\begin{itemize}
    \item \emph{Statistical similarity to the original data} --- for example, comparing distributions of variables or correlations; 
 %Statistical similarity metrics: A basic check is to compare summary statistics between the real and synthetic datasets. This includes means, standard deviations, and ranges of each attribute, as well as correlation or covariance between attributes. 
   \item  \emph{Predictive performance of models trained on synthetic data} --- for example, training a classifier on synthetic data and testing on real data; and
    \item  \emph{Specific analytic results} --- for example, hypothesis tests or effect size estimates derived from synthetic versus real data. 
\end{itemize}
 Additionally, privacy evaluation such as measuring re-identification risk or membership inference leakage is integral, but here we focus on utility evaluation.
%A variety of evaluation metrics and protocols have been proposed for both aspects, and we outline the most common approaches below. 
%\medskip

%Assessing the quality of differentially private synthetic data is a multi-faceted problem: one must evaluate utility (how useful or informative the synthetic data is for intended tasks) as well as privacy (to what extent the synthetic data protects against disclosure of the original data). A variety of evaluation metrics and protocols have been proposed for both aspects, and we outline the most common approaches below. 

%\paragraph{Utility Evaluation} 
%\subsection{Utility Evaluation: Statistical Similarity and Task Performance} 
Utility of synthetic data broadly means that analyses conducted on the synthetic data should yield results comparable to those performed on the real data. This can be assessed through different \emph{evaluation protocols}. \cite{Movahedi2024} outline two primary protocols for evaluating classifiers trained on differentially private (DP) synthetic data: \begin{itemize} 
\item \textbf{Protocol A}: A machine learning model is both trained and evaluated on synthetic data, typically using a held-out synthetic test set. 
\item \textbf{Protocol B}: A model is trained on synthetic data but evaluated on a real (non-synthetic) test set --- in practice, this might involve a data curator, who holds the real data and can perform the evaluation in a DP-compliant manner. \end{itemize}

Protocol B provides a more direct measure of how well a model trained on synthetic data would perform in real-world scenarios. To support consistent evaluation, \cite{Movahedi2024} also introduce a metric for assessing how closely the results from these protocols reflect actual predictive performance on real data.
%\medskip

Experimental findings strongly support the use of Protocol B whenever feasible. The authors found that relying solely on synthetic data for both training and testing (Protocol A) can yield misleading results --- either overly optimistic or failing to reveal poor generalization to real data. In contrast, evaluation on real data, even when access is privacy-limited, produces more reliable estimates of real-world performance. As they note, "\emph{the results of our study suggest that employing the second protocol is advantageous, particularly in biomedical health studies where the precision of the research is of utmost importance.}" In high-stakes domains such as clinical decision support, even small differences in model performance can have serious consequences. Thus, Protocol B offers a more trustworthy assessment and should be preferred whenever possible.%\medskip

This insight offers an important practical guideline: for critical applications of DP synthetic data, real-data evaluation should be incorporated if at all feasible, to avoid unwarranted confidence in models trained solely on synthetic data.
 %\medskip

Beyond model performance, another critical axis of evaluation is the validity of scientific or statistical conclusions drawn from analyses on synthetic data. \cite{Perez2024} provide an illuminating example by examining the reliability of hypothesis testing on DP synthetic data.%\medskip

In their study,  \cite{Perez2024}  generated differentially private synthetic datasets from biomedical sources --- including a prostate cancer cohort and a cardiovascular study --- using a variety of methods (e.g., DP histogram techniques, MWEM, Private-PGM, DP-GAN). They then applied a standard nonparametric test (the Mann–Whitney U test for group differences) to the synthetic datasets and assessed the frequency of Type I errors (false positives) and Type II errors (missed true findings). A striking result was that many DP synthetic data methods led to inflated Type I error rates --- that is, the tests often detected \emph{statistically significant differences} that were entirely artifacts of the privacy noise. This problem was particularly pronounced under strict privacy settings, where heavy noise can introduce spurious differences large enough to be mistaken for real effects. As the authors caution, "\emph{low p-values may be obtained in statistical tests simply as a byproduct of the noise added to protect privacy}."
%\medskip

These findings underscore the need to calibrate inference procedures when analyzing DP synthetic data. Conventional interpretations of p-values and confidence intervals may no longer be valid without adjustment.  \cite{Perez2024} identified only one approach --- a DP smoothed histogram method by \cite{WassermanLarry2010ASFf} --- that successfully maintained valid Type I error control across tested privacy levels. However, this method required relatively large sample sizes and moderate privacy budgets to also preserve statistical power (i.e., to keep Type II error at acceptable levels).
%\medskip
 
The broader lesson for practitioners is clear: when using DP synthetic data for inferential analysis or AI benchmarking, one must verify that the statistical procedures remain valid. This may involve adjusting significance thresholds, using inference-aware synthetic data methods such as those proposed by \cite{raisa2023}, or validating conclusions against a holdout real dataset when feasible.

\subsection{Limitations and Challenges} 
While privacy-preserving synthetic data is a powerful concept, several limitations and challenges have been identified in the literature. We discuss some key issues: privacy limitations, representation biases, and fairness concerns that arise in DP synthetic data, as well as the broader regulatory and ethical implications.

\paragraph{Privacy is not automatic:} As highlighted earlier, synthetic data are not automatically exempt from privacy risks or regulations. Simply stripping identifiers and generating new records does not guarantee that individuals cannot be re-identified or that sensitive information won’t be inferable.  
There are many subtle ways a synthetic data generator can leak information about the training data (\cite{Perez2024}). For instance, a complex model might memorize outliers or specific patterns unique to an individual. DP provides a formal safety net by bounding the influence of any single record, but DP synthetic data must be carefully configured --- choice of 
$\varepsilon$, model architecture, training procedure, and noise injection all affect the level of protection in practice. Moreover, as highlighted in recent empirical work by \cite{ganev2025}, even theoretically sound mechanisms like PATE-GAN can leak more privacy than intended if the implementation is flawed or privacy accounting is inaccurate. This underscores that achieving meaningful privacy protection depends not only on the theoretical mechanism but also on its correct and rigorous implementation.%\medskip

Furthermore, one must remember that neither the process of synthetic data generation nor the resulting datasets are automatically outside the scope of data protection laws (\cite{Pitkamaki2024}). In the European GDPR context, for example, synthetic data may still be considered personal data if there is any reasonable way to link it back to real individuals. This means organizations cannot assume that using synthetic data frees them from compliance obligations like lawful basis for processing or individuals’ rights; a risk assessment is still needed to judge if the synthetic data is sufficiently de-identified or not.
\bigskip

\noindent
\textbf{Disparate impact on subgroups:} A salient challenge discovered by \cite{ganev2022robinhoodmattheweffects} is that DP synthetic data generation can have uneven effects across different subpopulations. When data is imbalanced --- say, some classes or demographic subgroups are underrepresented --- the addition of DP noise and the constraints of the DP training process can distort the minority group data either by over-smoothing or under-representing them. \cite{ganev2022robinhoodmattheweffects}  applied the terms "\emph{Robin Hood}" and "\emph{Matthew}” effects to describe these two extremes. %\medskip 

In some cases, the DP mechanism reduced the gap between majority and minority --- a Robin Hood effect, essentially taking from the rich and giving to the poor: the synthetic data over-samples or amplifies the minority group relative to the original. In other cases, it increased the gap --- a Matthew effect, where the rich get richer: the majority group dominates even more in synthetic data. Unfortunately, both scenarios led to disparate impacts in downstream analysis --- for instance, classification models trained on the synthetic data had systematically different accuracy for the minority group versus the majority group (\cite{ganev2022robinhoodmattheweffects}).
%\medskip

Underrepresented groups tended to suffer more in terms of model performance, meaning the synthetic data did not faithfully preserve their information. This is a serious concern if synthetic data is used in contexts like healthcare or criminal justice, where fairness across subgroups is critical. It implies that DP might inadvertently worsen biases present in data or even create new biases due to the noise. Therefore, synthetic data generators should be evaluated not just on overall utility, but also on subgroup fidelity. Recent works (see, e.g., \cite{pmlr-v238-abroshan24a,kenfack2022repfairganmitigatingrepresentationbias,vanbreugel2021decafgeneratingfairsynthetic}) suggest possible mitigation (e.g., adjusting the training objective to maintain subgroup statistics, or post-processing synthetic data to correct biases), but it remains an area requiring attention to ensure equity in data sharing.%%\medskip
\bigskip

\noindent
\textbf{Utility-performance trade-offs:}  
By design, adding privacy-preserving noise degrades the fidelity of synthetic data to some extent. A common limitation across all DP synthetic data generation methods is a loss in utility, particularly under very strict privacy requirements. Practically, if the privacy parameter $\varepsilon$ is set to a very low value (indicating strong privacy), the resulting synthetic data may be too noisy to be useful. %\medskip

 \cite{Perez2024} observed that under tighter privacy budgets, certain DP synthetic data methods yielded unreliable hypothesis test results (many false positives). Similarly, other studies have reported that predictive models trained on high-noise synthetic data often fail to learn meaningful patterns {(\cite{GaboardiMarco2016DPCH,raisa2023})}. %\medskip
 
There is, therefore, a practical range of privacy settings within which DP synthetic data remains useful. This range is context-dependent and influenced by factors such as dataset size and complexity. For instance, larger datasets can tolerate stricter privacy (i.e., more noise), since the noise can be averaged out across more samples. In contrast, small datasets may be overwhelmed by the added noise, severely impairing downstream utility. %\medskip

This interplay suggests that DP synthetic data is often most viable when \emph{moderate privacy levels} (e.g., $\varepsilon$ in the range of 1–10, depending on application and risk tolerance, \cite{Perez2024}) are acceptable. In some cases, privacy budgets can also be strategically allocated --- for example, reserving a portion for \emph{real-data evaluation} using approaches like Protocol B, in which model performance is tested on real data rather than synthetic holdouts. %\medskip

 These trade-offs also motivate research into more efficient use of privacy budgets --- for example, by synthesizing only the most sensitive parts of a dataset and retaining non-sensitive real data where possible, or by using hybrid models that combine DP synthetic data with other privacy-preserving techniques. However, such strategies must be implemented carefully, as they can complicate privacy accounting and risk assessments. %\medskip

\bigskip

\noindent
\textbf{Reliability and user trust:} Another challenge is convincing data stakeholders to trust synthetic data. If a synthetic dataset yields erratic analytical results --- such as fluctuating summary statistics or unexpected correlations not present in the original data --- users may lose confidence in its utility. Validity enhancements, such as those proposed by \cite{raisa2023}, help ensure that uncertainty is appropriately represented, which may increase user trust by offering a more transparent picture (e.g., wider confidence intervals under strong privacy constraints). Without such adjustments, users may either over-trust synthetic data --- leading to false discoveries or overconfident decisions --- or under-trust it, dismissing it as useless if it fails to closely mirror real data. Managing this perception is partly an engineering challenge (developing tools to assess quality) and partly an educational one (training users to interpret results from synthetic data appropriately). The literature increasingly acknowledges this sociotechnical dimension; for example, the Finnish PRIVASA project emphasizes the importance of clearly communicating the limitations of synthetic data and treating it as a complement to, rather than a replacement for, real data in critical settings~(\cite{Pitkamaki2024}).

\medskip
In conclusion, while significant progress has been made in DP synthetic data generation, open challenges remain. Ensuring fairness, maintaining utility at stringent privacy levels, and navigating the regulatory definitions of personal data are all active areas of work. These limitations do not imply that synthetic data is futile; rather, they highlight where careful implementation and further research are needed. As with any privacy technology, understanding failure modes is crucial for safe deployment.
When used with care, DP synthetic data can significantly expand the range of safe and meaningful data use cases, enabling analysis where direct access to real data would be ethically or legally prohibited.

\section{Enhancing Differential Privacy}\label{cha_enhancing}
% Advanced Techniques in Differential Privacy
With the increasing demand for privacy-preserving machine learning, this section explores how {differential privacy (DP)} is combined with other technologies to meet both privacy and utility requirements. We focus on three key directions: the integration of {DP with cryptographic techniques}, the application of {DP in federated learning (FL)}, and hybrid approaches that jointly leverage {FL, homomorphic encryption (HE), and DP}. These combinations are increasingly explored in research and are beginning to see practical adoption in industry (e.g., in telemetry and federated analytics), though continued work is needed to improve efficiency and scalability.

\subsection{Expanding DP with Cryptography}\label{sec_crypto}
DP offers strong mathematical guarantees for protecting individual data contributions during statistical analysis. However, it traditionally assumes either a trusted curator (in the standard central DP model) or accepts a substantial loss of accuracy (in the local DP model, where individuals randomize their own data before sharing). In practical deployments 
--- such as in Google’s RAPPOR (\cite{rappor}) or Apple's telemetry systems (\cite{apple}) ---  
this trade-off between trust and utility often limits the effectiveness of DP (\cite{wagh2021}).%\medskip

Recent research has therefore turned toward cryptography to bridge this gap.
For a thorough review of the emerging intersection between DP and cryptography --- termed \emph{DP-cryptography} --- we refer to the work of \cite{wagh2021}.
As they note, DP and cryptographic techniques serve complementary purposes: DP bounds what can be learned from outputs, while cryptography controls who gets access to what inputs or computations. This intersection has led to two main paradigms:
\begin{itemize}
    \item \textbf{Cryptography for DP}, where encryption and secure computation are used to implement DP without a central trusted curator;

    \item \textbf{DP for Cryptography}, where DP is used to relax otherwise rigid cryptographic guarantees, improving efficiency while maintaining provable privacy.

\end{itemize}
Together, these approaches enable new architectures that enhance DP by protecting sensitive data both during computation (via cryptography) and in the output (via noise addition).
 
\subsubsection{Cryptography for DP}
Techniques like {secure multiparty computation (SMPC)}, {anonymous communication}, and {trusted execution environments (TEEs)} can be used to implement DP systems without relying on a central trusted curator. These designs aim to preserve utility --- often lost in local DP settings --- while maintaining realistic trust assumptions in practice.
%\medskip

One example is the work by \cite{agarwal2019encrypted}, who introduce the concept of \emph{private encrypted databases} that simultaneously provide cryptographic protection against untrusted servers and DP against untrusted analysts. Their framework, based on structured encryption and additively HE, supports encrypted operations for data management and private operations for query analysis.
%\medskip

The key components are:
\begin{enumerate}
    \item \emph{Structured encryption} — enabling encrypted search and updates;
    \item \emph{Private encrypted counters} — constructed using HE and the binary mechanism;
    \item \emph{HPX architecture} — a full encrypted database system that supports differentially private histogram queries with formal guarantees against various adversaries, including persistent and snapshot attacks.
\end{enumerate}

While \cite{agarwal2019encrypted} focus on private data management and encrypted querying, other systems emphasize anonymous data collection and aggregation. A prominent example is \emph{Prochlo} (\cite{prochlo,Erlingsson_amp2019}), a system developed at Google for privacy-preserving telemetry collection. Prochlo follows the \emph{Encode–Shuffle–Analyze} (ESA) architecture: clients locally perturb their data using DP, and a semi-trusted shuffler anonymizes the data before aggregation. This architecture provides stronger privacy than local DP alone, as the shuffler amplifies privacy guarantees while maintaining utility.%\medskip

In a different application domain, \emph{DJoin} (\cite{Narayan2012}) demonstrates how cryptographic protocols can be used to securely compute set joins across multiple data owners while satisfying central-model DP. It relies on SMPC to compute noisy intersections without requiring any party to reveal its raw data, assuming only semi-honest adversaries.
%\medskip

These approaches and systems illustrate the core idea behind the "\emph{cryptography for DP}" paradigm: using cryptographic techniques to eliminate or reduce trust in the central server, while enabling differentially private computation with improved utility.

    %applies secure multiparty computation (MPC) to compute differentially private set operations without a trusted party;

%Example: DJoin enables privacy-preserving computation of dataset intersections without a trusted curator.
%Combining DP with secure computation can achieve strong privacy with acceptable performance trade-offs.

\subsubsection{DP for Cryptography}
Conversely, DP can be used to \textit{relax} perfect cryptographic guarantees, enabling more efficient implementations that may leak limited information but remain provably private. Instead of aiming for strict indistinguishability, systems in this category introduce {DP noise} to outputs or internal states, accepting controlled leakage in exchange for significant gains in efficiency and scalability.
%Conversely, DP can be used to \emph{relax} perfect cryptographic guarantees, allowing for more efficient but "leaky" implementations that are provably private. For example, adding DP noise to decryption outputs or intermediate MPC states reduces padding and overhead in secure computation, as seen in systems like Shrinkwrap and Vuvuzela:%\medskip
%\begin{itemize}
%    \item \textbf{Shrinkwrap} \cite{Shrinkwrap} is a system for accelerating secure SQL query execution under DP. It reduces the performance cost of oblivious computation by applying DP noise to intermediate result sizes, allowing the system to avoid worst-case padding while still preserving privacy in the semi-honest model.
%    \item \textbf{Vuvuzela} \cite{Vuvuzela} is a scalable anonymous messaging system that uses DP to defend against traffic analysis attacks. It adds dummy messages and noise to communication metadata, ensuring that whether a user is idle or actively messaging is indistinguishable --- offering provable privacy at Internet scale.
%\end{itemize}
%\medskip

For example, \emph{Shrinkwrap} (\cite{Shrinkwrap}) accelerates secure SQL query execution by applying DP noise to intermediate result sizes. This allows the system to reduce the overhead of oblivious computation by avoiding worst-case padding, while still providing formal privacy guarantees in the semi-honest adversarial model. Similarly, \emph{Vuvuzela} (\cite{Vuvuzela}) is an anonymous messaging system that uses DP to protect against traffic analysis attacks. By injecting noise and dummy messages into communication metadata, it ensures that user activity patterns remain indistinguishable --- even at Internet scale.%\medskip

A different application of DP within cryptographic protocols is demonstrated by \cite{Li2022}, who propose a method for \textit{hardening approximate HE} schemes. In schemes such as CKKS (Cheon-Kim-Kim-Song), which support approximate arithmetic over encrypted real numbers, decryption errors can leak sensitive information despite standard IND-CPA security (i.e., secure against chosen plaintext attacks under standard cryptographic assumptions). To address this, the authors propose \emph{post-processing decrypted values} with a differentially private mechanism --- specifically, a Gaussian mechanism calibrated to the worst-case noise in the ciphertext.%\medskip
 
This modification elevates the security level to \emph{IND-CPAD} (IND-CPA with decryption oracle defense), offering stronger protection against adversaries who observe decryption outputs over time. \cite{Li2022} show that adding a modest amount of Gaussian noise --- proportional to the ciphertext error --- is enough to prevent leakage without significantly reducing output quality. 
This demonstrate the "\emph{DP for cryptography}" paradigm in action: DP not only protects data outputs but also enhances the security of cryptographic primitives in practice.

\subsubsection{HE Noise as an Implicit Source of Differential Privacy}
A novel perspective within the intersection of cryptography and DP is the idea that \textit{the inherent noise in approximate HE} might itself offer DP guarantees --- without the need to inject additional noise explicitly. \cite{Ogilvie2024} investigates this possibility using the CKKS scheme, which supports approximate arithmetic over encrypted real numbers and naturally accumulates noise during computation.%\medskip

The analysis shows that this idea is not merely theoretical. In practical settings such as ridge regression, the accumulated noise from CKKS operations can yield meaningful privacy guarantees (e.g., $\varepsilon \approx 2$ after 50 iterations, \cite{Ogilvie2024}). However, a significant limitation is that this \textit{noise is message-dependent}, meaning its magnitude varies with the input and may expose sensitive information through side-channel leakage.%\medskip

Despite this limitation, the approach opens the door to hybrid methods that reduce or even eliminate the need for externally added DP noise. This could improve computational efficiency while maintaining a quantifiable level of privacy. In this way, the work highlights a promising direction: aligning DP guarantees with the intrinsic properties of cryptographic schemes to build lighter-weight privacy-preserving systems.
 
%Cryptographic Primitives in Context:

%Enable privacy-preserving operations like secure multiparty computation (MPC) and fully homomorphic encryption (FHE).
%High computational overhead limits their practical deployment.

\medskip
Table~\ref{tab:dp_crypto} summarizes representative approaches at the intersection of DP and cryptography, highlighting their paradigms, key techniques, and application domains.

\begin{table}[ht]
\caption{Summary of Approaches at the Intersection of DP and Cryptography}\label{tab:dp_crypto}
\resizebox{\textwidth}{!}{\begin{tabular}{@{}llll@{}}
\toprule
\textbf{Paradigm} & \textbf{Example Systems} & \textbf{Key Idea} & \textbf{Use Case}\\
\hline
\begin{tabular}[c]{@{}l@{}}\textbf{Cryptography}\\
\textbf{for DP} \end{tabular}& 
\begin{tabular}[c]{@{}l@{}}Agarwal et al. \\ %\cite{agarwal2019encrypted} \\ 
Prochlo (ESA) \\ %\cite{prochlo,Erlingsson_amp2019} \\ 
DJoin \\ %\cite{Narayan2012}
\end{tabular} & 
\begin{tabular}[c]{@{}l@{}}Use SMPC, HE, TEEs, \\ or shuffling 
to enable \\ DP without full trust\end{tabular} & 
\begin{tabular}[c]{@{}l@{}}- Private analytics, \\ - Telemetry,\\ 
- Multi-owner data joins\end{tabular} \\
\midrule
\begin{tabular}[c]{@{}l@{}}\textbf{DP for } \\
\textbf{Cryptography}\end{tabular}  & 
\begin{tabular}[c]{@{}l@{}}Shrinkwrap \\ %\cite{Shrinkwrap} \\ 
Vuvuzela \\ %\cite{Vuvuzela} \\ 
Li et al. (CKKS) \\ %\cite{Li2022}
\end{tabular} & 
\begin{tabular}[c]{@{}l@{}}Use DP noise to relax \\ crypto  
guarantees for \\ efficiency or
side‑\\ channel resistance.\end{tabular} & 
\begin{tabular}[c]{@{}l@{}}- Efficient secure queries,\\ 
- Anonymous messaging,\\ 
- Hardened HE schemes\end{tabular} \\
\midrule
\textbf{HE Noise as DP} & 
\begin{tabular}[c]{@{}l@{}}Ogilvie (CKKS)\\ % cite{Ogilvie2024}
\end{tabular} & 
\begin{tabular}[c]{@{}l@{}}Leverage inherent HE \\ noise 
to provide DP \\ guarantees.\end{tabular} & 
\begin{tabular}[c]{@{}l@{}}- DP at low added cost \\ 
(e.g., ML training)\end{tabular} \\
\bottomrule
\end{tabular}}
\end{table}

\subsection{Differentially Private Federated Learning}
\label{sec_FL}
FL enables collaborative training of machine learning models without centralized data collection. In this paradigm, {\em clients} --- such as user devices or institutions --- train models locally on their private data and share updates with a central server, which aggregates them to build a global model. 
This general approach includes both \emph{cross-device FL}, 
where the clients are typically millions of personal devices (e.g., smartphones and other IoT devices), and \emph{cross-silo FL}, where the clients are a small number of organizations (e.g., hospitals or banks) with more stable and more structured data. These two settings differ in terms of scale, data characteristics, and system constraints, and therefore often require different design choices in practice (\cite{kairouz:hal-02406503}). %\medskip

While FL architectures avoid centralized raw data collection, model updates themselves may leak sensitive information, enabling attacks such as membership inference or model inversion (\cite{9308910,kairouz:hal-02406503,8835245}).  
To protect against such threats, DP is often integrated into FL systems, ensuring that individual contributions cannot be reverse-engineered from the global model. For a systematic review on the topic, we refer to the work of \cite{fu2024}. A broader overview of the FL landscape --- including privacy, optimization, system challenges, cross-device and cross-silo settings, and open problems --- can be found in the comprehensive survey by \cite{kairouz:hal-02406503}.

%where a large number of unreliable user devices (e.g., smartphones or sensors) participate, and \emph{cross-silo federated learning}, where a small number of stable clients (e.g., hospitals or banks) contribute structured datasets.
% Each setting presents different challenges and opportunities for privacy accounting, personalization, and communication efficiency.

\subsubsection{DP Settings, Granularity, and Accounting Mechanisms}
The DP literature in FL typically distinguishes between three privacy models: \emph{central DP}, \emph{local DP}, and the \emph{shuffle model}. These models differ in trust assumptions and where noise is introduced (see Section~\ref{sec_PrivacySettings}), and they are orthogonal to the FL setting itself (i.e., cross-device vs.\ cross-silo). Different combinations of privacy model and FL setting are used depending on deployment constraints and trust relationships.
%\medskip

%The DP literature in FL typically distinguishes between three privacy models: {central DP}, {local DP}, and the {shuffle model}. These differ in trust assumptions and where noise is introduced (see Section~\ref{sec_PrivacySettings}). 
In the context of FL, the \emph{shuffle model} --- where anonymized, randomized client messages are routed through a semi-trusted shuffler before aggregation --- can be viewed as a special form of {distributed DP}, since no single party observes both the raw data and final outputs. While some surveys (e.g., \cite{fu2024}) treat the shuffle model as a distinct category between local and central models, it fits naturally within our broader definition of distributed DP.
%\medskip

Within the central DP setting, privacy can be applied at different levels of granularity (unit of privacy, see Section \ref{sec_DPdefinition}):
\begin{itemize}
    \item \emph{Sample-level DP} protects individual samples in a client’s local dataset;
    \item \emph{Client-level DP} protects entire clients, ensuring that participation of a client cannot be inferred.
\end{itemize}
This distinction is specific to the central model; in local and shuffled settings, privacy is applied per user (client) by default, so separate notions of sample- or client-level DP are not required.
%Client-level DP is more stringent and better aligned with FL's threat model, where users may be concerned about participating at all.
%\medskip

To track cumulative privacy loss over many training rounds --- common in FL --- modern differentially private FL methods use advanced composition techniques that offer tighter accounting than the basic DP compositions given in Section \ref{sec_DPproperties}. These include Rényi DP (\cite{RDP2017}), Gaussian DP (\cite{DongJinshuo2022Gdp}), Moments Accountant (\cite{abadi2016deep}), and zCDP (\cite{BunMark2016CDPS}).
%, and are widely implemented in tools like TensorFlow Privacy and Opacus.
%\medskip %\enlargethispage{0.1\baselineskip}

A comprehensive summary of DP settings and mechanisms, protection levels, and accounting methods used in FL can be found in Table~3 of  \cite{fu2024}, which provides an excellent overview of the current landscape. Here, we discuss some of the works in each DP setting category as examples and give Table~\ref{tab:dp_fl} that summarizes the representative approaches discussed here, serving as an illustrative --- not exhaustive --- overview of differentially private FL.

\subsubsection{Federated Learning with Central Differential Privacy}\enlargethispage{\baselineskip}

In central DP setting, clients train models locally and send updates to a central server, which is assumed to be at least {semi-trusted}. The server adds noise to the aggregated updates to enforce DP. This model provides better utility than local DP because noise is only added once, after aggregation. However, it requires the server not to store or inspect individual updates unless combined with secure aggregation.%\medskip

The \emph{DP-FedAvg} algorithm, % % Fu luokittelee CDP ja client-level (tämän perusteella cross-devise), ChatGPT: originally cross-devise FL but used also cross-silo
introduced by \cite{McMahan2018b}, extends the standard Federated Averaging framework to provide user-level (client-level) central DP by clipping each client’s model update and adding Gaussian noise to the aggregated update on the server. Building on this foundation, \cite{andrew2021adaptive} % Fu luokittelee CDP ja client-level (tämän perusteella cross-devise)
propose a method for \emph{adaptive clipping}, which dynamically adjusts the clipping threshold during training based on a differentially private estimate of the quantile of client update norms. This approach reduces the need for manual hyperparameter tuning and improves model utility without compromising privacy guarantees. The method is compatible with other FL techniques like secure aggregation and is validated across several public FL benchmarks.
%\medskip

While DP-FedAvg and its refinements aim to improve utility under fixed or adaptive (but shared across all clients) privacy budgets, other work explores more flexible privacy models. \cite{10032626} propose  \emph{HDP-FL}, a FL framework that incorporates \emph{heterogeneous DP} (see Section \ref{sec_DPvariants}) to accommodate varying privacy preferences across clients. Unlike traditional approaches that assign a uniform privacy budget to all participants, HDP-FL allows clients to declare individualized privacy levels, balancing their privacy needs against the utility of the global model. % Fu luokittelee CDP ja sample level, clientit ei kuitenkaan ole esim sairaaloita tms. yhteneväisiä objekteja cross-devise???
In addition, it introduces a \emph{contract-theoretic incentive mechanism}, derived from economic theory, to ensure that clients truthfully declare their privacy sensitivity and adhere to the agreed DP noise levels. The framework supports both Laplace and Gaussian mechanisms and includes a theoretical convergence analysis. Experimental results demonstrate improved utility and truthful participation compared to homogeneous DP baselines.%\medskip
 
In addition to these cross-device settings, where clients are typically numerous and resource-constrained, recent work has also explored how DP can be integrated into {cross-silo FL}, where each client is an institution holding many user records (samples). 
 \cite{liu2024crosssilofederatedlearningrecordlevel} propose \emph{rPDP-FL}, a framework that supports sample-level personalized DP by combining client- and record-level sampling with Rényi DP accounting. Their approach accommodates heterogeneous privacy budgets even within the same client and shows improved utility under strict privacy constraints. In contrast, \cite{10.5555/3600270.3600699} advocate for \emph{silo-specific sample-level DP} (SSS-DP), where each silo sets its own sample-level privacy target. They evaluate the effectiveness of different personalization strategies under this privacy model and find that a simple mean-regularized multi-task learning (MR-MTL) baseline achieves strong performance with minimal privacy overhead. Together, these studies highlight the importance of adapting privacy granularity and personalization to the structure and needs of cross-silo federated settings.
%\medskip

These research developments reflect the increasing sophistication of DP mechanisms in FL and have informed the design of privacy-preserving systems at scale. One prominent example is Google's Gboard, which provides a large-scale, real-world deployment of FL with central DP. \cite{Gboard2023} train next-word prediction and other language models using the DP-FTRL algorithm, which applies noise centrally after secure aggregation of client updates. The system achieves formal zCDP guarantees, with all deployed neural language models now satisfying DP constraints. This work demonstrates the feasibility of training high-utility FL models with strong central privacy guarantees in a production setting.

\subsubsection{Federated Learning with Local Differential Privacy}%\enlargethispage{\baselineskip}

When combining FL with local DP, each client applies DP noise to their updates before sending them to the server. This setting removes the need to trust the server entirely, but generally leads to lower utility, since noise is added per client. It is particularly relevant when the threat model assumes a fully untrusted aggregator.%\medskip

\cite{10091486} propose \emph{PPeFL}, an edge-based FL framework that provides strong local DP guarantees by introducing a set of novel data perturbation mechanisms. Each participant perturbs selected model parameters locally using techniques such as filtering and screening with the exponential mechanism (FS-EM), strong privacy perturbation (DPM-SP), and enhanced utility perturbation (DPM-EU) before sharing with edge nodes. The architecture leverages intermediate edge aggregation to reduce communication with the central server and improve efficiency. The system supports varying privacy needs and achieves competitive model accuracy under strict privacy budgets. Experimental results show that PPeFL offers both strong privacy protection and practical utility, particularly in edge computing environments.
While PPeFL is designed to accommodate different privacy requirements across clients, it does not explicitly implement or analyze heterogeneous privacy budgets. We now turn to methods that formally define and enforce heterogeneous DP, assigning personalized privacy levels to each participant.
%\medskip
 
\cite{LingJie2024Eflp} address the utility degradation often associated with applying local DP in FL by proposing a method that combines DP-SGD with heterogeneous privacy budgets. Each client adds locally calibrated DP noise to their gradients before sending them to the server, with the noise level determined by the client's chosen privacy budget. To mitigate the impact of high-noise updates, the server employs a privacy-weighted aggregation mechanism, assigning lower weights to clients with stricter privacy. The system also incorporates preprocessing techniques, such as feature extraction using HOG (histogram of oriented gradients) and CNNs, to improve learning efficiency. Experiments on benchmark datasets (e.g., MNIST and CIFAR-10) demonstrate that this approach significantly improves model accuracy and reduces communication overhead, particularly in scenarios with many clients and diverse privacy requirements.
%\medskip

In addition, \cite{9253545} propose a FL framework for IoT systems that enforces heterogeneous local DP through gradient perturbation. Their method, called \emph{LDP-FedSGD}, allows each client to add calibrated noise to its gradient updates based on an individually selected privacy budget. The server aggregates only the noised updates, ensuring privacy even under an honest-but-curious threat model. To balance privacy and utility, the authors introduce four local mechanisms --- \emph{Three-Outputs}, PM-OPT, PM-SUB, and HM-TP (a hybrid of Three-Outputs and PM-SUB) mechanism --- and show through experiments on real-world datasets that their approach improves model accuracy and communication efficiency compared to baseline local DP methods.

%%%%%%%%%%%%%%%%
% authors!!! citation!!!
Unlike heterogeneous DP methods that tailor privacy budgets to client preferences, \emph{ProxyFL} introduced by \cite{KalraShivam2023Dflt} addresses a different challenge of training differentially private federated models when clients have \emph{limited local data}. Each participant maintains two models: a \emph{proxy model}, trained under DP-SGD and shared with the server, and a \emph{local model}, updated privately for inference.  This separation enables local DP without the need for secure aggregation. ProxyFL reduces update dimensionality, improves convergence under strict privacy budgets, and is 
particularly effective in federated settings with many clients, strict privacy requirements, or limited per-client training data. Experimental results demonstrate enhanced training stability and performance compared to standard DP-FL baselines.

\subsubsection{Federated Learning with Distributed Differential Privacy}\label{sec_distributedFL}
In distributed DP settings, privacy is achieved without trusting a single central server to add noise. This includes techniques like anonymizing shufflers (as in the shuffle model) and secure aggregation protocols combined with local noise addition. These approaches ensure that no party observes both raw data and the final output, while often providing stronger utility than pure local DP.%\medskip

 \cite{KairouzPeter2021TDDG} present a privacy-preserving FL framework that combines {local noise} addition with {secure aggregation}. Each client quantizes its model updates and adds discrete Gaussian noise locally before transmitting them. These privatized updates are then securely aggregated using a cryptographic protocol by \cite{Bonawitz2017} that ensures the server learns only the aggregated sum, not any individual update. The authors provide a rigorous privacy analysis under zCDP, accounting for the effects of quantization, modular clipping, and noise discretization. This approach exemplifies a distributed DP setting in which privacy is preserved without requiring trust in a central server to add noise.%\medskip

While the Kairouz approach relies on local noise addition combined with secure aggregation to hide individual updates, another prominent strategy in the distributed DP setting is the \emph{shuffle model}, where privacy is amplified not by cryptographic aggregation but by routing locally randomized messages through a semi-trusted shuffler that anonymizes and mixes them before aggregation.
For example, \cite{pmlr-v130-girgis21a} introduce \emph{CLDP-SGD}, a communication-efficient FL algorithm that provides privacy guarantees in the shuffled DP model. The method combines local DP, client and data subsampling, and a secure shuffler to amplify privacy while maintaining utility. A key contribution is the integration of compressed and private gradient updates, which drastically reduces communication cost per client while preserving DP. The authors provide a detailed privacy analysis that composes sampling and shuffling, and show that their approach matches known lower bounds for centralized private empirical risk minimization under convex loss. %Preliminary experiments on MNIST support the theoretical guarantees.
%\medskip

In addition, \cite{Liu_Cao_Chen_Guo_Yoshikawa_2021} propose \emph{FLAME}, a FL framework in the shuffle model. FLAME avoids reliance on a trusted analyzer by inserting a semi-trusted shuffler between clients and server, allowing privacy amplification through message anonymization. The framework includes three protocols --- \emph{SS-Simple}, \emph{SS-Double}, and \emph{SS-Topk} --- which progressively improve privacy-utility trade-offs using subsampling and gradient sparsification. Notably, SS-Topk achieves higher accuracy than both local and curator-based DP-FL baselines under the same central privacy budget, demonstrating that shuffle-based methods can provide strong privacy with minimal accuracy loss in practice.%\medskip
\medskip

Table~\ref{tab:dp_fl} summarizes the representative approaches of differentially private FL discussed here. 
%For a more comprehensive overview of DP in FL, see Table 3 in %\cite{fu2024}.

\begin{table}[ht]
\caption{Summary of Differentially Private FL Approaches by DP Setting}\label{tab:dp_fl}
\resizebox{\textwidth}{!}{\begin{tabular}{@{}llll@{}}
\toprule
\textbf{DP Setting} & \textbf{Example Systems} & \textbf{Key Idea} & \textbf{Use Case}\\
\midrule
\textbf{Central DP} & 
\begin{tabular}[c]{@{}l@{}}DP-FedAvg \\%\cite{McMahan2018b} \\ 
Adaptive Clipping \\%\cite{andrew2021adaptive} \\
HDP-FL \\%\cite{10032626} \\
DP-FTRL \\%\cite{Gboard2023} \\
rPDP-FL \\%\cite{liu2024crosssilofederatedlearningrecordlevel} \\
SSS-DP \\%\cite{10.5555/3600270.3600699}
\end{tabular} & 
\begin{tabular}[c]{@{}l@{}}Noise added at \\ the server after \\ aggregating client \\ updates; supports \\ sample- or client- \\level DP.
\end{tabular}  & 
\begin{tabular}[c]{@{}l@{}}- Mobile keyboard models \\ (e.g., Google Gboard), \\
- Adaptive privacy in FL, \\
- Cross-silo collaboration \\
(e.g., healthcare), \\
- Large-scale deployments\end{tabular} \\
\midrule
\textbf{Local DP} & 
\begin{tabular}[c]{@{}l@{}}PPeFL \\%\cite{10091486} \\ 
Heterogeneous \\ \quad LDP-FL \\%\cite{LingJie2024Eflp,9253545} \\
ProxyFL \\%\cite{KalraShivam2023Dflt}
\end{tabular} & 
\begin{tabular}[c]{@{}l@{}}Clients perturb \\ updates locally,  \\ supports personalized \\
budgets, client-level \\DP by default, \end{tabular}
&\begin{tabular}[c]{@{}l@{}}- Edge computing, \\ - IoT systems, \\ - Small-data clients\end{tabular} \\
\midrule
\begin{tabular}[c]{@{}l@{}}\textbf{Distributed DP/}\\
\textbf{Shuffle} \end{tabular} & 
\begin{tabular}[c]{@{}l@{}}zCDP + Secure \\ \quad Aggregation \\%\cite{KairouzPeter2021TDDG} \\ 
CLDP-SGD \\%\cite{pmlr-v130-girgis21a} \\
FLAME \\%\cite{Liu_Cao_Chen_Guo_Yoshikawa_2021}
\end{tabular} & 
\begin{tabular}[c]{@{}l@{}}Combines local noise \\ with secure aggre- \\ 
 gation or shuffler \\anonymization, \\ 
provides client-level \\ DP with privacy \\ amplification\end{tabular} & 
\begin{tabular}[c]{@{}l@{}}- Privacy without trusting \\ the server, \\ - Communication-efficient \\ FL in multi-party settings.\end{tabular} \\
\hline
\end{tabular}}
\end{table}

\ignore{
\subsection{Conclusion}
In summary, DP has become a central component of FL systems, addressing the privacy risks posed by sharing model updates instead of raw data. Depending on the trust assumptions and deployment constraints, privacy can be enforced using central, local, or distributed DP models --- each with distinct trade --- offs between utility, communication, and implementation complexity. Recent advances include methods with adaptive clipping, personalized/heterogeneous privacy budgets, and lightweight architectures designed for low-data or resource-constrained clients. Together, these developments demonstrate the growing maturity and versatility of DP-FL approaches, paving the way for scalable, privacy-preserving machine learning across diverse application domains.
}

\ignore{
The hybrid DP and compression approach by Jiang et al. (2023)  \cite{Jiang2023} targets industrial edge computing – they compress local model gradients to reduce communication, then apply an adaptive DP noise mechanism to protect those gradients during transmission, effectively preventing inference attacks on the industrial data. These applications demonstrate that across industrial IoT and control systems, integrating DP often requires tailoring the mechanism (feature transformation, adaptive noise, etc.) to preserve the physical or statistical relationships that indicate anomalies.}

\subsection{Integrating DP, Cryptography, and FL}
Recent advances have explored how DP and cryptographic techniques can be combined within FL to strengthen security guarantees and support deployment in adversarial environments. While FL already avoids centralizing raw data, sensitive information can still be leaked through model updates. DP addresses this by bounding the influence of any single client’s data, and cryptographic methods --- especially HE --- offer protection against curious or malicious servers that might inspect updates before or after aggregation.

%\medskip  
%%%%%%%%%%%%%%%%%%%%
\cite{grivetsebert:cea-04485721} propose a rigorous framework that combines DP and HE in FL settings. To address the incompatibility between HE and DP noise, it introduces a novel \emph{Poisson-based stochastic quantization operator}, 
which allows encrypted computations while retaining formal DP guarantees. In their system, clients generate and clip updates locally, add {Gaussian noise}, and then {encrypt the noised updates using the BFV (Brakerski/Fan-Vercauteren) HE scheme}. A key contribution is showing that the quantization required for HE can be treated as a post-processing step, preserving the original DP guarantees. Their experiments on the {FEMNIST dataset} demonstrate that this combined approach can achieve state-of-the-art privacy–utility trade-offs with only a marginal increase in computation time due to encryption.%\medskip 

\cite{9448383} present a system for the Industrial Internet of Things (IIoT) that incorporates DP, HE, FL, and \emph{blockchain}. They design a {data protection aggregation scheme} where $k$-means clustering, random forests, and AdaBoost models are trained across distributed clients under both {DP and HE protections}. Their scheme incorporates {blockchain} as a coordination and integrity layer, enhancing auditability and trust across devices. HE ensures that computations on encrypted updates are possible without decryption, while {DP mechanisms} control information leakage from the aggregated results. Their evaluation demonstrates improved accuracy and privacy, as well as system robustness under constrained industrial scenarios.%\medskip

Together, these works illustrate a promising direction for FL systems where multiple layers of protection --- statistical, cryptographic, and architectural --- are integrated. While challenges remain in terms of communication efficiency, computational cost, and threat modeling, these hybrid frameworks offer a strong foundation for secure and privacy-preserving machine learning in real-world deployments.

\section{Use Cases of Differential Privacy} \label{chaUseCases}
Differential Privacy (DP) has emerged as a foundational concept for protecting individual privacy in data analysis and machine learning. While its theoretical underpinnings are well established, practical applications across various domains continue to expand and evolve. This section explores a range of use cases where DP plays a critical role in enabling privacy-preserving data sharing and computation. We begin with applications in cybersecurity, focusing on DP in cyber-physical systems (CPS), anomaly detection, and face recognition, highlighting how DP mitigates risks in sensitive and high-stakes environments. The section then turns to healthcare and finance --- two data-intensive sectors where privacy concerns are paramount --- and examines how DP facilitates secure analytics while complying with regulatory and ethical standards. Together, these case studies illustrate the versatility and growing importance of DP in real-world systems.

\subsection{Cybersecurity Applications of DP}
\label{sec_cybersecurity}
%Privacy-Preserving Vulnerability Analysis: Organizations might be hesitant to share detailed SBOMs because they reveal sensitive information about software dependencies, potentially exposing security risks. Applying DP can help anonymize this data, enabling secure sharing without revealing proprietary details.
Cybersecurity applications often involve the analysis of sensitive data such as network traffic logs, user authentication records, malware samples, and system telemetry. Applying DP in this domain allows organizations to derive security insights or share threat-related information without exposing personal or proprietary details. In this section, we explore three use cases in cybersecurity: CPS, anomaly detection, and face recognition. CPS are increasingly deployed in critical infrastructure and industrial environments, where DP helps protect operational data and support secure collaboration. Anomaly detection leverages behavioral patterns for identifying attacks or misconfigurations, and DP enables the sharing of detection models or insights without revealing underlying user behavior. Face recognition systems process highly identifiable visual data, and integrating DP helps mitigate the risks of biometric data leakage. For each use case, we examine how DP can be applied and highlight technical approaches or case studies that demonstrate its practical benefits in safeguarding privacy while supporting robust security analytics.

%Cybersecurity applications often involve analyzing sensitive data: network traffic logs, user authentication records, malware samples, threat intelligence reports, etc. Applying DP in this domain enables organizations to gain security insights or share information with partners without revealing personal or proprietary details. In this section, we explore three key use cases in cybersecurity: anomaly detection, face recognition, and CTI. For each, we discuss how DP can be employed, and highlight technical approaches or case studies demonstrating the benefits of DP.

\subsubsection{Differential Privacy in Cyber-Physical Systems}
CPSs integrate computational intelligence with physical processes through interconnected sensors, actuators, embedded controllers, and  communication networks. Examples include smart grids, electric vehicle infrastructure, industrial automation, smart healthcare, cyber-physical metaverse systems (CPMS), and resource-constrained edge environments. These systems continuously generate vast amount of sensitive telemetry and behavioral data, which are crucial for tasks such as control, anomaly and intrusion detection, predictive maintenance, and stability assessment in critical infrastructures. However, this data often contains private or safety-critical information, and its exposure --- whether through shared analytics platforms, machine learning models, or cloud services --- can pose severe risks.%\medskip
 
DP offers a principled solution to this privacy challenge by introducing controlled randomness to statistical computations and machine learning pipelines, making it difficult for adversaries to infer sensitive details about individuals or devices. Compared to traditional methods like encryption and $k$-anonymity, DP is particularly well-suited to CPS environments due to its lightweight nature and ability to provide formal privacy guarantees without needing to modify raw datasets (\cite{8854247}). This is critical for CPSs, which often operate under real-time constraints and resource limitations.
 A comprehensive survey of DP techniques tailored for CPSs is given by \cite{8854247}. % survey
%\medskip
\bigskip

\noindent
\textbf{Centralized CPS:}
%In centralized CPS monitoring, DP is used to sanitize real-time data streams or control signals before aggregation, enabling privacy-preserving status reporting, fault detection, or performance monitoring. For instance, DP can be applied to noisy state estimation in smart grid monitoring or to anonymize device-level telemetry in smart manufacturing.\medskip
In centralized CPSs such as smart buildings or energy systems, DP is applied to protect real-time data reporting and query results, often through mechanisms like Laplace or Gaussian noise addition. Techniques such as point-wise privacy by \cite{Eibl2017} and demand-response perturbation by \cite{987893e673a24eaa8bb63c5d0011c47b} help safeguard user behaviors without compromising grid efficiency or automation performance. For example, DP has been employed to protect smart meter readings (\cite{BARBOSA2016355}), real-time stream processing in Internet of Things (IoT) (\cite{10.1145/3133956.3134102}), and fog node aggregation in smart grid scenarios (\cite{CaoHui2019Adpa}). \cite{YogiManasKumar2025Anuc} present a user-centric privacy framework based on user-level DP and query schedulers to protect access patterns and individual privacy during CPS database interactions, and \cite{10.1145/3659582} address task privacy in CPMS using DP-based task obfuscation in mobile crowd sensing. While the last is not purely centralized, the scheme assumes a central data processing center for query coordination. 
\bigskip

\noindent
\textbf{Distributed CPS:}
In distributed CPS and Industrial IoT (IIoT) settings, DP is increasingly integrated with federated learning, secure aggregation, and edge computing. These hybrid approaches enable collaborative model training or event detection without transmitting raw sensor data, thereby reducing privacy risks while preserving functionality. For instance, \cite{9875039} develop a federated intrusion detection system (IDS) for electric vehicle charging infrastructure using utility-optimized local DP and reinforcement learning-based budget allocation. \cite{JIANG2024103108} introduce a resource-aware federated learning framework that combines local DP with model parameter selection to reduce bandwidth use in constrained CPS environments, and  \cite{ZHANG202217} propose PEMFL that uses DP with momentum-optimized federated learning to secure industrial CPS model updates.%\medskip 

%Mechanisms such as dual variable perturbation, differentially private load monitoring, and fog-based aggregation show promise in enabling privacy-preserving operations across vast, heterogeneous CPS deployments \cite[148, 160, 284]{DP_CPS_Survey}. \medskip

For intrusion detection, \cite{FRIHA2023103097} propose a fully decentralized federated IDS with DP for IIoT, and \cite{BASAK2023108661} present a DP-augmented federated intrusion detection model tailored for healthcare CPS. \cite{LIANG2024108862} propose PB-fdGAN, a permissioned blockchain-enabled decentralized federated GAN framework with local DP for intrusion detection in industrial CPSs. It addresses issues like model homogenization and aggregation quality using federated distillation and QoS-aware blockchain consensus. %(Quality of Service) 
%\medskip  

Recent work on smart grid systems further refines DP integration. \cite{EFedDSA,SecFedSA} proposed EFedDSA and SecFedSA both implement DP-based federated learning to enable privacy-preserving grid stability assessments, improving efficiency and reducing communication overhead.
%\medskip

Beyond centralized and federated architectures, \cite{Nawshin2024} explore a decentralized approach, applying DP to deep learning on-device for malware detection in Android-based IoT systems. Their work demonstrates how DP can support privacy-preserving threat detection in CPS environments under a zero-trust security model.
\medskip
%\bigskip

\noindent
\textbf{Conclusion: } Overall, DP acts as a critical enabler for balancing operational utility and confidentiality in CPS. It supports scalable, decentralized data processing architectures while mitigating risks of linkage, differencing, and correlation attacks --- threats particularly relevant in real-time and spatiotemporal data environments. As CPS deployments continue to grow in scale, complexity, and societal importance, the role of DP in preserving trust and regulatory compliance is set to become increasingly central.

\subsubsection{DP in Anomaly Detection} 
\label{sec_anomaly}
\emph{Anomaly detection} is a technique used to identify unusual or unexpected patterns in data that deviate from normal behavior. These anomalies may indicate critical events such as security breaches, fraud, system failures, or medical conditions, but they can also arise from benign causes such as data entry errors or rare but legitimate activity. Anomaly detection methods can be statistical, model-based, or learned from data, and they are widely used across domains, including cybersecurity, fraud detection, network monitoring, healthcare, and industrial maintenance. In addition, \emph{malware detection} can be viewed as a specialized form of anomaly detection when it relies on identifying behavioral deviations from benign software norms.%\medskip

When applied to sensitive datasets, however, anomaly detection raises serious privacy concerns. DP offers a formal framework to address these concerns by ensuring that the output of an analysis does not reveal sensitive information about any individual in the data. Yet, applying DP to anomaly detection is particularly challenging, as the noise required for privacy can obscure the very anomalies one aims to find. Recent research has introduced novel insights, mechanisms, and frameworks to navigate this tension and support effective anomaly detection under DP guarantees.
\bigskip

\noindent
\textbf{Fundamental Challenges:}
 Classical DP poses inherent limitations for anomaly detection. The core DP requirement --- that output distributions remain nearly unchanged with the modification of any single record --- conflicts with the nature of outliers, which are by definition rare and highly sensitive to individual data changes. As a result, preserving DP often demands the addition of excessive noise to obscure the influence of such data points, severely compromising detection accuracy. In fact, \cite{9623509} has shown that achieving both meaningful privacy guarantees and useful accuracy for data-dependent outlier detection is essentially infeasible under standard DP without relaxing its assumptions.
%\medskip

Moreover, the level of noise typically required to satisfy DP can suppress small but meaningful anomalies, diminishing detection performance and potentially concealing malicious activity.
\cite{7963194} explore this issue in the context of integrity attacks on control systems, showing that DP-induced noise can inadvertently hide the presence of an attacker. To reduce false positives, systems must elevate their detection thresholds, which inadvertently widens the window for stealthy attacks. In essence, while DP is effective at concealing sensitive user behavior, it may also unintentionally shield adversarial actions if not carefully designed.

%In these cases, designing a mechanism with adaptive noise --- one that takes into account the local data density or anomaly score --- can provide a better trade-off between privacy and utility.
\bigskip

\noindent
\textbf{Mechanisms and Techniques:}  
To address the above mentioned challenges, researchers have proposed variants and extensions to DP that adjust the privacy guarantee, and designed new mechanisms to better suit for anomaly detection.
 One promising approach developed by
 \cite{9623509} is {\em Sensitive Privacy} (SP) that generalizes DP by providing privacy levels based on each record’s "outlierness". 
SP allows accurately flagging an anomalous record while still giving most records a strong DP-like protection. In essence, SP adds an extra knob to tune the privacy-utility trade-off for outliers: genuine anomalies might receive less noise (weaker privacy) so they can be detected, while non-outliers are protected with stronger noise. This approach --- together with $n$-{\em step lookahead mechanisms} that offers theoretical guarantees under the SP (\cite{9623509}) --- 
yields practical anomaly detection accuracy with rigorous (though slightly relaxed) privacy guarantees.
%when dealing with data-dependent anomalies. 

%\medskip 
Beyond noise addition, new DP mechanisms are emerging to better integrate with anomaly detection models. For example, 
\cite{9268470} propose a {\em privatized communication protocol} for distributed anomaly detection in control systems that avoids the classic Laplace noise addition altogether. Instead of perturbing raw data, their method shares randomized summary sets of data (solutions of a chance-constrained optimization) that satisfy DP with high confidence. This guarantees privacy while explicitly linking the detection threshold’s design to the desired privacy level. Such approaches illustrate that DP need not always mean simplistic noise injection; it can be achieved through clever aggregation or bounding techniques that preserve more signal relevant to anomalies. 

%\medskip 

%Another emerging idea is using machine learning to maintain utility under DP: Cui et al. design a DP mechanism based on Generative Adversarial Networks (GANs) to synthesize or perturb data during federated model training, achieving higher accuracy for IoT anomaly detection while still satisfying privacy. These novel mechanisms underscore a trend toward hybrid solutions that combine DP with model-based methods to retain anomaly-related information.
%
%Cui, L., Qu, Y., Xie, G., Zeng, D., Li, R., Shen, S., & Yu, S. (2022). Security and Privacy-Enhanced Federated Learning for Anomaly Detection in IoT Infrastructures. IEEE Transactions on Industrial Informatics, 18(5), 3492-3501.
In the context of malware detection, \cite{Nawshin2024} propose a novel approach called DP-RFECV-FNN, which integrates \dps with a \emph{Feedforward Neural Network} (FNN) to detect and classify Android malware in IoT networks. The method supports \emph{zero trust security} by verifying and authenticating applications while preserving user privacy. Their framework leverages both static features (e.g., permissions, code structure) and dynamic features (e.g., runtime behavior), achieving high classification accuracy --- ranging from 97.78\% to 99.21\% for static features and 93.49\% to 94.36\% for dynamic features --- even under strict privacy budgets ($\varepsilon = 0.1$ to $1.0$). This work illustrates that deep learning-based anomaly detection can maintain strong DP guarantees while remaining effective in adversarial environments.
\bigskip
 
\noindent
\textbf{Integration with Federated Learning:}
FL has become an attractive solution for privacy-sensitive anomaly detection scenarios, particularly in distributed settings like IoT, edge computing, and cross-silo environments (see also Section \ref{sec_FL}). It allows local training on private data without central aggregation. However, when FL is combined with DP, new challenges emerge in preserving data privacy without degrading detection performance or system convergence. Several recent works explore this intersection and propose techniques to integrate DP with FL while maintaining utility, robustness, and scalability of anomaly detection.%\medskip

\cite{cui2022security} address this challenge by proposing a decentralized, blockchain-enabled FL framework for IoT anomaly detection. Their system augments FL with a novel GAN-based DP mechanism (DP-GAN) that generates synthetic local model parameters satisfying DP guarantees, without significantly compromising detection accuracy. The approach includes an added discriminator --- the DP identifier --- that ensures generated updates meet
formal privacy guarantees while preserving anomaly-relevant signals.
Moreover, a staleness-aware asynchronous FL protocol improves convergence efficiency in heterogeneous IoT networks. Blockchain consensus replaces the centralized aggregator, further enhancing resilience against poisoning and single-point failures. The proposed numerical experiments demonstrate competitive detection accuracy and improved convergence compared to standard FL models. %\medskip

%%%%%%%%%%%%%%%%%%%%%%%%%%%
 \cite{Zhou2023} focus on edge computing contexts vulnerable to poisoning attacks. They propose a differentially private FL framework for detecting poisoned model updates (a form of intrusion at the model level). They first perform anomaly detection on client-uploaded model parameters to filter out malicious updates. Then they dynamically adjust the privacy budget to strike a balance between model accuracy and robustness. The numerical experiments confirm that their system successfully mitigates model poisoning while preserving anomaly detection capabilities, making it a practical contribution to FL security. % \medskip
 %%%%%%%%%%%%%%

In addition, adaptive DP strategies have been combined with robust aggregation in FL to handle the non-fully trusted model with malicious clients and honest-but-curious servers: 
\cite{Le2023}
 present an {\em adaptive privacy-preserving FL} (Ada-PPFL) scheme where the server applies client-level DP noise adaptively (saving ~35\% noise budget) and a detection algorithm called DPAD to pinpoint anomalous model updates even though they are DP-noised. Such approach simultaneously tackle privacy (against an honest-but-curious server) and security (poisoning attacks by adversarial clients), demonstrating the practical utility of DP in safeguarding collaborative anomaly detection.
 %\medskip
%Taken together, these contributions demonstrate that integrating DP into FL for anomaly detection is a rich and multifaceted research area. Whether through secure hyperparameter tuning, adversarial robustness, decentralized consensus, or communication-aware compression, each approach targets a different layer of the FL pipeline. Collectively, they advance the goal of privacy-preserving anomaly detection in federated environments where sensitive data and diverse system constraints coexist.

%%%%%%%%%%%%%%%%%%%%%%%%%%%
\bigskip

\noindent
\textbf{Conclusion:}
The landscape of DP in anomaly detection is rapidly evolving. Emerging techniques show that by adapting privacy mechanisms to the anomaly context --- whether through new privacy definitions, smarter noise addition, or integration with federated architectures --- we can detect irregularities without compromising individual privacy. The convergence of DP with federated learning, blockchain, and adaptive algorithms is particularly promising for IoT and cybersecurity applications, enabling collaborative defense systems that are both privacy-preserving and resilient to attacks. Nonetheless, open challenges remain in fine-tuning the privacy-utility balance and rigorously proving the security of DP-empowered anomaly detectors. Ongoing research is likely to yield even more refined methods (e.g., combining DP with explainable AI to understand anomalies without revealing sensitive data, see, e.g., \cite{Ezzeddine2024}) that will shape the next generation of secure and private anomaly detection across diverse domains.%\medskip

\subsubsection{Differential Privacy  in Face Recognition Systems}  
\emph{Face recognition} is widely used for authentication, surveillance, and personal device unlocking, yet it raises serious privacy concerns. Facial images contain sensitive biometric features that can uniquely identify individuals, and if misused or leaked, they can compromise one’s identity and other personal data. Unlike passwords, faces cannot be easily changed if stolen. Traditionally, highly accurate face recognition models are trained and deployed on powerful servers, meaning user face data is often transmitted to or stored by third-party providers. This creates a risk of \mybf{uncontrolled information release} --- an adversary or untrusted service could access raw face images and link them to identities or other sensitive records. Therefore, protecting facial features during both the training and inference of face recognition systems is critical.%\medskip

DP has emerged as a key technique to mitigate privacy risks in biometric systems by adding carefully calibrated noise to data or computations. In the context of face recognition, applying DP means introducing randomness that masks or perturbs facial features in a controlled way, so that an individual’s presence or exact image has little influence on the system’s outputs. This provides a theoretical guarantee that an adversary cannot confidently distinguish whether any particular face was in the data, thereby protecting sensitive details. Crucially, this privacy protection must be achieved while \mybf{preserving the utility} of the face recognition --- that is, the system should still recognize or verify faces with high accuracy. 
%The challenge is balancing noise and accuracy: too much noise could render faces unrecognizable, while too little fails to protect privacy.  
In the following, we review how DP is applied to face recognition to protect sensitive facial features, discuss several state-of-the-art approaches, and {examine the role of federated learning} in privacy-preserving face recognition. %\medskip
\bigskip

\noindent
\textbf{Challenges:}
The main challenge in face recognition is balancing noise and accuracy. Too much noise makes faces unrecognizable, while too little fails to protect privacy. DP introduces noise to protect privacy, but it inevitably reduces accuracy, especially in complex tasks like low-resolution or in-the-wild images. Strong privacy guarantees (low $\varepsilon$) can result in high false match/non-match rates, while larger $\varepsilon$ values weaken privacy. Finding the right $\varepsilon$ often requires policy decisions, domain expertise, and legal guidance, as biometric privacy is increasingly regulated. Despite these challenges, there is growing consensus that formal privacy protection is essential to maintain public trust and comply with privacy standards.%\medskip

In addition, interpreting the DP guarantee in a biometric context can be non-intuitive. While DP makes it less likely that a person’s exact identity (face) can be extracted from the data, it does not eliminate the risk of revealing other personal attributes. A DP-sanitized face image may still disclose characteristics like rough age or hair color, as these broader features can survive the noise. Additionally, the high dimensionality and structure of facial data make it vulnerable to sophisticated attackers who could exploit prior knowledge or multiple DP outputs. Proper privacy budgeting is essential to prevent sequential attacks, where multiple noisy samples are collected over time.

\bigskip

\noindent
\textbf{Differential Privacy for Protecting Facial Features:}
DP can be applied at various stages of a face recognition pipeline to protect facial data. A common approach is to perturb either the \myem{input images} or their \myem{feature representations} before they are exposed to potential attackers or external systems. By injecting noise into facial data (either in pixel space or in some transformed domain), the system ensures that each individual’s face cannot be exactly recovered or distinguished by unauthorized parties. 
In face recognition, the goal is often to \mybf{hide identifiable features} (e.g. exact pixel values or high-resolution details of a face) while \mybf{retaining discriminative features} that allow the model to correctly match or classify faces. %\medskip

% Täällä voisi ottaa ne lyhemmät myös ja kirjoittaa jotain siltä väliltä.

% Uses LDP!!!
\cite{chamikara2020peep} % (2020) 
propose the \emph{Privacy using EigEnface Perturbation} (PEEP) protocol, which applies LDP (see Definition \ref{localDP}) to the classic eigenface technique for face recognition. Instead of storing or transmitting raw face images, the faces are first projected into a lower-dimensional subspace --- the "eigenface" space. In this space, noise is added to the eigenface coefficients using LDP, ensuring that even if the perturbed data is intercepted, the original face image cannot be reconstructed. This approach protects user privacy while still enabling face recognition through a standard eigenface algorithm operating on the noisy data. One of the key advantages of PEEP is its ability to thwart \mybf{membership inference} and \mybf{model memorization attacks}, as the noise prevents the model from learning or recalling specific faces. Despite the noise added for privacy, PEEP maintains a reasonable classification accuracy (typically between 70–90\%) and allows tuning the privacy–accuracy trade-off by adjusting the privacy budget.%\medskip

Another line of research applies DP in the \emph{frequency domain} to protect facial images.  \cite{ji2022privacy} % (2022) 
introduce a method that transforms face images using a \emph{Block Discrete Cosine Transform} (BDCT) and adds noise to the frequency coefficients. The idea is that not all frequency components are equally important for recognition: low-frequency components capture the overall appearance, while high-frequency components contain finer details. Instead of applying a uniform noise level to all components, \cite{ji2022privacy}  propose a \emph{learnable privacy budget allocation} scheme that adjusts the noise for each frequency band based on its relevance to recognition. This allocation is learned automatically by observing its impact on the recognition model's performance, ensuring an optimal balance of privacy and accuracy. Because the noise injection follows the DP framework, the system guarantees that the original face cannot be reliably recovered. \cite{ji2022privacy} report that their method provides strong privacy protection with minimal accuracy loss, outperforming previous privacy-preserving techniques. It is lightweight, integrates easily into existing face recognition models, and resists both white-box and black-box reconstruction attacks. This means that even if an attacker knows the noise mechanism (white-box attack) or uses a neural network to try to reverse-engineer the face (black-box attack), the original face remains protected.
%\medskip

% Ranked DP 
 \cite{ou2025rdp} %(2025) 
introduce \emph{Ranked Differential Privacy} (Ranked DP), a novel approach for protecting sensitive facial features in multi-scale subspaces. Ranked DP operates by decomposing face images into multiple scales (e.g., through wavelet transforms or multi-level feature extraction) and sparsifying the less significant components. The remaining feature coefficients are then ranked based on their importance to identity recognition and image quality. The overall privacy budget ($\varepsilon_0$) is allocated across these ranked coefficients, with Laplacian noise added in proportion to their importance. This \emph{selective noise injection} optimizes privacy protection, focusing on the most sensitive features while maintaining the visual quality of the image.
To further enhance performance, \cite{ou2025rdp} formulate an optimization problem to maximize the utility of the perturbed images under the given $\varepsilon_0$ budget. They propose two solutions for noise scale optimization: an analytical normalization approximation, suitable for real-time applications, and a gradient descent approach for more accurate offline use. Experiments on real-world datasets show that Ranked DP significantly outperforms previous methods in the privacy-utility trade-off. 
 At a strong privacy setting ($\varepsilon_0 = 0.2$), Ranked DP achieves a {\em peak signal-to-noise ratio} $10$ dB higher --- roughly a tenfold improvement --- over existing techniques, indicating substantially less distortion in the protected faces.
 This ensures better recognition accuracy while effectively hiding sensitive facial features, making Ranked DP a promising solution for privacy-preserving biometric systems.
%\medskip

Beyond these specific approaches, other literature also explores \dps for face images in creative ways. For instance, \cite{CroftWilliamL.2021Ooiv}  investigated \emph{direct obfuscation} of facial images via DP, treating the image pixels with noise to anonymize faces while aiming to keep them useful. \cite{LUO2025104434} propose \emph{Differential Privacy Obfuscation} (DPO)-\emph{Face}  focusing on facial sensitive regions, which means applying stronger perturbation to particular facial regions (like eyes or mouth) that are highly distinctive, while leaving other regions more intact. This region-based approach attempts to ensure both privacy and image usability by noising the most identifying features of the face, essentially masking what most readily identifies a person. \cite{XieYun2023Ppff} 
propose a general privacy protection framework for \emph{edge-based face recognition} (EFR) 
systems that addresses privacy concerns throughout the entire life cycle of face data. In particular, they introduce a LDP-algorithm based on feature proportion differences and use identity authentication and hashing to ensure the legitimacy of the terminal device and the integrity of the face image. 

In all these cases, a common insight is that transforming the face data to a suitable domain (frequency, PCA subspace, feature space, etc.) before adding noise can yield a better privacy-utility trade-off than adding noise naively to raw pixels. By targeting perturbations to the parts of the data most reflective of identity, the sensitive information is masked while leaving enough signal for the \ml model to do its job. 

An alternative application of \dps in face recognition occurs at the \emph{model training stage} rather than on the input data. Techniques like DP-SGD (see Section \ref{sec_DP-SGD}) allow face recognition models to be trained on sensitive datasets while limiting the leakage of individual’s information from the model’s parameters. In such schemes, noise is added to the model weight updates or gradients during training, ensuring that the final model parameters satisfy \dpns. This protects against attacks that try to infer if a particular person’s face was used in the training data (a form of membership inference). However, training deep face recognition models with DP-SGD often results in a greater accuracy cost, particularly since face recognition typically requires learning fine-grained features. For example, \cite{mao2018privacy} %(2018) 
explored an \emph{edge-computing} scenario where a \emph{deep CNN} for face recognition is split between a user device and a server, with differentially private noise added to the intermediate activations  sent to the server. This ensures that the user’s raw data never leaves the device without DP protection. \cite{mao2018privacy} showed that it is possible to train and fine-tune face recognition models in such a scheme with an acceptable accuracy loss. Generally, DP in training provides strong guarantees 
(preventing even powerful attackers from extracting training images from the model), but the privacy-accuracy trade-off tends to be harsher than with data/input-level perturbation methods. As a result, input-level perturbation --- as in PEEP (\cite{chamikara2020peep}), frequency-DP (\cite{ji2022privacy}), and  Ranked DP (\cite{ou2025rdp}) --- remains the preferable choice, as it is often more tractable to preserve accuracy while still guaranteeing privacy for each input.\medskip
\bigskip

\noindent
\textbf{Integration with Federated Learning:}
FL is a complementary approach that addresses privacy by keeping data decentralized. In a federated face recognition system, individual user devices (e.g. smartphones) train the model locally on their face data, and only share model updates (not raw images) with a central server or aggregator. This setup means that the sensitive facial images never leave the devices, mitigating the risk of data exposure from transmission or a central database. FL has been proposed for face recognition to allow collaborative training on a large number of users’ data without violating privacy. The integration of DP into FL further enhances privacy by ensuring that individual data points cannot be reconstructed from the model updates.
%\medskip 

 One such approach is \emph{PrivacyFace} by \cite{MengQiang2022IFLF},  %(2022) 
 which combines FL with a \emph{Differentially Private Local Clustering} (DPLC) method that enforces \dps during model training. In this framework, face images are first grouped into \emph{privacy-agnostic clusters}, ensuring that the data is anonymized before being processed by the global model. The clustering approach ensures that private features of individual users’ data are protected by sanitizing the class centers shared between clients. Additionally, a \emph{consensus-aware face recognition loss} is employed, which aligns the local models with the global model, minimizing the impact of privacy protection on overall recognition accuracy. This mechanism helps mitigate the risk of membership inference attacks, where an adversary might infer whether a particular user’s data was used in training, all while maintaining high model accuracy.
 
%\medskip
%DAFL
%95.6% and 97.9% accuracy in face recognition
%and speaker recognition tasks, respectively, under medium privacy settings. Compared with existing work, it
%can save 56.0% of training time with a 3.1% accuracy loss in the best case

Another notable method is \emph{Differentially Private Domain Adaptation-based Federated Learning} (DAFL) by \cite{WangZhousheng2024DDaf},  %(2024), 
which focuses on improving the privacy-utility trade-off in FL for face recognition. DAFL combines \dps with \emph{domain adaptation techniques} to enhance the model's performance across different data distributions from client devices. It introduces a \emph{dropout mechanism} to handle label imbalance in biometric identification systems during training, further improving the model's robustness while ensuring privacy. The use of \dps ensures that even strong adversaries cannot extract private data from the trained model, and the system achieves an effective balance between privacy and face recognition accuracy.
%\medskip

These approaches highlight the potential of integrating DP with FL to protect privacy in face recognition tasks, with different methods focusing on privacy-preserving clustering, domain adaptation, and balancing privacy costs with model performance.
\bigskip

\noindent
\textbf{Conclusion:} 
DP provides strong guarantees by ensuring that any single face image or attribute has limited impact on the data or trained model, making it difficult for attackers to infer if a specific person’s data was included. For face recognition systems, \dps mitigates risks such as:
%DP offers a tunable and computationally lightweight approach to protecting biometric data in face recognition systems. By bounding the influence of any single individual’s data, DP mitigates key privacy risks, including:

\begin{itemize}
    \item \textbf{Reconstruction attacks}: \DP adds noise to fine-grained features, making it harder to reconstruct recognizable faces from sanitized data.
    \item \textbf{Membership inference attacks}: 
    DP ensures model outputs do not change significantly whether a person's data is included or not.
    \item \textbf{Linkage attacks}: DP makes it difficult to link noisy face data to identities through external databases.
\end{itemize}
\medskip
Additionally, \dpns-based approaches are lightweight and flexible, with noise injection being computationally inexpensive compared to alternatives like encryption. Methods like learnable privacy budget allocation by \cite{ji2022privacy} and selective noise injection by \cite{ou2025rdp} allow for privacy with minimal accuracy loss, often maintaining high performance close to non-private models. This makes \dps a practical solution for deployment, as systems can ensure strong privacy protection with little impact on accuracy.
As threats and regulations continue to evolve, DP provides a foundational layer of protection. However, its integration with other security and fairness mechanisms will be essential to ensure face recognition systems remain both effective and aligned with ethical and legal standards.

\subsection{DP in Healthcare}
\label{sec_health}
DP has emerged as a principled approach to protecting patient information while enabling data analysis and sharing in healthcare. At a high level, DP ensures that the outcome of an analysis does not significantly change when any single individual's data is added or removed, thereby limiting what an adversary can infer about that individual. A key parameter in this framework is the privacy budget $\varepsilon$, which quantifies the privacy loss: smaller values of $\varepsilon$ correspond to stronger privacy guarantees, typically achieved by injecting more noise, but at the cost of reduced data utility. Choosing an appropriate $\varepsilon$ in practice remains challenging, as its abstract nature makes it difficult to interpret or to set clear guidelines for acceptable privacy in medical contexts. 

DP mechanisms can be applied in two main paradigms. In the \emph{interactive setting}, a trusted curator responds to user queries --- such as statistical summaries or machine learning model evaluations --- with noise added to ensure privacy. In contrast, the \emph{non-interactive setting} involves releasing a sanitized version of the data, such as differentially private synthetic data or summary tables, that can be analyzed freely without further interaction with the original database~(\cite{Dankar2013}).

A scoping review by \cite{Ficek2021} identified several key areas of health research where DP has been applied, including genomics, neuroimaging, and health surveillance using personal devices. The majority of studies focused on algorithm development, particularly for tasks related to data release and predictive modeling. 
This highlights the methodological emphasis of the field, while relatively fewer studies address system-level implementation or systematic evaluation of privacy-utility tradeoffs.

In what follows, we organize this section by key application areas. We begin with statistical analysis, including applications in \emph{Genome-Wide Association Studies} (GWAS), then discuss approaches to data sharing through differentially private synthetic data, and conclude with predictive modeling techniques, including federated learning in health care settings.

\subsubsection{Statistical Analysis}\label{sec_health1} A classical application of DP in health research is in releasing statistical findings or conducting epidemiological analyses on sensitive databases. Health agencies and researchers often need to share summary statistics (incidence rates, query results, contingency tables, etc.) from patient registries or surveillance systems. DP provides a way to add just enough noise to these statistics to protect individual identities while still allowing useful insight. For example, queries like "\emph{How many patients had condition X last year?}" can be answered with a DP mechanism (e.g., adding Laplace noise to the count). If done interactively, a privacy budget is decremented with each query to limit cumulative disclosure. In a non-interactive scenario, one might release an entire table of DP-safe statistics or even a DP regression model parameters (\cite{Dankar2013}). 
%\medskip

An important area using statistical analysis in healthcare is GWAS. GWAS are essential tools in identifying genetic variants associated with diseases. These studies typically examine \emph{single nucleotide polymorphisms} (SNPs) --- positions in the genome where individuals differ by a single DNA base --- which may be associated with specific traits or diseases.
However, genomic data is inherently identifiable and shared among relatives, making it particularly vulnerable to privacy breaches --- even aggregate statistics such as allele frequencies can risk re-identification. As \cite{HomerNils2008RICT} famously showed, an individual's presence in a GWAS cohort can be inferred from population-level summary statistics. %\medskip 

One influential approach to applying DP in GWAS is the method by  \cite{GWAS2}, who refined neighbor distance mechanisms to privately select the top-associated SNPs. Their technique combined input perturbation with convex optimization to improve both accuracy and computational efficiency over Laplace- and score-based alternatives. This work demonstrated that it is possible to release differentially private GWAS results without relying on full data encryption or synthetic data generation.

Subsequent developments have tackled deeper challenges specific to genomics. \cite{GWAS3} highlighted that standard DP assumes independence between individuals --- an assumption that fails in genomic data due to familial relationships. They proposed a modified privacy framework that accounts for dependency between tuples, providing more realistic privacy guarantees in settings like GWAS.

Building on these conceptual foundations, more recent systems integrate DP with cryptographic techniques. \cite{Raisaro2018} extended the i2b2 framework to support privacy-preserving genomic cohort exploration using homomorphic encryption (HE) combined with DP. Their system allows researchers to compute encrypted aggregate statistics (e.g., allele frequencies, mutation counts) across pseudonymized genomic datasets, with query outputs perturbed according to a per-user privacy budget. Similarly, the MedCo platform enables federated exploration of distributed clinical-genomic databases using a combination of HE and DP, facilitating secure multi-institutional GWAS 
(\cite{RaisaroJeanLouis2019MESa}).

Together, these developments demonstrate a rich and evolving landscape for private GWAS. From early statistical DP mechanisms to dependency-aware privacy models and cryptographic infrastructures, each stage has brought GWAS closer to being both analytically powerful and ethically responsible.

\subsubsection{Data Sharing} \label{sec_health2}
Healthcare is often cited as a prime beneficiary of synthetic data generation. Patient data (\emph{electronic health records} (EHR), clinical trial data, genomic data, etc.) are highly sensitive and protected by laws (like GDPR in Europe and HIPAA in the U.S.). This makes it difficult to share data between hospitals, researchers, and pharmaceutical companies, slowing down research and model development as noted in \cite{Pitkamaki2024}. 
Synthetic health data offers a way to bypass some of these hurdles by creating a shareable dataset that mimics real patients. For instance, a hospital might generate a differentially private synthetic dataset of its patient records and allow external researchers to use it to develop predictive models or to perform exploratory analyses. Because the data is synthetic and comes with a privacy guarantee, this could be done without violating patient confidentiality.% \medskip

A survey by \cite{MURTAZA2023100546} reviews various synthetic data generation approaches in the healthcare domain, covering both DP-based methods and alternative techniques. Here, we briefly review five representative approaches to differentially private synthetic data generation in healthcare (see also Section \ref{sec_synthetic_data}).
%\medskip
 
One such method is DP-CTGAN introduced by \cite{Fang2022}. DP-CTGAN integrates DP into the \emph{Conditional Tabular GAN} (CTGAN) architecture for generating synthetic medical datasets. It introduce gradient clipping and noise injection into the training of the discriminator, allowing the generator to learn realistic tabular data while providing formal DP guarantees. This approach outperforms previous models like DPGAN and PATE-GAN on multiple benchmark datasets and can also be adapted for federated learning to protect data across institutions without centralizing it. %\medskip

Another notable contribution is the RDP-CGAN framework by \cite{torfi2022}, which employs {R\'enyi DP} for generating synthetic medical data using convolutional GANs and autoencoders. Their approach is designed to handle both continuous and discrete data types, and to capture temporal and correlated structures common in real-world health records. By incorporating privacy-preserving training at both the autoencoder and discriminator stages, and using 1D convolutional architectures, the model achieves high fidelity in synthetic data generation across several benchmark datasets. The authors demonstrate that RDP-CGAN outperforms prior DP-based models such as DPGAN and PATE-GAN under the same privacy budgets, especially in terms of predictive performance on downstream tasks. Their work highlights the importance of architectural choices and robust privacy accounting in achieving strong utility-privacy trade-offs for synthetic health data. %\medskip

A complementary strategy is proposed by \cite{LeeDongha2020Gseh}, who developed a \emph{Dual Adversarial Autoencoder} (DAAE) model for generating synthetic sequences of EHRs. Their architecture combines a recurrent autoencoder with two adversarial critics --- one operating in the latent space and one on the output space --- to improve the realism and diversity of generated sequences. DP is incorporated during training using a differentially private version of the {Adam optimizer}. The resulting synthetic data maintains high predictive utility while providing provable privacy guarantees. %\medskip

Another example, from \cite{ImtiazSana2021SaPS}, focuses on smart healthcare data collected from wearable devices. They propose a BGAN-based architecture that supports different DP integration strategies, such as adding Laplacian noise to input data (noisy input), directly to generated outputs (noisy output), or during training. Their results show that adding noise to the input is particularly effective for preserving privacy while maintaining the statistical fidelity of the synthetic health data. %\medskip

Finally, \cite{KossenTabea2022TSBI} apply DP to the generation of labeled medical images in the neuroimaging domain. They use a differentially private WGAN to generate synthetic \emph{Time-of-Flight Magnetic Resonance Angiography} (TOF-MRA) image patches along with segmentation labels for brain vessel segmentation. Their study explores the trade-off between privacy and model utility, showing that even with moderate privacy budgets (e.g., $\varepsilon=7.4$), the synthetic data can be used to train segmentation networks that perform reasonably well on real data. This work is notable as one of the first, dating back to 2022, to apply DP-GANs for labeled 2D brain images in a downstream 3D medical application. %\medskip

While these approaches show strong potential, careful validation by domain experts is essential before synthetic health data is used. As highlighted by \cite{Pitkamaki2024}, the practical use of synthetic data --- particularly in privacy-sensitive domains like healthcare --- requires clear definitions, careful consideration of regulatory constraints, and realistic expectations about its capabilities and limitations. Clinical plausibility checks are critical --- implausible records, such as a male patient with a pregnancy code or a procedure following recorded death, can undermine trust. Ensuring logical consistency is an added challenge beyond statistical fidelity. Domain-specific constraints may need to be hardcoded into the generation process, or post-processing used to filter out errors. Purely data-driven models may fall short here, and hybrid approaches --- combining generative models with rule-based corrections --- can be more effective. Despite these challenges, synthetic data is becoming an integral part of privacy-preserving health data infrastructures, enabling safer and more scalable data sharing.
 
\subsubsection{Predictive Modeling} \label{sec_health3}
Machine learning models have shown great potential in healthcare applications, such as predicting disease outcomes or learning population-level patterns from patient data. However, when trained on sensitive health data, these models can inadvertently leak personal information. DP has been employed to mitigate this risk by enabling model training procedures that provide formal privacy guarantees. In the central DP model (where a trusted server has access to raw data), this is typically achieved through techniques such as DP-SGD (see Section \ref{sec_DP-SGD}), which adds noise to gradient updates, or through output perturbation applied to the learned model parameters. %\medskip

For example, \cite{ding2022} contribute theoretical and practical improvements to differentially private empirical risk minimization  for healthcare systems using analytic Gaussian mechanisms. They propose two algorithms --- OPERA, based on output perturbation, and GRPUA, based on gradient perturbation --- which achieve tighter utility bounds in both convex and non-convex settings. Their methods outperform previous approaches on standard healthcare datasets, particularly in high-privacy regimes. %\medskip

In addition, \cite{orabe2025} incorporate DP-SGD into deep learning for cardiovascular risk prediction from multi-hospital electrocardiogram (ECG) data. In their study, a neural network was trained to classify 17 cardiac conditions; with a moderate privacy budget ($\varepsilon=10$ in a central DP setting), the model’s overall performance degradation was small --- only a 0.03--0.04 reduction in AUC occurred. However, they observed that the impact of DP was not uniform across all outcomes: in particular, certain less prevalent conditions experienced noticeably larger drops in AUC than more common ones. This highlights that utility trade-offs under DP can vary by subgroup, as the added noise may disproportionately affect patterns that are already more difficult to learn or distinguish.%\medskip

Indeed, \cite{Suriyakumar2021} present an in-depth empirical evaluation of differentially private models for clinical prediction tasks using MIMIC-III time-series and NIH Chest X-Ray datasets. Their findings emphasize the steep utility trade-offs when applying DP-SGD, particularly in \emph{long-tail} clinical data distributions (i.e., where a small number of conditions or patient subgroups are common, while many others are rare and sparsely represented). They also show that DP can reduce the influence of minority group data in model training, raising fairness concerns. These results underscore the importance of carefully evaluating both the privacy and equity implications of DP in clinical applications.%\medskip

% Although they did not find DP to systematically improve robustness to dataset shifts (e.g., year-to-year shifts in hospital data), their work offers important insights into the broader limitations of DP in real-world predictive modeling.
In addition, federated learning has gained traction in healthcare as a way for multiple institutions to collaboratively train models without exchanging raw data. This is especially attractive given the strict data sharing regulations in the field. However, federated learning alone does not provide formal privacy guarantees, as model updates can still be susceptible to inference attacks. To mitigate this, several approaches integrate DP into either the aggregation process or the local training phases.
%\medskip

\cite{Khanna2022} propose a differentially private federated learning framework for predicting breast cancer status from gene expression data. By applying noise to model weights during server-side aggregation, their system maintains high prediction accuracy even under strong privacy constraints. Their implementation also includes a practical mechanism for converting desired $\varepsilon$ values into corresponding noise multipliers, making it more accessible for clinical users.%\medskip

 \cite{khan2023differentialprivacyadaptiveweight} present the DP-SimAgg method, which introduces DP into a similarity-weighted aggregation scheme for brain tumor segmentation using MRI data. Noise is added during the server-side model aggregation, allowing collaborators to remain lightweight. Their method demonstrates that even with strict privacy budgets (e.g., $\varepsilon=0.1$), high segmentation performance can be preserved, highlighting the feasibility of privacy-preserving federated learning in imaging applications.%\medskip

An alternative distributed approach is presented by \cite{lamp2024dpruldifferentiallyprivaterulelearning}, who focus on distributed rule learning for clinical decision support. Their DP-RuL framework uses local DP (randomized response) to privately aggregate logical rules from multiple client devices or sites. 
 The algorithm employs \emph{Monte Carlo Tree Search} (\cite{Browne2012MCTS}) to discover population-level rules, and introduces an adaptive privacy budget allocation mechanism to optimize the trade-off between privacy and utility.
  This approach is particularly suited for mobile or edge-based health monitoring applications.%\medskip
 
Finally, \cite{ADHIKARY2024106577} propose \emph{PrivLet}, a privacy-preserving framework for gait classification in hemiplegic patients using smartphone IMU sensors. Their method integrates local DP into the wavelet-domain representation of time-series data by adding noise directly to the wavelet coefficients before classification. This approach is computationally lightweight and well-suited for edge devices such as smartphones. The authors demonstrate that PrivLet maintains competitive classification accuracy across a range of privacy budgets, making it a viable solution for real-time clinical prediction tasks in resource-constrained environments.

%\medskip
 
Together, these studies demonstrate the diversity of approaches to privacy-preserving predictive modeling in healthcare. Whether through centralized DP-SGD, FL with DP-weight aggregation, or local DP protocols for rule-based systems, all aim to balance model performance with provable privacy guarantees—though achieving this balance remains a core challenge.
 
%\medskip
\bigskip

\noindent
\textbf{Conclusion:} % Tämä voisi olla myös subsection, jos edellisissäkin sectioneissa on tällainen
DP offers a principled framework for protecting sensitive health data while enabling meaningful data analysis, model development, and data sharing. Across statistical data release, synthetic data generation, and predictive modeling, a growing body of research demonstrates both the potential and limitations of DP in health care contexts. While recent methods show promising utility even under strong privacy guarantees, challenges remain in preserving performance for minority subgroups, ensuring clinical plausibility in synthetic data, and supporting practical deployment across institutions. Continued interdisciplinary research is needed to bridge the gap between formal privacy mechanisms and the complex realities of medical data use.

%%%%%%%%%%%%%%%%%%%%%%%%%%%%%%%%%%%%%%%%%%%%%%%%%%%%%%%%%%%%%%%%%%%%%%%%%%%%%%%%%%%%%%%%%%%%
\subsection{DP in Finance}\label{sec:dp-finance}
The financial sector is increasingly adopting DP to enable secure data analysis, sharing, and model development while complying with strict confidentiality regulations. Applications range from fraud detection and credit risk modeling to auditing and secure financial market operations. This section surveys key use cases grouped by functional area.

\subsubsection{Fraud Detection} \label{sec_finance1}
One of the most prominent applications of DP in finance is \emph{fraud detection}. Financial firms (banks, payment networks) use DP to collaboratively detect fraudulent transactions without revealing customer-level details. The goal is to identify anomalous spending patterns or unauthorized activities across large, distributed datasets. A common approach is local DP, where each institution perturbs or encodes its transaction data before sharing. 
%\medskip 

For instance, \cite{perez2024locallydifferentiallyprivateembedding} explore the use of local DP in collaborative fraud detection. They propose a mechanism where financial institutions publish transaction embeddings with local DP guarantees. These embeddings can then be used by downstream models without exposing sensitive patterns or identities. Empirical results show robustness against inference, inversion, and membership attacks across multiple fraud detection datasets.
%\medskip 

In addition, federated learning combined with DP allows financial institutions to collaboratively train models on transaction data without revealing sensitive customer information. \cite{Ma2024} demonstrate this approach by developing a credit card fraud detection system in which multiple banks train a \emph{convolutional neural network} (CNN) collaboratively using federated learning. Their system enables the detection of fraudulent transactions across institutions while keeping data decentralized and private, illustrating the potential of distributed learning frameworks for privacy-preserving financial analytics. 
%\medskip 

Further, also federated learning combined with DP and \emph{secure multiparty computation }(SMPC) allows financial institutions to collaboratively train fraud detection systems.
\cite{Byrd2021} propose a DP + SMPC framework that enables institutions to train a shared logistic regression model on credit card fraud data. Their protocol adds Laplace noise to model updates and aggregates them securely, ensuring that neither individual model weights nor customer-level data are exposed. This approach achieved strong performance on a highly imbalanced real-world dataset, suggesting that privacy need not come at the cost of utility.
%\medskip  

Finally, a bank might generate a \emph{synthetic version of its transaction data} to share with fintech partners or researchers for developing fraud detection algorithms, thus avoiding direct exposure of customer-level information. Synthetic data can also support broader collaboration, such as enabling insurers to share claim data for stress testing and risk modeling without breaching confidentiality. In fraud detection, where datasets are typically highly imbalanced, synthetic data offers a promising way to simulate rare fraudulent behaviors for model training. However, preserving important statistical features --- such as heavy tails or extreme outliers that are critical for stress testing --- remains a significant challenge. \cite{Assefa2021} note that while DP provides a strong framework for privacy protection, it often struggles to preserve the tail behavior and extremal dependencies that are essential for realistic financial modeling and decision-making.

\subsubsection{Credit Scoring} 
\emph{Credit scoring} is the process of evaluating a borrower’s \emph{likelihood of default} --- the probability that the borrower will fail to repay a loan within a specified time frame. This probability is inferred from financial and demographic information such as income, repayment history, and outstanding debts. Credit scoring models support key lending decisions, including loan approvals, credit limits, and interest rates. Because these models rely on sensitive personal data, protecting individual privacy during data analysis and model development is essential.
%\medskip 

DP techniques help mitigate the risks of disclosing sensitive attributes such as income, liabilities, or repayment history. \cite{luo2023} evaluate how different DP mechanisms affect the utility of credit card datasets when training predictive models, offering insights into the privacy-utility trade-offs relevant to default prediction. However, recent advances go further by integrating DP into collaborative model training across multiple institutions. For instance, \cite{WANG2024104051} propose a vertical federated learning framework for imbalanced credit scoring that combines DP with adaptive gradient clipping and cost-sensitive learning. Their \emph{interpretable logistic regression model} (IVLR-ACS) enables financial institutions to jointly train a credit scoring model without revealing raw data, while also addressing class imbalance --- a key challenge in default risk modeling. Experiments across eight real-world datasets demonstrate that their approach achieves strong performance and protects customer privacy, showing the potential of DP-enhanced collaborative scoring systems in regulated financial environments. 

\subsubsection{Financial Auditing}
\emph{Financial auditing} is the process of independently verifying an organization’s financial records to ensure accuracy, regulatory compliance, and integrity. Modern audits increasingly rely on machine learning models trained on digital accounting records, such as journal entries from ERP systems. A key challenge is that these datasets contain highly sensitive, client-specific information that auditors are legally and contractually bound to protect. Traditional data‑sharing approaches are often infeasible due to strict regulatory requirements (such as GDPR) and client confidentiality obligations. At the same time, building robust anomaly detection models often requires pooling knowledge across multiple clients or industries, which risks exposing private financial details.  
%\medskip

\cite{Schreyer2022} address this challenge by proposing a federated learning framework for training audit anomaly detection models across clients, enhanced with DP and split learning. Their system allows auditors to develop industry‑specific models using decentralized journal entry data without exchanging raw accounting records between entities. 
The models --- such as deep autoencoders --- can detect both global anomalies (e.g., rare posting times) and local anomalies (e.g., suspicious co‑occurrences of accounting fields). By applying gradient clipping and injecting DP noise into model updates, the framework ensures that no client’s data can be reconstructed from the learned model, offering a practical, privacy‑preserving approach to collaborative audit analytics.%\medskip

In this context, DP is particularly valuable: it enables collaborative model training across multiple clients while providing formal guarantees that no individual organization’s records can be inferred, even indirectly.

\subsubsection{Secure Market Operations}\label{sec_finance3}
Financial markets are particularly sensitive to information leakage, where revealing individual trading intentions or strategies can lead to front-running, price manipulation, or strategic disadvantage. Here, DP offers a principled alternative to the complex, ad hoc privacy strategies commonly used in electronic markets.%\medskip

\cite{Diana2020} propose differentially private call auction mechanisms that obscure the identity and quantity of submitted bids while preserving market efficiency. Their approach introduces calibrated noise into the auction's pricing and matching logic, providing per-share privacy guarantees based on participation rate. Simulation results show that the privacy-preserving auctions perform competitively with their non-private counterparts, and could offer a viable replacement for mechanisms such as dark pools or hidden orders, with formal privacy guarantees.%\medskip 

In a complementary direction, researchers have explored the generation of synthetic market microstructure data --- such as \emph{limit order book} (LOB) data --- with privacy preservation in mind. This data is essential for developing and testing trading algorithms, but is rarely shared due to its strategic sensitivity. \cite{Assefa2021} highlight that while traditional LOB datasets capture valuable patterns like spread dynamics and trade volume clustering, publishing them directly risks revealing individual strategies or institutional behavior. Synthetic data generation, possibly combined with DP or agent-based simulation, can offer a safe alternative for training and benchmarking without exposing proprietary trading signals.%\medskip

Together, these approaches demonstrate that DP can support secure market operations both in real-time mechanism design and in offline data provisioning for financial research and infrastructure development.
\bigskip

\noindent
\textbf{Conclusion:}
%In summary, 
DP is emerging as a versatile tool for addressing privacy challenges across a range of financial applications. From fraud detection and credit scoring to financial auditing and secure market operations, recent research demonstrates that DP can enable meaningful data analysis and machine learning while preserving confidentiality in settings where privacy is both ethically and legally mandated. These developments show that privacy preservation need not come at the cost of utility --- especially when combined with techniques like federated learning, cost-sensitive modeling, and adaptive noise injection. Beyond the core application areas covered in this section, DP also holds promise for enhancing regulatory reporting and compliance processes, where aggregate disclosures must be accurate yet safe from reverse engineering. As the financial sector continues to digitalize and integrate advanced analytics into decision-making, the role of DP is likely to expand, offering a principled framework for balancing innovation with the fundamental need for trust and confidentiality.

\section{Practicalities and User Expectations}\label{chaUserExpectations}

As differential privacy (DP) evolves from a theoretical ideal to a practical tool for privacy protection, a growing body of research has focused on the human, organizational, and systemic factors that shape its adoption and effectiveness. This section surveys key empirical findings related to user expectations, communication strategies, practitioner challenges, the risk of privacy theater, and regulatory uncertainty. It highlights how the gap between formal guarantees and real-world deployment affects not only usability and implementation but also public trust and policy alignment.

Specifically, we synthesize results from empirical user studies, communication experiments, practitioner interviews, and policy analyses, organizing the discussion around four recurring themes: user expectations, communication challenges, practitioner usability, and institutional adoption. To promote responsible communication in practice, we conclude the section by introducing a practical DP workflow and  a concise DP disclosure checklist that operationalizes these considerations.

\subsection{User Expectations} 
Although DP is often described as offering strong, mathematically grounded guarantees, public understanding of what those guarantees entail is limited. \cite{Cummings2023} found that while users generally care about privacy --- especially in relation to specific risks such as access by hackers, law enforcement, or employers --- their willingness to share data depends more on whether these concrete threats are addressed than on formal guarantees provided by DP. %\medskip

Importantly, users often encounter DP through general-purpose explanations in settings like product announcements or privacy policies and these explanations often fail to address users' core privacy concerns. \cite{Cummings2023} found that the specific framing of DP --- whether it emphasizes noise injection, anonymization, algorithmic process, or a guarantee of safety --- can shape users’ expectations in different ways. However, none of these framings consistently improve willingness to share data unless they address users' specific fears. Moreover, many users form inaccurate expectations based on such explanations, assuming stronger protections than actually exist.%\medskip

This issue is compounded by overconfidence: \cite{10.1145/3555762} showed that users frequently believe they understand DP, while struggling with key concepts such as probabilistic guarantees and the meaning of $\varepsilon$. These misunderstandings create a risk that users consent to data sharing based on incomplete or misleading impressions of how their privacy is protected. To address this, we will consider several approaches for more effectively communicating DP to users and practitioners  in Section \ref{secX}.

\subsection{The Risk of Privacy Theater}
DP is widely regarded as a gold standard for privacy protection due to its formal mathematical guarantees. However, when descriptions of DP omit or obscure key parameters --- particularly the magnitude or meaning of $\varepsilon$ --- users may be misled into assuming stronger privacy protections than are actually provided. In such cases, real-world deployments of DP risk devolving into privacy theater: privacy becomes a performative claim rather than a substantive safeguard. %\medskip

 \cite{10.1145/3555762}  explicitly address this 
 concern in their study of user perceptions. They show that low-transparency explanations can foster overconfidence, leading users to assume stronger privacy protections than are actually provided. Moreover, their experiments reveal that even when users are presented with explanations of varying transparency, most do not change their willingness to share data. This suggests that technical guarantees are often overshadowed by factors such as institutional trust or the belief that one has "nothing to hide."
%\medskip

\cite{10.1145/3433638} add that misapplications of DP --- such as high-$\varepsilon$ deployments, opaque parameter disclosure, and flawed sequential composition practices --- further compound this risk. They point out that companies like Apple and Google have used $\varepsilon$ values as high as 14, while failing to explain the implications of such settings to users. According to the authors, these deployments may still retain the label of DP but ultimately deviate from its foundational aims. It is worth noting, however, that some guidelines and deployments suggest even higher values (e.g., $\varepsilon \approx 20$) may be reasonable in practice, depending on the application context and privacy--utility trade-offs (\cite{NIST2022,Ponomareva2023}, see also Section \ref{section_privacy_budget}).%\medskip

\cite{weiss2024shareshareriskslaypeople}  further highlight the mismatch between user expectations and the parameters typically used in real-world DP deployments. In a behavioral study 
involving realistic NLP scenarios, they found that most participants were only willing to share sensitive text data under substantially lower $\varepsilon$ values than are commonly used in practice --- values in the range of $3$ to $10$ exceed what most users find acceptable. Importantly, acceptable thresholds varied with both data sensitivity and dataset size, underscoring that privacy guarantees are only meaningful when they align with users' contextual perceptions of risk. These findings reinforce the need for participatory or adaptive approaches to setting $\varepsilon$, rather than relying solely on developer-driven decisions. Without such alignment, DP risks becoming a symbolic privacy label lacking substance. %\medskip

Finally, recent work by \cite{gomez2025varepsilondeltaconsideredharmful} critiques the whole widespread practice of reporting DP guarantees as a single  $(\varepsilon, \delta)$-DP pair with a fixed $\delta$. The authors argue this approach is incomplete, arbitrary, and often misleading, making comparisons across algorithms difficult and privacy risks hard to interpret. Instead, they recommend replacing $(\varepsilon, \delta)$-DP reporting with Gaussian DP (GDP, \cite{DongJinshuo2022Gdp}) using its single parameter $\mu$. 
As they write, 

\begin{quote}
    \emph{We advocate for the DP community to move away from reporting $\varepsilon$ at fixed $\delta$ as the standard for reporting privacy parameters. Instead, we propose to analyze the DP algorithms in terms of privacy profiles (\cite{Balle2}) or, equivalently, trade-off curves (\cite{DongJinshuo2022Gdp}), using numerical accountants, and converting the result to a conservative $\mu$-GDP guarantee for reporting. If the full privacy profile or trade-off curve is sufficiently close to that of a Gaussian mechanism, the $\mu$-GDP guarantee will give an essentially complete picture of the privacy of the algorithm.}
\end{quote}
Further, if GDP does not provide a sufficiently tight fit, Gomez et al.\ recommend publishing the full privacy profile (trade-off curve)\footnote{A privacy profile or trade-off curve describes the full set of $(\varepsilon, \delta)$ pairs a mechanism satisfies, often interpreted via the success rates of optimal membership inference attacks at different false positive rates.} along with code to reproduce it.

\subsection{Communicating DP to End Users} 
Explanations of DP that focus on mathematical rigor often alienate non-experts. Many users struggle to interpret abstract parameters like $\varepsilon$ or understand how added noise affects the accuracy of results. To address this, several studies have investigated communication strategies aimed at improving end users understanding of DP’s core trade-offs.%\medskip

\cite{281238} replicated the influential study by \cite{9152658} to assess how communication strategies affect user comprehension and data sharing decisions in a DP context. \cite{9152658} found that explaining the implications of DP and  local DP (e.g., who needs to be trusted) rather than their formal definitions significantly improved user comprehension and willingness to share sensitive data. Their results also revealed that participants found data perturbation concepts particularly difficult to understand. \cite{281238} confirmed many of these findings in a German context, observing that while willingness to share was generally higher than in the original U.S.-based study, users still struggled to understand the privacy guarantees offered by DP. Both studies emphasize that merely stating that DP is used is insufficient: users need context-sensitive explanations that clarify the implications of the technique to make informed data-sharing decisions.%\medskip

\cite{291162} and \cite{WEN2023118799} show that visual aids, odds-based framing, and analogies such as lotteries can help users better grasp the privacy-utility trade-off. However they also note that improved understanding does not always lead to greater willingness to share.
Similarly, \cite{10646708} find that interactive visualizations that depict trade-offs between accuracy and privacy improve users’ ability to make more expert-aligned decisions about $\varepsilon$ values. However, they also caution that overemphasis on accuracy loss may backfire, pushing users toward riskier privacy choices. \cite{281270} further suggest that metaphors --- like noisy images or spinner-based decision tools --- can make DP’s mechanisms more approachable, but must be carefully aligned with the actual guarantees to avoid misleading mental models.
%\medskip

\cite{Franzen1,Franzen2} designed and tested privacy decision interfaces that use both text and visual elements --- such as frequency-based risk illustrations adapted from medical settings --- to explain the trade-off between privacy and data utility. Their findings show that such interfaces help users better understand what DP protects and how different levels of noise (controlled by $\varepsilon$) affect their privacy. However, they also found that not all users benefit equally: factors like a person's numeracy (comfort with numbers) and cognitive style influence how well they can interpret quantitative risk information. Therefore, they emphasize the need for privacy interfaces that are not only accurate but also accessible to users with diverse backgrounds and abilities. %\medskip

Taken together, these findings suggest that no single explanation strategy suffices. Effective communication of DP requires layered, role-sensitive approaches that balance accuracy with accessibility. These should combine visual metaphors, interactive trade-off tools, and contextual framing to match users’ privacy concerns and decision heuristics.

\subsection{Challenges for Data Practitioners}
It is not only the end users who may find DP difficult to grasp.  Even among technically literate users --- such as developers and analysts --- implementing and understanding DP can be challenging. \cite{10.1145/3687011} and \cite{ngong2024evaluatingusabilitydifferentialprivacy} show that users of open-source DP libraries often struggle with core concepts like the composition of privacy budgets and the interpretation of $\varepsilon$, particularly when tools lack clear documentation, usable examples, or intuitive error handling. In both studies, users frequently relied on external tutorials or handouts to understand library behavior, and even then, encountered errors that hindered task completion or led to privacy violations.%\medskip

\cite{Sarathy2023} reinforce these findings through qualitative interviews with non-expert practitioners using the DP Creator tool (\cite{DPCreator}). They identify four recurring pain points:
\begin{enumerate} 
    \item Confusion over metadata and privacy budget settings.
    \item Frustration with black-box tools that obscure raw data.
    \item New governance responsibilities placed on analysts and depositors.
    \item Poor compatibility of DP tools with exploratory data analysis.
\end{enumerate}
Participants expressed both optimism about DP’s potential and hesitation due to usability limitations, lack of guidance, and the risks of incorrect decision-making.

While privacy awareness has increased since the introduction of the GDPR, most developers remain unfamiliar with formal privacy engineering frameworks such as ISO/IEC 27550 (\cite{iso_27550_2019}) and the NIST Privacy Framework (\cite{nist_privacy_framework}). \cite{Iwaya2023} show that the adoption of privacy engineering practices is often inconsistent and largely reactive --- shaped by organisational culture, leadership commitment, and external regulatory pressures. As a result, privacy measures are frequently implemented in response to compliance demands rather than being systematically embedded into the software development lifecycle. Practitioners often feel underqualified to make privacy-related design decisions and express concerns about making incorrect choices or facing accountability. This leads to  hesitation, delegation, or avoidance of DP and other privacy-enhancing techniques.

\subsection{Regulatory and Policy Considerations}
%\section{Enabling Institutional Adoption: Regulatory and Policy Considerations}
DP offers strong formal guarantees and has generated growing interest from policymakers and public institutions. However, realizing its full potential in these contexts requires addressing regulatory complexities, operational constraints, and evolving legal interpretations.

\cite{DrechslerJorg2023DPfG} notes that public sector agencies, such as national statistical offices, face unique challenges in adopting DP --- particularly due to requirements for survey design, data weighting, and long-term reproducibility. These factors complicate the implementation of privacy-preserving mechanisms that are commonly used in industry but must be adapted to meet the strict legal and methodological standards of official statistics.

\cite{cummings2024centeringpolicypracticeresearch} emphasize that aligning DP with institutional goals requires more than technical soundness. Policymakers and data stewards often evaluate privacy in terms of public accountability, fairness, and group-level risk --- dimensions not directly captured by DP’s individual-level guarantees. Clearer guidance on how to interpret and communicate these guarantees in policy-relevant terms would help build trust and support responsible adoption.%\medskip

The regulatory treatment of DP-generated outputs, such as synthetic data (see Section \ref{sec_synthetic_data}), is also evolving. As shown by \cite{Pitkamaki2024}, synthetic data is not automatically considered anonymous under laws like the GDPR or Finland’s Secondary Use Act. Determining whether synthetic datasets meet privacy standards depends on context and implementation (e.g., a hospital may release a DP-synthetic admissions dataset only after confirming low re-identification risk). However, with careful design and risk assessment, synthetic data can serve as a useful bridge between privacy protection and data availability.%\medskip

As legal frameworks mature and guidance becomes more refined, institutions are increasingly well-positioned to leverage DP in ways that uphold both privacy and data utility. Targeted support --- such as domain-specific standards, evaluation tools, and use-case-driven policy models --- can further enable the responsible and effective adoption of DP across sectors.
For example:
\begin{itemize}
    \item \textbf{Healthcare:} Standardized risk‑assessment protocols for releasing differentially private health statistics could help hospitals share aggregate patient data for public‑health research while complying with GDPR and similar regulations.
    \item \textbf{Finance:} DP frameworks could enable banks to publish industry‑wide risk indicators or fraud‑detection benchmarks without exposing client‑level financial data.
    \item \textbf{Education:} Schools and education authorities could release anonymized, differentially private student performance metrics to support policy evaluation and academic research without compromising individual privacy.
    \item \textbf{Smart cities:} Municipalities could share differentially private mobility or energy‑usage data to inform urban planning and sustainability initiatives without exposing individuals’ travel or consumption patterns.
\end{itemize}
These kinds of sector‑specific tools and policies would provide practical pathways for responsibly implementing DP in real‑world settings.

%\section{Recommendations for Usability and Governance}
\subsection{Recommendations and Best Practices}\label{secX}
To bridge the gap between the theoretical foundations of DP and its practical adoption, recent literature offers a range of actionable recommendations. These span user communication, developer support, and deployment practices:\bigskip

\noindent
\textbf{For Communicating with End Users}
\begin{itemize}
\item \textbf{Reformulate DP explanations} to address the specific types of information disclosures that users fear and care about, such as access by third parties or misuse by institutions (\cite{Cummings2023}).
\item \textbf{Prefer $\boldsymbol{\mu}$-GDP over $(\boldsymbol{\varepsilon},\boldsymbol{\delta})$-DP} for reporting and comparing privacy guarantees, as it avoids arbitrary $\delta$ choices and is easier to interpret (\cite{gomez2025varepsilondeltaconsideredharmful}).
\item \textbf{Use visual aids and interactive tools} --- such as frequency-based risk displays, odds-based framing, or metaphor-driven interfaces --- to explain probabilistic guarantees and $\varepsilon$-related trade-offs (\cite{10646708,281270,291162,WEN2023118799}).
\item \textbf{Design accessible privacy interfaces} that communicate risk and trade-offs effectively to users with diverse backgrounds, abilities, and levels of numeracy (\cite{Franzen1,Franzen2}).
\item \textbf{Apply adaptive (personalized) approaches to setting $\boldsymbol{\varepsilon}$}, based on contextual sensitivity and user preferences, rather than relying solely on fixed, developer-defined values (\cite{boenisch2024wayindividualizedprivacyassignment,Edabi2015,Jorgensen2015,weiss2024shareshareriskslaypeople}).
\end{itemize}
\medskip

\noindent
\textbf{For Supporting Developers and Data Practitioners}
\begin{itemize}
\item \textbf{Align software tools and libraries} with familiar data science workflows, while maintaining rigorous DP guarantees. This includes integrating guidance into code libraries and providing intuitive interfaces (\cite{10.1145/3687011}).
\item \textbf{Provide educational resources and helpful error messages} to support non-expert developers in understanding DP concepts and common implementation pitfalls (\cite{ngong2024evaluatingusabilitydifferentialprivacy}).
\item \textbf{Optimize hyperparameters} --- such as learning rate, batch size, clipping thresholds and noise scales --- in a principled way to maximize utility without weakening privacy guarantees (\cite{Ponomareva2023}).
\item \textbf{Ensure auditability and transparency} 
by adopting practices such as open-source code, clear documentation, and independent audits \cite{gomez2025varepsilondeltaconsideredharmful}. Even theoretically sound DP mechanisms can be misimplemented in practice (e.g., PATE-based methods, \cite{ganev2025}), 
%as illustrated by issues in PATE-based systems \cite{ganev2025}. More broadly, 
{and recent grey-box studies by \cite{cebere2026privacytheorybugspractice} have uncovered multiple privacy violations in widely used DP libraries
%, underscoring the value of external review.
%
%by adopting practices such as open-source code, clear documentation, and independent audits \cite{gomez2025varepsilondeltaconsideredharmful}. Past work has shown that even theoretically sound DP mechanisms, like PATE-based methods, can be misimplemented in practice \cite{ganev2025}.
%(e.g., PATE-based methods, \cite{ganev2025}). 
%{\color{red} More broadly, recent grey-box audits have uncovered multiple privacy violations in widely used DP libraries \cite{cebere2026privacytheorybugspractice}, 
highlighting the importance of} external review and empirical auditing.
%\item \textbf{Ensure auditability and transparency} by adopting practices such as open-source code, clear documentation, and independent audits (\cite{gomez2025varepsilondeltaconsideredharmful}). Past work (e.g., PATE-based methods, \cite{ganev2025}) has shown that even theoretically sound approaches can be misimplemented, so enabling external review helps detect and correct errors.

\end{itemize}
\medskip

\noindent
\textbf{For Deployment, Governance, and Policy Transparency}
\begin{itemize}
\item \textbf{Fully report and document privacy parameters and decisions} (see DP Disclosure Checklist below), including $\varepsilon$, $\delta$, \emph{or preferably the GDP parameter} $\mu$, along with unit of privacy, accounting methods, and the rationale behind key trade-offs. Using $\mu$-GDP for reporting provides a single, comparable privacy parameter that better reflects the true privacy profile of many mechanisms (\cite{gomez2025varepsilondeltaconsideredharmful}). Clear documentation helps prevent privacy theater and fosters trust among users and stakeholders (\cite{cummings2024centeringpolicypracticeresearch,Ponomareva2023,10.1145/3555762}).

\item \textbf{Foster cross-functional collaboration} between privacy engineers, legal teams, and User Experience (UX) designers to ensure DP implementations are compliant, understandable, and aligned with organizational goals (\cite{Iwaya2023}).

\item \textbf{Develop sector-specific guidelines} (e.g., for healthcare, education, or official statistics) to ensure that DP methods are tailored to the legal, technical, and social requirements of different application domains (\cite{Pitkamaki2024}).
\end{itemize}

\noindent
The above recommendations highlight many complementary aspects of deploying DP systems.
To provide an at-a-glance overview of the considerations discussed throughout this article, we summarize a typical workflow for designing and deploying DP systems.
\medskip

\noindent\textbf{Practical DP Workflow} 
\begin{enumerate}
\item \textbf{Define privacy objective and threat model.}  
Clarify the risks to be mitigated, the potential adversaries, and the type of information that must be protected.

\item \textbf{Choose the deployment and trust model.}  
Decide whether the system operates under central, local, or distributed/federated DP and where DP protection is applied (input, training, output).

\item \textbf{Specify the protected unit.}  
Determine the unit of privacy (e.g., record-level or user-level), define the neighboring relation, and establish contribution bounds that determine sensitivity.

\item \textbf{Choose a privacy-preserving method.}  
Choose appropriate DP mechanisms or training approach (e.g., Laplace, Gaussian, DP-SGD)  together with an accounting method and assumptions required to track cumulative privacy loss.

\item %\textbf{Set privacy parameters and evaluate utility.}  
\textbf{Set privacy parameters, validate utility, and audit the system.}  
Determine suitable privacy parameters (e.g., $\varepsilon$, $\delta$, or $\mu$ in GDP), evaluate the resulting privacy?utility trade-off, and where possible validate the implementation through empirical privacy auditing or testing.

\item \textbf{Document and communicate the guarantee.}  
Report assumptions, parameters, mechanisms, and limitations transparently (see the DP Disclosure Checklist below).
\end{enumerate}
In practice, this process is often iterative: if the resulting utility is insufficient or deployment constraints change, practitioners may need to revisit earlier decisions, such as the choice of mechanisms, contribution bounds, or even the protected unit.

Building on the above workflow, transparent reporting is an essential part of responsible DP deployment. 
To translate these considerations into practice
and reduce the risk of overclaiming, 
% practitioners should both validate their systems (e.g., via privacy auditing) and communicate guarantees transparently. 
%To support the latter, 
we propose a concise \textbf{DP Disclosure Checklist} for practitioners communicating privacy guarantees (see Table~\ref{tab:dp-checklist2}). In addition, we provide a brief illustrative disclosure example demonstrating how these elements may be communicated in practice.
%\enlargethispage{\baselineskip}

\bigskip
{\small
\begin{mdframed}\vspace{-1mm}
 \begin{example}\label{ex_disclosure}
 {
 \textbf{Illustrative Disclosure Example} \smallskip

\noindent 
This model was trained under central $(\varepsilon = 3, \delta = 10^{-6})$-DP with a trusted service provider having access to raw data. Protection is provided at the user level, where neighboring datasets differ by the addition or removal of a single user?s data. Gradients were clipped at a fixed threshold $C$ to bound per-user contribution, and Gaussian noise was added during training using DP-SGD. The reported privacy budget accounts for both training and hyperparameter tuning, and was computed with an RDP accountant assuming Poisson subsampling. The guarantee applies to the released model parameters only and reflects a single training run. It protects against inference of an individual user?s participation but does not address fairness concerns, distribution shift, or side-channel attacks outside the DP mechanism.}
\end{example}
\end{mdframed}
}

\begin{landscape}
\begin{table*}[t!h]
{
\caption{DP Disclosure Checklist}
\label{tab:dp-checklist2}
%\resizebox{\textwidth}{!}{
{\small
\begin{tabular}{@{}llllll@{}}
\toprule
\textbf{Item} & \textbf{What to disclose} & \textbf{Examples} \\
\midrule
\begin{tabular}[c]{@{}l@{}} DP setting \& trust model \end{tabular}
& \begin{tabular}[c]{@{}l@{}} Specify DP setting and trust assumptions (who can access\\ raw data, intermediate computations, or model updates).\end{tabular}
& \begin{tabular}[c]{@{}l@{}} Central DP; local DP; federated/distributed \\ DP; secure aggregation.\end{tabular} \\ \midrule

DP coverage
%DP insertion point \& data access
&  \begin{tabular}[c]{@{}l@{}} State where DP is applied, which stages are included in \\ accounting, and what outputs are protected. \end{tabular}
& \begin{tabular}[c]{@{}l@{}} Input or output perturbation; \\ training-time DP (e.g., DP-SGD);\\ training + tuning; final model release only.\end{tabular} \\ \midrule

Protected unit \& adjacency
%Protected unit, adjacency \& contribution bounds
&  \begin{tabular}[c]{@{}l@{}} Specify privacy unit, neighboring relation, and contribution \\ limits that determine sensitivity.\end{tabular}
& \begin{tabular}[c]{@{}l@{}} Example-level; replace-one;\\ clipping threshold $C$; per-user limits;\\ group-privacy factor $k$.\end{tabular}
\\ \midrule

DP definition \& parameters 
&  \begin{tabular}[c]{@{}l@{}} Report the formal privacy notion and final disclosed \\ parameters; provide an interpretable explanation of the \\ guarantee, and when feasible present guarantees in a form \\ that supports intuitive interpretation (e.g., via $\mu$-GDP).\end{tabular}
& \begin{tabular}[c]{@{}l@{}} $(\varepsilon,\delta)$-DP; pure DP; RDP; $\mu$-GDP; \\explanation of $\delta$; $\mu$ value.\end{tabular} \\ \midrule

Composition \& accounting 
&  \begin{tabular}[c]{@{}l@{}} Describe how cumulative privacy loss is computed, accountant \\  used, internal conversions, and required assumptions.\end{tabular}
& \begin{tabular}[c]{@{}l@{}} RDP or PLD accountant; advanced \\ composition; RDP $ \rightarrow (\varepsilon,\delta)$; Poisson \\subsampling  for amplification.\end{tabular} \\ \midrule

\begin{tabular}[c]{@{}l@{}} Mechanism \& \\ \qquad implementation\end{tabular}
&  \begin{tabular}[c]{@{}l@{}} Disclose mechanism type, noise distribution, actual \\ sampling/batching strategy, and technical conditions affecting \\ privacy guarantees.\end{tabular}
& \begin{tabular}[c]{@{}l@{}} Exponential mechanism; fixed size \\mini batching vs.\ Poisson sampling;\\ seed management.\end{tabular} \\ \midrule

Scope \& limitations 
&  \begin{tabular}[c]{@{}l@{}} Define protection scope, adversary model, and 
explicitly \\ state risks outside the DP guarantee.\end{tabular}
& \begin{tabular}[c]{@{}l@{}} Auxiliary information risks; population \\ inference; fairness/bias; side-channel threats.\end{tabular} \\ \midrule

\begin{tabular}[c]{@{}l@{}} Documentation, auditing, \& \\ \qquad transparency\end{tabular}
&  \begin{tabular}[c]{@{}l@{}} Provide accessible material supporting verification, \\ auditing, and interpretation.\end{tabular}
& \begin{tabular}[c]{@{}l@{}} Public documentation; configuration details;\\ code references; auditing results or reports;\\ use of audited libraries (e.g., Opacus, \\ TensorFlow Privacy, JAX Privacy).\end{tabular} \\
\bottomrule
\end{tabular}}}
\end{table*}
\end{landscape}

To conclude, the practical impact of DP depends not only on its formal guarantees, but on how those guarantees are communicated, interpreted, and embedded within real-world organizational workflows. Empirical studies indicate that users' concerns are typically framed around concrete risks --- such as misuse by organizations or third-party access --- rather than abstract mathematical definitions. Misinterpretations of privacy parameters, particularly $\varepsilon$, can therefore lead either to misplaced trust or unwarranted skepticism.

For practitioners, implementation challenges extend beyond algorithm design to parameter selection, usability of tools, and clarity of documentation. Effective deployment requires layered, accessible, and context-sensitive communication strategies that acknowledge diverse levels of technical literacy while maintaining formal rigor.

At the governance level, legal interpretations of DP --- including its status under regulations such as the GDPR --- continue to evolve. In the absence of settled case law, organizations may hesitate to rely on DP in high-stakes domains. Nonetheless, when implemented transparently and documented rigorously, DP offers one of the strongest currently available mechanisms for reconciling data utility with quantifiable privacy protection. Continued development of sector-specific standards, evaluation frameworks, and regulatory guidance will be essential to support responsible adoption.

\section{Future Trends and Research Directions}\label{chaFuture}
%Differential Privacy (DP) has evolved from a theoretical construct into a widely studied and increasingly deployed framework for data protection. However, real-world implementation remains challenging, and several key research directions are emerging to bridge the gap between formal guarantees and practical needs. This chapter synthesizes anticipated future trends and open research questions in the DP landscape, with an emphasis on making DP-training more robust, flexible, and accessible.
%
Differential privacy (DP) has emerged as a leading paradigm for privacy‑preserving data analysis, particularly when dealing with sensitive data in machine learning training processes. Once largely confined to academic research, DP is increasingly becoming mainstream, with high‑profile deployments by Apple, Google, Microsoft, and the U.S. Census Bureau, as well as growing recognition in regulatory frameworks such as GDPR guidance and the NIST 2025 recommendations. DP‑training integrates noise systematically into training procedures, ensuring the trained models generalize well without compromising individual privacy. Despite considerable progress, the rapid evolution of technology and the increasing complexity of datasets continue to highlight critical challenges and opportunities for future research.
\bigskip

\noindent
\textbf{Efficient and Scalable DP-Training: } % OK
 A crucial direction for future research involves developing scalable DP-training methods capable of efficiently handling large-scale datasets and complex neural network architectures. Current DP approaches, particularly gradient perturbation techniques like DP-SGD, often incur high computational and privacy overheads. Novel strategies leveraging advanced optimization methods, such as adaptive clipping, dynamic noise adjustment, and improved gradient approximation, are necessary to enhance efficiency and scalability (\cite{Papernot2021Making}). 
Moreover, designing machine learning models and architectures specifically tailored for DP --- rather than retrofitting privacy mechanisms onto existing ones --- may lead to more efficient and inherently private systems (\cite{Papernot2021Making}). Interestingly, DP has also been observed to improve generalization in some settings, acting as a form of regularization (\cite{anil-etal-2022-large,de2022unlockinghighaccuracydifferentiallyprivate}). Future work may further explore this dual role of DP, tuning privacy not only to protect individual data but also to improve model robustness and mitigate overfitting.
\bigskip

\noindent
\textbf{Federated and Distributed DP Learning: } 
Another significant trend involves integrating DP with federated and distributed learning frameworks. Federated learning allows models to be trained locally on user devices or institutional silos, sharing only model updates rather than raw data, which already provides some privacy benefits.  Incorporating DP into this paradigm further strengthens privacy guarantees, especially in scenarios where sensitive data cannot be centrally aggregated. Notably, research distinguishes between \emph{cross-device federated learning}, where a large number of unreliable user devices (e.g., phones, sensors) participate, and \emph{cross-silo federated learning}, where a small number of stable clients (e.g., hospitals, banks) contribute structured datasets. Each setting presents different challenges and opportunities for privacy accounting, personalization, and communication efficiency. This makes DP-enhanced federated learning a rich area for future work on balancing utility, privacy, and scalability across diverse deployment environments (see, e.g., \cite{KairouzPeter2021TDDG,kairouz:hal-02406503} and Section~\ref{sec_FL}).
\bigskip

\noindent
\textbf{Integration with Cryptography: }
In addition, integrating DP with cryptographic techniques presents a promising research direction. Combining DP with secure multi-party computation (MPC), homomorphic encryption (HE), and secure aggregation methods can significantly enhance data security and privacy guarantees. Exploring these intersections further will help design systems capable of secure, privacy-preserving computations across diverse applications, particularly those involving multiple stakeholders and sensitive distributed datasets (see, e.g., \cite{Bonawitz2017,wagh2021}, and Section \ref{sec_crypto}).
\bigskip

\noindent
\textbf{Advanced Mechanisms for Improved Utility-Privacy Trade-offs:} %ok
Current DP methods often face trade-offs between privacy guarantees and model utility. Future research should prioritize innovative DP mechanisms, such as adaptive composition strategies, optimized noise allocation, and refined privacy accounting methods, to achieve better utility at stringent privacy budgets. Techniques like Rényi DP (\cite{RDP2017}), Gaussian DP (\cite{DongJinshuo2022Gdp}), and privacy loss distribution (PLD) accounting (\cite{KoskelaAntti2019CTDP,KoskelaAntti2022IPAw}) have already shown promise and warrant further exploration.
\bigskip

\noindent
\textbf{Adaptive and Personalized Privacy:} %OK
Moreover, traditional DP methods often apply uniform privacy guarantees across datasets. Future research need to further develop adaptive and personalized privacy frameworks that dynamically tailor privacy guarantees to the sensitivity of individual data points or user preferences  (\cite{AlagganMohammad2017HDP,boenisch2024wayindividualizedprivacyassignment,Edabi2015,Jorgensen2015}). Such adaptive approaches would improve utility by allocating privacy budgets more effectively, thus better addressing varying user privacy needs and real-world contexts.
\bigskip

\noindent
\textbf{DP in Emerging Machine Learning Models: } % ok
The rapid advancement of emerging machine learning architectures, particularly transformers, graph neural networks (GNNs), and large language models (LLMs), presents distinct challenges for DP-training (see, e.g., \cite{anil-etal-2022-large,li2022largelanguagemodelsstrong,Ponomareva2022,yu2024differentially}). Future studies must address how DP techniques can effectively handle the intrinsic characteristics of these models, such as extensive parameter sets, complex attention mechanisms, and intricate graph structures, without substantial degradation of performance.
\bigskip

\noindent
\textbf{Privacy Auditing and Real-World Attacks: } % ok
Formal DP guarantees are worst-case by design, but this often leaves practitioners uncertain about actual information leakage (\cite{Cummings2024Advancing}). 
Therefore robustness of DP mechanisms against real-world adversarial settings is a critical area for future research. Developing systematic privacy auditing techniques and evaluating DP-training under realistic threat models, including membership inference, reconstruction attacks, and attribute inference attacks, is vital. Comprehensive auditing frameworks will help practitioners confidently deploy DP technologies by providing clarity about their resilience against practical privacy threats (\cite{CarliniNicholas2018TSSE,Jayaraman2019,8835245}).
\bigskip

\noindent
\textbf{Estimating Empirical Privacy Loss: } %\enlargethispage{0.1\baselineskip}
While differential privacy provides formal worst-case guarantees, these bounds can be overly conservative for many real-world threat models. This has motivated research on \emph{empirical estimation of privacy loss}, where the effective privacy budget $\hat{\varepsilon}$ is measured against specific adversaries, often using membership inference attacks. \cite{pmlr-v202-zanella-beguelin23a} recently proposed a Bayesian framework for estimating $\hat{\varepsilon}$ that models the joint distribution of false positive and false negative rates, yielding much tighter credible intervals and reducing sample complexity by up to two-thirds compared to traditional methods. Looking forward, these kinds of techniques --- Bayesian estimation and other statistically grounded approaches --- are expected to complement theoretical privacy accounting, giving practitioners clearer insights into real-world privacy risks and helping guide budget selection and hyperparameter tuning.\bigskip

\noindent
\textbf{Transparent and Explainable DP: } % ok
As DP becomes mainstream, transparency and explainability in DP-training methods will become increasingly critical. Users and practitioners alike will need clearer insights into privacy trade-offs, the impact of noise introduction, and the implications for model interpretability and fairness. Research on integrating explainability frameworks with DP, including visualization techniques and interpretable privacy accounting, will be essential for wider acceptance and deployment of DP-based technologies (\cite{10646708,Cummings2023,281270,291162,WEN2023118799}).
\bigskip

%\noindent
%\textbf{Conclusion: }
DP continues to evolve as both a theoretical foundation and a practical tool for data protection. Future research stands at a pivotal point, with growing emphasis on scalability, utility, and integration with distributed and federated systems. Addressing challenges such as computational efficiency, adaptability to new model architectures, and support for complex data modalities will be critical to enabling real-world adoption. In parallel, advancing fairness-aware DP methods will be essential to ensure that privacy protections do not unintentionally amplify existing biases or disproportionately affect vulnerable groups. Finally, efforts in transparency, explainability, and personalized privacy will help align DP with broader societal expectations for trust and accountability. Together, these directions aim to make DP a robust, equitable, and widely accepted framework for privacy-preserving technologies in an increasingly data-driven world.

\section{Conclusion}
% Tarvitaanko vielä erikseen? Joka luvussa on oma ja tuo future trends kyllä jossain määrin concludeaa kaiken

Differential Privacy (DP) is a mathematically robust approach to privacy, designed to ensure that the inclusion or exclusion of any single individual’s data does not significantly affect the outcome of an analysis. This property makes DP resistant to auxiliary information attacks while providing mathematically guaranteed privacy protection and composable privacy guarantees across multiple analyses. %\medskip

This review has comprehensively explored various aspects and applications of DP. Initially, foundational concepts, definitions, and theoretical properties were discussed to establish a robust understanding of DP's core principles and mechanisms. Following this, the integration of DP into machine learning, emphasizing privacy-preserving training procedures, notably differentially private stochastic gradient descent (DP-SGD), was examined to illustrate practical methodologies and their associated challenges.%\medskip

The report subsequently delved into recent advancements and specialized techniques for enhancing differential privacy, such as the intersection of DP with cryptographic methods and federated learning. The generation and evaluation of privacy-preserving synthetic data were also covered, highlighting how DP can facilitate secure data sharing and analytics without compromising individual privacy.%\medskip

Furthermore, the review included an extensive survey of real-world applications and use cases of differential privacy across diverse domains, including cybersecurity, healthcare, and finance, demonstrating DP's broad relevance and impact. Practical considerations were examined to understand user expectations, regulatory frameworks, communication challenges, and best practices for implementing DP.%\medskip

While sections covering foundational definitions and machine learning methodologies provided essential background, subsequent sections particularly emphasized recent developments, emerging techniques, and current literature, illustrating ongoing innovation and active research directions within the DP community.%\medskip

This report underscores the increasing importance and effectiveness of DP as a critical tool in balancing the need for data utility with privacy preservation. Although achieving an optimal balance between privacy guarantees and data utility remains a central challenge in DP research and applications, significant advancements and innovative research promise continued progress. The ongoing exploration into adaptive mechanisms, personalized privacy frameworks, and robust defenses against privacy attacks holds great potential, ensuring DP will continue to effectively address future data privacy challenges.%\medskip

In summary, DP is a foundational tool for privacy-preserving data science, combining strong theoretical guarantees with growing practical adoption.

\section*{Acknowledgements}
The authors would like to thank Jenni Lampainen for carefully reading and commenting the original draft of this survey.
The authors acknowledge the use of OpenAI's ChatGPT for support in refining the writing of this paper. 

The work was supported by {the Research Council of Finland, Projects No.\ \#340182 and \#340140}, and
 the European Union's Horizon Europe project Privacy Preserving Identity Management for Digital Wallet and Secure Data Sharing and Processing for Cyber Threat Intelligence Data (PRIVIDEMA, Grant Agreement No.\ 101167964). Views and opinions expressed are however those of the authors only and do not necessarily reflect those of the European Union or the European Cyber-security Competence Centre. Neither the European Union nor the European Cybersecurity Competence Centre can be held responsible for them.

 Tapio Pahikkala acknowledges the research environment provided by ELLIS Institute Finland.

%The work was supported by the European Union's Horizon Europe project Privacy Preserving Identity Management for Digital Wallet and Secure Data Sharing and Processing for Cyber Threat Intelligence Data (PRIVIDEMA, Grant Agreement No.\ 101167964).
\bigskip

%\pagebreak 
%\bibliographystyle{abbrv}
%\bibliographystyle{plainnat}   % or abbrvnat, unsrtnat, etc.
%\bibliography{d_bib}    
%\bibliography{sn-bibliography}% common bib file

\begin{appendices}
\section{Recommended Literature}\label{chaLiterature}
% Tämä voisi olla ehkä appendixina ennen kirjastoja

We now present a selection of key publications that have significantly contributed to the development and understanding of differential privacy (DP), serving as valuable resources for readers seeking to deepen their expertise or explore specific topics in more detail.\bigskip\bigskip\

% General
\noindent
\textbf{Title:} Calibrating Noise to Sensitivity in Private Data Analysis \newline
\textbf{Authors:} Cynthia Dwork, Frank McSherry, Kobbi Nissim, and Adam Smith \newline \textbf{Reference:} \cite{Dwork2006}\newline 
\textbf{Year:} 2006 \newline

\noindent
This seminal paper introduces the formal definition of DP and establishes a rigorous mathematical framework for privacy-preserving data analysis. It defines the core $\varepsilon$-DP model and presents the Laplace mechanism, which adds calibrated noise based on a query's global sensitivity to achieve privacy guarantees. The authors prove key properties such as composability and post-processing invariance, laying the theoretical foundation for virtually all future work in DP. This paper is essential reading for understanding both the motivation and the formal guarantees underlying modern approaches to data privacy.

%\bigskip
\bigskip
\bigskip

\noindent
\textbf{Title:} The Algorithmic Foundations of Differential Privacy\newline
\textbf{Authors:} Cynthia Dwork and Aaron Roth \newline \textbf{Reference:} \cite{DworkRoth2014}\newline 
\textbf{Year:} 2014 \newline

\noindent
This monograph provides the definitive foundation for understanding DP, presenting both its theoretical underpinnings and practical mechanisms. Dwork and Roth introduce DP as a formal privacy guarantee that ensures an individual's participation in a dataset has a negligible impact on the outcome of any analysis. The work thoroughly develops the mathematical formulation of DP, explores key mechanisms such as the Laplace and exponential mechanisms, and presents powerful composition theorems which allow for cumulative privacy accounting across multiple analyses. It also covers advanced techniques like the sparse vector method, as well as applications in linear query answering, mechanism design, and machine learning. With its rigorous treatment of both privacy and utility trade-offs, the book is essential reading for anyone seeking a deep understanding of privacy-preserving data analysis. 
\bigskip
\bigskip
%\bigskip\newpage

\noindent
\textbf{Title:} How to DP-fy ML: A Practical Guide to Machine Learning with Differential Privacy \newline
\textbf{Authors:} Natalia Ponomareva, Hussein Hazimeh, Alex Kurakin, Zheng Xu, Carson Denison, H. Brendan McMahan, Sergei Vassilvitskii, Steve Chien, and Abhradeep Thakurta \newline \textbf{Reference:}  \cite{Ponomareva2023}
\newline 
\textbf{Year:} 2023 \newline
 
\noindent
This survey paper provides a comprehensive guide for applying DP to Machine Learning (ML) models. It addresses the challenges in adopting DP for modern complex ML models, such as privacy-utility trade-offs, architectural modifications, hyperparameter tuning, and privacy accounting. The paper is aimed at both researchers and practitioners, providing foundational theory, practical algorithms, and detailed steps for implementation. The paper offers step-by-step guides for implementing DP in real-world ML systems and suggests privacy budgets ($\varepsilon\leq 10$) for practical deep learning applications. It highlights privacy amplification techniques, including sampling and multi-device considerations, for enhancing privacy guarantees. The guide aspires to enable broader adoption of DP ML models with good accuracy and privacy, tailored to both academic research and industry deployment.
\bigskip
\bigskip
%\bigskip

\noindent
\textbf{Title:} Guidelines for Evaluating Differential Privacy Guarantees\newline
\textbf{Authors:} Joseph P.\ Near, David Darais, Naomi Lefkovitz, and Gary S.\ Howarth \newline \textbf{Reference:} \cite{NIST2025} \newline
\textbf{Year:} 2025 \newline

\noindent
This comprehensive publication from the U.S. National Institute of Standards and Technology provides guidance for understanding, evaluating, and comparing DP guarantees in real-world systems. It is aimed at both technical and non-technical stakeholders and covers formal definitions, privacy parameters ($\varepsilon,\delta$), algorithmic mechanisms, and practical deployment issues. Key contributions include a conceptual "privacy pyramid" for framing evaluation, extensive discussion on the interpretation of $\varepsilon$ in different contexts, and an exploration of empirical auditing methods. It also offers practical hazards and flowcharts to help guide implementation decisions and includes supplementary Jupyter notebooks for hands-on exploration

%\bigskip
\bigskip\bigskip

\noindent
\textbf{Title:} Advancing Differential Privacy: Where We Are Now and Future Directions for Real-World Deployment\newline
\textbf{Authors:} Rachel Cummings, Damien Desfontaines, David Evans, Roxana Geambasu, Yangsibo Huang, Matthew Jagielski, Peter Kairouz, Gautam Kamath, Sewoong Oh, Olga Ohrimenko, Nicolas Papernot, Ryan Rogers, Milan Shen, Shuang Song,  Weijie Su, Andreas Terzis, Abhradeep Thakurta, Sergei Vassilvitskii, Yu-Xiang Wang, Li Xiong, Sergey Yekhanin, Da Yu, Huanyu Zhang, and Wanrong Zhang \newline \textbf{Reference:} \cite{Cummings2024Advancing} \newline
\textbf{Year:} 2024 \newline

\noindent
This comprehensive article provides a detailed review of current practices and state-of-the-art methodologies in the field of DP, with a focus of advancing DP's deployment in real-world applications. It delves into the theoretical foundations of DP, explores practical implementations across various domains, and discusses the challenges faced in balancing data utility with privacy guarantees. The authors highlight recent advancements in DP techniques, including algorithmic innovations and applications in real-world scenarios. Additionally, the article addresses policy considerations and outlines potential directions for future research, making it an essential read for those interested in the evolving landscape of data privacy.

%\bigskip
\bigskip\bigskip
%\newpage

% Special techniques

\noindent
\textbf{Title:} Deep Learning with Differential Privacy\newline
\textbf{Authors:}
Martin Abadi, Andy Chu, Ian Goodfellow, H. Brendan McMahan, Ilya Mironov, Kunal Talwar, and Li Zhang
\newline \textbf{Reference:}  \cite{abadi2016deep} \newline
\textbf{Year:} 2016 \newline

\noindent
This influential paper introduced the first practical framework for training deep neural networks with DP. Abadi et al.\ adapt stochastic gradient descent (SGD) to incorporate privacy-preserving mechanisms through a carefully designed algorithm known as DP-SGD. Key innovations include per-example gradient clipping, calibrated noise addition, and a novel \emph{moments accountant} technique, which provides significantly tighter tracking of cumulative privacy loss compared to traditional composition theorems. This work marks a turning point in privacy-preserving machine learning, showing that large, non-convex models can be trained with formal DP guarantees while maintaining utility.
\bigskip
\bigskip
%\bigskip

\noindent
\textbf{Title:}  R\'enyi Differential Privacy \newline
\textbf{Authors:} Ilya Mironov \newline \textbf{Reference:} \cite{RDP2017}
\newline
\textbf{Year:} 2017\newline

\noindent
This foundational paper introduces Rényi Differential Privacy (RDP), a relaxation of traditional DP based on Rényi divergence. RDP retains many of the favorable properties of standard DP --- such as robustness to auxiliary information, post-processing invariance, and composition --- but offers significantly tighter bounds, especially for heterogeneous or repeated queries. The paper presents a clear definition of RDP, proves key theoretical properties, and demonstrates its effectiveness in analyzing mechanisms like Gaussian noise and DP-SGD. RDP also provides a convenient framework for privacy accounting and enables accurate, efficient tracking of cumulative privacy loss in practice.

\bigskip
\bigskip
%\bigskip

\noindent
\textbf{Title:} Amplification by Shuffling: From Local to Central Differential Privacy via Anonymity \newline
\textbf{Authors:} \'Ulfar Erlingsson, Vitaly Feldman, Ilya Mironov, Ananth Raghunathan, Kunal Talwar, and Abhradeep Thakurta \newline \textbf{Reference:} \cite{Erlingsson_amp2019} \newline
\textbf{Year:2019} \newline

\noindent
This paper presents a general privacy amplification result showing that shuffling locally differentially private (LDP) reports can substantially strengthen privacy guarantees when viewed in the central model. The authors prove that for many practical settings, shuffling LDP data achieves central DP with privacy loss reduced by a factor proportional to the square root of the population size. This finding provides a theoretical foundation for architectures like Encode-Shuffle-Analyze (ESA) and highlights that anonymity and unlinkability in LDP systems can be leveraged for significantly improved privacy-utility trade-offs. The paper also introduces a longitudinal monitoring protocol under LDP with polylogarithmic privacy cost per user change.

\bigskip
\bigskip
%\bigskip

%\newpage

% Enchanging DP
\noindent
\textbf{Title:}  DP-Cryptography: Marrying Differential Privacy and Cryptography in Emerging Applications  \newline
\textbf{Authors:} Sameer Wagh, Xi He, Ashwin Machanavajjhala, and Prateek Mittal \newline \textbf{Reference:} \cite{wagh2021}\newline
\textbf{Year:} 2021 \newline

\noindent
This paper presents a comprehensive review of the emerging field at the intersection of DP and cryptography, which they term DP-cryptography. The paper outlines two major research directions: using cryptographic primitives (such as secure multiparty computation and anonymous communication) to enable differentially private data analysis without relying on a trusted curator, and conversely, using DP to relax and speed up traditional cryptographic protocols. Through system case studies like Prochlo, DJoin, and Shrinkwrap, the authors illustrate how combining these two paradigms can yield practical, scalable, and privacy-preserving data processing architectures that neither approach achieves alone. This paper provides both foundational context and concrete guidance for designing systems that reconcile privacy, trust, and utility. %\medskip

%Differential privacy (DP) and cryptography are powerful tools for preserving privacy in data-driven systems. This paper explores the interplay between these two fields to address privacy and utility challenges in practical applications. DP provides a robust privacy framework, while cryptographic primitives add secure computation and communication capabilities. Combining the two can bridge gaps in trust assumptions, utility, and scalability. \medskip

%The integration of differential privacy and cryptography opens a new frontier for designing systems with strong privacy guarantees while addressing practical limitations. It lays the foundation for future research on scalability, trust models, and advanced use cases.
\bigskip
\bigskip{}
%\bigskip

\noindent
\textbf{Title:} Differentially Private Federated Learning: A Systematic Review\newline
\textbf{Authors:} Jie Fu, Yuan Hong, Xinpeng Ling, Leixia Wang, Xun Ran, Zhiyu Sun, Wendy Hui Wang, Zhili Chen, and Yang Cao \newline \textbf{Reference:} \cite{fu2024}\newline
\textbf{Year:} 2024 \newline

\noindent
Fu et al.\ provide a comprehensive survey of DP in federated learning, covering key privacy models (central, local, shuffle), protection levels (sample- vs. client-level), and modern composition techniques such as zCDP and RDP. The article reviews core mechanisms like DP-SGD and highlights recent advances including heterogeneous privacy budgets, adaptive noise, and communication-efficient designs. It serves as an up-to-date reference for both foundational concepts and current research directions in privacy-preserving federated learning.
%\bigskip{}
\bigskip \bigskip

\noindent
\textbf{Title:} Advances and Open Problems in Federated Learning \newline
\textbf{Authors:} Peter Kairouz, Brendan H. Mcmahan, Brendan Avent, Aurélien Bellet, Mehdi
Bennis, Arjun Nitin Bhagoji, Kallista Bonawitz, Zachary Charles, Graham
Cormode, Rachel Cummings, et al. \newline \textbf{Reference:} \cite{kairouz:hal-02406503} \newline
\textbf{Year:} 2021 \newline

\noindent
 Kairouz et al.\ present a comprehensive and influential survey covering the full landscape of FL research. The work outlines fundamental concepts such as cross-device and cross-silo FL, explores practical challenges like communication bottlenecks, system heterogeneity, and privacy preservation, and highlights open problems in optimization, personalization, fairness, and robustness. Particular attention is given to DP, secure aggregation, and system design under realistic threat models. With contributions from leading researchers across academia and industry, the survey offers both a rigorous overview and a forward-looking roadmap, making it an essential reference for anyone studying or working in federated learning. 
 
\bigskip
\bigskip
%\bigskip
% 2024 
%\newpage

% Applications
\noindent
\textbf{Title:} Differential Privacy Techniques for Cyber Physical
Systems: A Survey\newline
\textbf{Authors:} Muneeb Ul Hassan, Mubashir Husain Rehmani, and Jinjun Chen \newline \textbf{Reference:} \cite{8854247} \newline
\textbf{Year:} 2020 \newline

\noindent
 Hassan et al.\ provide a comprehensive survey of DP techniques tailored for Cyber-Physical Systems (CPSs), with a focus on practical challenges and system-specific constraints. The survey categorizes existing work into centralized and distributed settings and examines how DP mechanisms—such as Laplace and Gaussian noise—are applied to protect real-time sensor data, control signals, and user behaviors in domains like smart grids, healthcare, and industrial automation. Special attention is given to lightweight, resource-aware implementations suitable for embedded and edge devices. The paper also highlights limitations of traditional privacy models in CPS and outlines open research directions, including adaptive noise calibration, secure multi-party learning, and integration with edge-cloud architectures. This survey is a valuable resource for researchers seeking a structured overview of DP applications in safety-critical, real-time environments.\bigskip{}
\bigskip
%\bigskip
%\bigskip

\noindent
\textbf{Title:} Differential Privacy in Health Research: A Scoping Review \newline
\textbf{Authors:} Joseph Ficek, Wei Wang, Henian Chen, Getachew Dagne, and Ellen Daley \newline \textbf{Reference:} \cite{Ficek2021} \newline
\textbf{Year:} 2021 \newline

\noindent
This comprehensive scoping review maps the application of DP in health-related research. The authors classify 54 studies into categories such as algorithm development, data system design, and privacy-utility appraisal. Key application areas include genomics, neuroimaging, health surveillance, and medical phenotyping. While progress has been made in data release and predictive modeling, the review highlights major gaps in support for inferential statistics, correlated data, and real-world deployments. The paper calls for further development of DP methods tailored to health data and encourages experimental case studies to evaluate privacy-utility trade-offs in practical settings.

\bigskip
\bigskip
%\bigskip

\noindent
\textbf{Title:} Finnish Perspective on Using Synthetic Health Data for Privacy (PRIVASA Project) \newline
\textbf{Authors:} Tinja Pitk\"am\"aki, Tapio Pahikkala, Ileana Montoya Perez, Parisa Movahedi, Valtteri
Nieminen, Tom Southerington, Juho Vaiste, Mojtaba Jafaritadi, Muhammad Irfan Khan,
Elina Kontio, Pertti Ranttila, Juha Pajula, Harri P\"ol\"onen, Aysen Degerli, Johan Plomp, and Antti Airola \newline \textbf{Reference:} \cite{Pitkamaki2024} \newline
\textbf{Year:} 2024 \newline

\noindent
The paper examines how synthetic health data can be utilized to enhance data privacy while supporting research, development, and decision-making, focusing on the Finnish regulatory and healthcare context. It is part of the PRIVASA (Privacy-preserving AI for Synthetic and Anonymous Health Data) project, which integrates interdisciplinary insights and applies synthetic data to privacy-sensitive healthcare scenarios. This document highlights the importance of synthetic data in reconciling privacy preservation and data accessibility. It concludes with a call for multi-disciplinary research, improved regulatory guidance, and ethical stewardship.
\bigskip
\bigskip
%\bigskip

% User expectations
\noindent
\textbf{Title:} "I need a better description": An Investigation Into User Expectations For Differential Privacy \newline
\textbf{Authors:} Rachel Cummings, Gabriel Kaptchuk, and Elissa M.\ Redmiles \newline \textbf{Reference:} \cite{Cummings2023} \newline
\textbf{Year:} 2023 \newline

\noindent
This paper presents a rigorous user study examining how people perceive and interpret DP. Using two large-scale surveys, the authors explore whether users care about the types of data disclosures DP protects against, how in-the-wild descriptions of DP affect their privacy expectations, and whether such descriptions influence willingness to share data. The study finds a mismatch between user concerns and the actual protections provided or described, and introduces a framework for aligning DP communication with user expectations. An essential resource for understanding the human factors in DP adoption.
\bigskip
\bigskip
%\bigskip

\noindent
\textbf{Title:} Understanding Risks of Privacy Theater with Differential Privacy \newline
\textbf{Authors:} Mary A.\ Smart, Dhruv Sood, and Kristen Vaccaro \newline \textbf{Reference:} \cite{10.1145/3555762} \newline
\textbf{Year:} 2022 \newline

\noindent
This study critically examines how users respond to explanations of DP, especially when key details --- like the privacy parameter $\varepsilon$ --- are withheld. Across two large-scale user studies involving decisions about sharing browser history, the authors explore whether opaque descriptions of DP lead to "privacy theater": the illusion of protection without substance. Surprisingly, they find that most participants decide whether to share before seeing any explanation, and that many do not understand what DP actually protects against. The paper highlights the risks of superficial DP messaging and stresses the need for transparent, tailored communication to ensure informed user consent.
\bigskip
\bigskip
%\bigskip

\noindent
\textbf{Title:} What Are the Chances? {E}xplaining the Epsilon Parameter in Differential Privacy \newline
\textbf{Authors:} Priyanka Nanayakkara, Mary Anne Smart, Rachel Cummings, Gabriel Kaptchuk, and Elissa M. Redmiles\newline \textbf{Reference:}  \cite{291162} \newline
\textbf{Year:} 2023 \newline

\noindent
This paper investigates how to effectively explain the probabilistic nature of DP, particularly the role of the privacy parameter $\varepsilon$, to end users. The authors develop three explanation methods --- textual, visual, and sample-based --- and evaluate them in a vignette-based survey with 963 participants. Their findings show that odds-based explanations (text and visual) improve users’ objective understanding of DP and increase self-efficacy compared to state-of-the-art methods that omit $\varepsilon$. Importantly, participants' willingness to share data was sensitive to the communicated value of $\varepsilon$, underscoring the need for transparent and accessible DP communication.

\section{DP Libraries}
We now present a list of most popular DP libraries for various machine learning frameworks.%\bigskip

\begin{enumerate}

\item \textbf{TensorFlow Privacy}\vspace{-1mm}
\begin{itemize}
    \item \url{https://github.com/tensorflow/privacy}

    \item Developed and maintained by Google.

    \item Integrates directly with TensorFlow/Keras.

    \item Offers ready-to-use DP optimizers, including DP-SGD. 

    \item Actively maintained and well-documented.
    
    \item Can be used in combination with \emph{TensorFlow Federated} to facilitate federated learning with differential privacy.

\end{itemize}

    \item \textbf{Opacus}\vspace{-1mm}
\begin{itemize}
    \item \url{https://github.com/pytorch/opacus}
    \item Developed by Meta AI.
    \item Designed specifically for PyTorch.
    \item Focuses on performance and scalability in DP training.
    \item Includes tools for privacy accounting and per-example gradient clipping.
    \item Suitable for integration into existing PyTorch workflows with minimal modification.
\end{itemize}

\item \textbf{JAX Privacy}\vspace{-1mm}
\begin{itemize}
    \item \url{https://github.com/google-deepmind/jax_privacy}
    \item Developed by Google DeepMind.
    \item Implements DP-SGD and other privacy-preserving algorithms using JAX.
    \item Provides components for end-to-end training pipelines and reproducing results from recent research on differentially private image classification.
    \item Designed with a focus on transparency, modularity, and reproducibility in research settings.
\end{itemize} %\bigskip

\item \textbf{fastDP}\vspace{-1mm}
\begin{itemize}
    \item \url{https://github.com/awslabs/fast-differential-privacy}
    \item Developed by AWS Labs.
    \item Designed for PyTorch.
    \item Focuses on scalability and efficient training under differential privacy constraints.
    \item Implements DP-SGD with optimized per-example gradient computation.
    \item Supports distributed training for large-scale workloads.
\end{itemize}
\item \textbf{Diffprivlib} (IBM Differential Privacy Library)\vspace{-1mm}
\begin{itemize}
    \item \url{https://github.com/IBM/differential-privacy-library}
    \item Developed by IBM Research.
    \item Built on top of scikit-learn.
        \item Primarily focused on classical machine learning models (e.g., linear regression, SVMs).
    \item Includes DP-SGD-like mechanisms for certain estimators.
\end{itemize}
\end{enumerate}

\begin{landscape}
\begin{table}[ph!]
\caption{Comparison of Libraries Supporting Differential Privacy} 
\label{table_libraries} 
\begin{tabular}{@{}lp{2.5cm}ccccp{3.2cm}p{5.7cm}@{}}
\toprule
\textbf{Library} & \textbf{Framework} & \textbf{DP-SGD } & \textbf{Privacy} & \textbf{Use Case} & \textbf{URL} & \textbf{Notes}  \\
 & & \textbf{Support} & \textbf{Accounting} & \textbf{Focus} & &  \\
\midrule
\textbf{TensorFlow Privacy} & TensorFlow/Keras & Yes & Yes & Deep learning & \href{https://github.com/tensorflow/privacy}{TensorFlow} & Well-documented and maintained by Google. \\
\textbf{Opacus} & PyTorch & Yes & Yes & Deep learning & \href{https://github.com/pytorch/opacus}{PyTorch} & High-performance training by Meta;\newline widely used in \newline PyTorch ecosystem. \\
\textbf{JAX Privacy} & JAX & Yes & Yes & Research (JAX) & \href{https://github.com/google-deepmind/jax_privacy}{JAX} & DP-SGD tools for reproducible training pipelines. \\
\textbf{fastDP} & PyTorch & Yes & Yes & Scalable DP training & \href{https://github.com/awslabs/fast-differential-privacy}{fastDP} & Optimized for speed and scalability;\newline supports various\newline optimizers and distributed training. \\
\textbf{Diffprivlib} & scikit-learn & Partial & Yes & Classical ML & \href{https://github.com/IBM/differential-privacy-library}{Diffprivlib} & Strong support for traditional ML models, less focused on deep learning. \\
\bottomrule
\end{tabular}
\end{table}
\end{landscape}

\end{appendices}
 \end{sloppy}

\end{document}